%% file: main.tex
\documentclass[article,onecolumn,superscriptaddress,nofootinbib,floatfix,onecolumn]{revtex4-2}

\usepackage{amsmath, latexsym, amssymb, hyperref, graphicx, color, multirow, makecell, diagbox}
\usepackage[table]{xcolor}
\usepackage{tabularx}
\usepackage[T1]{fontenc}

\usepackage{dcolumn}

\usepackage{bm}
\usepackage{xspace}

\usepackage{float}
\usepackage{epsfig}
\usepackage{longtable}[htpb!]
\usepackage{rotating}
\usepackage{ulem}
\usepackage{amssymb}
\usepackage{amsmath}
\usepackage{academicons}
\usepackage{txfonts}
\usepackage{upgreek}
\usepackage{mathtools}
\usepackage{booktabs}
\usepackage{cases}

\usepackage{soul} 

\usepackage[capitalize]{cleveref}
\definecolor{nicered}{rgb}{.7,.1,.1}
\definecolor{nicegreen}{rgb}{.1,.5,.1}
\definecolor{darkblue}{rgb}{0,0,.5}
\hypersetup{colorlinks, citecolor=nicegreen,linkcolor=nicered, urlcolor=darkblue}
\usepackage{multirow}
\usepackage{slashed}
\usepackage{verbatim}
\usepackage{siunitx}
\usepackage{longtable}[htpb!]

\usepackage{catchfilebetweentags}
\usepackage{siunitx, bbding, pifont}
\usepackage{longtable}[htpb!]

\newcommand{\mli}[1]{\mathit{#1}}

\newcommand{\qbar}    {{\overline{q}}}
\providecommand{\QQbar}{\mathrm{Q}\overline{\mathrm{Q}}}
\providecommand{\uubar}{\mathrm{u}\overline{\mathrm{u}}}
\providecommand{\ddbar}{\mathrm{d}\overline{\mathrm{d}}}
\providecommand{\qqbar}{\mathrm{q}\overline{\mathrm{q}}}

\providecommand{\ccbar}{\mathrm{c}\overline{\mathrm{c}}}
\providecommand{\bbbar}{\mathrm{b}\overline{\mathrm{b}}}
\providecommand{\bbarq}{\mathrm{b}\overline{\mathrm{q}}}

\providecommand{\ttbar}{\mathrm{t}\overline{\mathrm{t}}}
\providecommand{\ppbar}{\mathrm{p}\mbox{-}\overline{\mathrm{p}}}
\providecommand{\lele}{\ell^{+}\ell^{-}}
\providecommand{\nunubar}{\nu\overline{\nu}}

\newcommand{\epem}{\mathrm{e}^+\mathrm{e}^-}
\newcommand{\mumu}{\mu^+\mu^-}
\newcommand{\tautau}{\tau^+\tau^-}
\newcommand{\gaga}{\gamma\gamma}

\newcommand{\sqrts}{\sqrt{s}}
\newcommand{\alphas}{\alpha_\text{s}}

\newcommand{\jpsi}{\mathrm{J}/\psi}
\newcommand{\alphasmZ}{\alphas(m_{_\mathrm{Z}})}
\newcommand{\lqcd}{\Lambda_\text{QCD}}
\newcommand{\lew}{\Lambda_\mathrm{EW}}

\newcommand{\BR}{\mathcal{B}}

\newcommand{\fbinv}{fb$^{-1}$}
\newcommand{\abinv}{ab$^{-1}$}

\newcommand{\LumiInt}{\mathcal{L}_{\mathrm{int}}}

\newcommand{\madgraph}{\textsc{MadGraph5\_aMC@NLO}}
\newcommand{\mgshort}{\textsc{MG5\_aMC}}

\usepackage{xspace}
\newcommand*{\eg}{e.g.,\@\xspace}
\newcommand*{\ie}{i.e.,\@\xspace}

\newcommand{\orcid}[1]{\href{https://orcid.org/#1}{\hspace*{0.1em}\raisebox{-0.45ex}{\includegraphics[width=1em]{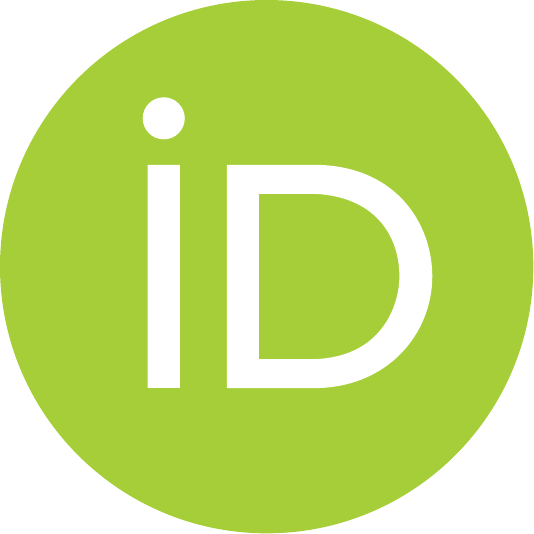}}}}

\renewcommand\arraystretch{1.5}

\begin{document}

\title{Rare and exclusive few-body decays of the Higgs, Z, W bosons, and the top~quark}

\author{David~d'Enterria\orcid{0000-0002-5754-4303}}\email{david.d'enterria@cern.ch}
\affiliation{CERN, EP Department, CH-1211 Geneva, Switzerland}
\author{V\u{a}n D\~{u}ng L\^{e}\orcid{0009-0006-2306-1000}}\email{dunglvht@gmail.com}
\affiliation{University of Science - Ho Chi Minh National University, Vietnam}

\begin{abstract}
\noindent 
We perform an extensive survey of rare and exclusive few-body decays ---defined as those with branching fractions $\BR \lesssim 10^{-5}$ into two to four final particles--- of the Higgs, Z, W bosons, and the top quark. Such~rare decays can probe physics beyond the Standard Model (BSM), constitute a background for exotic decays into new BSM particles, and provide precise information on quantum chromodynamics factorization with small nonperturbative corrections. We tabulate the theoretical $\BR$ values for about 200 rare decay channels of the four heaviest elementary particles, indicating the current experimental limits in their observation. Among those, we have computed for the first time ultrarare Higgs boson decays into photons and/or neutrinos, H and Z radiative decays into leptonium states, radiative H and Z quark-flavour-changing decays, and semiexclusive top-quark decays into a quark plus a meson, while updating predictions for a few other rare H, Z, and top quark partial widths. The feasibility of measuring each of these unobserved decays is estimated for p-p collisions at the High-Luminosity Large Hadron Collider (HL-LHC), and for $\epem$ and p-p collisions at the Future~Circular~Collider~(FCC).
\end{abstract}

\maketitle

\vspace*{-2\baselineskip}

\tableofcontents

\section{Introduction}

\begin{figure}[htpb!]
\centering
\includegraphics[width=0.85\textwidth]{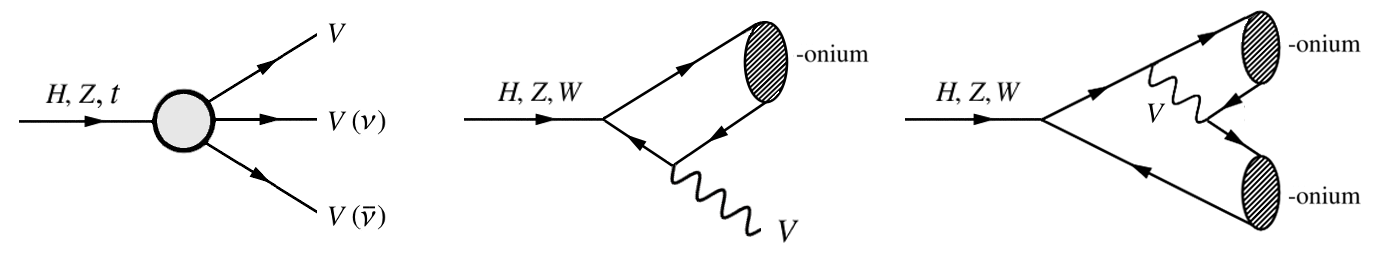}
\caption{Examples of schematic diagrams of rare and exclusive two- and three-body decays of the Higgs, Z, and W bosons, and of the top quark. The leftmost diagram shows a rare decay into two or three gauge bosons V = Z, W, $\gamma$ (or into a gauge boson plus two neutrinos $\nu$) through virtual loops (grey circle). The center and rightmost diagrams show, respectively, typical exclusive decays into a gauge boson (mostly a photon) plus an onium bound state, and into two onium states (dashed blobs).\label{fig:rare_decays_diags}}
\end{figure}

With the discovery of the Higgs boson at the CERN Large Hadron Collider (LHC) about ten years ago~\cite{ATLAS:2012yve,CMS:2012qbp}, the full particle content of the Standard Model (SM) of particle physics has become fully fixed. Among the 17 existing elementary particles (6 quarks, 6 leptons, 4 gauge bosons, and the scalar boson), the top quark, the Higgs and the electroweak (W, Z) bosons are the most massive ones. Studying in detail the properties of the four heaviest elementary particles, with masses around the electroweak scale $\lew\approx\mathcal{O}(100$~GeV), is an important priority in precision SM studies and in searches for new physics beyond it (BSM). At the LHC, the large center-of-mass energies (up to $\sqrt{s} = 14$~TeV) and integrated luminosities (up to $\LumiInt = 3$~\abinv\ at the end of the high-luminosity, HL-LHC, phase)~\cite{Azzi:2019yne} available in proton-proton (p-p) collisions, as well as the many \abinv\ to be integrated in the very clean ``background-free'' $\epem$ collision environment of the next planned lepton collider facilities, such as the Future Circular Collider (FCC-ee)~\cite{FCC:2018evy} or the CEPC~\cite{CEPCStudyGroup:2018ghi}, will allow the collection of very large H, W, Z, and top-quark data samples. In addition, and despite much more complicated background conditions than at $\epem$ colliders, p-p collisions at $\sqrt{s}=100$~TeV planned at FCC-hh~\cite{FCC:2018vvp} will produce unprecedented numbers of  W, Z, Higgs, and top particles. The very large data samples expected at these machines will make it possible to measure many of their rare few-body decays ---understood here as decays into two to four final-state particles, with branching fractions below $\BR \approx 10^{-5}$---, which remain unobserved to date. Broadly speaking, in this work we consider three types of rare few-body decays shown in Fig.~\ref{fig:rare_decays_diags}: (i) decays into three (or four) lighter gauge bosons, or into one (or two) gauge boson plus two neutrinos, (ii) decays into a lighter gauge boson plus a single hadronic (or leptonic) bound system in the form of a quarkonium (or leptonium) state, and (iii) exclusive decays into two quarkonium bound states. The reason for the rarity of the first type of decays is the fact that they proceed through suppressed (heavy particle) loops, whereas the second and third decay modes occur very scarcely because the probability to form a single (let alone double) -onium bound state, out of the outgoing quarks or leptons of the primary decay, is very small.

The incentives for the study of such decays are multiple, with some specificities depending on the decaying ``mother'' particle. A first general motivation for their study is the possibility that new physics phenomena alter some very rare partial decay widths. Precision tests of suppressed (or forbidden) processes in the SM ---such as flavour changing neutral currents (FCNC), or processes violating lepton flavour (LFV) or lepton flavour universality (LFUV)--- are powerful probes of BSM physics. While competitive studies exist at low energies (\eg\ in rare kaon decays, LFV searches with muons or taus, or electric dipole moment measurements), most investigations at multi-GeV mass scales have so far been mostly carried out exploiting b-quark decays at colliders~\cite{LHCb:2018roe,Belle-II:2018jsg}. Unlike in the latter case where, due to the relatively low masses of the B hadrons involved, large power corrections to the decay rates lead to sizable theoretical uncertainties, power corrections in decays of the electroweak and Higgs bosons and top quark are under better theoretical control thanks to the large boost of the final-state hadrons. More concrete motivations are succinctly mentioned next for each particle. 
In the case of the Higgs boson, very rare decays with photons and/or neutrinos (Fig.~\ref{fig:rare_decays_diags}, left) lead to experimental signatures that are identical to potential exotic BSM Higgs decays~\cite{Curtin:2013fra,LHCHiggsCrossSectionWorkingGroup:2016ypw}, and therefore need to be well estimated as background(s) for the latter.
Exclusive Higgs decays such as those shown in Fig.~\ref{fig:rare_decays_diags} (center) are sensitive to the Yukawa couplings of the charm and lighter quarks~\cite{Bodwin:2013gca,Isidori:2013cla,Kagan:2014ila,Konig:2015qat,Perez:2015lra}, as well as to FCNC $\rm H \to qq'$ decays, which are otherwise very difficult to access experimentally due to the smallness of the quark masses involved, and/or their heavily suppressed loop-induced rates~\cite{LHCHiggsCrossSectionWorkingGroup:2016ypw}. 
In the Z boson case, specific decay modes allow probing FCNC couplings (which are loop- and Glashow--Iliopoulos--Maiani (GIM)~\cite{Glashow:1970gm} suppressed processes in the SM) in a model-independent way~\cite{Guberina:1980dc,Perez:2003ad}. In addition, rare decays of the H and Z bosons shown in Fig.~\ref{fig:rare_decays_diags} where the onium state decays into diphotons or dileptons, constitute a background for different searches for exotic BSM decays~\cite{Agrawal:2021dbo}, such as \eg\ $\mathrm{Z}\to\gamma\,a(\gaga)$ with $a$ being an axion-like particle~\cite{dEnterria:2021ljz} or a massive graviton~\cite{dEnterria:2023npy} decaying into photons; or $\rm H \to A'(\ell^+\ell^-)+X$ where a dark photon A$'$ further decays into a $\ell^+\ell^-$ lepton pair~\cite{Curtin:2014cca}. 
The measurements of exclusive decays of the W boson~\cite{Arnellos:1981gy,Mangano:2014xta,Jones:2020bvu} and of the top quark~\cite{dEnterria:2020ygk} have also been suggested \eg\ as alternative means to determine the W boson and top quark masses and widths via two- or three-body invariant mass analyses, free of invisible neutrinos or (messy) jets involved in the inclusive decay modes. In the top quark case, the study of its suppressed radiative decay rates, induced by the offdiagonal parts of the quark dipole moments, has been of interest for many decades because they also provide an experimentally clean probe for new physics~\cite{Beneke:2000hk}. Of particular importance are precision studies of rare FCNC decays such as $\rm t \to Zc$, $\rm t \to \gamma c$, and $\rm t \to cg$, for which many BSM extensions can enhance their branching ratios by orders of magnitude, thereby yielding compelling phenomenology~\cite{Beneke:2000hk,Aguilar-Saavedra:2004mfd}.

Theoretically, the calculation of the partial widths of the few-body decays schematically shown in Fig.~\ref{fig:rare_decays_diags} (left) are carried out through an expansion of the underlying virtual loops in the electroweak (EW) and/or quantum chromodynamics (QCD) couplings ($\alpha$ and $\alphas$, respectively), at a given order of perturbative accuracy. First calculations were computed at leading order (LO), but more recent results exist at next-to-leading-order (NLO) accuracy. With regards to the center and right diagrams of Fig.~\ref{fig:rare_decays_diags}, the formalism of QCD factorization~\cite{Lepage:1979zb,Lepage:1980fj,Efremov:1978rn,Efremov:1979qk,Chernyak:1983ej} is a well-established approach to study and compute rates for hard exclusive processes with individual hadrons in the final state. Within this framework, the production of the final state occurs in two stages: first, a pair of quarks is produced in a short-distance partonic process; second, the quarks form a given hadronic bound state. The short-distance physics at the energy scale of the initial heavy particle is appropriately separated from the long-distance dynamics governing the formation of the final hadron(s), and the decay amplitudes can be therefore obtained from the convolution of hard-scattering functions calculable in perturbative QCD (pQCD) and nonlocal hadronic matrix elements. These latter objects, called light-cone distribution amplitudes (LCDAs), are nonperturbative and scale-dependent functions that encode the infrared physics of the final hadron formation.
The decay amplitudes are formally given as expansions in the ratio of the two (hard and soft) scales in the problem, given by the large energy released $E = m_\mathrm{X}/2$ in the process and the final hadron mass, respectively. In the cases of interest in this work, the hard scale is set by the heavy mass of the decaying bosons or top quark, $\mu_\text{hard}\approx \lew$ rendering the impact of nonperturbative power corrections at typical hadronization energy scales $\lqcd\approx 0.2$~GeV, $\mathcal{O}(\lqcd/\lew)\lesssim 10^{-3}$, under control. 
The studies of exclusive decays of the H, Z, W bosons, and top quark, therefore not only provide a sensitive test of the SM but, in particular, also stringent tests of the QCD factorization formalism, including constraints on poorly known aspects of the nonperturbative formation of hadronic bound states. In the case of final states with charm and bottom quarks, they can bring forward valuable insights into partially conflicting mechanisms of heavy quarkonium production~\cite{Lansberg:2019adr,Chapon:2020heu}.
\vspace{0.15cm}

The main purpose of this work is to present a comprehensive summary of the current theoretical and experimental status of rare and exclusive few-body decays of the four heaviest SM particles. We have first collected all calculations and experimental upper limits for rates of rare and exclusive decays existing in the literature, revised them, and complemented them with $\sim$60 additional channels theoretically estimated here for the first time. In total, we provide a list of about 200 predicted rare decays branching ratios, and identify those that are potentially observable at the HL-LHC, or at FCC, and those with negligible rates unless some BSM physics enhances them. This document should therefore help guide and prioritize future experimental and theoretical studies of the different channels. 
The paper is organized as follows. Section~\ref{sec:compil} provides an overall description of the theoretical and experimental details of the branching ratios compiled for each decay channel, as well as an explanation of how estimates of future experimental upper limits are derived. Sections~\ref{sec:Higgs}, \ref{sec:Zboson}, \ref{sec:Wboson}, and \ref{sec:top_quark} present, respectively, a detailed list of all rare decays of the Higgs, Z, and W bosons, and top quark, including revised branching fractions for a few channels, as well as ultrarare H decays, H and Z decays into leptonium states, radiative H and Z quark-flavour-changing decays, and semiexclusive top-quark decays into a quark plus a meson, computed here for the first time. The paper is closed with a summary of the main findings.

\vspace{-0.19cm}

\section{Results}
\label{sec:compil}

The rare decays results covered in this work are organized in tables per decaying particle listing their theoretical branching fractions, their current experimental upper limits, and future bounds expected at HL-LHC, FCC-ee, and FCC-hh colliders. Here, we discuss first the theoretical models employed to compute them, followed by an explanation of the method used to estimate future experimental limits. The numerical values of the relevant SM parameters used in the (re)calculations of a few branching fractions are also provided.

\vspace{-0.15cm}
\subsection{Theoretical predictions}

For all rare decays collected here, we indicate the theoretical framework used to calculate their corresponding branching fractions. All channels with hadronic final states (Fig.~\ref{fig:rare_decays_diags}, center and right) have been obtained through various implementations of the QCD factorization formalism, using different prescriptions to describe the meson form-factors and to evolve them using the renormalization group equations. We provide here a succinct description of each model so that their acronyms listed below are understandable. The concrete models employed mostly depend on the identity (mass) of the underlying quarks and associated final hadron(s), which defines their kinematics and relevant energy scales. For example, the form-factors of heavy (charm and/or bottom) mesons also include short-distance physics at energy scales of the heavy-quark mass ($m_\mathrm{Q}\gg\lqcd$, for $\rm Q =c,b$), which are often described in terms of long-distance matrix elements (LDMEs) rather than LCDAs

\subsubsection{Light-cone (LC) factorization}

A common method for computing exclusive light-hadron production in high-energy decays is given by the so-called amplitude expansion in the light cone (LC), where 
the perturbatively small expansion parameter is the chirality factor $m_\mathrm{q}/E$ (with $m_\mathrm{q}$ and $E$ being the mass of the produced quark and its typical energy, respectively), and the meson form-factors are nonperturbative objects described by LCDAs~\cite{Chernyak:1983ej} that encode the internal motion of the quark-antiquark pair inside their bound state, constrained by QCD sum rules~\cite{Shifman:1978bx,Colangelo:2000dp}. 
The LC approach implements directly QCD factorization by describing the production of a given final state as a convolution of a hard scattering kernel times the LCDA object. As an example, for the case of the exclusive decay of a heavy particle X into a meson M plus a photon ($\rm X\to M+\gamma$), the decay amplitude can be written schematically as 
\begin{align}
A(\rm  X\to M+\gamma)&\sim f_\mathrm{M}\int_0^1 \mathrm{d}x\, H_\mathrm{M}(x,\mu)\phi_\mathrm{M}(x,\mu),
\label{eq:LC}
\end{align}
where $f_\mathrm{M}$ is the meson decay constant, $x$ is the difference in the momentum fractions carried out by the quark and the antiquark inside the outgoing meson, $H_\mathrm{M}(x,\mu)$ is the hard scattering kernel, calculable in pQCD from the relevant matrix-element diagrams, and evaluated at the energy scale $\mu$ of the decay, and $\phi_\mathrm{M}(x)$  is the LCDA object, which 
at LO can be interpreted as the amplitude for finding a quark-antiquark pair with longitudinal momentum fraction difference $x$ inside the meson. The $f_\mathrm{M}$ constant is a parameter that describes the strength of the interaction between a meson and its decay products. It is connected with the form factor of the SM transition current sandwiched between the vacuum and the meson states, which 
for pseudoscalar (P) and vector meson (VM) mesons (with masses $m_\mathrm{P},m_\mathrm{VM}$ and wavefunctions  $P(k)$ and $\mli{VM}(k)$ at four-momentum $k$) read, respectively, 
\begin{equation}
\begin{aligned}
\label{eq:f_M-theory}
\langle P(k)|\qbar_1(0) \gamma ^\mu \gamma^5 q_2(0) |0\rangle &=-i f_\mathrm{P}k^\mu,\\
\langle \mli{VM}(k) |\qbar_1(0)\gamma^\mu q_2(0) |0\rangle &=-i f_\mathrm{VM}m_\mathrm{VM}\varepsilon^{*\mu}_\mathrm{VM}.
\end{aligned}
\end{equation}
Here above, $\gamma^{\mu,5}$ are Dirac gamma matrices, and $\varepsilon^{*\mu}_\mathrm{VM}$ is the polarization vector of the meson, $q(\overline{q})(x)$ is the creation (annihilation) operator of the fermion field $q$ with spacetime coordinate $x$. The $f_\mathrm{M}$ numerical values are obtained from experimental measurements of electromagnetic meson decays widths into leptons (of mass $m_\mathrm{\ell}$) or photons 
\begin{equation}
\begin{aligned}
 \Gamma(\mathrm{VM} \to \ell^+\ell^-)&=  \frac{4\pi\alpha^2(m_\mathrm{VM})}{3}Q^2_{\rm q}  \frac{f^2_\mathrm{VM}}{m_\mathrm{VM}},\\
  \Gamma(\mathrm{P} \to \gamma \gamma)&=  4\pi\alpha^2(m_\mathrm{P})Q^4_{\rm q}  \frac{f^2_\mathrm{P}}{m_\mathrm{P}},\quad\quad \\
 \Gamma(\mathrm{P}^\pm \to \ell^\pm\, \nu_\ell) &= \frac{\pi\ \alpha^2(m_\mathrm{P}) f^2_{\mathrm{P}}\ m_{\rm P}\  m_\ell^2}{16 m_{\rm Z}^4\cos^4 \theta_\mathrm{w} \sin^4 \theta_\mathrm{w}} \left[1-\left(\frac{m_\ell}{m_{\rm P}}\right)^2\right]^2 |V_{\rm q_1q_2}|^2,
\label{eq:f_M}
\end{aligned}
\end{equation}
where $\alpha$ is the QED coupling, $Q_{\rm q}$ the meson quark electric charge, $\theta_\mathrm{w}$ is the weak mixing angle, $m_\mathrm{Z}$ the Z boson mass, and $|V_{\rm q_1q_2}|$ the Cabibbo--Kobayashi--Maskawa (CKM) matrix element of the two constituent quarks.

Higher-order pQCD corrections can be included in Eq.~(\ref{eq:LC}) through the evolution of $H_\mathrm{M}(x,\mu)$ from $m_\mathrm{X}$ down to the meson mass scale. In the simplest approximation where one ignores the internal motion of quarks inside the mesons, the LCDA takes the form $\phi(x,\mu)=\delta(x)$ or $\phi(x,\mu)=x\delta(x)$ for parity-even and odd states, respectively, where $\delta(x)$ is the Dirac delta function, and the whole meson information in Eq.~(\ref{eq:LC}) is encoded in the $f_\mathrm{M}$ parameter. In such a delta-approximation scheme, one can then write \eg\ the partial decay width of a Z boson into an exclusive $\jpsi +\gamma$ final state as~\cite{Luchinsky:2017jab} 
\begin{align}\label{delta appr}
\Gamma_\delta(\rm Z\to  \jpsi + \gamma)&=\frac{\pi \alpha^2(0) f_\mathrm{ \jpsi}^2Q^2_\mathrm{c}}{6 \cos^2 \theta_\mathrm{w} \sin^2 \theta_\mathrm{w}m_\mathrm{Z}}.
\end{align}

\subsubsection{Soft-Collinear Effective Theory (SCET)}

As an alternative to the LC approach, many decay modes involving light mesons have been calculated using Soft-Collinear Effective Theory (SCET)~\cite{Bauer:2001yt}, where QCD factorization is rephrased in the language of effective field theory (EFT)~\cite{Georgi:1993mps} to properly address the problem of the multiple scales appearing in the calculations. The SCET framework provides a systematic expansion of $\rm X \to M + Y$ decay amplitudes in powers of a small expansion parameter $\lambda = \lqcd/E$ (with $E=m_\mathrm{X}/2$ the meson energy), and allows proper resummation of the large logarithms appearing in the ratios of scales $\log^n(m_\mathrm{M}/m_\mathrm{X})$, given by the mass of the mother particle and that of the mesonic final state, which spoil the convergence of the perturbative expansion. 
Most often, SCET is combined with LCDA. At the scale of the large energies released in EW and Higgs boson decays, even charm and bottom quarks can be treated as light quarks, and hence heavy-quark mesons can be described by LCDAs. In a few other calculations, the EFT approach is combined with form-factors based on the nonrelativistic quark model (NRQM)~\cite{Manohar:1983md}. Final-state mesons containing one light and one heavy quark can be instead described in Heavy-Quark Effective Theory (HQET) where the LCDA describes the hadronic physics at two distinct scales, $m_\mathrm{Q}$ and $\lqcd$, of which the former should be tractable by perturbative methods~\cite{Beneke:2000ry}. 
In HQET, the hadronic matrix elements are expressed as a combination of perturbatively computable coefficients and new, suitably defined, hadronic matrix elements that exploit the constraints provided by heavy quark symmetries~\cite{Neubert:1993mb} on the nonperturbative matrix elements at low scales.

At leading order in the SCET expansion, the $\rm X \to M + \gamma$ decay amplitude can be written in a factorized form as~\cite{Konig:2018wuf}:
\begin{align}
A&= \sum_i \int \mathrm{d}t\, C_i(t,\mu) \,\langle M(k)|\bar{q}(t\bar n)\frac{\bar{n}\mkern-7.5mu/}{2} \Gamma_i[t\bar n,0]q(0) |0\rangle  + 
\text{power corrections},
\label{eq:SCET}
\end{align}
where $i$ runs over different combinations of gamma matrices $\Gamma_i$ that have the appropriate spin structure to form the meson~M with wavefunction $M(k)$ from a suitable quark-antiquark pair $(\qqbar)$ combination; 
$n,\bar{n}$ are four-momenta in the lightcone basis\footnote{In the rest frame of a decaying particle X (with mass $m_\mathrm{X}\gg m_\mathrm{M}$), the 4-momenta of the two final-state particles are $k_1^\mu = n^\mu E$ and $k_2^\mu = \bar{n}^\mu E$, where $E = m_\mathrm{X}/2$ is the energy of the final particles in the X rest frame, and the light-cone vectors satisfy $n\cdot\bar{n}=2$.}; the Wilson coefficients $C_i(t,\mu)$ are calculable hard-scattering EFT coefficients that depend on the process and are evaluated at a factorization scale $\mu$ that can be determined in a matching computation from perturbation theory. The meson matrix element in Eq.~(\ref{eq:SCET}) has a direct connection with the meson form factor 
given by Eqs.~(\ref{eq:f_M-theory}), through~\cite{Grossman:2015cak}:
\begin{equation}
\begin{aligned}
\langle M(k)|\bar{q}(t\bar n)\frac{\bar{n}\mkern-7.5mu/}{2} \Gamma_i[t\bar n,0]q(0) |0\rangle &= -i f_\mathrm{M} E\int_0^1 \mathrm{d}x e^{ixt\bar{n}\cdot k}\phi_\mathrm{M}(x,\mu); \quad\rm  M=P, VM_{||} \\
\langle \mli{VM}(k)_\perp|\bar{q}(t\bar n)\frac{\bar{n}\mkern-7.5mu/}{2} \Gamma_i[t\bar n,0]q(0) |0\rangle &= -i f^\perp_\mathrm{VM}(\mu) E \varepsilon_{\mathrm{VM}}^{\perp *\mu} \int_0^1 \mathrm{d}x e^{ixt\bar{n}\cdot k}\phi^{\perp}_\mathrm{VM}(x,\mu),
\label{eq:ME_scet}
\end{aligned}
\end{equation}
where $E = \bar{n}\cdot k/2$; and $\perp,||$ stand for transverse and longitudinal polarizations of the meson. These last expressions can be thought of as the generalization of the relation of matrix elements to decay constants in the nonlocal case:
in the local limit $t\to0$, the matrix elements of Eq.~(\ref{eq:ME_scet}) translate into Eq.~(\ref{eq:f_M-theory}).
To make the connection to the LC approach clearer, we can further take the Fourier transform of $C_i(t,\mu)$ and call it hard scattering kernel $H_\mathrm{M}(x,\mu)\equiv \int \mathrm{d}t\, C_\mathrm{M}(t,\mu) \,e^{ixt\bar{n}\cdot k}$. We then have a form of the factorized scattering amplitude 
\begin{align}\label{scet-amp}
A=-if_\mathrm{M} E\int_0^1 \mathrm{d}x H_\mathrm{M}(x,\mu)\phi_\mathrm{M}(x,\mu) + \,\text{power corrections},
\end{align}
which looks like the LC decay amplitude given by the convolution of hard-scattering kernels with meson LCDA of Eq.~(\ref{eq:LC}). One can also expand the LCDAs in terms of Gegenbauer polynomials as~\cite{Grossman:2015cak}
\begin{align}\label{delta}
   \phi_\mathrm{M}(x,\mu) = 6x(1-x) \left[ 1 + \sum_{n=1}^\infty a_n^\mathrm{M}(\mu)\,C_n^{3/2}(2x-1) \right] ,
\end{align}
with coefficients $a_n^\mathrm{M}(\mu)$, moments of the LCDA, that can be obtained from nonperturbative approaches such as lattice QCD, or QCD sum rules. At the electroweak scale, the LCDAs are close to the asymptotic $6x(1-x)$ form. Thus for the asymptotic case, one has \eg\ that the $\rm Z \to  \jpsi+\gamma$ partial decay width amounts to~\cite{Konig:2018wuf}:
\begin{align}
\Gamma(\rm Z \to  \jpsi+\gamma)&=\frac{9}{4}\frac{\pi \alpha^2(0) f_\mathrm{ \jpsi}^2}{6 \cos^2 \theta_\mathrm{w} \sin^2 \theta_\mathrm{w}m_\mathrm{Z}}Q_\mathrm{c}^2\left[ 1-\frac{10}{3}\frac{\alphasmZ}{\pi}\right].
\end{align}
This asymptotic expression is similar to that of the delta approximation in Eq.~\eqref{delta appr}, and includes an extra term with higher-order QCD coupling, $\alphasmZ$, corrections coming from a better approximation of the LCDA. 

\subsubsection{Non-Relativistic QCD (NRQCD)}

The formation of heavy quarkonium in the decay of a massive particle X is a multi-scale problem involving the hard scale given by $m_\mathrm{X}$, the heavy-flavour quark mass $m_\mathrm{Q}$, the relative momentum of the heavy quark pair $m_\mathrm{Q}v$, and the binding energy of the heavy quark pair $m_\mathrm{Q}v^2$,  where $v$ is the typical relative velocity of the heavy quarks inside the meson. Since $m_\mathrm{X} \gg mv \gg mv^2 \gg \lqcd$, the simple static limit applied for light quarks is not sufficient for these systems. Therefore, for heavy quark-antiquark bound states such as charmonium, bottomonium, and bottom-charm (B$_\mathrm{c}$) mesons, a third class of models is often used where in addition to the usual expansion in powers of $\alphas$, the interactions are organized as an expansion in $v$. Since $v^2\approx0.3,0.1$ for charmonium and bottomonium, respectively, the heavy quarks are nonrelativistic and described in a Non-Relativistic QCD (NRQCD) approach~\cite{Bodwin:1994jh}.  In the NRQCD framework, the decay width of a heavy particle X into a given quarkonium meson M$(n)$ with quantum numbers $n =\, ^{2s+1}L_{J}^{[1,8]}$ (corresponding to a state with spin $s$, orbital angular momentum $L$, total angular momentum $J$, in the $[1]$ or $[8]$ colour singlet or octet representations, respectively), $\mathrm{X} \to \mathrm{M}(n) + \mathrm{Y}$, factorizes as
\begin{equation}
\mathrm{d}\Gamma[\mathrm{X} \to \mathrm{M}(n) + \mathrm{Y}] = \sum_{n} \mathrm{d}\hat{\Gamma}[\mathrm{X}\to (\QQbar)_n + \mathrm{Y}]\,\langle \mathcal{O}^\mathrm{M}(n)\rangle,
\end{equation}
where the amplitude $\sum_{n} \mathrm{d}\hat{\Gamma}[\mathrm{X}\to (\QQbar)_n + \mathrm{Y}] \propto \frac{1}{2m_\mathrm{X}}|\mathcal{A}|^2 d\Phi_3$ can be computed perturbatively, and 
$\langle \mathcal{O}^\text{M}(n)\rangle$ are LDMEs which, for the colour singlet (CS) states, are related to the radial quarkonium wavefunction at the origin $|R^{\mathrm{M}}(0)|$. For S-wave quarkonia at LO in $v$, the $(^1S_0^{[1]})$ and $(^3S_1^{[1]})$ states are the sole to contribute, and 
their LDMEs  read
\begin{eqnarray}
\langle \mathcal{O}^{\eta_c}(^1S_0^{[1]})\rangle =\frac{N_\mathrm{c}}{2\pi}|R^{\eta_c}(0)|^2, \quad
\langle \mathcal{O}^{\jpsi}(^3S_1^{[1]})\rangle = \frac{3N_\mathrm{c}}{2\pi}|R^{\jpsi}(0)|^2, 
\label{eq:LDME}
\end{eqnarray}
for the singlet charmonium ground states, where $N_\mathrm{c} = 3$ is the number of colours. The LDMEs for the CS contributions can be obtained from the electromagnetic decays of quarkonia into dileptons or diphotons~\cite{Eichten:1995ch, Bodwin:2007fz}, or from potential models. The NRQCD encapsulates two different mechanisms 
to describe the evolution of the heavy-flavour quark pair into a quarkonium meson, such as the colour-singlet model (CSM), and the colour-octet model (COM). In the CSM, the $\QQbar$ pairs are produced in CS states at the hard-scattering scale $m_\mathrm{X}$, and their quantum numbers are conserved during hadronization. 
The COM requires the extra radiation of gluons, and its contributions are more suppressed than in inclusive reactions, 
but has also been considered in a few cases. Beyond the non relativistic description of NRQCD, the relativistic quark model (RQM)~\cite{Ebert:2002pp} has been also employed in a few exclusive decay processes involving heavy quarkonia. The NRQCD calculations of heavy particle decays into quarkonia have been performed at LO, NLO, or next-to-NLO (NNLO) accuracy, resumming in some cases also leading (LL) or next-to-leading (NLL) logarithms. 

As an example, the production of a bottom meson $\rm B \equiv (\bbarq)_n = B_\mathrm{u,d}, B_\mathrm{s}, B_\mathrm{c}$ in the heavy particle decay process $\rm X\to \overline{\rm B} + Y$, can be described in a NRQCD-inspired heavy-quark recombination model~\cite{Braaten:2001bf} where the width takes the form
\begin{align}
    \mathrm{d}\Gamma[\mathrm{X} \to \overline{\rm B} + \mathrm{Y}] = \sum_{n}{\mathrm{d}\hat{\Gamma}[\mathrm{X}\to (\bbarq)_n + \mathrm{Y}]\;\rho[(\bbarq)_n\to \overline{\rm B}]},
\end{align}
with $(\bbarq)_n$ representing the Fock state of the bottom b and accompanying $\rm \overline{q}$ quark, and the factor $\rho[(\bbarq)_n\to \overline{\rm B}]$ is a nonperturbative probability for $(\bbarq)_n$ to evolve into the $\rm \overline{\rm B}$ meson. In the case of $\rm B_c^\pm$ production, the nonperturbative transition $\rm (\bbarq)_n\to \overline{\rm B}_c$ follows a definite velocity power-counting because the relative velocity of the charm and bottom quarks in the rest frame of the bottom meson (like that of the two bottom quarks in the $\Upsilon$ case) are small. This latter assumption is not fulfilled in the $\rm B_\mathrm{u,d,s}$ cases, where the Fock state $(\bbarq)_n$ contributions to the $\rm \overline{\rm B}$ meson with different quantum numbers (\eg\ in colour or angular momentum) are not necessarily suppressed, as aforementioned 
(although for practical purposes, the only relevant states are  $n=\!^1S_0^{[1]},^3S_1^{[1]},^1S_0^{[8]},^3S_1^{[8]}$~\cite{Braaten:2001bf}). By using the heavy-quark spin symmetry~\cite{Bodwin:1994jh}, one can further reduce the nonperturbative transition probabilities $\rho$ from four to two as~\cite{dEnterria:2020ygk}
\begin{eqnarray}
\rm \rho_1^{\overline{\rm B}} \equiv \rho[(\bbarq)_{^1S_{0}^{[1]}}\to \overline{\rm B}], \quad 
\rho_8^{\overline{\rm B}} \equiv \rho[(\bbarq)_{^1S_{0}^{[8]}}\to \overline{\rm B}], \quad 
3\rho_{1,8}^{\overline{\rm B}} = \rho[(\bbarq)_{^3S_{1}^{[1,8]}}\to \overline{\rm B}], 
\end{eqnarray}
where the $\rho_1$ and $\rho_8$ probabilities can then be related to the standard LDMEs, $\langle \mathcal{O}^{\overline{\rm B}}(^{2s+1}L_{J}^{[c]})\rangle$, via 
\begin{eqnarray}
\langle \mathcal{O}^{\overline{\rm B}}(^1S_0^{[1]})\rangle = 2N_\mathrm{c}\frac{4m_\mathrm{b}m_\qbar^2}{3}\rho_{1,8}^{\overline{\rm B}}, \quad 
\langle \mathcal{O}^{\overline{\rm B}}(^1S_0^{[8]})\rangle = \left(N_\mathrm{c}^2-1\right)\frac{4m_\mathrm{b}m_\qbar^2}{3}\rho_{1,8}^{\overline{\rm B}}, \quad  
\langle \mathcal{O}^{\overline{\rm B}}(^3S_1^{[1,8]})\rangle = 3\langle \mathcal{O}^{\overline{\rm B}}(^1S_0^{[1,8]})\rangle, 
\label{eq:rho1_rho8}
\end{eqnarray}
where $m_\mathrm{b}$ is the bottom quark mass, $m_\mathrm{\qbar} = m_{\rm \bar d}, m_{\rm \bar s}\approx 0.3$~GeV are the constituent light quark masses.
The concrete application of the formulas above for the semiexclusive decay of a top quark into a B-meson plus an up-type quark, $\rm t\to \overline{\rm B}_{(s)}^0+q$ (discussed in Section~\ref{sec:top_semiexcl}), results in the following widths at LO in the QCD coupling $\alpha_\mathrm{s}$~\cite{dEnterria:2020ygk}:
\begin{eqnarray}
\label{eq:Gamma_Bjet}
\Gamma(\rm t\to \overline{\rm B}^0+q) &=& \left(\rho_1^{\overline{\rm B}}+8\rho_8^{\overline{\rm B}}\right)|V_\mathrm{qd}|^2\frac{\alpha^2 \pi}{9}\frac{m_{\bar{\mathrm{d}}}^2(m_\mathrm{t}^2-m_\mathrm{b}^2)^2(m_\mathrm{t}^2+m_\mathrm{b}^2)}{m_\mathrm{t}^3m_\mathrm{W}^4\sin^4{\theta_\mathrm{w}}}\left[1+\mathcal{O}\left(\frac{m_\mathrm{q}}{m_\mathrm{b}}\right)\right], \nonumber\\
\Gamma(\rm t\to \overline{\rm B}_{s}^0+q) &=& \frac{|V_\mathrm{qs}|^2}{|V_\mathrm{qd}|^2}\frac{m_{\bar{\rm s}}^2}{m_{\bar{\mathrm{d}}}^2}\Gamma(\rm t\to \overline{\rm B}^0+q)\,,
\end{eqnarray}
with similar (yet longer) expressions available in Ref.~\cite{dEnterria:2020ygk} for the $\rm t\to \Upsilon+q$ decays.

\subsection{Experimental limits}

As we will see below, predictions for rare decays of the electroweak and Higgs bosons, and the top quark, have branching fractions in the $10^{-5}$ to $10^{-15}$ range (or even down to $10^{-22}$ for positronium\,+\,photon final states). Such tiny branching fractions are very challenging experimentally, and no such decays have yet been observed for any of the particles. To provide an idea of the size of the data samples of W$^\pm$, Z, H, and t particles discussed in this work, Table~\ref{tab:data_samples} collects their total number produced in past, current, and future colliders. 
The LEP numbers of W and Z bosons are obtained from the LEP-I and LEP-II electroweak summaries~\cite{ALEPH:2005ab,ALEPH:2013dgf}, and indicate that no rare decay mode with rates below $10^{-7}$ and $10^{-5}$ for the Z and W bosons, respectively, has ever been probed in the clean experimental conditions of an $\epem$ collider. We note, somehow curiously, that the last LEP-II operation, which integrated 2.46~fb$^{-1}$ over $\sqrts = 189$--209~GeV~\cite{LEPWorkingGroupforHiggsbosonsearches:2003ing}, featured a Higgs cross section of $\sigma_\mathrm{H}\approx 3$~fb (adding Higgstrahlung and weak-boson-fusion production processes)~\cite{dEnterria:2017jmj}, and therefore LEP-II \textit{did} produce a few Higgs(125 GeV) boson counts (however, the Higgs searches at the time were optimized for the associated production of a $m_\mathrm{H}= 115$~GeV scalar boson plus an onshell Z boson, for which the H(125~GeV) signal would not have been visible). For the next $\epem$ machine, we focus on FCC-ee because it is the planned facility with the largest data samples expected to be collected~\cite{FCC:2023}. The FCC-ee numbers in Table~\ref{tab:data_samples} cover the time span of the baseline 15-year program with four interaction points and four dedicated runs under consideration: Z-pole, WW threshold, HZ Higgstrahlung, and $\ttbar$ threshold. The largest data sample of all will be for Z bosons, for which $6\cdot 10^{12}$ particles will be produced. The number of H bosons at FCC-ee includes those produced in the HZ ($1.45\times 10^6$) and $\ttbar$~($+330$k events) runs, plus via $\rm WW\to H$ at all runs ($+125$k). Although, in much more complicated background conditions than in $\epem$ collisions, the HL-LHC (with 3~\abinv of p-p collisions at $\sqrts = 14$~TeV collected per ATLAS/CMS experiment) and the FCC-hh (p-p at $\sqrt{s}=100$~TeV with 30~\abinv) will truly serve as W, Z, Higgs, and top factories. We also list the number of the four massive particles produced at the Tevatron in $\ppbar$ collisions at $\sqrts = 1.96$~TeV, as there exist still a couple of competitive rare decay limits from the CDF experiment. The hadron collider numbers in Table~\ref{tab:data_samples} are obtained from the corresponding production cross sections for each particle multiplied by the integrated luminosity quoted (plus the contribution from $\rm \ttbar\to W^+W^-+X$ decays for the W$^\pm$ counting). The cross sections at hadron colliders have been either obtained from the existing literature~\cite{Mangano:2016jyj} or, when not readily available or not fully up-to-date, have been recomputed at next-to-next-to-leading-order (NNLO) accuracy with \textsc{mcfm}~v.8.0~\cite{Boughezal:2016wmq} with the NNPDF3.1\_NNLO parton distribution functions (PDFs)~\cite{NNPDF:2017mvq}. Theoretical (scale, PDF) uncertainties are at most 10\% and not quoted.

\tabcolsep=1.5mm
\begin{table}[htbp!]
\centering
\caption{Total number of Higgs, Z, and W bosons, and top quarks produced (or expected to be produced) in $\epem$ collisions at LEP and FCC-ee, as well as in $\ppbar$ at Tevatron, and in p-p collisions at HL-LHC, and FCC-hh. For $\epem$ colliders, the Z-pole, Higgstrahlung and Higgs weak-fusion, and pair production (WW, $\ttbar$) cross sections (without ISR) are indicated. At LEP and FCC-ee, the numbers of W bosons and top-quarks consider two bosons and two quarks produced per collision around the $\epem\to \rm W^+W^-,\,\ttbar$ thresholds. For hadron machines, the integrated luminosities and the NNLO production cross section for each particle are indicated. All top quark numbers are for pair production (\ie\ the number of events are multiplied by two), and the number of W bosons is the (W$^+$\,+\,W$^-$) sum including also the contributions from $\ttbar$ decays.
\label{tab:data_samples}}
\vspace{0.15cm}
\begin{tabular}{l|ccccccccc}\hline
Collider                   & \multicolumn{2}{c}{W$^\pm$ bosons} & \multicolumn{2}{c}{Z bosons} & \multicolumn{2}{c}{H bosons} & \multicolumn{2}{c}{top quarks}\\
& $\sigma$(W) & $N$(W) & $\sigma$(Z) & $N$(Z) & $\sigma$(H) & $N$(H)  & $\sigma$($\ttbar$) & $N$(top) \\\hline
LEP & 2--10~pb & $0.8\times 10^5$ & 59 nb & $2\times 10^{7}$ & $\sim$2,\,1 fb & $\sim$5 & -- & -- \\
FCC-ee & 2--10~pb & $5\times 10^8$ & 59~nb & $6\times 10^{12}$ & 200,\,30~fb & $1.9\times 10^6$ & 0.5~pb & $3.8\times 10^6$ \\[2pt]
\textit{Increase factor} LEP$\,\mapsto\,$FCC-ee & 1 & 6250 & 1 & $300,000$ & 70,\,30 & $400,000$ & -- & --\\[2pt]
\hline
Tevatron (1.96~TeV, 10~\fbinv) & 25.3 nb & $2.5\times 10^{8}$ & 7.6 nb & $7.6\times 10^{7}$ & 1.1 pb & $1.1\times 10^{4}$ & 7.1 pb  & $1.4\times 10^{5}$ \\
HL-LHC  (14~TeV, $2\times 3$~\abinv) & 200 nb & $1.2\times 10^{12}$ & 62.5 nb & $3.8\times 10^{11}$ & 58~pb & $3.5\times 10^{8}$ & 1 nb & $1.2\times 10^{10}$ \\
FCC-hh (100~TeV, 30~\abinv) & 1300 nb & $4.1\times 10^{13}$ & 415 nb & $1.2\times 10^{13}$ & 0.93~nb & $2.8\times 10^{10}$ & 35 nb & $2.1\times 10^{12}$  \\ [2pt]
\textit{Increase factor} Tevatron$\,\mapsto\,$HL-LHC & 8 & 4800 & 8.2 & 5000 & 52.7 & 31\,800 & 141 & 86\,000\\
\textit{Increase factor} HL-LHC$\,\mapsto\,$FCC-hh & 6.5 & 34 & 6.7 & 32 & 16 & 80 & 35 & 175\\ [2pt]
\hline
\end{tabular}
\end{table}

The largest yield increases across colliders are factors of $6\cdot10^3$, $3\cdot10^5$, and $4\cdot10^5$ for W, Z, and H bosons, respectively, going from LEP to FCC-ee, and a factor of $9\cdot10^4$ for top quarks from Tevatron to HL-LHC.
As of today (end of 2024), measurements at LEP, Tevatron, and LHC have been able to set upper limits at 95\% confidence level (CL) in about one-fourth of the $\sim$200 rare channels considered here. The most stringent experimental rare decays limits are listed in the tables below, including in some cases new results which are not yet available in the 2024~PDG decays listings~\cite{ParticleDataGroup:2024cfk}. One key goal of our work is to provide reasonable expectations of the achievable upper bounds at the HL-LHC, FCC-ee, and FCC-hh. The corresponding extrapolations are obtained through three different means:
\begin{enumerate}
    \item For a few decay channels, there exist dedicated ATLAS and/or CMS studies that have determined the expected limits at the end of the HL-LHC phase~\cite{TheATLAScollaboration:2013nbo,ATLAS:2015xkp,CMS:2022kdd,Cerri:2018ypt,Liu:2020bem,ATLAS:2016qxw}. For such cases, we provide directly the expected limits (divided by $\sqrt{2}$ to account for the statistical combination of two experiments, ATLAS\,+\,CMS) with their bibliographical reference.
    \item For those channels where LHC limits exist today in measurements with a given integrated luminosity (labelled \eg\ ``$\LumiInt$(13\,TeV)'' next), but for which no dedicated studies exist for HL-LHC, we estimate the latter 
    by assuming that they will be statistically improved by the size of the final data sample, namely by the ratio of squared-root integrated luminosities $\sqrt{\smash{2\times 3~\mathrm{ab}^{-1}}/\smash{\LumiInt(\text{13\,TeV})}}$, where the factor of two assumes an ATLAS\,+\,CMS combination. For individual LHC measurements carried out with the full Run-2 integrated luminosity of around 140~\fbinv, this translates into an expected HL-LHC improvement of more than a factor of six. Bounds estimated that way are conservative, as they ignore improvements in the data analyses (\eg\ optimization of the event selection, enlarged categorization depending on the production mode, adoption of more advanced statistical profiling methods, etc.), increased particle production cross sections (from collision energies rising from 13 to 14~TeV), and possible multi-experiment combinations (\eg\ adding results from LHCb) for some channels. We have checked that our simple statistical approximation here gives limits comparable to those obtained with the method 1) above whenever dedicated studies are available.
    \item For those channels where the best limits today are from Tevatron, the corresponding HL-LHC extrapolation is obtained also statistically, by multiplying the current bound by the square-root of the ratio of the number of decaying particles X between Tevatron and HL-LHC, $\sqrt{\smash{N_\mathrm{X}\text{(HL-LHC)}}/\smash{N_\mathrm{X}\text{(Tevatron)}}}$. According to the increase factors quoted in Table~\ref{tab:data_samples}, this implies an improvement of about a factor of 70 in the expected W and Z bosons upper bounds.
    Such an estimate is conservative for the same reasons stated in point 2.\ above and because, in general, the overall acceptance/efficiency of the ATLAS/CMS detectors is larger than that of CDF.
\end{enumerate}

Finally, for the FCC-ee and FCC-hh cases, whenever a given decay channel has a branching fraction commensurate with the expected number of particles produced (\ie\ whenever $\BR\times N_\mathrm{X} \gtrsim 1$, where $N_\mathrm{X}$ are the FCC numbers given in Table~\ref{tab:data_samples} for any given particle), it will appear listed as ``producible at FCC-ee/FCC-hh''. Assuming that potential backgrounds are small and/or well under control, as is often the case at $\epem$ colliders, and even more so at FCC-ee where very precise detector performances are being required~\cite{FCC:2023}, observation can be achieved with a handful of events. Such an assumption is, however, unwarranted at hadron colliders, such as FCC-hh, where at least 100 signal events are typically required to suppress backgrounds in searches for rare prompt decays of Higgs and Z bosons~\cite{CMS:2013rmy,ATLAS:2015rsn}. Future detector and experimental analysis progresses at the FCC-hh could make it less critical than for the LHC. It is conceivable that much more advanced data analysis techniques will be available by the time this new energy-frontier collider starts to operate, which can render observable some producible decays.


\subsection{Input parameters of the calculations}

For the numerical evaluation of various theoretical expressions for branching fractions computed in this work, the input parameters listed in Table~\ref{tab:params} are used. This include leptonium and quark and boson masses and widths (left column), couplings and CKM matrix elements (center column), and hadron masses and form-factor parameters: decay constants $f_\mathrm{M}$ and HQET first inverse moments $\lambda_\mathrm{M}$ (right column).

\tabcolsep=3.5mm
\begin{table}[htbp!]
\renewcommand\arraystretch{1.05}
\centering
\caption{Numerical values of SM parameters used here in theoretical calculations of various branching fractions. Most of the values are from the PDG~\cite{ParticleDataGroup:2024cfk} (the leptonium masses, $m_{\ell\ell}$, are given as twice the single lepton masses, ignoring tiny binding energies), except for those where a specific reference is provided. \label{tab:params}}
\vspace{0.15cm}
\begin{tabular}{cc|cc|ccccc}
\toprule
\multicolumn{2}{c}{masses and widths}  &   \multicolumn{2}{c}{couplings and CKM elements}& \multicolumn{4}{c}{meson masses and form-factor parameters (in GeV)} \\\midrule
$m_\mathrm{ee}$ &  $1.02991$~MeV &  $\alpha(0)$ &  1/137.036 &   $m_\mathrm{\pi^\pm}$ &  0.13957 &  $f_\mathrm{\pi }$ &  0.1304~\cite{Grossman:2015cak}\\
$m_{\mu\mu}$ &  0.21126~GeV &   $\alpha(m_\mathrm{Z})$ &  1/128.943 &   $m_\mathrm{K^\pm}$ &  0.493677 &  $f_\mathrm{K}$  &  0.1562~\cite{Grossman:2015cak}\\
$m_{\tau\tau}$ &  3.5537~GeV &  $\alphasmZ$    &  0.1180 &&&  $\lambda_\mathrm{K}$ &  0.255~\cite{Puhan:2023ekt} \\
$m_\mathrm{c}$ &  1.50~GeV &  $\sin^2{\theta_\mathrm{w}}$ &  0.2351 &   $m_\mathrm{\rho^\pm }$ &  0.77526  &  $f_\mathrm{\rho }$ &  0.212~\cite{Grossman:2015cak}\\
$m_\mathrm{b}$ &  4.18~GeV  &  $\rm G_F$ &  $1.1664\cdot 10^{-5}$~GeV$^{-2}$ &   $m_\mathrm{K^{*\pm}}$ &  0.89166  &  $f_\mathrm{K^{*}}$ &  0.203~\cite{Grossman:2015cak}\\
$m_\mathrm{t}$ &  172.69~GeV &  $|V_\mathrm{ud,cs}|$ &  0.974 &   $m_\mathrm{\phi}$ &  1.0194&  $f_\mathrm{\phi}$ &  0.223~\cite{Konig:2015qat}\\
$m_\mathrm{W}$ &  80.377~GeV &  $|V_\mathrm{us,cd}|$ &  0.225 &   $m_\mathrm{D^\pm}$ &  1.86966  &  $f_\mathrm{D}$ &  0.212~\cite{FlavourLatticeAveragingGroupFLAG:2021npn}\\
$m_\mathrm{Z}$ &  91.188~GeV  &  $|V_\mathrm{cb,ts}|$ &  0.041  &&&  $\lambda_\mathrm{D}$ &  0.354~\cite{Lu:2021ttf} \\
$m_\mathrm{H}$ &  125.25~GeV  &  $|V_\mathrm{ub}|$ &  0.00382 &   $m_\mathrm{D_s^\pm}$ &  1.96835  &  $f_\mathrm{D_s}$ &  0.2499~\cite{FlavourLatticeAveragingGroupFLAG:2021npn}\\
$\Gamma_\mathrm{W}$ &  2.085~GeV  &  $|V_\mathrm{tb}|$ &  0.999 &  $ m_\mathrm{D^{*0}}$ &  2.007  &  $f_\mathrm{D^{*0}}$ &  1.28  $f_\mathrm{D^{}}$~\cite{Becirevic:2012ti}\\
$\Gamma_\mathrm{W\to \mu\nu}$&   $0.1063~\Gamma_\mathrm{W}$ &  $|V_\mathrm{td}|$ &  0.0086 &   $m_\mathrm{D^{*\pm}}$ &  2.01027  &  $f_\mathrm{D^{*}}$ &  1.28  $f_\mathrm{D^{}}$~\cite{Becirevic:2012ti}\\
$\Gamma_\mathrm{Z}$ &  2.4955~GeV  &&&   $m_\mathrm{D_s^{*\pm}}$ &  2.1123  &  $f_\mathrm{D_s^{*}}$ &  1.26  $f_\mathrm{D_s}$~\cite{Becirevic:2012ti}\\
$\Gamma_\mathrm{Z\to\ell\ell}$ &  $0.0337~\Gamma_\mathrm{Z}$ &&&   $m_\mathrm{B^\pm}$ &  5.27934  &  $f_\mathrm{B}$ &  0.190~\cite{FlavourLatticeAveragingGroupFLAG:2021npn}\\
$\Gamma_\mathrm{H}$ &  4.1~MeV~\cite{Djouadi:2018xqq} &&&&&  $\lambda_\mathrm{B}$ &  0.338~\cite{Mandal:2023lhp} \\
$\Gamma_\mathrm{H\to \gaga}$ &$2.5 \cdot 10^{-3}~\Gamma_\mathrm{H}$ &  &  &   $m_\mathrm{B^{*\pm,0}}$ &  5.32471  &  $f_\mathrm{B^*}$ &  0.941 $f_\mathrm{B}$~\cite{Colquhoun:2015oha}\\
$\Gamma_\mathrm{t}$ &  1.331~GeV~\cite{Chen:2022wit}  &  &  &&&  $f_\mathrm{B_\mathrm{s}}$ &  0.2303~\cite{FlavourLatticeAveragingGroupFLAG:2021npn} \\
&& & &&&  $\lambda_\mathrm{B_s}$ &  0.438~\cite{Khodjamirian:2020hob} \\
& & & &  $ m_\mathrm{B_s^{*0}}$ &  5.415  &  $f_\mathrm{B_\mathrm{s}^*}$ &  0.953 $f_\mathrm{B_\mathrm{s}}$~\cite{Colquhoun:2015oha}\\
&&&&   $m_\mathrm{B_\mathrm{c}^\pm}$ &  6.27447  &  $f_\mathrm{B_\mathrm{c}}$ &  0.427~\cite{Colquhoun:2015oha} \\
&&&&   $m_\mathrm{B_\mathrm{c}^{*\pm}}$ &  6.32877~\cite{Colquhoun:2015oha}  &  $f_\mathrm{B_\mathrm{c}^{*}}$ &  0.988 $f_\mathrm{B_\mathrm{c}}$~\cite{Colquhoun:2015oha} \\
\bottomrule
\end{tabular}
\end{table}

\clearpage
\section{Rare Higgs boson decays}
\label{sec:Higgs}

\subsection{Rare two-, three-, and four-body Higgs boson decays}

\begin{figure}[htpb!]
\centering
\includegraphics[width=0.95\textwidth]{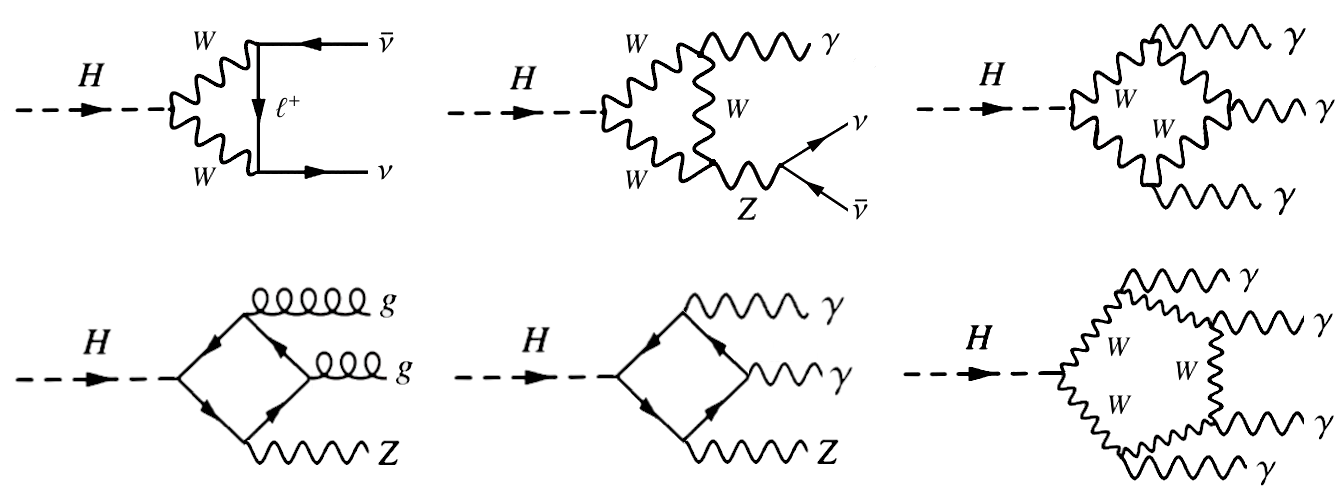}
\caption{Representative diagrams of rare 2-, 3-, and 4-body decays of the H boson into photons and/or neutrinos, and into Z bosons plus gluons or photons.}
\label{fig:H_rare}
\end{figure}

Figure~\ref{fig:H_rare} shows rare decays of the H boson into two neutrinos, a photon plus two neutrinos, three or four photons, and a Z boson plus two gluons or two photons. Calculations of rates for these processes, which proceed via virtual loops in the SM, do not exist in many cases to our knowledge. Since they are very suppressed and have no obvious phenomenological impact, they are not included in the standard Higgs decay codes~\cite{Djouadi:2018xqq}, although they are intriguing for diverse reasons exposed below. We have therefore computed them with the \madgraph\ program~\cite{Alwall:2014hca} (called MG5\_AMC, hereafter) with QCD and EW loop corrections up to NLO accuracy~\cite{Hirschi:2015iia,Frederix:2018nkq}, including massive fermions in the loops (where needed), using the input parameters of Table~\ref{tab:params}. The corresponding results are listed in Table~\ref{tab:H_decays_V_V_V}.

Our first calculation is that of the invisible two-body Higgs boson decay into a neutrino-antineutrino pair, which is infinitesimal in the SM ($\BR\approx 10^{-26}$) compared to the standard invisible (four-neutrino) $\rm H \to ZZ^\star \to 4\nu$ decay ($\BR\approx 0.1\%$)~\cite{Djouadi:2018xqq}. However, since the SM assumption of massless $\nu$'s is incorrect, the $\rm H \to \nunubar$ decay can receive extra contributions depending on the mechanism of neutrino mass generation actually realized in nature. Therefore, it is a process worth keeping an eye on, when considering invisible Higgs decays and accounting for massive neutrinos~\cite{dEnterria:2025rjj}. The second diagram of Fig.~\ref{fig:H_rare} shows the photon-plus-neutrinos decay, which proceeds through Z and W loops and, since the neutrino pair goes undetected, it appears experimentally as an unbalanced monophoton decay of the Higgs boson. Such a final state is shared by many exotic BSM Higgs decays~\cite{Curtin:2013fra}, and it is worth computing its SM rate. This decay is actually the least rare considered in this whole survey, and has a branching fraction of $\BR = 3.74\cdot 10^{-4}$ that is about 20\% larger than the naive estimate given by $\BR(\rm H\to Z\gamma) = 1.54\cdot 10^{-3}$~\cite{Djouadi:2018xqq} combined with $\BR(\rm Z \to \nunubar)=0.20$~\cite{ParticleDataGroup:2024cfk}. This is so because extra W-induced channels (not shown in Fig.~\ref{fig:H_rare}) contribute to the amplitude.

The third ultrarare Higgs decay of interest here is $\rm H \to 3\gamma$. At variance with the naive assumption that it is forbidden for a scalar particle to decay into three photons, because such a process would violate charge-conjugation (C) symmetry, $\rm C(H)=+1 \neq C(3\gamma) = (-1)^3 = -1$, such a decay is possible, albeit extremely suppressed, for EW-induced decays. In a situation akin to the case of the triphoton decay of the scalar positronium state (para-positronium)~\cite{Bernreuther:1981ah,Pokraka:2017ore} or of the $\pi^0$ meson~\cite{Dicus:1975cz}, which is forbidden in QED but not in the EW theory, such a Higgs decay can proceed through the W box shown in the top-right panel of Fig.~\ref{fig:H_rare}. Since $\rm H \to 3\gamma$ violates C-symmetry, it must also violate parity in order to conserve CP and, therefore, the final state must be composed of three spatially symmetric photons with vanishing total angular momentum. As a consequence, the partial width of this channel features an utterly small $\mathcal{O}(10^{-40})$ probability. The box diagram is however not the only contribution to this decay, which can also proceed via $\rm H \to Z(\gaga)\gamma$. Estimates of this latter contribution are of the same size as the box ones~\cite{dEnterria:2025rjj}.

\begin{table}[htpb!]
\caption{Compilation of rare Higgs decays to two neutrinos, a photon plus two neutrinos, three or four photons, and a Z boson plus two photons or two gluons. For each decay, we provide the theoretical branching fraction computed with \mgshort. The last two columns indicate whether the decay can be produced at FCC-ee/FCC-hh.
\label{tab:H_decays_V_V_V}}
\resizebox{\textwidth}{!}{%
\input{tables/exclusive_H_decays_ultrarare.tex}
}
\end{table}
Two other Higgs boson decays into three gauge bosons are possible: $\rm H\to Zgg$ and $\rm H\to Z\gaga$. They have been considered before in the literature, albeit in the $m_\mathrm{t}\to\infty$ limit and/or with old SM parameters~\cite{Kniehl:1990yb,Abbasabadi:2004wq,Abbasabadi:2008zz}. Both decays proceed through a fermion box (mostly, a top quark box) as shown in the bottom of Fig.~\ref{fig:H_rare} (left and center diagrams). The $\rm H\to Z\gaga$ decay rate is much more suppressed than the naive expectation of it being similar to $\BR(\rm H \to Z\gamma)\approx 1.5\cdot10^{-3}$ times the QED coupling for the extra photon emission because, due to the charge conjugation properties of the gauge boson couplings, there are no W bosons in the virtual box. It is worth noting that the similar box diagram of the $\rm H\to \gamma g g$ decay, which results by changing the Z boson by an emitted photon, is forbidden by the generalized Furry's theorem~\cite{Nishijima:1951} that applies for the photon case but is evaded by the axial coupling of the Z boson to fermions in the $\rm H\to Zgg$ case (this decay can however go through  $\rm H \to Z(gg)\gamma$ with an estimated rate of $\mathcal{O}(10^{-40})$~\cite{dEnterria:2025rjj}).
We have calculated both processes here with \mgshort\ --including top-, bottom-quarks, and tau leptons in the box-- finding branching fractions of $\BR(\rm H\to Zgg) = 6.3\cdot 10^{-7}$ and $\BR(\rm H\to Z\gaga) = 2.4\cdot 10^{-9}$, respectively. With such low $\mathcal{B}$ values, one expects short of one $\rm H\to Zgg$ event at the FCC-ee, and both very rare channels are only producible at a machine such as FCC-hh, where they offer a relatively clean signal of two energetic gluons or photons plus a back-to-back lepton-antilepton pair.

\begin{figure}[htpb!]
\centering
\includegraphics[width=0.43\textwidth]{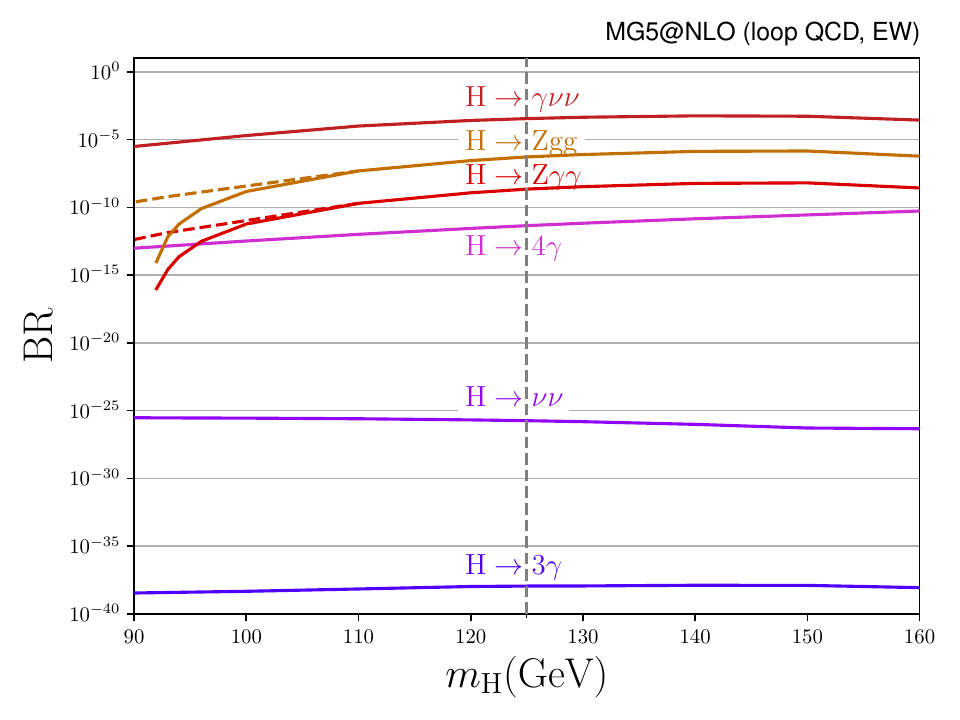}
\includegraphics[width=0.56\textwidth]{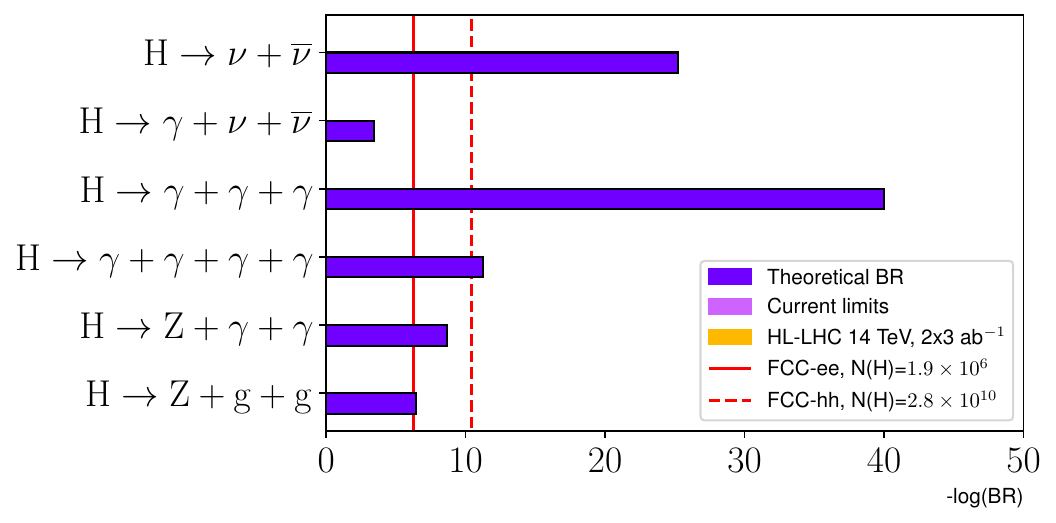}
\caption{Theoretical decay branching fractions of the Higgs boson into rare two-, three-, or four- gauge bosons and/or neutrinos, computed in this work shown as a function of the Higgs boson mass (left) and as blue bars in negative log scale (right). In the left panel, the dashed lines for $m_\mathrm{H}<m_\mathrm{Z}$ show the $\rm H \to Z^*gg,\,Z^*\gaga$ decay rates with offshell Z bosons. In the right panel the red vertical lines indicate the FCC-ee (solid) and FCC-hh (dashed) reach based only on the total number of H bosons to be produced at both facilities.
\label{fig:H_rare_BR}}
\end{figure}

Finally, it is interesting to compute the 4-photon decay of the Higgs boson as it may constitute a background for exotic decays into a pair of light axion-like or scalar particles, each of which further decays into two photons, $\mathrm{H} \to a(\gaga)\,a(\gaga)$. The Higgs $4\gamma$ decay has a branching fraction of $\BR = 4.56\cdot 10^{-12}$, which is 28 orders-of-magnitude larger than the 3-$\gamma$ box-mediated decay, as C and P conservation is not an issue here, and the rate is only suppressed due to the presence of heavy charged-particle loops. Among all rare channels discussed in this section, LHC searches have been performed to date for the $4\gamma$ final state alone~\cite{ATLAS:2015rsn,CMS:2022fyt,CMS:2022xxa,ATLAS:2023ian} (searches for $3\gamma$ decays have focused on Z' resonances off the Higgs peak~\cite{ATLAS:2015rsn}), but unfortunately, the limits have been set only on the process $\rm H \to a(\gaga)a(\gaga)$ with two intermediate ALPs decaying into photons. It would not be too difficult for ATLAS/CMS to recast these searches into upper bounds on the $\rm H\to 4\gamma$ ``continuum'' decay. For reference, Fig.~\ref{fig:H_rare_BR} shows the branching fractions for the six very rare decays of the Higgs boson into two-, three-, and four particles, computed here, as a function of the Higgs boson mass $m_\mathrm{H}$ (left), and of negative log branching fraction for the SM Higgs mass (right).


\subsection{Exclusive Higgs decays into a gauge boson plus a meson}
\label{sec:H_gauge_meson}

\begin{figure}[htpb!]
\centering
\includegraphics[width=0.85\textwidth]{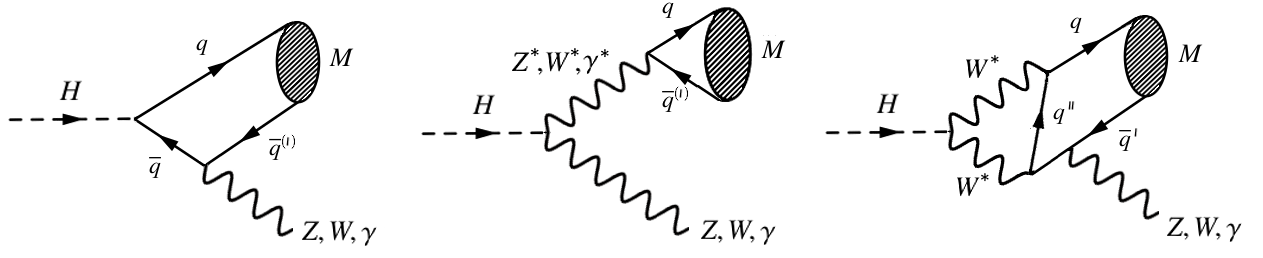}
\caption{Schematic diagrams of exclusive decays of the H boson into a meson plus a gauge boson: direct (left), indirect (center), and W-loop (right) processes. The solid fermion lines represent quarks, and the gray blob represents the mesonic bound state.}
\label{fig:H_gauge_meson_diags}
\end{figure}

Since it is extremely difficult to experimentally access the Yukawa couplings to first-generation ($\rm q=u,d,s$) and second-generation ($\rm q=s, c$) quarks, due to the smallness of the $\rm H \to \qqbar,\ccbar$ decay widths, and the very large QCD dijet backgrounds at the LHC, it has been proposed to constrain those couplings via rare exclusive decays into a photon (or a gauge boson) plus a vector meson~\cite{Bodwin:2013gca,Isidori:2013cla,Kagan:2014ila,Konig:2015qat,Perez:2015lra}. The relevant Feynman diagrams for the $\rm H \to V+ M$ (where $\rm V=Z,W^{\pm},\gamma$, and M~=~meson) decays in the SM are shown in Fig.~\ref{fig:H_gauge_meson_diags}. The first diagram represents LO amplitudes where the Higgs boson directly couples to a quark-antiquark pair that radiates a gauge boson and forms the mesonic bound state. The second diagram depicts indirect contributions to the decay amplitude, where the scalar boson decays first into two gauge bosons, one of which transforms (offshell) into a hadron state via $\rm V^* \to \qqbar$. 
The third diagram shows the radiative FCNC decay into a gauge boson plus a flavoured meson, through a double W loop. The indirect diagram provides the dominant contribution to the decay rates due to the smallness of the Yukawa couplings to the first- and second-generation quarks, and the largest sensitivity to the Higgs-quark coupling comes from the (destructive) interference between the two amplitudes. Thus, for example, the $\rm H \to \gamma+\jpsi,\gamma+\phi$ modes allow direct access to the flavour-diagonal coupling of the Higgs boson to the charm and strange quarks, respectively, while the $\rm H \to \gamma+\rho,\gamma+\omega$ decays can probe the Higgs couplings to up and down quarks. The radiative decays $\rm H \to \gamma+M$, where $\rm M =  K^{*0}, D^{*0}, B^{*0}, B_s^{*0}$ -- which in the SM can only proceed through 
a virtual W boson because the photon (and Z) splitting preserves flavour, or through the direct process in BSM scenarios with flavour-changing Higgs decays -- provide possibilities to probe BSM flavour-violating q-q' Higgs couplings~\cite{Kagan:2014ila,Kamenik:2023hvi}.

The indirect amplitude, in which the virtual photon or Z boson couples to the vector meson through the matrix element of a local current, can be parameterized in terms of a single hadronic parameter: the vector-meson decay constant $f_\mathrm{M}$ (see Table~\ref{tab:params}). This quantity can be directly obtained from the experimental measurements of the vector-meson leptonic partial decay width (which proceeds through the EW annihilation $\qqbar\to\gamma^*,\mathrm{Z}^*\to\ell^+\ell^-$), given by
\begin{equation}
\Gamma(\mathrm{M \to \ell^+\ell^-}) = \frac{4\pi Q_\mathrm{q}^2 f_\mathrm{M}^2}{3m_\mathrm{M}}\alpha^2(m_\mathrm{M}),
\label{eq:VM_ll_width}
\end{equation}
where $Q_\mathrm{q}$ is the electric charge of the constituent quarks, and $m_\mathrm{M}$ the mass of the meson. Corrections due to the offshellness of the photon and to the contribution of the $\rm H \to \gamma\,Z^*$ process are suppressed by $m^2_\mathrm{M}/m^2_\mathrm{H}$ and $m^2_\mathrm{M}/(m^2_\mathrm{Z}-m^2_\mathrm{M})$, respectively, and hence very small~\cite{Konig:2015qat}.
The direct amplitudes, which are much smaller than the indirect ones except for the heaviest $\rm H \to \gamma+\Upsilon$ decay, have been calculated theoretically by different groups. The hierarchy $m_\mathrm{M}\ll m_\mathrm{H}$ implies that the vector meson is emitted at very high energy $E_\mathrm{M}\gg m_\mathrm{M}$ in the Higgs boson rest frame. The constituent partons of the vector meson can thus be described by energetic particles moving collinear to the direction of M, and QCD factorization, either in the SCET or NRQCD incarnation, can be employed to compute them.

\subsubsection{Higgs decays into a photon plus a vector meson}

\begin{figure}[htpb!]
\centering
\includegraphics[width=0.6\textwidth]{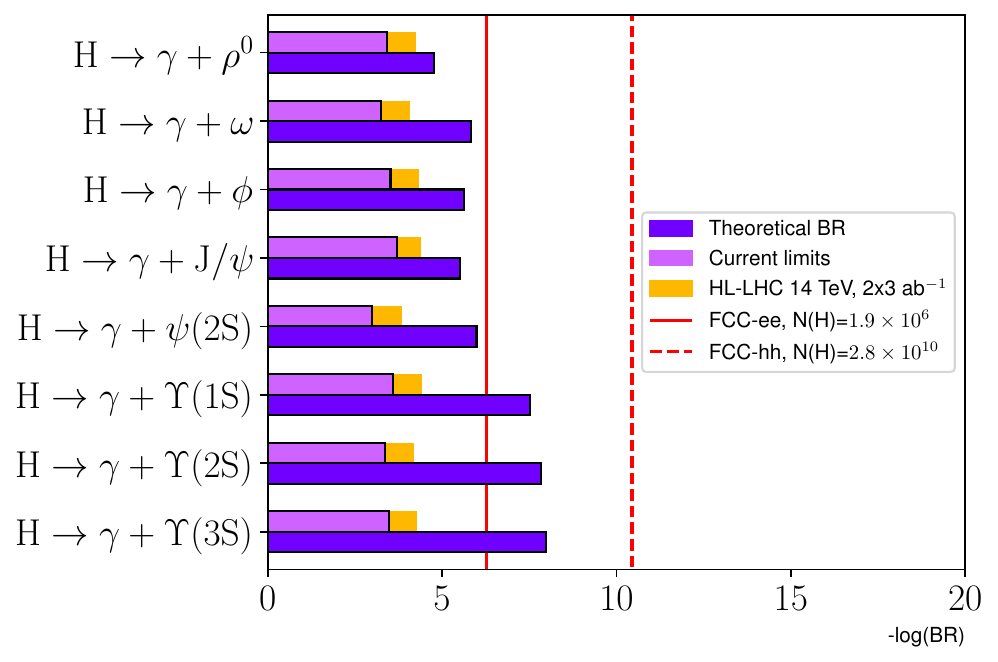}
\caption{Branching ratios (in negative log scale) of exclusive $\rm H \to \gamma +vector$-meson decays. Most recent theoretical predictions (blue bars) compared to current experimental upper limits (violet) and expected conservative HL-LHC bounds (orange). The red vertical lines indicate the FCC-ee (solid) and FCC-hh (dashed) reach based only on the total number of H bosons to be produced at both facilities.
}
\label{fig:H_gamma_meson_limits}
\end{figure}

Table~\ref{tab:H_decays_gamma_meson} lists the corresponding theoretical predictions and experimental limits for Higgs decays into a photon plus a vector meson, and Fig.~\ref{fig:H_gamma_meson_limits} displays them in graphical form. Theoretical $\BR$ values are in the range of $\mathcal{O}(10^{-5}$--$10^{-9})$, with larger rates for decays into light-quark and charm vector mesons compared to bottomonium due to stronger cancellation of direct-indirect contributions for the latter. Experimental upper bounds have been set for all decays at the LHC~\cite{ATLAS:2017gko,ATLAS:2018xfc,CMS:2019wch,ATLAS:2023alf,CMS:2024hhg}, and HL-LHC could observe $\rm H\to \gamma+\rho$ and $\rm H\to \gamma+\jpsi$. Most of these decays will be visible at FCC-ee, but heaviest bottomonium radiative decays will require the number of Higgs bosons produced in a machine like FCC-hh. 


\begin{table}[htpb!]
\caption{Compilation of exclusive Higgs decays to a photon plus a meson. For each decay, we provide the predicted branching fraction(s) and the theoretical approach used to compute it, as well as the current experimental upper limit and that conservatively estimated for HL-LHC. The last two columns indicate whether the decay can be produced at FCC-ee/FCC-hh.
\label{tab:H_decays_gamma_meson}}
\resizebox{\textwidth}{!}{%
\input{tables/exclusive_H_decays_gamma_meson.tex}
}
\end{table}

\subsubsection{Higgs decays into a Z boson plus a meson}

Higgs decays of the form $\rm H \to Z + M$ involve neutral currents and are very similar to the radiative decays just discussed. However, an important difference with respect to the $\gamma\,+\,$meson decays, is that now since the massive final-state gauge boson can be in a longitudinal polarization state, both pseudoscalar and vector mesons can be produced. Also, the interference terms between the direct and indirect amplitudes are smaller. Although these decays are less useful for measuring the Higgs couplings to light quarks, they probe the important effective H-$\gamma$-Z coupling, and provide independent constraints on the meson distribution amplitudes. In addition, they constitute a background for exotic BSM decays of the Higgs boson into, \eg\ a Z boson plus an ALP ($\mathrm{H} \to \mathrm{Z} + a$)~\cite{CMS:2023eos}. 

\begin{figure}[htpb!]
\centering
\includegraphics[width=0.6\textwidth]{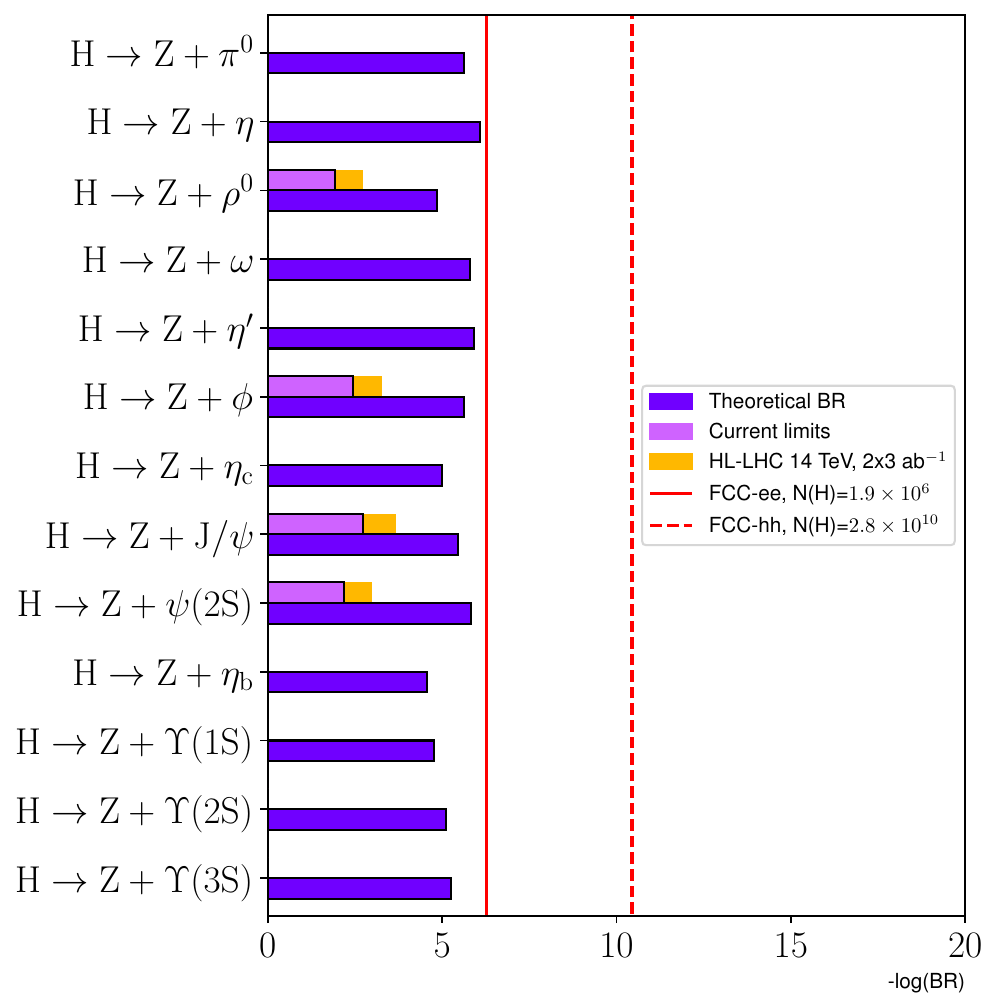}
\caption{Branching ratios (in negative log scale) of exclusive $\rm H \to Z + meson$ decays. The most recent theoretical predictions (blue bars) are compared to current experimental upper limits (violet) and conservatively expected HL-LHC bounds (orange). 
The red vertical lines indicate the FCC-ee (solid) and FCC-hh (dashed) reach based only on the total number of H bosons to be produced at~both~machines.}
\label{fig:H_Z_meson_limits}
\end{figure}

Table~\ref{tab:H_decays_Z_meson} compiles the theoretical predictions and experimental limits for Higgs decay branching ratios into a Z boson plus a meson, and Fig.~\ref{fig:H_Z_meson_limits} presents them in graphical form. All decays have rates in the $10^{-5}$--$10^{-6}$ range with small differences across models for the same channel, except for the $\rm H \to Z+\rho$ case where the calculations of~\cite{Isidori:2013cla} and~\cite{Alte:2016yuw} differ by a factor of six, seemingly because the former had neglected the indirect contributions. 
Experimental upper bounds have been set for a few channels listed in Table~\ref{tab:H_decays_Z_meson}, and they are in the $\mathcal{O}(10^{-2}$--$10^{-3})$ range\footnote{We note that one could a priori also derive limits for $\BR(\rm H \to Z +\eta_\mathrm{c})$ and $\BR(\rm H \to Z + \eta_\mathrm{b})$ from a recent 95\% CL upper limit on the cross section for the $\rm H \to Z(\ell\ell) a(\gaga)$ process over $m_a \approx 1$--30~GeV performed by the CMS Collaboration~\cite{CMS:2023eos}. However, the upper bound cross sections measured at the $m_a = m_{\eta_\mathrm{c},\eta_\mathrm{b}} \approx 2.9,~9.4$~GeV mass points: $\sigma \lesssim 7,$~2.1~fb, corresponding to $\mathcal{B(\rm H \to Z + a)} \lesssim 0.002,~0.62\cdot 10^{-3}$ after normalizing by the Higgs production cross section and correcting for the dilepton decays of the Z boson) are too weak to place any sensible constrain on these $\BR$ (namely, they would exclude the nonphysical region $\BR(\rm H \to Z + \eta_\mathrm{c,b})\gtrsim 10$, dividing them by the corresponding $\BR(\eta_\mathrm{c,b}\to\gaga)$ values).}.
In general, the experimental $\rm H \to Z + M$  limits are about a factor of ten worse than their $\rm H \to \gamma + M$ counterparts because of the extra events lost from the requirement of Z boson identification via dilepton decays (with $\BR\approx 3\%$ for each charged lepton pair). Conservatively, the HL-LHC will be able to set upper bounds about 100 times above their expected SM branching fractions, and it appears that experimental evidence will only be possible at FCC-ee for all of them. Bottomonia-plus-Z and $\eta_\mathrm{c}$-plus-Z decays have the largest rates, but no upper-limit has been set to date for them.


\begin{table}[htpb!]
\caption{Compilation of exclusive Higgs decays to a Z boson plus a meson. For each decay, we provide the predicted branching fraction(s) and the theoretical approach used to compute it, as well as the current experimental upper limit and that estimated for HL-LHC. The last two columns indicate whether the decay can be produced at FCC-ee/FCC-hh.
\label{tab:H_decays_Z_meson}}
\resizebox{\textwidth}{!}{%
\input{tables/exclusive_H_decays_Z_meson.tex}}
\end{table}

\subsubsection{Higgs decays into a gauge boson plus a flavoured meson}

The third diagram of Fig.~\ref{fig:H_gauge_meson_diags} shows the radiative decay into a photon or Z boson plus a flavoured meson. In the SM, this FCNC process can only proceed through a double W loop, and we are not aware of any theoretical calculation of these rates to date. We estimate here the branching fractions for $\rm H \to \gamma/Z+VM$ with $\rm VM =  K^{*0}, D^{*0}, B_s^{*0}, B_d^{*0}$, which are all excited vector states, as explained below. 

On the one hand, calculations for the inclusive flavour-changing $\rm H \to qq'$ rates exist~\cite{Benitez-Guzman:2015ana,Aranda:2020tqw,Kamenik:2023hvi}, but without the photon (or Z boson) emission and the exclusive final-state meson formation. These predicted FCNC Higgs branching fractions, $\BR(\rm H\to qq') \equiv \BR(\rm H\to \qqbar' +  q'\overline{q})$, amount to \footnote{The $\rm H\to ds$ branching fraction has been updated as per Ref.~\cite{Tammaro2025}.}
\begin{eqnarray}
\BR(\rm H\to uc) &=& (2.7 \pm 0.5) \cdot 10^{-20}, \quad \BR(\rm H\to db) = (3.8 \pm 0.6) \cdot 10^{-9}, \nonumber \\
\BR(\rm H\to sb) &=& (8.9 \pm 1.5) \cdot 10^{-8}, \;\,\quad \BR(\rm H\to ds) = (1.9 \pm 0.3)\cdot 10^{-15}.
\label{eq:BR_H_qqprime}
\end{eqnarray}
On the other hand, the work of~\cite{Kagan:2014ila} determined the $\rm H\to \gamma+VM$ branching fractions, and the work~\cite{Alte:2016yuw} provided the $\rm H\to Z+VM$ decay widths assuming (arbitrary) $\mathcal{O}(1)$ flavour-changing Yukawa couplings. From the results of~\cite{Kagan:2014ila} (resp.,~\cite{Alte:2016yuw}), the branching ratios of Higgs decaying into a photon (resp., a Z boson) plus a flavoured neutral meson can be determined through the following EFT\,+\,LCDA-based expressions
\begin{align}
\BR\left(\mathrm{H\to \gamma+VM(qq')}\right)&=\frac{ \alpha(0)}{2\ m_\mathrm{H}} \left( \frac{f_\mathrm{VM} m_\mathrm{VM}}{2\,\lambda_\mathrm{VM}(\mu)} Q_\mathrm{q} \right)^2 \frac{|\kappa_{\rm qq'}|^2+|\kappa_{\rm q'q}|^2}{ \Gamma_\mathrm{H}},\label{eq:BR_H_gamma_m}\\
\mathcal{B}(\rm H\to Z+VM(qq'))&=\frac{9m_{\rm H}}{8 \pi v^2}\left(f^\perp_{\rm VM}(\mu_{\rm HZ})v_{\rm q}\right)^2\frac{|\kappa_{\rm qq'}|^2+|\kappa_{\rm q'q}|^2}{2\ \Gamma_\mathrm{H}}\frac{r_{\rm Z}}{(1-r_{\rm Z})^3}(1-r_{\rm Z}^2+2r_{\rm Z} \ln r_{\rm Z})^2,\label{eq:BR_H_Z_m}
\end{align}
where $\kappa_{\rm qq'}$ are the off-diagonal Higgs coupling to fermions, $Q_{\rm q}$ is the quarks charge, $\varv\approx 246$~GeV is the Higgs vacuum expectation value, $v_{\rm q}=T_3^{\rm q}/2-Q_{\rm q}\sin^2\theta_{\rm w}$ is the vector current coupling of the $\mathrm{Z}$ boson, $T_3^\mathrm{q}$ is the  third component of weak isospin, and $r_{\rm Z}=m_{\rm Z}^2/m_{\rm H}^2$ is the squared Z-to-H mass ratio. In Eq.~(\ref{eq:BR_H_gamma_m}), for the numerical evaluation of the meson HQET first inverse moments $\lambda_\mathrm{VM}(\mu)$, which play an important role when working out rates for exclusive decays involving charm and bottom mesons~\cite{Lee:2005gza,Lu:2021ttf}, we have used the values quoted in Table~\ref{tab:params} with $\rm \lambda_{B_s^{*0}}=\lambda_{B_d^{*0}}=\lambda_B$, $\rm \lambda_{D_s^{*0}}=\lambda_D$, and $\rm \lambda_{K^{*0}}=\lambda_K$. In Eq.~(\ref{eq:BR_H_Z_m}), the $f^\perp_{\rm VM}(\mu_{\rm HZ})$ objects are the transverse decay constants of the vector meson, evaluated at the hard scale $\mu_{\rm HZ} = (m^2_{\rm H}-m^2_{\rm Z})/m_{\rm H}\approx 58.8 \text{ GeV}$. These transverse decay constants are coupling constants of the vector meson to the tensor current defined by
\begin{align}
    \langle 0|\overline{q}'(0)\sigma^{\mu\nu}q(0)|VM(p,\lambda)\rangle = if^\perp_{\rm VM}(\mu)(p^\mu\varepsilon^{\nu}-p^\nu\varepsilon^{\mu}),
\end{align}
where $p,\ \lambda$ are the momentum and polarization of the vector meson, respectively, and $\mu$ is the factorization scale. The corresponding numerical values used here are~\cite{Pullin:2021ebn,RBC-UKQCD:2008mhs}: $f^\perp_{\rm B^{*0}}(3\text{ GeV})= 0.2 \text{ GeV}$, $f^\perp_{\rm B^{*0}_{\rm s}}(3\text{ GeV}) = 0.236 \text{ GeV}$, $f^\perp_{\rm K^{*0}}(2 \text{ GeV}) =0.717 f_{\rm K^{*0}}$, $f^\perp_{\rm D^{*0}}(2\text{ GeV})= 0.202 \text{ GeV}$. These transverse decay constants can be evolved to the higher scale $\mu_{\rm HZ}$ using the renormalization group equation~\cite{Konig:2015qat}, thereby giving: $f^\perp_{\rm VM}(\mu_{\rm HZ})=0.86\ f^\perp_{\rm VM}(3 \text{ GeV})=0.83\ f^\perp_{\rm VM}(2 \text{ GeV})$, using $\alpha_\mathrm{s}(\mu_{\rm HZ})=0.126$, $\alpha_\mathrm{s}(3\text{ GeV})=0.25$, $\alpha_\mathrm{s}(2\text{ GeV})=0.29$, and $N_\mathrm{f}=5$ number of flavours.

Finally, the last ingredient needed to compute the branching fractions of the Higgs decays into a photon or Z boson plus a flavoured meson, are the off-diagonal Higgs coupling to fermions $\kappa_{\rm qq'}$, which appear in the Lagrangian density
\begin{align}
    \mathcal{L}_{H\to\overline{q}q'} = \sum_{q}\kappa_{\rm qq'} h\overline{q}_Lq'_R + \text{h.c.}\ ,
\end{align}
and can be determined by matching the effective tree-level expressions from the Lagrangian with the NLO branching fractions given by Eq.~(\ref{eq:BR_H_qqprime}):
\begin{align}
    \BR(\mathrm{ H\to qq')\times \Gamma_{\mathrm{H}}  = \Gamma(\rm H\to \overline{q}q'+ \overline{q'}q)}  = \frac{N_\mathrm{c}}{8\pi}\left(|\kappa_{\rm qq'}|^2+|\kappa_{\rm q'q}|^2\right)m_\mathrm{H}.
\end{align}
The final results for the off-diagonal Higgs couplings read:
\begin{align}
    \begin{cases}
        |\kappa_{\rm bs}|^2+ |\kappa_{\rm sb}|^2 = 2.4 \times 10^{-11},\\
        |\kappa_{\rm bd}|^2+ |\kappa_{\rm db}|^2 = 1.0 \times 10^{-12},\\
        |\kappa_{\rm cu}|^2+ |\kappa_{\rm uc}|^2 = 7.4 \times 10^{-24},\\
        |\kappa_{\rm ds}|^2+ |\kappa_{\rm sd}|^2 = 5.2 \times 10^{-19}.
    \end{cases}\label{eq:kappa_h}
\end{align}
Combining Eq. \eqref{eq:kappa_h} with \eqref{eq:BR_H_gamma_m} or \eqref{eq:BR_H_Z_m}, we arrive at the theoretical branching fractions listed in Table~\ref{tab:H_decays_boson_flavouredmeson}. Such exclusive FCNC radiative decays are extremely suppressed in the SM, in the range of $\mathcal{O}(10^{-14}$--$10^{-30})$. Therefore, radiative flavoured meson decays of the Higgs boson appear utterly rare to be visible at any current or future collider, and therefore a very clean probe of any BSM physics that may enhance Higgs FCNC decays. Experimental limits have been recently set for the $\rm H \to \gamma+ K^{*0}, \gamma + D^{*0}$ channels at $\mathcal{O}(10^{-4}\mbox{--}10^{-3})$~\cite{ATLAS:2023alf,ATLAS:2024dpw}, to be compared with our $\mathcal{O}(10^{-23}\mbox{--}10^{-27})$ SM predictions.

\begin{table}[htpb!]
\caption{Compilation of exclusive Higgs decays to a gauge boson plus a flavoured meson. For each decay, we provide the branching fraction predicted using Eqs.~(\ref{eq:BR_H_gamma_m}), (\ref{eq:BR_H_Z_m}), as well as the current experimental upper limit and that estimated for HL-LHC. The last two columns indicate whether the decay can be produced at FCC-ee/FCC-hh.
\label{tab:H_decays_boson_flavouredmeson}}
\resizebox{\textwidth}{!}{%
\input{tables/exclusive_H_decays_boson_flavouredmeson.tex}
}
\end{table}

\begin{figure}[htpb!]
\centering
\includegraphics[width=0.6\textwidth]{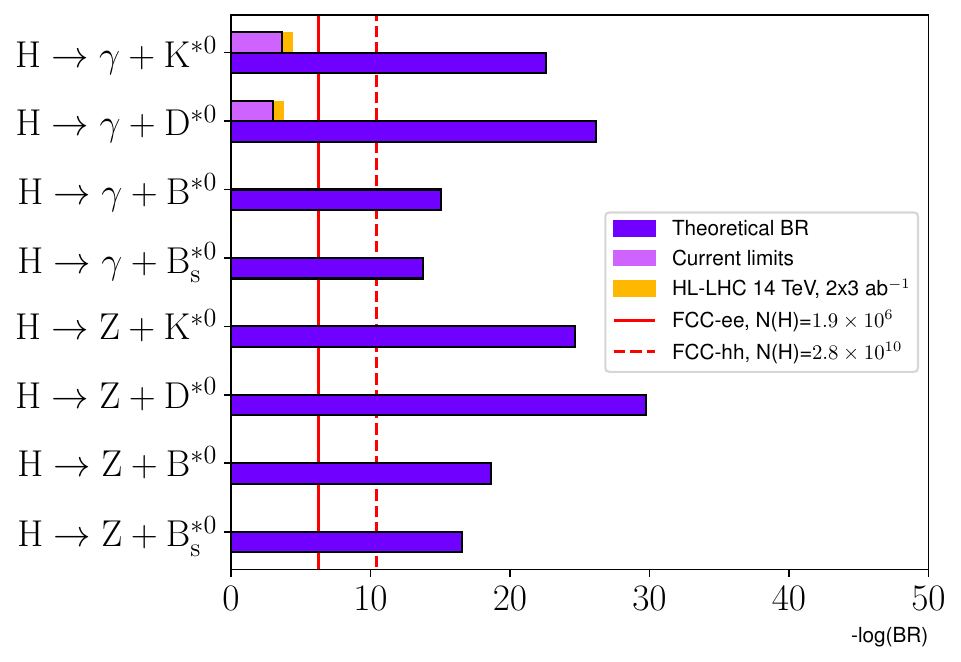}
\caption{Branching fractions (in negative log scale) of exclusive $\rm H \to \gamma/ Z+$~flavoured-meson decays. Our predictions via Eqs.~(\ref{eq:BR_H_gamma_m}), (\ref{eq:BR_H_Z_m}) (blue bars) are compared with current experimental upper limits (violet) and expected conservative HL-LHC bounds (orange). The vertical red~lines indicate the FCC-ee (solid) and FCC-hh (dashed) reach based only on the total number of H bosons to be produced at both~facilities.
}
\label{fig:H_boson_flavouredmeson_limits}
\end{figure}

\subsubsection{Higgs decays into a W boson plus a meson}

\begin{figure}[htpb!]
\centering
\includegraphics[width=0.6\textwidth]{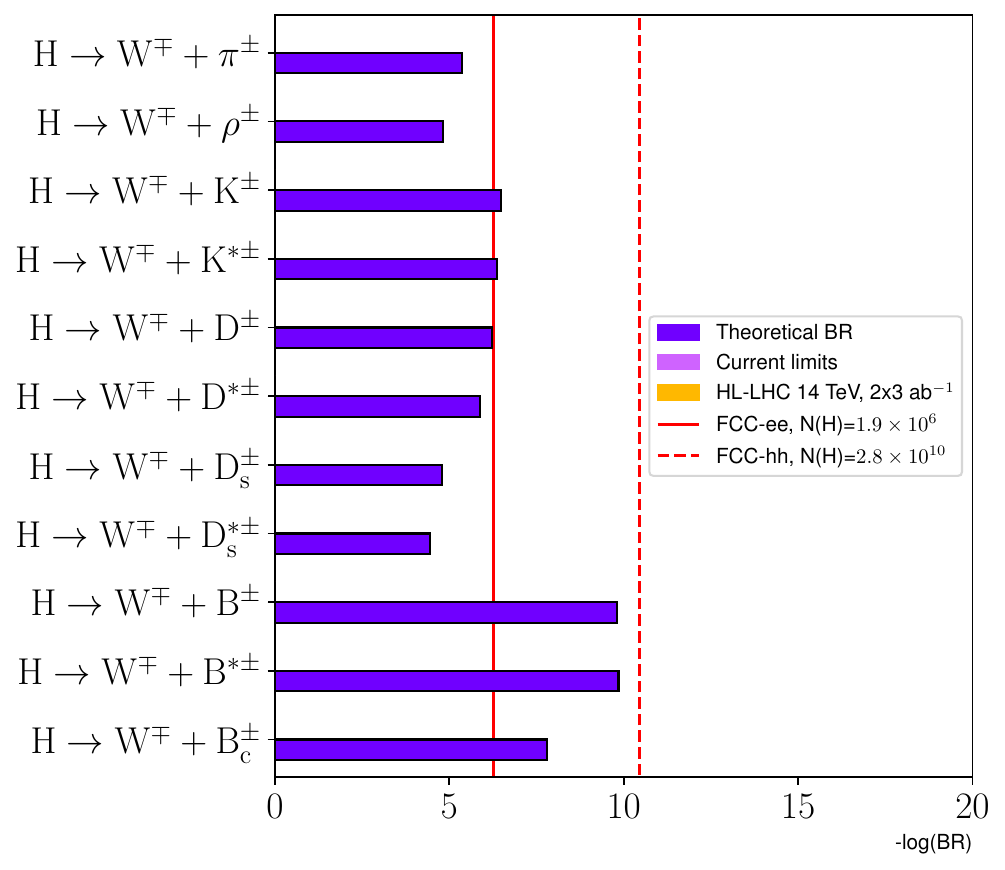}
\caption{Branching fractions (in negative log scale) of exclusive $\rm H \to W^{\pm} + meson$ decays. The theoretical predictions are shown as blue bars. The red vertical lines indicate the FCC-ee (solid) and FCC-hh (dashed) reach based only on the total number of H bosons to be produced at both facilities.}
\label{fig:H_W_meson_limits}
\end{figure}

The charged $\rm H \to W^\pm M^\mp$ decays (corresponding to the diagrams of Fig.~\ref{fig:H_gauge_meson_diags} with W bosons in the final state) differ qualitatively from the neutral radiative decays discussed above, because the W attaches itself to a charged current, and one can probe flavour-violating couplings of the Higgs boson. The theoretical complication is that the W has both transverse and longitudinal polarizations, yielding lengthier analytical expressions~\cite{Isidori:2013cla,Alte:2016yuw}. Table~\ref{tab:H_decays_W_meson} lists the theoretical predictions for Higgs decays into a W$^\pm$ boson plus a charged meson, and Fig.~\ref{fig:H_W_meson_limits} presents them in graphical form. The rates for these processes range roughly between $10^{-5}$ and $10^{-10}$ with small differences between models for the same final state. The EFT\,+\,NRQM theoretical predictions of Ref.~\cite{LHCHiggsCrossSectionWorkingGroup:2016ypw}, which update the $\BR$ numerical values computed in~\cite{Isidori:2013cla} (but have likely a typo in the $\BR(\rm H\to B^{*\pm}W^\mp) = 10^{-5}$ rate quoted, which should be $10^{-10}$), agree with the alternative EFT\,+\,LCDA rates of Ref.~\cite{Alte:2016yuw}. There has been no experimental search to date at the LHC for any of these 11 decays. The most promising channel, with a $\mathcal{O}(10^{-5})$ branching fraction, is $\rm H \to W^\pm + \rho^\mp$ with the rho meson decaying into two pions with almost 100\% probability, but seemingly experimental evidence for 7 (all) of them will only be possible at FCC-ee (FCC-hh).


\begin{table}[htpb!]
\caption{Compilation of exclusive Higgs decays to a W boson plus a meson. For each decay, we provide the predicted branching fraction(s) and the theoretical approach used to compute it. The last two columns indicate whether the decay can be produced at FCC-ee/FCC-hh.
\label{tab:H_decays_W_meson}}
\input{tables/exclusive_H_decays_W_meson.tex}
\end{table}

\subsection{Radiative Higgs leptonium decays}

\begin{figure}[htpb!]
\centering
\includegraphics[width=0.6\textwidth]{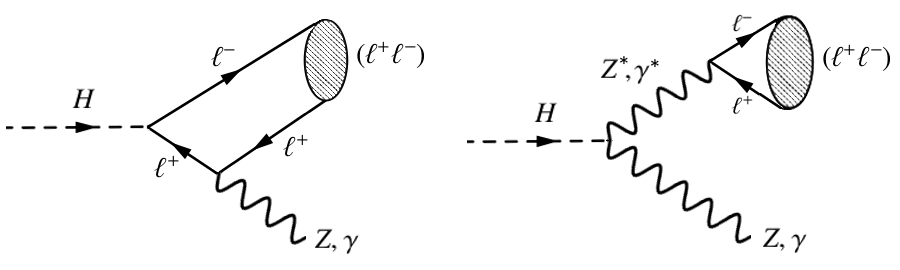}
\caption{Schematic diagrams of exclusive decays of the H boson into a photon or a Z boson plus a leptonium state in the direct (left) and indirect (right) processes. The solid fermion lines represent leptons, the gray blob represents the leptonium bound state.}
\label{fig:H_leptonium_diags}
\end{figure}

Figure~\ref{fig:H_leptonium_diags} shows the diagrams of the decay of the Higgs boson into a photon or a Z boson plus a leptonium state $(\lele)$, where $(\lele) = (\epem),\,(\mumu),\,(\tautau)$ represent positronium, dimuonium, and ditauonium bound states, respectively. Depending on the accompanying gauge boson (Z or $\gamma$), leptonium can be produced in spin triplet (ortho) and/or spin singlet (para) states. Such decays are similar to the exclusive radiative decays into mesons shown in Fig.~\ref{fig:H_gauge_meson_diags}, but changing the quark lines for lepton lines. Of course, in the SM there is no unlike-flavour leptonic decay possible, at variance with the $\rm H \to W^{\pm} + M$ case. However, in the presence of BSM physics leading to LFV Higgs decays, $(\ell\ell')$ bound states could be formed directly in exclusive radiative decays. We provide here an estimate of the decay rates for the SM processes by considering the similar $\rm H \to \gamma + M,\,Z + M$ processes, and changing the quarkonium form-factors by leptonium ones. Apart from probing lepton Yukawa couplings, if such decays were to be produced with relatively large rates, they would provide interesting cross checks for any LFUV effects potentially observed with the ``open'' leptons~\cite{Fael:2018ktm}.

The overall probability for forming an onium bound state is determined by one single parameter: its radial wavefunction at the origin of the space coordinate, which for leptonium bound states (of principal quantum number $n$) reads~\cite{dEnterria:2022alo},
\begin{equation}
|\phi_{n,(\ell\ell)}(r=0)|^2=\frac{\left(m_{\ell} \alpha(0)\right)^3}{8 \pi n^3}.
\label{eq:wf_origin_ll}
\end{equation}
Since the QED coupling is much smaller than the QCD one, $\alpha(m_{\ell\ell})\ll C_F\alphas(m_{\qqbar})$, and since the charged lepton masses are smaller than the quark masses of the same generation, $m_{\ell\ell}\ll m_{\qqbar}$, the ratio $[\alpha(0) m_{\ell\ell}/(\alpha_\mathrm{s}(m_{\qqbar})m_{\qqbar})]^{3}$ is very small, and
one can anticipate that those decays will be orders-of-magnitude more suppressed than the $\rm H \to \gamma + M(\qqbar)$ ones discussed in Section~\ref{sec:H_gauge_meson}.

We start by calculating the branching fraction for the $\rm H \to \gamma+(\lele)$ process in which,  due to charge-conjugation conservation, the leptonium can only be in the ortho $(\lele)_1$ state. The partial width of this decay can be derived from the similar expressions for quarkonium vector mesons~\cite{Bodwin:2013gca}, namely
\begin{align}
	\BR(\rm H\to \gamma+(\lele)_1) = \frac{1}{8 \pi}\frac{m_\mathrm{H}^2-m_{\ell\ell}^2}{m_\mathrm{H}^2\,\Gamma_\mathrm{H}}\left|\mathcal{A}_\text{dir}+\mathcal{A}_\text{ind}\right|^2,
 \label{eq:Gamma_H_ll_gamma}
\end{align}
with the direct and indirect amplitudes corresponding to the diagrams of Fig.~\ref{fig:H_leptonium_diags} left and right, respectively, and where the  $\rm H\to \gamma+Z^*$ contribution followed by the $\rm Z^* \to (\lele)_1$ transition is negligible compared to the corresponding $\rm H\to \gamma+\gamma^*$ one. These two amplitudes can be written as a function of the wavefunction at the origin $\phi_\mathrm{n,(\ell\ell)}(0)$, as follows
\begin{align}
\mathcal{A}_\text{dir}&=2 Q_{\ell} \sqrt{4\pi \alpha(0)} \left(\sqrt{2}\mathrm{G_F}\, m_{\ell\ell}\right)^{1/2} \frac{m_\mathrm{H}^2-m_{\ell\ell}^2}{\sqrt{m_\mathrm{H}}(m_\mathrm{H}^2-m_{\ell\ell}^2/2-2m_\mathrm{\ell}^2)}\,\phi_\mathrm{n,(\ell\ell)}(0),\\
\mathcal{A}_\text{ind} &=  \frac{Q_{\ell}\sqrt{4\pi \alpha(0)}\, f_{\ell\ell}}{m_{\ell\ell}}  \left(16\pi\Gamma_\mathrm{H\to\gaga}\right)^{1/2} \frac{m_\mathrm{H}^2-m_{\ell\ell}^2}{m_\mathrm{H}^2} = -\frac{2Q_{\ell}\sqrt{4\pi \alpha(0)}}{m_{\ell\ell}^{3/2}}  \left(16\pi\Gamma_\mathrm{H\to\gaga}\right)^{1/2} \frac{m_\mathrm{H}^2-m_{\ell\ell}^2}{m_\mathrm{H}^2}\, \phi_\mathrm{n,(\ell\ell)}(0)\,,
\label{eq:H_ll_gamma_indirect}
\end{align}
where for the second equality of the indirect component, we have adopted the Van~Royen--Weisskopf formula~\cite{Azhothkaran:2020ipl,VanRoyen:1967nq} to relate the leptonium decay constant, $f_\mathrm{M}$, to its wavefunction at the origin:
\begin{align}
f^2_\mathrm{M} = 4 N_\mathrm{c} \frac{|\phi_\mathrm{M}(0)|^2}{m_\mathrm{M}}.
\label{eq:VanRoyen}
\end{align}
Here, the particle M can be either a pseudoscalar/vector meson or a para/ortho-leptonium state, and the number of colours $N_\mathrm{c}= 1$ is used for the latter. We choose $\phi_\mathrm{M}(0)$ to be real, and then the decay constant $f_\mathrm{M}$ comes with a phase that decides the interference between indirect and direct amplitudes. In this work, we choose the (positive) phase that yields a destructive interference similar to the quarkonium case~\cite{Bodwin:2013gca}. 
Since $m_{\ell\ell}\ll m_\mathrm{H}$, plugging Eq.~(\ref{eq:wf_origin_ll}) into (\ref{eq:H_ll_gamma_indirect}), one can see that the indirect amplitudes are independent of the leptonium masses. Using Eqs.~(\ref{eq:Gamma_H_ll_gamma})--(\ref{eq:H_ll_gamma_indirect}) with the numerical values of Table~\ref{tab:params}, we determine the radiative ortholeptonium $(n=1)$ branching fractions listed in the first three rows of Table~\ref{tab:H_decays_boson_leptonium}. 
The radiative ortholeptonium decays of the Higgs boson have all numerically  similar $\mathcal{O}(10^{-12})$ rates for the three leptons, they have not been searched-for to date at the LHC, and only a future machine such as the FCC-hh can try to set upper limits on them at about 10 times their SM values. The experimental search has a very clean signature characterized by a secondary vertex from the boosted $(\lele)_1$ decay, which leads to significantly displaced triphotons (for positronium and dimuonium), $\epem$ pairs (for dimuonium and ditauonium), or $\mumu$ pairs (for ditauonium).\\

We determine next the rates of the $\rm H \to Z+(\lele)$ decays. At variance with the photon case, these decays have a massive final-state gauge boson that can be in a longitudinal polarization state. As a consequence, both scalar (para-) and vector (ortho-) leptonium states can be produced, whereas in the $\rm H\to \gamma + (\lele)_1$ case, the leptonium could only be a (transversely polarized) ortho state. We consider first the Higgs decays into Z-boson-plus-ortholeptonium by properly adapting the theoretical expressions for the similar $\rm H \to Z + VM$ decays, where VM are quarkonia vector mesons such as $\jpsi,\Upsilon$~\cite{Gao:2014xlv}. The corresponding $\rm Z + (\lele)_1$ decay width can then be written as
\begin{align}
	\Gamma(\rm H \to Z + (\lele)_1) = \Gamma_1+\Gamma_2+\Gamma_3,\label{eq:Gamma_H_ll_Z}
\end{align}
where $ \Gamma_1,\,\Gamma_2,$ and $\Gamma_{3}$ are the contributions from $\rm H\to Z+Z^*\to Z+(\lele)_1$, $\rm H\to Z+\gamma^*\to Z+(\lele)_1$, and their interference, respectively. Using $r_\mathrm{Z}=m_\mathrm{Z}^2/m_\mathrm{H}^2$, $r_{\ell\ell}=m_{\ell\ell}^2/m_\mathrm{H}^2$, the K\"all\'en function $\lambda(a,b,c)=a^2+b^2+c^2-2 (ab+ac+bc)$, and the $\rm H\gamma Z$ effective coupling $C_{Z\gamma}\approx 5.54$~\cite{Korchin:2013ifa}, these individual partial widths can be expressed as follows 
\begin{align}
\Gamma_1&=\frac{m_\mathrm{H}^3(g_{\ell\ell} f_{\ell\ell})^2 }{16\pi \varv^4}\frac{\lambda^{1/2}(1,r_\mathrm{Z}, r_{\ell\ell})}{(1-r_{\ell\ell}/r_\mathrm{Z})^2} [(1-r_\mathrm{Z}-r_{\ell\ell})^2+8 r_\mathrm{Z} r_{\ell\ell}],
\nonumber \\
\Gamma_2&=\frac{\alpha(0)^3 f_{\ell\ell}^2 Q_{\ell}^2m_\mathrm{H}^3}{32\pi^2 \varv^2 \sin^2\theta_\mathrm{W}}\frac{C_{Z\gamma}^2}{m_{\ell\ell}^2}\lambda^{1/2}(1,r_\mathrm{Z}, r_{\ell\ell})[(1-r_\mathrm{Z}-r_{\ell\ell})^2+2 r_\mathrm{Z} r_{\ell\ell}],\nonumber \\
\Gamma_{3}&=\frac{3\alpha(0)^2 f_{\ell\ell}^2 g_{\ell\ell} Q_{\ell} m_\mathrm{H} C_{Z\gamma}}{8\pi \cos\theta_\mathrm{W} \sin^2\theta_\mathrm{W} \varv^2}\frac{\lambda^{1/2}(1,r_\mathrm{Z}, r_{\ell\ell})}{1-r_{\ell\ell}/r_\mathrm{Z}}(1-r_\mathrm{Z}-r_{\ell\ell}),\label{eqn:H_Z_ll}
\end{align}
with $\varv\approx 246$ GeV, and $g_{\ell\ell}=T_3^\ell -2Q_{\ell}\sin^2\theta_\mathrm{W}= 1/2+ 2 \sin^2\theta_\mathrm{W}$ (where $Q_{\ell}$ is the lepton charge, and $T_3^\ell$ its third component of the weak isospin).
Using the expressions above with the numerical values of Table~\ref{tab:params}, we determine the  branching fractions listed in the second three rows of Table~\ref{tab:H_decays_boson_leptonium}. These $\rm H \to Z + (\lele)_1$ decays have $\mathcal{O}(10^{-11}$--$10^{-13})$ rates, and they have not been searched-for to date at the LHC. 
A recent work~\cite{Martynenko:2024rfj} has computed higher-order corrections to the Higgs leptonium decays, finding  consistent results with ours within a factor of three\footnote{Note that the Higgs decay width into $\gamma+\,$positronium originally quoted in Ref.~\cite{Martynenko:2024rfj} had a typo that wrongly enhanced it by a factor of 25~\cite{MartynenkoPrivateComm}.}. In this case, only a High-Luminosity machine such as FCC-hh would be able to provide limits approaching the SM value with a particularly unique search for a hard photon accompanied by softer triphotons, or by an $\epem$ pair\footnote{The boosted positronium would have an extremely long path-length (its natural triphoton decay, with lifetime $c\tau = 142$~ns, would take place many km(!) away from the interaction point: $L = (c\tau)(\beta*\gamma) \approx 1450$~km, for $\beta\gamma \approx (m_\mathrm{H}-m_\mathrm{Z})/m_\mathrm{(ee)} \approx 3.4\,10^{4}$), but the bound state would be first potentially broken
into its individual $\epem$ components by the detector magnetic field~\cite{Nemenov:2001smx}, or by two-photon annihilation of the constituent positron in the detector material.}, issuing from a significantly displaced vertex from the secondary positronium decay.

Lastly, we compute the rates of the $\rm H \to Z+(\lele)_0$ decays, whose width can be written using similar expressions for the $\rm H \to Z+ M$ decays (where M is a pseudoscalar meson)~\cite{Alte:2016yuw}, as follows
\begin{align}
\Gamma(\rm H\to Z+({\lele})_0) &= \frac{m_\mathrm{H}^3}{4\pi \varv^4}\,\lambda^{3/2}(1,r_\mathrm{Z},r_{\ell\ell})\,\big| F^\mathrm{Z+\ell\ell_0} \big|^2 \;\mbox{ with } F^\mathrm{Z+\ell\ell_0}=F_\text{dir}^{Z+\ell\ell_0}+F_\text{ind}^{Z+\ell\ell_0}.
\label{eq:GammaHZll}
\end{align}
The contribution from the direct amplitude to the partial width (\ref{eq:GammaHZll}) amounts to
\begin{align}
	F_\mathrm{dir}^{Z+\ell\ell_0} = -f_{\ell\ell}\,a_\mathrm{\ell}\,\frac{m_\mathrm{\ell}^2}{m_\mathrm{H}^2} \frac{1-r_\mathrm{Z}^2+2r_\mathrm{Z}\ln r_\mathrm{Z}}{(1-r_\mathrm{Z})^3} , 
\end{align}
with $a_\mathrm{\ell}=T_3^\mathrm{\ell}/2=1/4$, which is suppressed by a factor $m_{\ell\ell}^2/m_\mathrm{H}^2$, that makes it completely negligible (the contribution to the final  $(\tau\tau)_0$ amplitude is of about 0.09\%). 
The contribution from the indirect amplitude to the width (\ref{eq:GammaHZll}) reads
\begin{align}
	F_\text{ind}^{Z+\ell\ell_0} &= f_{\ell\ell}\,a_\mathrm{\ell}.
\end{align} 
Using the expressions above with the numerical values of Table~\ref{tab:params}, we determine Z-plus-paraleptonium decay rates, $\rm H \to Z + (\lele)_0$, that amount to $10^{-12}$--$10^{-16}$ (three last rows of Table~\ref{tab:H_decays_boson_leptonium}).\\

\begin{figure}[htpb!]
\centering
\includegraphics[width=0.6\textwidth]{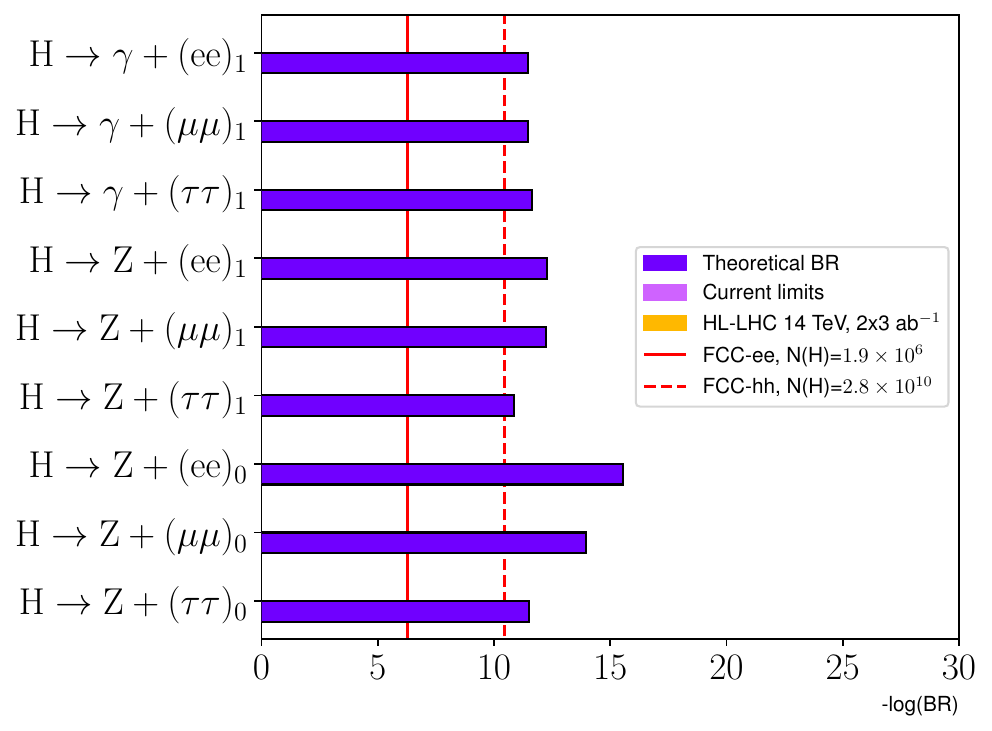}
\caption{Branching fractions (in negative log scale) of exclusive $\rm H \to \gamma,Z + leptonium$ decays: The theoretical predictions computed here are shown as blue bars. 
The red vertical lines indicate the FCC-ee (solid) and FCC-hh (dashed) reach based only on the total number of H bosons to be produced at both facilities.}
\label{fig:H_boson_leptonium_limits}
\end{figure}

All branching fractions computed here for Higgs decays into Z or $\gamma$ plus leptonium are listed in Table~\ref{tab:H_decays_boson_leptonium} and shown in graphical form in Fig.~\ref{fig:H_boson_leptonium_limits}. As aforementioned, the decay rates are minuscule, in the $\mathcal{O}(10^{-11}$--$10^{-16})$ range, and only BSM effects enhancing them can made them visible. If higher-order corrections increase the $\rm H \to \gamma + (\epem)_1$ channel by about 400 times as indicated in Ref.~\cite{Martynenko:2024rfj}, then a machine producing as many Higgs bosons as the FCC-hh can attempt to observe the $\rm H \to \gamma + (3\gamma)$ with a displaced triphoton vertex from the late orthopositronium $(ee)_1\to3\gamma$ decay, or $\rm H \to \gamma + \epem$ with a displaced dielectron from the magnetic-field-induced breaking of orthopositronium into its constituents. 

\begin{table}[htpb!]
\caption{Compilation of exclusive Higgs decays to a photon or a Z boson plus an ortho- $(\lele)_1$ or para- $(\lele)_0$ leptonium state (only the ground states, $n=1$, are considered). For each decay, we provide the prediction of its branching fraction computed here. The last two columns indicate whether the decay can be produced at FCC-ee/FCC-hh.
\label{tab:H_decays_boson_leptonium}}
\input{tables/exclusive_H_decays_boson_leptonium.tex}
\end{table}

\subsection{Exclusive Higgs decays into two mesons}

\begin{figure}[htpb!]
\centering
\includegraphics[width=0.99\textwidth]{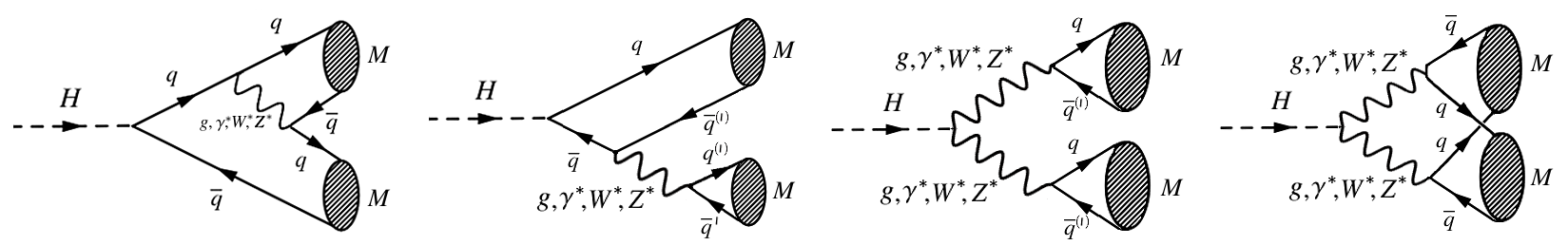}
\caption{Schematic diagrams of exclusive decays of the H boson into two mesons. The wavy lines indicate gauge bosons, the solid fermion lines represent quarks and the gray blobs are the meson bound states.}
\label{fig:H_2meson_diags}
\end{figure}

Figure~\ref{fig:H_2meson_diags} shows representative diagrams of the exclusive decay of the Higgs boson into two mesons, which can proceed through a multitude of intermediate states coupling to the scalar particle: quarks, gluons, virtual EW gauge bosons. First estimates of these processes were performed ignoring the internal motion of the produced quark-antiquark pairs~\cite{Doroshenko:1987nj}, and then further improved within different approaches for the meson-pair formation~\cite{Kartvelishvili:2008tz,Belov:2021toy, Faustov:2022jfk,Gao:2022iam, Faustov:2023phi,Belov:2023iop}. Except for a channel involving the $\phi$ meson, only exclusive decays involving charmonium and/or bottomonium final states have been computed to date, \ie\ no calculations of decays to a pair of light mesons exist to our knowledge (due to the difficulties explained in Section~\ref{sec:Z_2mesons} for the similar $\rm Z \to M+M$ case). 

Table~\ref{tab:H_2meson_decays} lists the corresponding theoretical predictions and experimental limits for concrete $\rm H \to 2(\QQbar)$ and $\rm H \to (\QQbar)(\QQbar')$ decay modes. 
The calculations are carried out in multiple frameworks (LC\,+\,LCDA, RQM, NRQCD/NRCSM, NRQCD\,+\,LDME) and predict rates in the $\mathcal{O}(10^{-9}$--$10^{-11})$ range\footnote{Semiexclusive decays $\rm H \to B_\mathrm{c}+Q,+Q',\,B_\mathrm{c}^* + Q + Q'$ have much larger rates~\cite{Jiang:2015pah} and, since they are not rare nor exclusive, they are not covered here.}, with some differences in the results for the same final state driven by the partial inclusion of the diagrams shown in Fig.~\ref{fig:H_2meson_diags}. As a matter of fact, the existing predictions have not fully included all processes shown in the figure with, \eg\ the W-induced and quark ``crossed'' decays often considered subleading and not added to the rates. The $\rm H \to \phi + \jpsi$ decay is the only process computed to date that includes the $\rm H\to W^*W^*$ intermediate diagram. The direct (quark-induced) contributions are only relevant for the heavier double bottomonia with larger Yukawa couplings, and calculations of final decays involving charmonium ignore them. 
\begin{table}[htpb!]
\caption{Compilation of exclusive Higgs decays to two mesons. For each decay, we provide the predicted branching fraction(s) and the theoretical approach used to compute it, as well as the current experimental upper limit and that estimated for HL-LHC. The last two columns indicate whether the decay can be produced at FCC-ee/FCC-hh.
\label{tab:H_2meson_decays}}
\resizebox{\textwidth}{!}{%
\input{tables/exclusive_H_decays_meson_meson.tex}}
\end{table}

\begin{figure}[htpb!]
\centering
\includegraphics[width=0.6\textwidth]{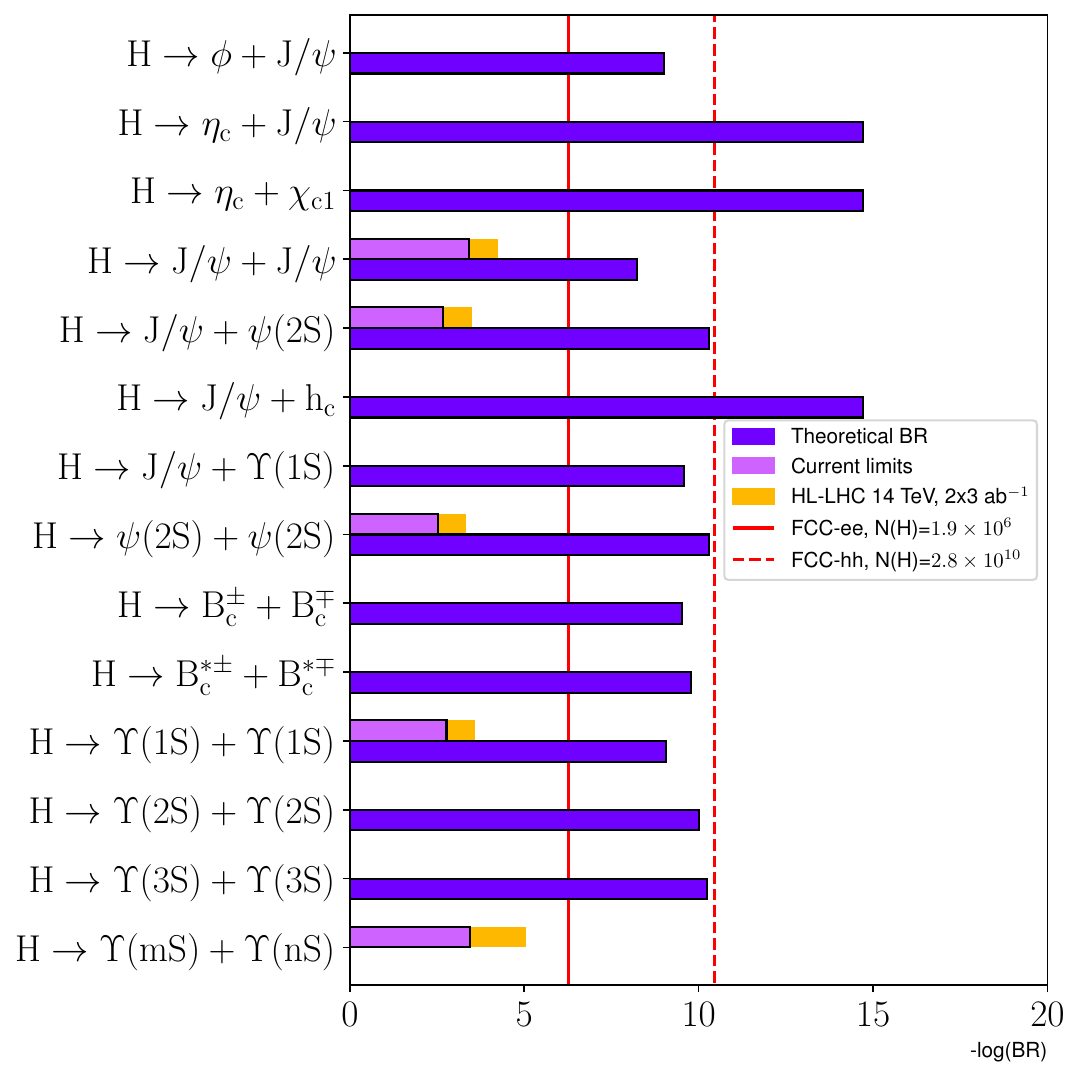}
\caption{Branching ratios (in negative log scale) of exclusive $\rm H \to meson + meson$ decays. Most recent theoretical predictions (blue bars) are compared to current experimental upper limits (violet) and expected conservative HL-LHC bounds (orange). The red vertical lines indicate the FCC-ee (solid) and FCC-hh (dashed) reach based only on the total number of H bosons to be produced at~both~machines.}
\label{fig:H_2meson_limits}
\end{figure}

Experimentally, the double-$(\QQbar)$ and $(\QQbar)(\QQbar')$ decays have been searched for at the LHC~\cite{CMS:2022fsq}, but the current $\mathcal{O}(10^{-3}$--$10^{-4})$ limits are 4--5 orders-of-magnitude larger than the predictions. Their production at the HL-LHC, as well as at any future lepton collider, appears unfeasible, and only a machine like FCC-hh will have the Higgs production rates required to reach those decay modes, as indicated in Fig.~\ref{fig:H_2meson_limits}.


\clearpage
\section{Rare Z boson decays}
\label{sec:Zboson}

\subsection{Rare three-body Z boson decays}

\begin{figure}[htpb!]
\centering
\includegraphics[width=0.99\textwidth]{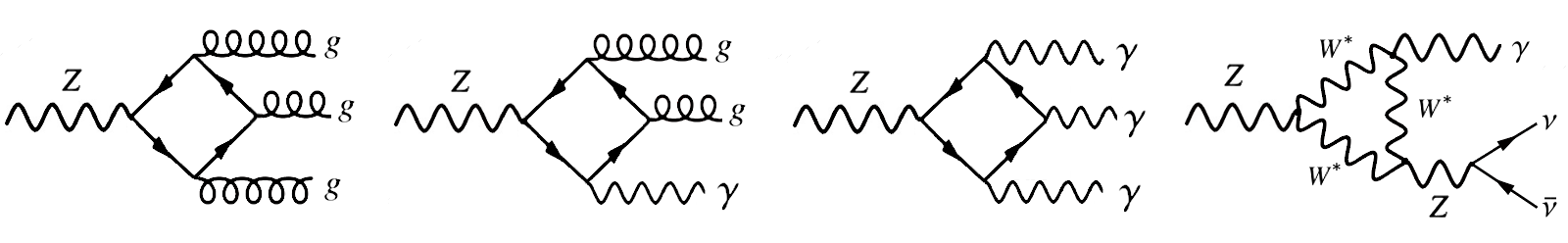}
\caption{Representative diagrams of rare 3-body decays of the Z boson into gluons and/or photons and into a photon plus neutrinos. The solid fermion lines represent quarks in the first and second diagrams, and quarks and leptons in the third one.}
\label{fig:Z_rare}
\end{figure}

The two-body decay of the Z boson into two massless vector particles ($\rm Z \to \gaga, gg$) is forbidden by the Landau--Yang theorem, which states that a spin-1 boson cannot decay into two massless bosons~\cite{Landau:1948kw,Yang:1950rg}, and of course the $\rm Z \to \gamma g$ decay is forbidden by conservation of colour. On the other hand, three-body decays into gauge bosons, or into a photon plus two neutrinos, are possible but can only be induced at the loop level in the SM (Fig.~\ref{fig:Z_rare}). 
Calculations for these virtual processes exist for many years~\cite{Laursen:1982rg,Glover:1993nv,Hopker:1993pb,Hernandez:1999xn} but they are not all always consistent among each other and, to our knowledge, have not been revised since then, although key underlying parameters, such as the $\alphasmZ$ coupling constant, hadronic Z boson partial widths, and top quark mass, have (noticeably) changed. For example, the original $\rm Z\to3\gamma$ calculations from~\cite{Glover:1993nv} used $m_\mathrm{t}=124$~GeV, whereas the calculation of the $\rm Z \to 3g$ decay of~\cite{Hopker:1993pb} used the $m_\mathrm{b}\to 0$, $m_\mathrm{t}\to \infty$ limits, and the prediction for $\rm Z\to g+g+\gamma$ from~\cite{Laursen:1982rg} used $\alphasmZ =0.17$, $m_\mathrm{t}=20$~GeV, and $\Gamma_\mathrm{Z} = \Gamma_\mathrm{Z\to \qqbar}\approx 2$~GeV. 
We note also that many of these older works, including the review of Ref.~\cite{Perez:2003ad}, provide contradictory values for some of these different branching fractions. We have therefore recalculated the rates for all these Z boson rare decays with \mgshort\ in the SM framework with QCD and EW loop corrections computed up to NLO accuracy, with massive fermions in the loops where needed, using the input parameters of Table~\ref{tab:params}.
The updated branching fractions (Table~\ref{tab:Z_decays_V_V_V}) change with respect to the existing ones (using the old formulas with updated SM parameters, where needed) by 10--25\% up or down, except for the $\rm Z \to \gamma + \nunubar$ case, where our result is six times smaller than that estimated in~\cite{Hernandez:1999xn}. This older work quoted a partial width for the triangle diagram which is smaller than our results for this contribution, whereas the relative impact of box diagrams was not clearly specified. So we suspect that a larger cancellation of contributions is present in our setup compared to that in this previous study.

\begin{table}[htpb!]
\caption{Compilation of exclusive Z decays to three gauge bosons (photons and/or gluons, with two additional channels requiring one soft photon, $E_\gamma <1$~GeV, as explained in the text) and a photon plus two neutrinos. For each decay, we provide the predicted branching fraction and the theoretical approach used to compute it, as well as the current experimental limit and that estimated for HL-LHC. The last column indicates whether the decay can be produced~at~FCC-ee.
\label{tab:Z_decays_V_V_V}}
\resizebox{\textwidth}{!}{%
\input{tables/exclusive_Z_decays_not_gamma_meson.tex}
}
\end{table}

\begin{figure}[htpb!]
\centering
\includegraphics[width=0.6\textwidth]{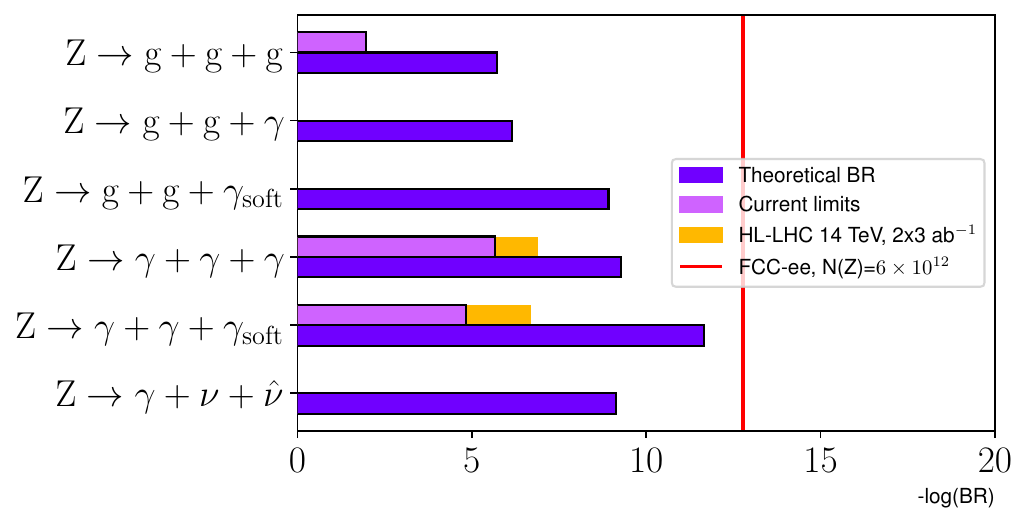}
\caption{Branching fractions (in negative log scale) of rare three-body Z boson decays. The most recent theoretical predictions (blue bars) are compared to current experimental upper limits (violet) and expected conservative HL-LHC bounds (orange). The red vertical line indicates the expected FCC-ee reach based only on the total number of Z bosons to be produced.}
\label{fig:Z_V_V_V_limits}
\end{figure}

Among the four rare Z boson decays shown in Fig.~\ref{fig:Z_rare}, those with final-state gluons have the largest rates, $\mathcal{O}(10^{-6})$, whereas the pure electroweak processes have much smaller, $\mathcal{O}(10^{-10})$, branching fractions. Experimentally, limits exist for $\rm Z\to 3g$ and $\rm Z\to 3\gamma$ of $\mathcal{O}(10^{-2})$ and $\mathcal{O}(10^{-6})$, respectively, although no searches have been yet performed to our knowledge for $\rm Z\to \gamma\,g\,g$ and $\rm Z\to \gamma\,\nunubar$ decays. The study of the 3-gluon decay appears hopeless at hadronic machines given the huge QCD trijet backgrounds, although the triphoton decay can be constrained to 100 times the expected SM value at the HL-LHC. Observation of all such decays appears only feasible at a lepton collider such as FCC-ee\footnote{In this whole section, we will be providing future Z-boson limits expectations for the FCC-ee alone because, although the FCC-hh will produce twice more Z bosons (Table~{\ref{tab:data_samples}}), the backgrounds are much larger for this collider.}, as indicated by the red vertical line in Fig.~\ref{fig:Z_V_V_V_limits}. We also provide estimates of the branching fraction for the $\rm Z \to gg\,\gamma_\text{soft}$ and $\rm Z \to \gaga\,\gamma_\text{soft}$ decays (where the soft photon has $E_\gamma < 1$~GeV) that can mimic the forbidden $\rm Z \to gg,\gaga$ channels if the low-energy photon goes experimentally undetected. The branching ratios for both such ``fake'' Landau--Yang-violating decay modes are $\mathcal{O}(10^{-9})$ and $\mathcal{O}(10^{-12})$, respectively.

\subsection{Exclusive Z boson decays into a gauge boson plus a meson}

\begin{figure}[htpb!]
\centering
\includegraphics[width=0.85\textwidth]{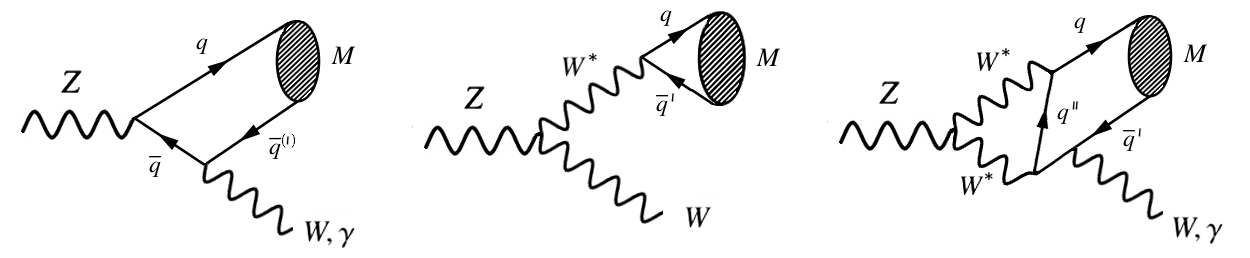}
\caption{Schematic diagrams of exclusive decays of the Z boson into a photon or W boson plus a meson. The solid fermion lines represent quarks, and the gray blobs the mesonic bound state.}
\label{fig:Z_decays_gamma_meson_diags}
\end{figure}

Figure~\ref{fig:Z_decays_gamma_meson_diags} shows the schematic diagrams of the exclusive decay of the Z boson into a photon (or a W boson) plus a meson. 
Due to the similarity of these diagrams to those of the Higgs boson (Fig.~\ref{fig:H_gauge_meson_diags}), and since the Z boson yields at colliders are about three to six orders-of-magnitude larger than for the scalar boson (Table~\ref{tab:data_samples}), the study of such processes provides valuable information on, both, theoretical elements (SCET and NRQCD validation, LCDAs/LDMEs constraints, etc.) and experimental aspects (optimization of search techniques) for the corresponding studies of exclusive Higgs boson decays. The observation of any such exclusive decays would provide unique opportunities to reconstruct the Z boson from isolated hadrons, and would improve our knowledge of meson transition form-factors, which describe the M\,$\to\gamma^*\gamma^{(*)}$ decays. In addition, exclusive Z boson decays into flavoured mesons 
probe FCNC processes.

\subsubsection{Z boson decays into a photon plus an unflavoured meson}

\begin{figure}[htpb!]
\centering
\includegraphics[width=0.6\textwidth]{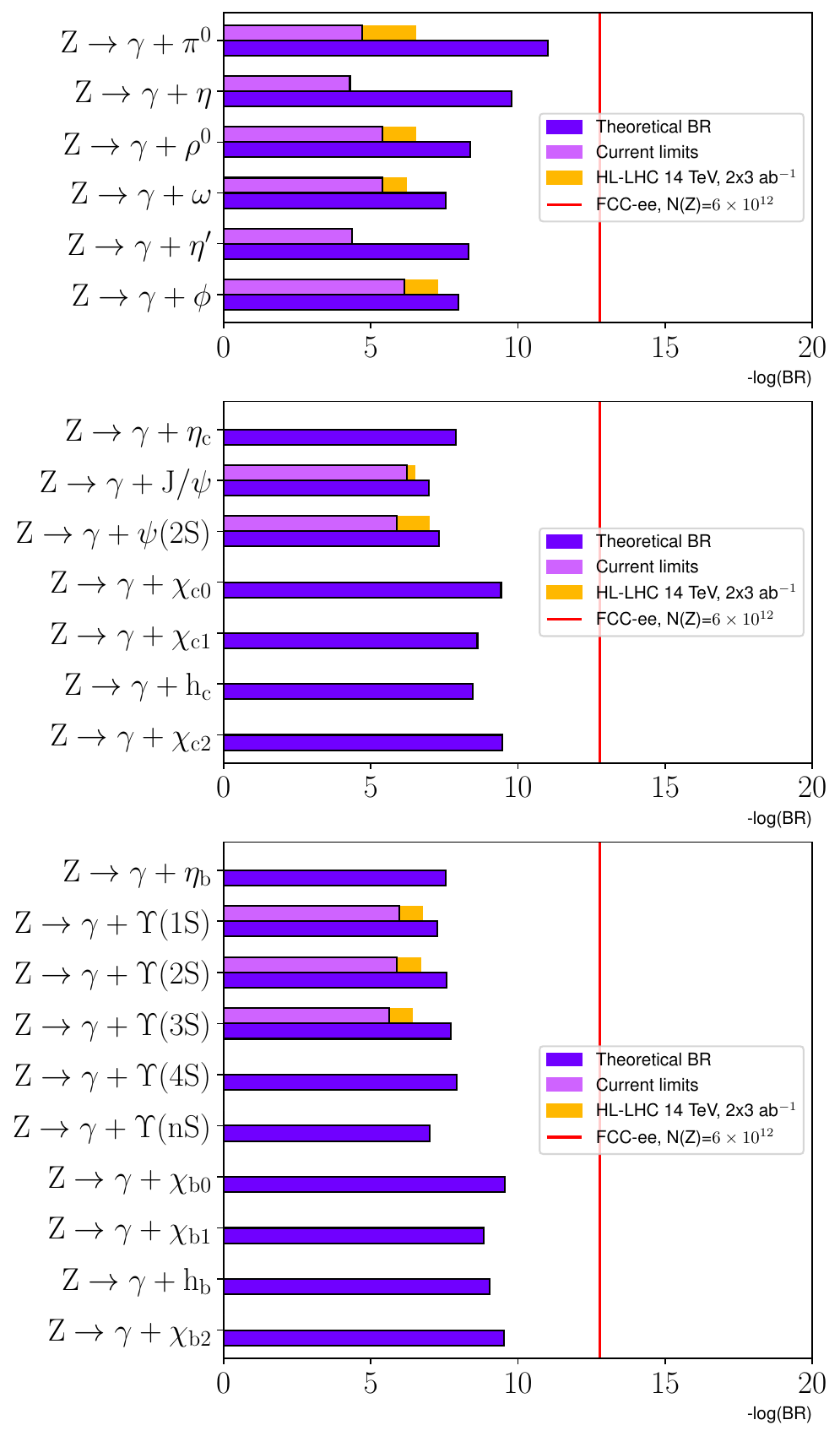}
\caption{Branching fractions (in negative log scale) of exclusive $\rm Z \to \gamma+M$ decays, with M being light (upper), charm (middle), and bottom (lower) unflavoured mesons. The most recent theoretical predictions (blue bars) are compared to current experimental upper limits (violet) and expected conservative HL-LHC bounds (orange). The red vertical line indicates the expected FCC-ee reach based only on the total number of Z bosons to be produced.}
\label{fig:Z_gamma_meson_limits}
\end{figure}

The theoretical predictions and experimental limits for the branching fractions of the radiative Z boson decay into light, charm, and bottom unflavoured mesons ($\rm Z \to \gamma + \qqbar, \ccbar, \bbbar$, proceeding through the left and center diagrams of Fig.~\ref{fig:Z_decays_gamma_meson_diags}) are listed in Table~\ref{tab:Z_decays_gamma_meson},~\ref{tab:Z_decays_gamma_ccbar}, and~\ref{tab:Z_decays_gamma_bbbar}, respectively, and shown in graphical form in Fig.~\ref{fig:Z_gamma_meson_limits}.
After forty years from their discovery, no hadronic radiative decay of the electroweak bosons has yet been observed, despite searches performed by the CDF~\cite{CDF:1998kzn,CDF:2013lma}, ATLAS~\cite{ATLAS:2017gko, ATLAS:2018jqi,ATLAS:2018zsq,ATLAS:2022rej,ATLAS:2023alf}, CMS~\cite{CMS:2018gcm,CMS:2019vaj,CMS:2019wch,CMS:2024hhg}, and LHCb~\cite{LHCb:2022kta} experiments. In some cases, such as for the $\rm Z \to \gamma\eta,\,\gamma\eta'$ channels, LEP measurements at the Z pole still provide the best upper limits~\cite{ALEPH:1991qhf}.\\

The exclusive Z-boson radiative decays into light unflavoured mesons ($\rm Z \to \gamma + M$), listed in Table~\ref{tab:Z_decays_gamma_meson}, have theoretical branching fractions in the range of $\mathcal{O}(10^{-8}$--$10^{-12})$, whereas the current experimental limits are in the $\mathcal{O}(10^{-5}$--$10^{-7})$ range with seven channels searched for by ATLAS (3), LHCb (1), CDF at Tevatron (1), and ALEPH at LEP (2). The $\rm Z\to \gamma\phi$ decay is close to being detected at HL-LHC because its predicted $\BR$ is about 1/5 of our projected conservative limit. Otherwise, all channels are producible at the FCC-ee.

\begin{table}[htpb!]
\caption{Compilation of exclusive Z decays to a photon plus a light unflavoured meson. For each decay, we provide the predicted branching fraction(s) and the theoretical approach used to compute it, as well as the current experimental upper limit and that estimated for HL-LHC. The last column indicates whether the decay can be produced at FCC-ee.
\label{tab:Z_decays_gamma_meson}}
\resizebox{\textwidth}{!}{%
\input{tables/exclusive_Z_decays_gamma_meson_light_quarks.tex}
}
\end{table}

The exclusive Z-boson radiative decays to charmonium mesons ($\rm Z \to \gamma + \ccbar$), collected in Table~\ref{tab:Z_decays_gamma_ccbar}, have rates in the $\mathcal{O}(10^{-8}$--$10^{-10})$ range as computed within multiple approaches (LC, SCET, NRQCD), which are in general consistent among themselves for the same process. 
Only one single channel has been searched for, $\rm Z \to \gamma + \jpsi$, with $\mathcal{O}(10^{-6})$ experimental upper bounds~\cite{CMS:2024hhg,CMS:2018gcm}. This decay may be visible at HL-LHC, because its predicted rate is about $1/3$ of our projected conservative limit. All channels are producible at FCC-ee.\\

\begin{table}[htpb!]
\caption{Compilation of exclusive Z decays to a photon plus a charmonium state. For each decay, we provide the predicted branching fraction(s) and the theoretical approach used to compute it, as well as the current experimental upper limit and that estimated for HL-LHC. The last column indicates whether the decay can be produced at FCC-ee.
\label{tab:Z_decays_gamma_ccbar}}
\resizebox{\textwidth}{!}{%
\input{tables/exclusive_Z_decays_gamma_meson_charmonium.tex}
}
\end{table}

The exclusive Z-boson radiative decays to bottomonium mesons ($\rm Z \to \gamma + \bbbar$), listed in Table~\ref{tab:Z_decays_gamma_bbbar}, have also branching fractions in the $\mathcal{O}(10^{-8}$--$10^{-10})$ range. Three $\rm Z \to \gamma + \Upsilon(nS)$ channels have been searched for at the LHC, setting limits in the $\mathcal{O}(10^{-6})$ range.
The $\rm Z\to \gamma +\Upsilon(1S)$ channel might be visible at HL-LHC, as the predicted branching fraction is about $1/4$ of our projected conservative limit. As for the photon-plus-charmonium case, all photon-plus-bottomonium channels are producible at FCC-ee. 
 
\begin{table}[htpb!]
\caption{Compilation of exclusive Z decays to a photon plus a bottomonium state. For each decay, we provide the predicted branching fraction(s) and the theoretical approach used to compute it, as well as the current experimental upper limit and that estimated for HL-LHC. The last column indicates whether the decay can be produced at FCC-ee.
\label{tab:Z_decays_gamma_bbbar}}
\resizebox{\textwidth}{!}{%
\input{tables/exclusive_Z_decays_gamma_meson_bottomonium.tex}
}
\end{table}

\clearpage
\subsubsection{Z boson decays into a photon plus a flavoured meson}
\label{sec:Z_gamma_M}

\begin{figure}[htpb!]
\centering
\includegraphics[width=0.6\textwidth]{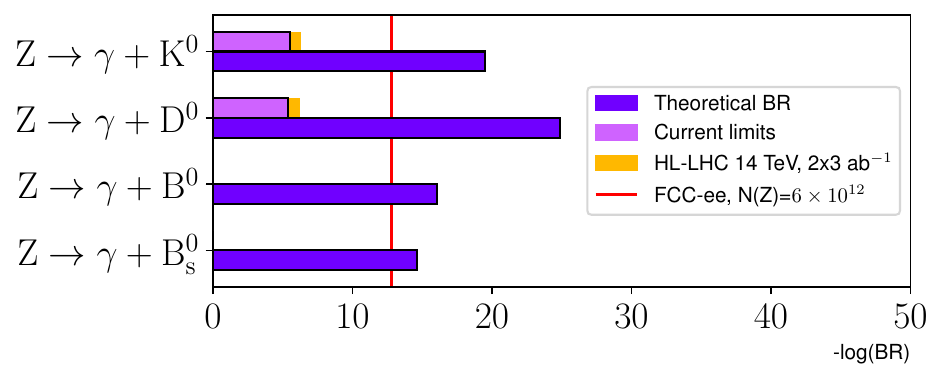}
\caption{Branching fractions (in negative log scale) of exclusive $\rm Z \to \gamma+M$ decays, with M being a flavoured meson. The theoretical predictions computed here (blue bars) are compared to current experimental upper limits (violet) and expected conservative HL-LHC bounds (orange). The red vertical line indicates the expected FCC-ee reach based only on the total number of Z bosons to be produced.}
\label{fig:Z_gamma_flavoured_meson_limits}
\end{figure}

In the SM, the absence of direct FCNC decays of the Z boson implies that the only possible exclusive radiative Z boson decays into flavoured mesons can proceed through the (extremely suppressed) right diagram of Fig.~\ref{fig:Z_decays_gamma_meson_diags}. We are not aware of existing evaluations of the branching fractions for such exclusive decays in the SM, but only of indirect limits for them determined in Ref.~\cite{Grossman:2015cak} using different SM constraints. Therefore, we estimate the numerical value of these exclusive FCNC channels here next. The standard Lagrangian density describing the Z boson coupling to quarks is given by the interaction term
\begin{align}
\mathcal{L}_\mathrm{Z\to \overline{q}q}\subset  -\frac{e}{\sin \theta_\mathrm{w} \cos \theta_\mathrm{w}} Z^\mu\bar{q} \gamma_\mu(v_{\rm q} -a_q \gamma_5) q,
\end{align}
with vector $v_{\rm q} \equiv T_3^q/2-Q_q\sin^2 \theta_\mathrm{w}$ and axial $a_q \equiv T_3^q/2$ couplings of a quark of electric charge $Q_q$ and third component of weak isospin $T^q_3$.
One can extend this Lagrangian to include interactions between quarks of different generations, as follows
\begin{align}
\mathcal{L}_\mathrm{Z\to \overline{q_i}q_j}\subset  -\frac{e}{\sin \theta_\mathrm{w} \cos \theta_\mathrm{w}} Z^\mu\bar{q_i} \gamma_\mu(v_{\overline{i}j} -a_{\overline{i}j} \gamma_5) q_j,
\end{align} 
where $i,j \in \{\rm u,d,c,s,t,b\}$. The offdiagonal interaction couplings $v_{\overline{i},j\neq i}$ and  $a_{\overline{i},j\neq i}$ are zero in the SM, but are used here as effective couplings to encode contributions from higher-order loop processes (such as those shown in Fig.~\ref{fig:Z_decays_gamma_meson_diags}, right), which can take on a very small value. The work~\cite{Grossman:2015cak} provides generic expressions for the calculation of partial widths of exclusive Z boson decays into a flavoured meson plus a photon in the SCET$\,+\,$LCDA framework, using offdiagonal quark couplings $v_{\overline{i}j},a_{\overline{i}j}$ as inputs. For arbitrary $\rm \overline{q_i}q_j$ combinations, the decay width for the $\rm Z\to \overline{q}q'$ process can be written at tree level as 
\begin{align}
\Gamma_\mathrm{Z\to \overline{q}q'} =\frac{\alpha(m_\mathrm{Z})(|v_\mathrm{\overline{q}q'}|^2+|a_\mathrm{\overline{q}q'}|^2)}{4  m_\mathrm{Z}^5  \cos^2\theta_\mathrm{w} \sin^2\theta_\mathrm{w}}\lambda^{1/2}\left(m_\mathrm{Z}^2,m_1^2,m_2^2\right) \left[ 2 m_\mathrm{Z}^4 - m_1^4 - m_2^4 - m_2^2 m_\mathrm{Z}^2  - m_1^2 (m_\mathrm{Z}^2 - 2 m_2^2 ) \right].
\end{align}
where $\lambda(a,b,c)$ is the K\"all\'en function, and $m_{1,2}$ the quark masses. 
Ignoring quark masses ($m_1=m_2=0$) and setting $v_\mathrm{\overline{q}q'}=a_\mathrm{\overline{q}q'}$, the equation above simplifies to
\begin{align}
\Gamma_\mathrm{Z\to \overline{q}q'} = \frac{\alpha(m_\mathrm{Z})  m_\mathrm{Z} |v_\mathrm{\overline{q}q'}|^2}{ \sin^2\theta_\mathrm{w} \cos^2\theta_\mathrm{w}}.
\end{align}
Using this last expression, the numerical values of the effective couplings $v_\mathrm{\overline{q}q'}=a_\mathrm{\overline{q}q'}$ 
can now be estimated from the offdiagonal Z boson quark decay widths computed at NLO accuracy in~\cite{Kamenik:2023hvi}: 
$\Gamma_\mathrm{Z\to \bar{b}s}= 5.2\times 10^{-8}$~GeV,
$\Gamma_\mathrm{Z\to \bar{b}d}= 2.3\times 10^{-9}$~GeV,
$\Gamma_\mathrm{Z\to \bar{c}u}= 1.7\times 10^{-18}$~GeV;
and amount to
\begin{equation}
\begin{cases}
\; |v_\mathrm{\overline{b}s}|^2=|a_\mathrm{\overline{b}s}|^2&\approx  6.6\times 10^{-9} \\
\; |v_\mathrm{\overline{b}d}|^2=|a_\mathrm{\overline{b}d}|^2&\approx 2.9\times 10^{-10} \\
\; |v_\mathrm{\overline{c}u}|^2=|a_\mathrm{\overline{c}u}|^2&\approx 2.2\times 10^{-19} .
\label{eq:v_ij}
\end{cases}
\end{equation}
The decay $\rm Z \to \overline{s}d$ was first computed in~\cite{Ma:1979px}, with vector and axial couplings equal since the photon contribution is zero, and with the offdiagonal coupling obtained via
\begin{align}
v_\mathrm{\overline{s}d} = a_\mathrm{\overline{s}d} = \frac{\alpha(m_\mathrm{Z}) }{16 \pi  \sin^2\theta_\mathrm{w}}\sum_{i=\rm u,c,t} V_{i\rm s}V^*_{i\mathrm{d}}\left[ -\frac{3 x_i}{1-x_i}+\frac{x_i^2}{2 (1-x_i)}-\frac{3 x_i^2 \log (x_i)}{2 (1-x_i)^2}-\frac{x_i \log (x_i)}{(1-x_i)^2}\right],
\end{align}
with $x_i = m_{\mathrm{q}_i}^2/m_\mathrm{W}^2$. Using the parameters of Table~\ref{tab:params}, this coupling squared can be numerically evaluated as
\begin{align}
 |v_\mathrm{\overline{s}d}|^2=|a_\mathrm{\overline{s}d}|^2&\approx 4.27\times 10^{-13} .
 \label{eq:v_sd}
\end{align}
Plugging in the values of the effective couplings (\ref{eq:v_ij}) and (\ref{eq:v_sd}) into the SCET$\,+\,$LCDA expressions of Ref.~\cite{Grossman:2015cak} enables us to determine the numerical branching fractions for Z-boson radiative decays to flavoured mesons listed in Table~\ref{tab:Z_decays_gamma_meson_flavoured} and graphically presented in Fig.~\ref{fig:Z_gamma_flavoured_meson_limits}. 
The quoted rates are for the sum of $\rm M = M^0+\overline{M^0}$ radiative decays. 
All such exclusive FCNC decays of the Z boson are extremely suppressed, amounting to $\mathcal{O}(10^{-15}\mbox{--}10^{-25})$. The experimentally negligible branching fractions for such decays make those interesting channels to search for BSM FCNC decays of the Z boson without any irreducible SM background. Two decays have been experimentally searched-for to date: $\rm Z \to \gamma + D^0$~\cite{LHCb:2022kta,ATLAS:2024dpw} and  $\rm Z \to \gamma + K_s^0$~\cite{ATLAS:2024dpw}, where experimental limits are at the $\mathcal{O}(10^{-6})$ level, and can be improved by a factor of $\sim$10 at the end of the HL-LHC phase as per our extrapolations here.

\begin{table}[htpb!]
\caption{Compilation of exclusive Z decays to a photon plus a flavoured meson. For each decay, we provide the predicted branching fraction(s) and the theoretical approach used to compute it, as well as the current experimental upper limits (if any) and those estimated for HL-LHC. The last column indicates whether the decay can be produced at FCC-ee.
\label{tab:Z_decays_gamma_meson_flavoured}}
\resizebox{\textwidth}{!}{%
\input{tables/exclusive_Z_decays_gamma_meson_flavoured.tex}
}
\end{table}

\clearpage
\subsubsection{Z boson triphoton decays}

\begin{figure}[htpb!]
\centering
\includegraphics[width=0.85\textwidth]{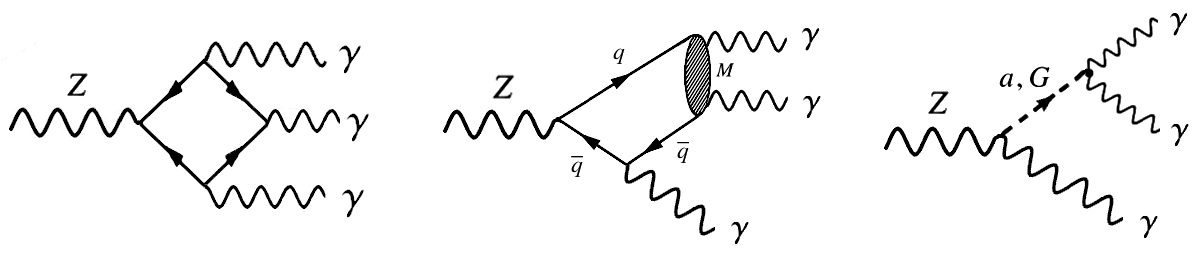}
\caption{Triphoton decays of the Z boson: SM direct decay (left), exclusive $\rm \gamma + meson(\gaga)$ decay (center), radiative decay into a BSM (ALP or massive graviton) diphoton system (right).}
\label{fig:Z_triphoton}
\end{figure}

As seen in Table~\ref{tab:Z_decays_V_V_V}, the $\rm Z \to 3\gamma$ decay is extremely suppressed in the SM ($\BR=6.4\cdot 10^{-10}$, as computed here) making of such a final state a particularly clean means to search for exotic Z decays into BSM particles with EW couplings. Many new physics scenarios contain new light (pseudo)scalar or tensor particles that couple to the electroweak (or Higgs) bosons and decay primarily to photons. Searches for $\mathrm{Z}\to\gamma\,a(\gaga)$, with $a$ being an ALP~\cite{Bauer:2018uxu,dEnterria:2021ljz} or a massive graviton~\cite{dEnterria:2023npy} decaying into photons, have attracted an increasing interest in the last years~\cite{Agrawal:2021dbo}. Figure~\ref{fig:Z_triphoton} shows possible diagrams leading to the triphoton decay of the Z boson. The direct decay (left diagram) and the exclusive radiative decays into C-even mesons followed by their diphoton decays (center) can mimic the BSM processes signals (right), and constitute backgrounds to the latter. The direct and mesonic backgrounds have been neglected so far in all ALP and massive-graviton upper limits set in studies based on current and future triphoton Z decay data. Such an assumption is justified so far given the null sensitivity to such rare SM processes, but it will not be the case at the FCC-ee facility where, as we see below, a few thousands such events are expected during the whole Z-pole run. Here, we quantify the triphoton branching fractions from meson decays by combining the exclusive $\rm Z\to\gamma+M$ results for spin-0,2 mesons of Tables~\ref{tab:Z_decays_gamma_meson}--\ref{tab:Z_decays_gamma_bbbar} with their corresponding two-photon branching fractions~\cite{ParticleDataGroup:2024cfk}. The results are listed in Table~\ref{tab:Z_3gamma}. The sum of all exclusive mesonic decay channels amounts to $\BR=1.8\cdot 10^{-10}$, representing an increase of about 30\% from the direct $3\gamma$ decay. The yields of SM triphoton decays will therefore amount to about 1\,000 events for the $N(\rm Z) = 6\cdot 10^{12}$ bosons expected to be collected during the full Z-pole operation at the FCC-ee. Searches for ALPs and new spin-2 particles will have to be carried out on top of such a relatively large number of background events that have been neglected so far in the derivation of ALP limits at future $\epem$ facilities~\cite{Bauer:2018uxu}.

\begin{table}[htpb!]
\caption{Compilation of exclusive Z decays into triphoton final states. The predictions of relevant $\rm Z \to \gamma + M$ radiative decays listed in Tables~\ref{tab:Z_decays_gamma_meson}--\ref{tab:Z_decays_gamma_bbbar} are combined with the diphoton branching fractions of the produced mesons, $\rm M \to \gaga$~\cite{ParticleDataGroup:2024cfk,Wang:2018rjg}.}
\label{tab:Z_3gamma}
\input{tables/exclusive_Z_decays_3_gammas.tex}
\end{table}

\subsubsection{Z boson decays into a W boson plus a meson}

\begin{figure}[htpb!]
\centering
\includegraphics[width=0.6\textwidth]{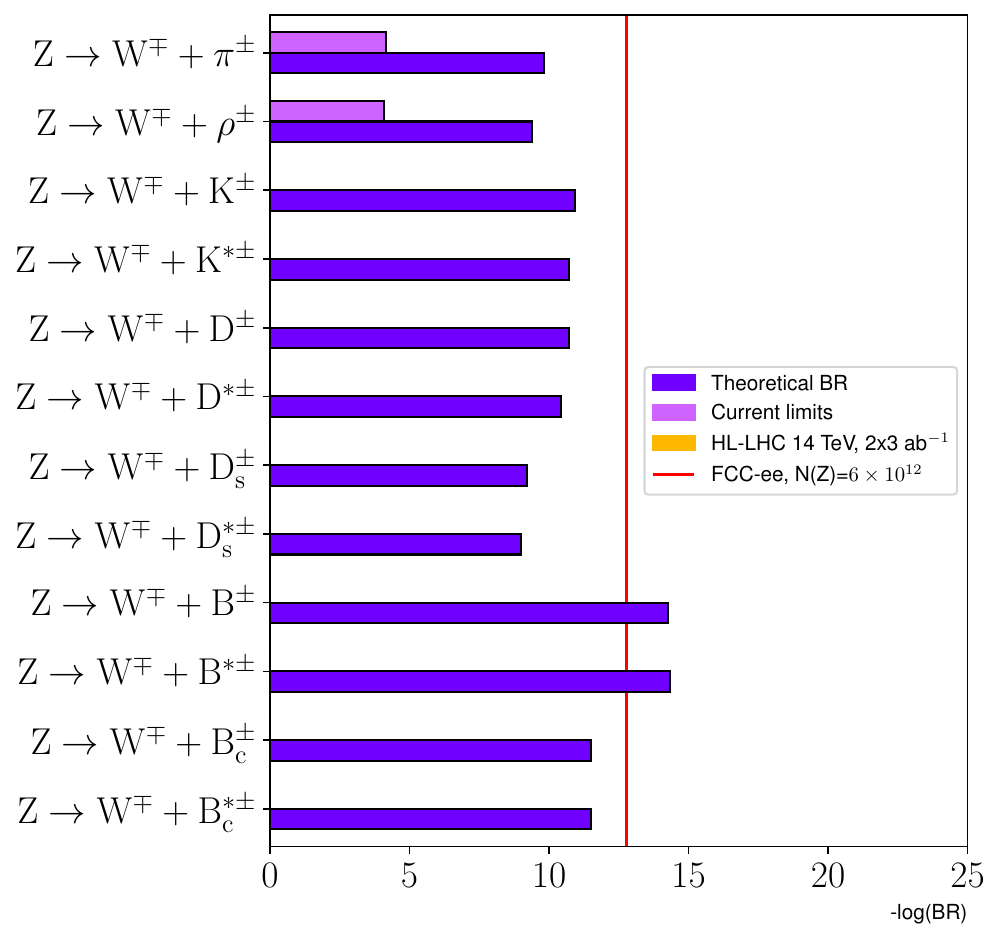}
\caption{Branching fractions (in negative log scale) of exclusive $\rm Z \to W^{\pm}+meson$ decays. The most recent theoretical predictions (blue bars) are compared to current experimental upper limits (violet) and expected conservative HL-LHC bounds (orange). The red vertical line indicates the expected FCC-ee reach based only on the total number of Z bosons to be produced.}
\label{fig:Z_W_meson_limits}
\end{figure}

The exclusive Z boson decays into $\rm W^\mp$ boson plus a charged meson $\rm M^\pm$ correspond to the diagrams of Fig.~\ref{fig:Z_decays_gamma_meson_diags} with W bosons in the final state. Theoretical estimations for such decays have been given in Ref.~\cite{Grossman:2015cak} for light-quark and $\rm D^\pm$ mesons. We estimate here the rest of missing exclusive decays: $\rm Z\to W^\mp + \{D^{*\pm},D^{*\pm}_s,B^\pm,B^{*\pm},B^{*\pm}_c,B^{\pm}_c\}$, using the expression
\begin{align}
\rm \Gamma(Z\to  W^\mp + M^\pm ) &= \frac{\alpha(m_\mathrm{Z})^2 f^2_{\rm M}}{48 m_\mathrm{Z}}|V_{\rm qq'}|^2 \frac{\sin^2 \theta_{\rm w}}{\cos^2 \theta_{\rm w}}\left(\frac{3}{2}+\frac{3}{2}\sin^2\theta_{\rm w} + \frac{227}{180}\sin^4\theta_{\rm w}\right),
\end{align}
where, to simplify the evaluations, we have neglected the small contributions from the Gegenbauer moments. Here $f_{\rm M}$ is the decay constant of the meson, and $V_{\rm qq'}$ is the relevant CKM matrix element.
Table~\ref{tab:Z_decays_W_meson} lists the theoretical predictions for all branching fractions, which are in the $10^{-10}$--$10^{-15}$ range. Only two channels have been searched for in the Z-pole run at LEP. Experimental observation will only be possible at FCC-ee for all of them. The same results compiled in Table~\ref{tab:Z_decays_W_meson} are shown in graphical form in Fig.~\ref{fig:Z_W_meson_limits}.

\begin{table}[htpb!]
\caption{Compilation of exclusive Z decays to a W boson plus a meson. For each decay, we provide the predicted branching fraction(s) and the theoretical approach used to compute it, as well as the current experimental upper limit and that estimated for HL-LHC. The last column indicates whether the decay can be produced at FCC-ee.
\label{tab:Z_decays_W_meson}}
\resizebox{\textwidth}{!}{%
\input{tables/exclusive_Z_decays_W_meson.tex}
}
\end{table}

\subsection{Radiative Z boson leptonium decays}

Figure~\ref{fig:Z_decays_gamma_meson_diags} shows the diagrams for the decay of the Z boson into a photon plus a leptonium state. Depending on the relative helicities of the outgoing leptons, their bound state can be in the spin singlet (paraleptonium, $(\ell\ell)_0$) or triplet (ortholeptonium, $(\ell\ell)_1$) states. No calculation for such processes exists to our knowledge, but the key ingredients can be obtained from the determination of the $\rm \epem \to \gamma+(\tautau)_0$ cross section at the Z pole provided in the ditauonium studies of Ref.~\cite{dEnterria:2023yao}. The motivations for their study are the same as for the similar Higgs decays, namely, the search for LFV or LFUV phenomena.

\begin{figure}[htpb!]
\centering
\includegraphics[width=0.7\textwidth]{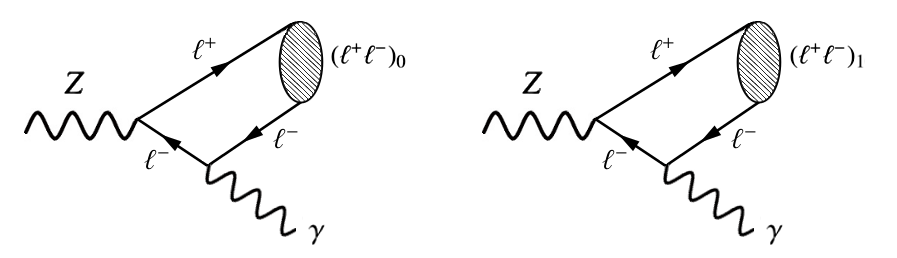}
\caption{Schematic diagrams of exclusive decays of the Z boson into para- (left) and ortho- (right) leptonium plus a photon. The solid fermion lines represent leptons, the gray blobs represent the leptonium bound state.}
\label{fig:Z_leptonium_diags}
\end{figure}

\begin{figure}[htpb!]
\centering
\includegraphics[width=0.6\textwidth]{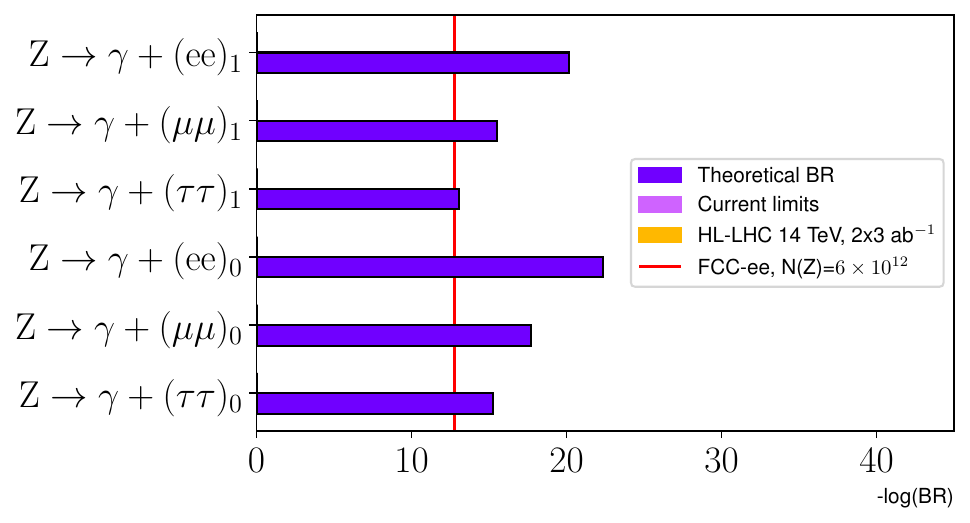}
\caption{Branching ratios (in negative log scale) of exclusive $\rm Z \to \gamma+leptonium$ decays. Theoretical predictions computed here are shown as blue bars. The red vertical line indicates the expected FCC-ee reach based on the total number of Z bosons to be produced.}
\label{fig:Z_gamma_leptonium_limits}
\end{figure}

The leading-order partial decay width of a Z boson into a leptonium-plus-a-photon can be derived from the ratio of the $\sigma(\epem\to \gamma+(\lele))$ cross section at the Z pole over the total resonant Z boson cross section in $\epem$ collisions, and amounts to
\begin{align}
\BR\left(\rm Z\to \gamma+(\lele)_0\right)&=\frac{\alpha(0)^4 \alpha(m_\mathrm{Z})^2}{9\cdot 256\,n^3} \frac{m_{\ell\ell}^2}{m_\mathrm{Z}^2}  \frac{\left(1-4 s_w^2\right)^2 \left(8 s_w^4-4 s_w^2+1\right) \left(m_\mathrm{Z}^2-m_{\ell\ell}^2\right)}{\Gamma_\mathrm{ee} \Gamma_\mathrm{Z} s_w^4 c_w^4}
\label{eq:Zgammall0}
\end{align}
for the paraleptonium case, and to
\begin{align}
\BR\left(\rm Z\to \gamma+(\lele)_1\right)&=\frac{\alpha(0)^4 \alpha(m_\mathrm{Z})^2}{9\cdot 256\,n^3} \frac{m_{\ell\ell}^2}{m_\mathrm{Z}^4} \frac{\left(8 s_w^4-4 s_w^2+1\right) \left(m_\mathrm{Z}^4-m_{\ell\ell}^4\right)}{\Gamma_\mathrm{ee} \Gamma_\mathrm{Z} s_w^4 c_w^4}
\label{eq:Zgammall1}
\end{align}
for the ortho-leptonium one, where $n$ is the principal quantum number of the leptonium resonance, and $s_w\equiv \sin\theta_\mathrm{W}$, and $c_w\equiv \cos\theta_\mathrm{W}$ are the sine and cosine of the Weinberg angle (Table~\ref{tab:params}). In both expressions above, $\alpha(0)$ and $\alpha(m_\mathrm{Z})$ are the electromagnetic coupling evaluated at zero\footnote{The leptonium wavefunction is proportional to $\alpha(0)^3$, Eq.~(\ref{eq:wf_origin_ll}), which combined with the emission of a final-state onshell photon, gives the $\alpha(0)^4$ dependence shown in Eqs.~(\ref{eq:Zgammall0}) and~(\ref{eq:Zgammall1}).} and at the Z pole mass, respectively, that, together with the rest of parameters, are listed in Table~\ref{tab:params}.
The obtained branching fractions are tabulated in Table~\ref{tab:Z_decays_gamma_leptonium}, and plotted in Fig.~\ref{fig:Z_gamma_leptonium_limits}. The dependence of these Z-boson radiative rates on $\alpha^6\,m_{\ell\ell}^2$, leads to vanishingly small branching fractions, $\mathcal{O}(10^{-13}$--$10^{-23})$, with ditauonium (positronium) featuring the largest (smallest) values. No experimental search has been conducted to date, although the relatively long lifetime of the leptonium objects (significantly boosted in the decay of the much heavier Z boson) would lead to a clean signature characterized by a displaced vertex from secondary decays of the $(\lele)$ into photons, $\rm e^\pm$, and/or $\mu^\pm$~\cite{dEnterria:2022alo}, akin to many BSM long-lived particles (but with an invariant mass at the corresponding known leptonium mass). Given the negligible rates, only an FCC-ee run at the Z pole would be able to provide limits approaching the SM values for the $\rm Z \to \gamma + (\tautau)_1$ case.

\begin{table}[htpb!]
\caption{Compilation of exclusive Z decays into a photon plus a ground state ($n=1)$ of ortho- or para-leptonium. For each decay, we provide the prediction of its branching fraction computed via Eq.~(\ref{eq:Zgammall0}) or (\ref{eq:Zgammall1}). The last column indicates whether the decay can be produced at FCC-ee.
\label{tab:Z_decays_gamma_leptonium}}
\input{tables/exclusive_Z_decays_gamma_meson_leptonium.tex}
\end{table}

\subsection{Exclusive Z boson decays into two mesons}
\label{sec:Z_2mesons}

\begin{figure}[htpb!]
\centering
\includegraphics[width=0.9\textwidth]{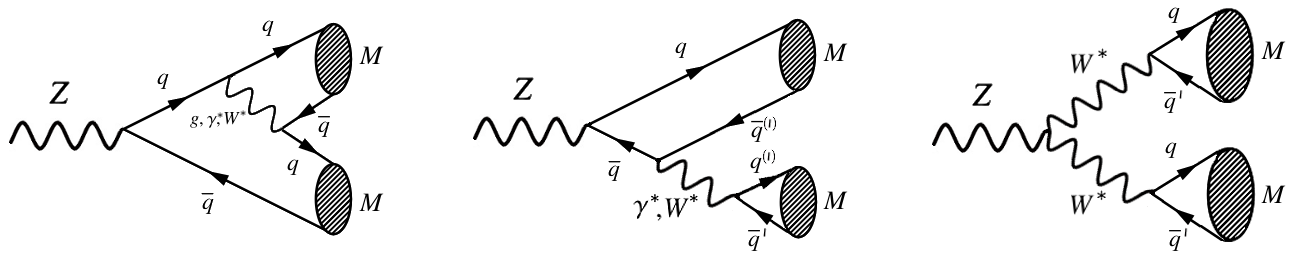}
\caption{Schematic diagrams of exclusive decays of the Z boson into two mesons. The solid fermion lines represent quarks, and the gray blobs the mesonic bound state.}
\label{fig:Z_2meson_diags}
\end{figure}

Figure~\ref{fig:Z_2meson_diags} displays the diagrams of exclusive Z boson decays into two mesons, with contributions from direct quark decays (left and center), as well as from indirect $\rm V^*\to M$ transitions (right). Due to the coincidence of diagrams with those of the Higgs boson (Fig.~\ref{fig:H_2meson_diags}), and since the Z boson yields at colliders are $10^{3}$--$10^{6}$ times larger (Table~\ref{tab:data_samples}), the study of such processes provides valuable information for the corresponding Higgs rates. The two-body decays into the same pair of pseudoscalar mesons, $\rm Z \to M+M$, are quantum-mechanically forbidden due to the presence of two identical final-state particles and conservation of angular momentum (thus violating the spin-statistics theorem), but the mesons M can be in a vector state. The decays of Z bosons into double quarkonia, first studied in 1990~\cite{Bergstrom:1990bu}, have been investigated as a means to provide clean information on the quarkonium bound-state dynamics at large momentum transfers. As explained below, there is still a relatively large uncertainty in the theoretical predictions, but this is still the most promising place to search for exclusive double-charmonia/bottomonia decays.

To our knowledge, there are no calculations of exclusive decays of the Z boson involving light vector mesons, but only into charmonium and/or bottomonium states. We have evaluated here for the first time the double light vector meson decay $\rm Z \to VM+VM$ via the two direct (left and center) diagrams of Fig.~\ref{fig:Z_2meson_diags}, using the EFT\,+\,NRQM formalism of~\cite{Gao:2022mwa} in which the width can be written as
\begin{align}
\Gamma_\mathrm{Z\to VM\,VM} = \Gamma^1_\mathrm{VM\,VM}(1+ 2 R+R^2 ),
\label{eq:Gamma_Z_MM}
\end{align}
where $\Gamma^1_\mathrm{VM\,VM}$ is the partial width corresponding to the left diagram of Fig.~\ref{fig:Z_2meson_diags}, amounting to
\begin{align}
\Gamma^1_\mathrm{VM\,VM}=\frac{2048 \pi^2 \alpha(0)\alphasmZ^2}{27 c_w^2 s_w^2 m_\mathrm{Z}^5} |\phi_\mathrm{VM}(0)|^4 \left(1-\frac{4m_\mathrm{VM}^2}{m_\mathrm{Z}^2}\right)^{5/2} ,
\label{eq:Gamma1_Z_MM}
\end{align}
and where $R$ is the ratio of the direct amplitudes $\mathcal{A}_1$ and $\mathcal{A}_2$ corresponding, respectively, to the left and center diagrams of Fig.~\ref{fig:Z_2meson_diags}, given by
\begin{align}
R = \frac{\mathcal{A}_2}{\mathcal{A}_1} = -\frac{9Q_\mathrm{q}^2\alpha(0)}{8\alphasmZ}\frac{m_\mathrm{Z}}{m_\mathrm{VM}}.
\label{eq:R_Z_MM}
\end{align}
For the light meson wavefunction at the origin $\phi_\mathrm{M}(0)$ in Eq.~(\ref{eq:Gamma1_Z_MM}), we derive it from its decay constant using the Van~Royen--Weisskopf expression, Eq.~\eqref{eq:VanRoyen}, as follows
\begin{align}
 |\phi_\mathrm{VM}(0)|^2 = \frac{m_\mathrm{VM} f^2_\mathrm{VM}}{ 4 N_\mathrm{c}},
\label{eq:f_phi}
\end{align}
for $N_\mathrm{c} =3$ colours, and without higher-order QCD corrections. We can only estimate the rate for the $\rm Z\to \phi+\phi$ channel because it is the only pure $|\qqbar\rangle$ state among the light mesons. Using Eqs.~(\ref{eq:Gamma_Z_MM})--(\ref{eq:f_phi}) and the parameters of Table~\ref{tab:params}, we obtain a $\BR = 2.1\cdot 10^{-12}$ rate (Table~\ref{tab:Z_2meson_light_decays}). There is no $\rm Z\to \phi+\phi$ experimental search performed to date, and only a machine like FCC-ee can provide enough Z bosons to start producing the decay. The table also quotes the forbidden $\rm Z\to \pi^0+\pi^0$ channel, for which the CDF upper bound could be improved by a factor of 100 at the HL-LHC. To provide a reasonable evaluation for exclusive double decays into other light mesons, rotation from flavour eigenstates $|\uubar\rangle,|\ddbar\rangle$ to physical eigenstates such as $\rho,\omega,\eta...$ would need to be performed. One can, however, anticipate branching fractions of the same order as the double-$\phi$ channel, $\mathcal{O}(10^{-12})$, since they only differ by a rotation and are enhanced by the second and third diagram of Fig.~\ref{fig:Z_2meson_diags} involving extra $\rm \gamma^*,W^*$ contributions. The scenario where Z decays into light pseudoscalar plus vector mesons suffers from the same complications of mixed states, but their rates are expected to be of the same order-of-magnitude as the double vector meson case because they have the same extra contribution from the photon propagator (Fig.~\ref{fig:Z_2meson_diags}, center)~\cite{Gao:2022mwa}.

\begin{table}[htpb!]
\caption{Compilation of exclusive Z decays to a pair of light mesons. For the forbidden two-pion decay, we provide the current experimental upper limit. For the double-$\phi$ decay, we provide the branching fraction predicted using Eqs.~(\ref{eq:Gamma_Z_MM})--(\ref{eq:f_phi}). The last column indicates whether the decay can be produced at FCC-ee.
\label{tab:Z_2meson_light_decays}}
\resizebox{\textwidth}{!}{%
\input{tables/exclusive_Z_decays_light_meson_meson}
}
\end{table}

There is a recent calculation of exclusive Z decays into a pair of charged pseudoscalar mesons, $\rm Z \to \pi^+\pi^-, K^+K^-$~\cite{Cheng:2018khi}. Also, the branching fractions $\rm Z\to B_c^{(*)+}+B_c^{(*)-}$ can be estimated to be in the $\mathcal{O}(10^{-8})-\mathcal{O}(10^{-11})$ range~\cite{Berezhnoy:2016etd,Liao:2024sgv,Liao:2022tqx}, where the variation of results within a factor of 20 depends on the different potential models~\cite{Liao:2024sgv} or initial quark masses~\cite{Liao:2022tqx} used.

\begin{figure}[htpb!]
\centering
\includegraphics[width=0.6\textwidth]{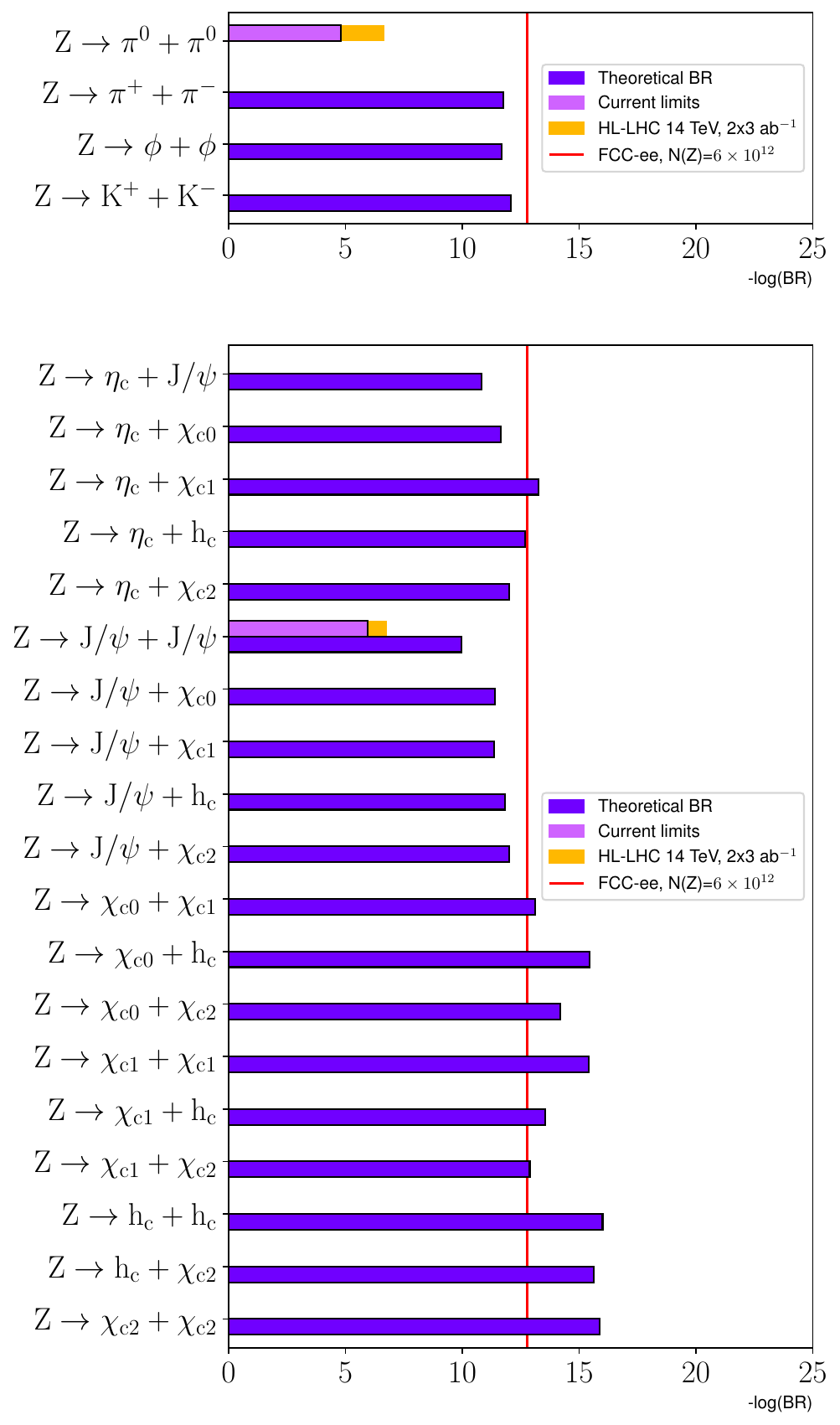}
\caption{Branching fractions (in negative log scale) of exclusive Z boson decays into two light mesons (upper) or two charmonium mesons (lower). The most recent theoretical predictions (blue bars) are compared to current experimental upper limits (violet) and expected conservative HL-LHC bounds (orange). The red vertical line indicates the expected FCC-ee reach based only on the total number of Z bosons to be produced.
\label{fig:Z_2meson_c_limits}}
\end{figure}

\begin{figure}[htpb!]
\centering
\includegraphics[width=0.6\textwidth]{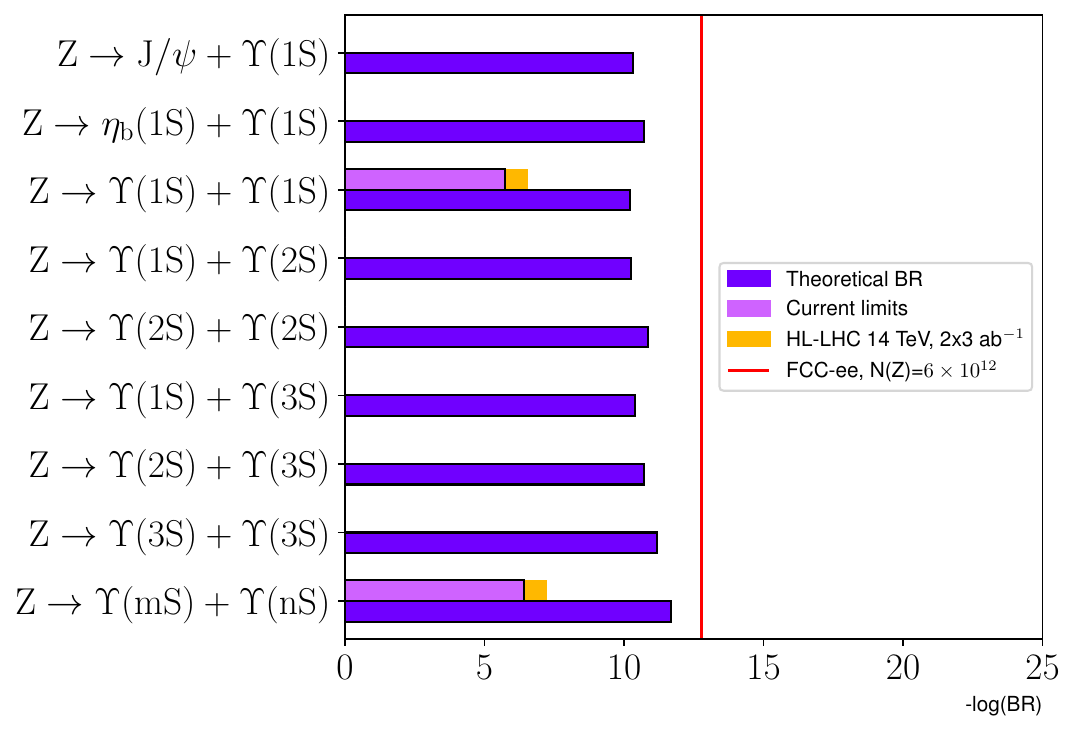}
\caption{Branching fractions (in negative log scale) of exclusive $\rm Z \to \Upsilon(nS)\,+\,$M decays. The most recent theoretical predictions (blue bars) are compared to current experimental upper limits (violet) and expected conservative HL-LHC bounds (orange). The red vertical line indicates the expected FCC-ee reach based only on the total number of Z bosons to be produced.
\label{fig:Z_2meson_b_limits}}
\end{figure}

Tables~\ref{tab:Z_2meson_c_decays} and~\ref{tab:Z_2meson_b_decays} list the theoretical predictions and experimental limits for concrete two-meson decay modes with pairs of charmonium mesons 
and bottomonium plus other mesons, respectively. The same results are shown in graphical form in Figs.~\ref{fig:Z_2meson_c_limits} and~\ref{fig:Z_2meson_b_limits}. 
The decay rates are very small, $\BR(\rm Z\to M+M)\lesssim 10^{-10}$, and quite sensitive to the model details and ingredients. All heavy-quarkonium channels have rates obtained within the NRQCD\,+\,LDMEs formalism. Alternative LC calculations show a good agreement with the latter~\cite{Likhoded:2017jmx} but, unfortunately, they can only provide predictions for final states with different mesons or with mesons in different excited states. The NRQCD predictions from the work~\cite{Likhoded:2017jmx} show discrepancies with other later works, because they did not consider diagrams involving a photon propagator (Fig.~\ref{fig:Z_2meson_diags}, center) that can enhance the rates by more than two orders-of-magnitude. In contrast, this latter study provides results for some excited states that have not been calculated elsewhere (although their results for channels such as $\rm Z\to \chi_{c1}+\chi_{c1},\,\chi_{c2}+\chi_{c2},\,\chi_{c0}+\chi_{c2}$,..., which cannot be predicted using LC formalism, may need to be revisited for the same reasons stated above). In addition, recent calculations including higher-order QED~\cite{Luo:2022ugd} and QCD\,+\,QED~\cite{Li:2023tzx} corrections yield significant enhancements in the branching ratios for double-quarkonia production. All that said, and within the relatively large theoretical uncertainties for those processes, the largest branching fractions expected are for double-charmonium with $\mathcal{O}(10^{-10})$ rates. Despite a few searches carried out at LEP (and one at the LHC), experimental observation will only be possible at FCC-ee for about half of them. 





\begin{table}[htpb!]
\caption{Compilation of exclusive Z decays to two charmonium mesons. For each decay, we provide the predicted branching fraction(s) and the theoretical approach used to compute it, as well as the current experimental upper limit and that estimated for HL-LHC. The last column indicates whether the decay can be produced at FCC-ee.
\label{tab:Z_2meson_c_decays}}
\resizebox{\textwidth}{!}{%
\input{tables/exclusive_Z_decays_charmonium_meson.tex}
}
\end{table}

\begin{table}[htpb!]
\caption{Compilation of exclusive Z decays to one bottomonium meson plus another meson. 
For each decay, we provide the predicted branching fraction(s) and the theoretical approach used to compute it, as well as the current experimental upper limit and that estimated for HL-LHC. The last column indicates whether the decay can be produced at FCC-ee.
\label{tab:Z_2meson_b_decays}}
\resizebox{\textwidth}{!}{%
\input{tables/exclusive_Z_decays_bottomonium_meson.tex}
}
\end{table}

\clearpage
\section{Rare W boson decays}
\label{sec:Wboson}

\subsection{Exclusive W decays into a photon plus a meson}

\begin{figure}[htpb!]
\centering
\includegraphics[width=0.6\textwidth]{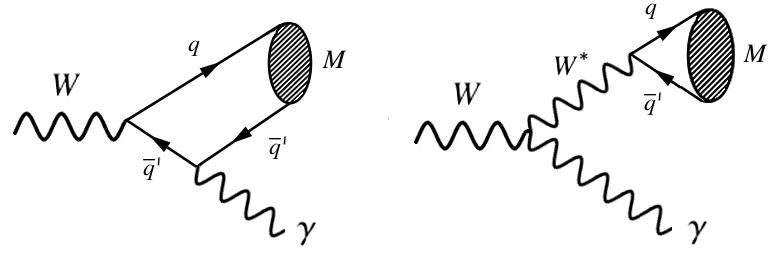}
\caption{Schematic diagrams of exclusive decays of the W bosons into a photon plus a meson. The solid fermion lines represent quarks, and the gray blobs represent the mesonic bound state.} 
\label{fig:W_gamma_meson_diags}
\end{figure}

\begin{figure}[htpb!]
\centering
\includegraphics[width=0.6\textwidth]{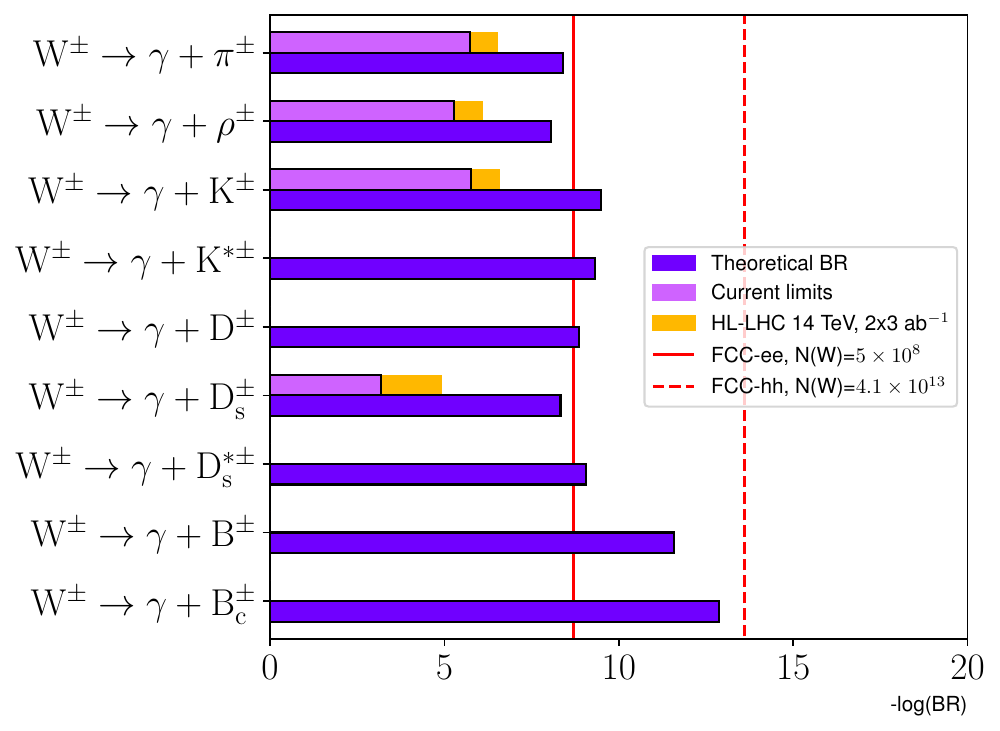}
\caption{Branching ratios (in negative log scale) of exclusive $\rm W^{\pm} \to \gamma+meson$ decays. Most recent theoretical predictions (blue bars) are compared to current experimental upper limits (violet) and expected conservative HL-LHC bounds (orange). The red vertical lines indicate the FCC-ee (solid) and FCC-hh (dashed) reach based only on the total number of W bosons to be produced at~both~machines.}
\label{fig:W_gamma_meson_limits}
\end{figure}

Figure~\ref{fig:W_gamma_meson_diags} displays the two generic diagrams contributing to exclusive decays of a W boson into a photon (the only gauge boson kinematically accessible for the least massive weak boson) plus a charged meson. 
The partial widths for these decays have been evaluated within the SCET framework in Ref.~\cite{Grossman:2015cak}, and can be written as
\begin{align}
    \Gamma(\rm W^\pm\to M^\pm\gamma)&=\frac{\alpha(0) m_\mathrm{W}f_\mathrm{M}^2}{48v^2}|V_{\mathrm{qq'}}|^2\left(|F_1^\mathrm{M}|^2+|\pm F_2^\mathrm{M}|^2\right),
    \label{eq:W_Mgamma_decays}
\end{align}
where, at leading logarithm, the real parts of the $F_1^\mathrm{M},\ F_2^\mathrm{M}$ objects can be written as 
\begin{align}
\text{Re } F_1^\mathrm{M}&=0.94-1.65\ a_1^\mathrm{M}(1 \text{ GeV})+0.42\ a_2^\mathrm{M}(1 \text{ GeV}) +\dots\nonumber \\
\text{Re } F_2^\mathrm{M}&=0.85-0.55\ a_1^\mathrm{M}(1 \text{ GeV})+1.25\ a_2^\mathrm{M}(1 \text{ GeV}) +\dots,
\label{eq:Gegenbauer_moms}
\end{align}
with $a_i^\mathrm{M}(\mu)$ being the $i$th Gegenbauer moment of the meson $\rm M$, evaluated at the renormalization scale $\mu$. Table~\ref{tab:W_decays_gamma_meson} compiles the corresponding theoretical predictions and experimental limits for branching ratios of the eight radiative exclusive decay modes considered in Ref.~\cite{Grossman:2015cak}, to which we have added the result for the  $\rm B_c^\pm$ meson. The Gegenbauer moments for this latter meson, with one-loop QCD radiative corrections and relativistic corrections~\cite{Wang:2017bgv}, are: $a_1^\mathrm{B_c}(\mu_{\rm B_c})=0.32,\ a_2^\mathrm{B_c}(\mu_{\rm B_c})=-0.34$ at scales $\mu_{\rm B_c} = 5.6\text{ GeV}$. These values can be evolved to lower scales using the LO renormalization group equation for $N_\mathrm{f}=5$~\cite{Grossman:2015cak}, yielding $a_1^{\rm B_c}(1 \text{ GeV})=1.35\ a_2^{\rm B_c}(\mu_{\rm B_c}),\ a_2^{\rm B_c}(1 \text{ GeV})=1.6\ a_2^{\rm B_c}(\mu_{\rm B_c})$, for $\alpha_\mathrm{s}(\mu_{\rm B_c})=0.21,\ \alpha_\mathrm{s}( 1 \text{ GeV})=0.4$, which we use in Eqs.~(\ref{eq:W_Mgamma_decays})--(\ref{eq:Gegenbauer_moms}) to obtain the result quoted in the last row of Table~\ref{tab:W_decays_gamma_meson}.
The theoretical branching fractions are in the $\mathcal{O}(10^{-8}$--$10^{-13})$ range, whereas current experimental upper bounds are three to five orders-of-magnitude larger, $\mathcal{O}(10^{-4}$--$10^{-6})$. Four channels have been searched for to date at the LHC (3 by CMS, 1 by ATLAS, and 1 by LHCb), and the HL-LHC will improve the existing limits by at least one order-of-magnitude, but still far from any possible observation. The FCC-ee will be able to produce events for three decay modes, and the FCC-hh will produce all of them. The results listed in Table~\ref{tab:W_decays_gamma_meson} are presented in graphical form in Fig.~\ref{fig:W_gamma_meson_limits}.

\begin{table}[htpb!]
\caption{Compilation of exclusive W boson decays to a photon plus a meson. For each decay, we provide the predicted branching fraction(s) and the theoretical approach used to compute it, as well as the current experimental upper limit and that estimated for HL-LHC. The last two columns indicate whether the decay can be produced at FCC-ee/FCC-hh.
\label{tab:W_decays_gamma_meson}}
\resizebox{\textwidth}{!}{%
\input{tables/exclusive_W_decays_gamma_meson.tex}
}
\end{table}

\subsection{Exclusive W decays into two mesons}

\begin{figure}[htpb!]
\centering
\includegraphics[width=0.9\textwidth]{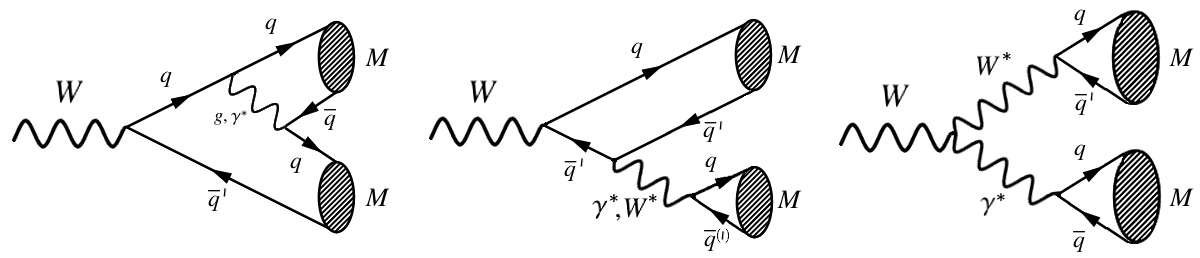}
\caption{Schematic diagrams of exclusive decays of the W boson into two mesons. The solid fermion lines represent quarks, and the gray blobs the mesonic bound state.}
\label{fig:W_2meson_diags}
\end{figure}

Figure~\ref{fig:W_2meson_diags} shows the schematic diagrams of the exclusive decay of the W boson into two mesons proceeding through quark decays (left and center), or through intermediate $\rm W^*,\,\gamma^* \to M$ transitions (right). Table~\ref{tab:W_2meson_decays} lists the corresponding theoretical predictions and experimental limits for concrete decay modes, all of them involving charm and/or bottom quarks mesons.  We are not aware of any study of exclusive W-boson decays involving one or two light mesons in the literature\footnote{The exclusive three-pion decay, $\rm W^\pm \to \pi^+\pi^-\pi^\pm$, estimated to have a $\BR \approx 10^{-7}$~\cite{Melia:2016knk}, has also been searched-for by the CMS experiment at the LHC~\cite{CMS:2019vaj}.}. The expected rates are truly tiny, of order $\mathcal{O}(10^{-11}\text{--}10^{-14})$, with relatively large uncertainties. The work of~\cite{Luchinsky:2021amp} compared the branching fractions predicted within two approaches (LC and NRCSM), which differ by factors of 5--10, whereas Ref.~\cite{Bagdatova:2023etj} computed those within the NRQCD framework. For decays with a small ratio $m_\mathrm{M}/m_\mathrm{W}$, the results from the LC approach should be preferable. Differences among predictions for the same process appear in some cases as due to not consistently considering all diagrams of Fig.~\ref{fig:W_2meson_diags}. The effect of the center diagram with a virtual photon in Fig.~\ref{fig:W_2meson_diags} was considered in~\cite{Bagdatova:2023etj}, and its contribution is equal or larger than the direct (left) diagram in Fig.~\ref{fig:W_2meson_diags}. None of the two theoretical works that have so far studied these processes, has considered the middle diagram of Fig.~\ref{fig:W_2meson_diags} with the $\rm W^*$ boson exchange nor the rightmost diagram of Fig.~\ref{fig:W_2meson_diags}. 
Using the $\delta$-approximation expression for the decay widths given in Ref.~\cite{Luchinsky:2021amp}, we can see that the $\rm W \to M1 + M2$ decay widths are proportional to $|V_{\rm q_1q_2}|^2\times f^2_{\rm M_1}\times f^2_{\rm M_2}$, where $f_{\rm M_1}$ and $f_{\rm M_2}$ are the decay constants of the two mesons. Using such a dependence, we can estimate the order-of-magnitude of other double-meson decay rates not computed to date. The most probable group of decays is the CKM-favoured $\rm W\to c+s\to \psi(\textit{n}S)+D^{(*)}_s,B^{(*)}_c+B^{(*)}_s$, already listed in Table~\ref{tab:W_2meson_decays}. The next most probable group of decays involving quarkonia is $\rm W\to c+d\to (c q)+(q d)$, that leads to the $\rm \BR(W^\pm\to \jpsi+D^\pm)\approx \mathcal{O}(10^{-13})$ estimate. Following that, we have the decays $\rm W\to b+c\to (b q)+(q c)$, that result in $\rm \BR(W^\pm\to \Upsilon(1S)+B_c^\pm)\approx \mathcal{O}(10^{-14})$. Finally, the least probable decay group involving bottom mesons is the CKM-suppressed $\rm W\to b+u\to (b q)+(q u)$, yielding $\rm \BR(W\to \Upsilon(1S)+B)\approx \mathcal{O}(10^{-16})$. 

\begin{figure}[htpb!]
\centering
\includegraphics[width=0.6\textwidth]{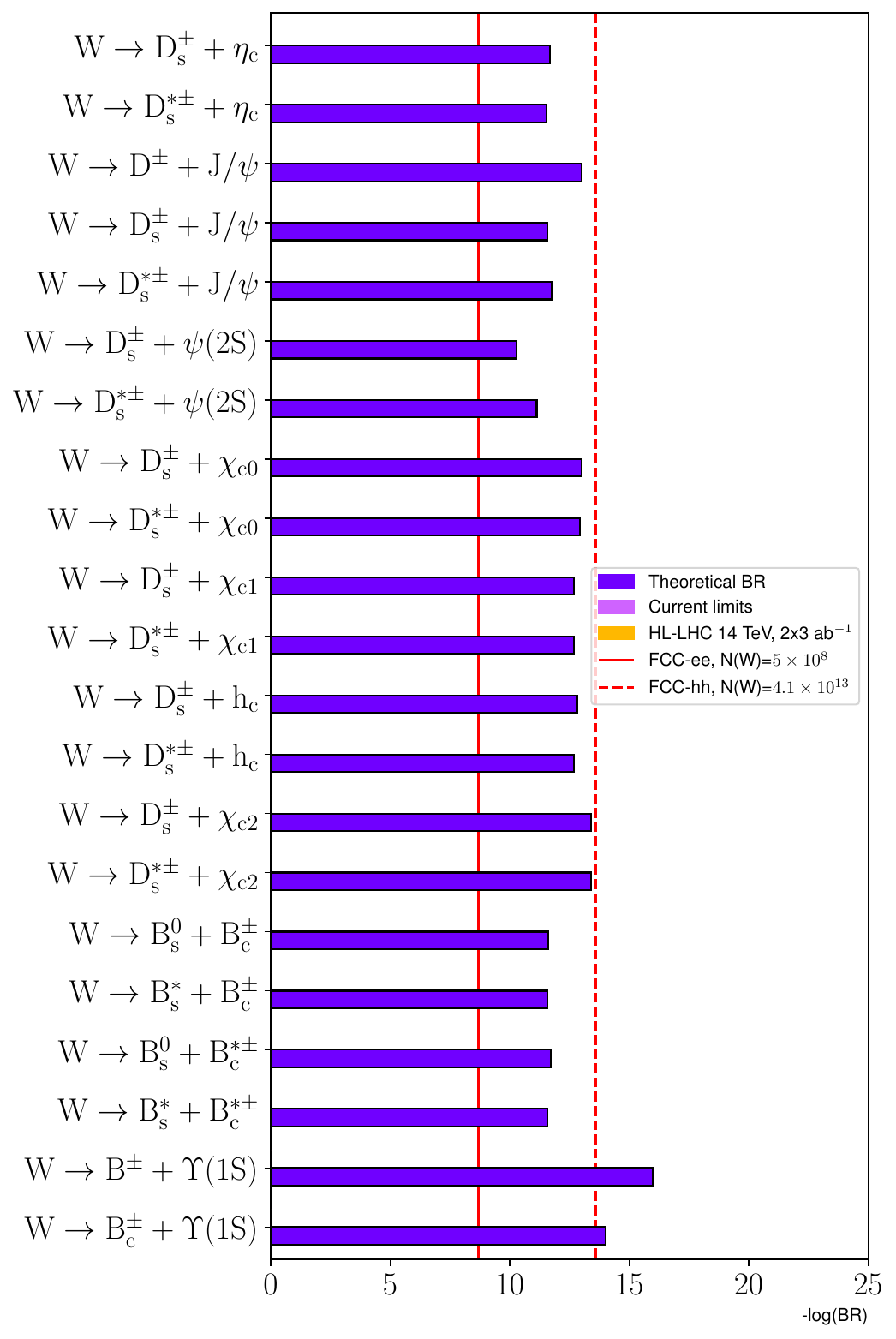}
\caption{Branching fractions (in negative log scale) of exclusive $\rm W^{\pm} \to meson+meson$ decays: Most recent theoretical predictions are shown as blue bars. The red vertical lines indicate the FCC-ee (solid) and FCC-hh (dashed) reach based only on the total number of W bosons expected at both facilities.
}
\label{fig:W_2meson_limits}
\end{figure}

Experimentally, none of the listed decays has been searched for at LEP, Tevatron, or the LHC to date. Only FCC-hh has possibilities to produce the 18 decay channels (Fig.~\ref{fig:W_2meson_limits}), but the expected tiny rates seemingly preclude their observation in the complex hadron-collider environment. 



\begin{table}[htpb!]
\caption{Compilation of exclusive W boson decays into two mesons. For each decay, we provide the predicted branching fraction(s) and the theoretical approach used to compute it. The last two columns indicate whether the decay can be produced at FCC-ee/FCC-hh.
\label{tab:W_2meson_decays}}
\resizebox{\textwidth}{!}{%
\input{tables/exclusive_W_decays_meson_meson.tex}
}
\end{table}

\clearpage
\section{Rare top quark decays}
\label{sec:top_quark}

\subsection{Two-body top quark decays}

\begin{figure}[htpb!]
\centering
\includegraphics[width=0.95\textwidth]{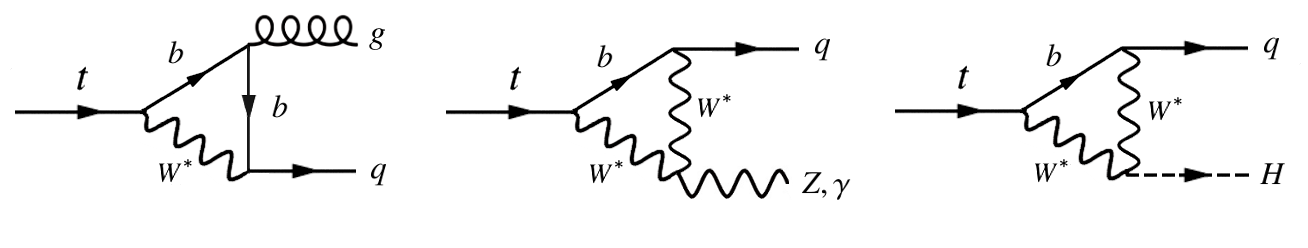}
\caption{Schematic diagrams of rare FCNC two-body decays of the top quark into a gauge (g, Z, or $\gamma$) boson or a Higgs boson plus an up-type quark ($\rm q = c,u$).
\label{fig:top_2body_diags}}
\end{figure}

The top quark decays to a W boson and a bottom quark, $\rm t \to W\,b$, with a branching fraction of nearly 100\%, with the other tree-level decays $\rm t \to W\,s$ and $\rm t \to W\,d$ comparatively suppressed by factors of $10^{-3}$ and $10^{-4}$, respectively, as per the CKM element hierarchy $V_\mathrm{tb}\gg V_\mathrm{ts} > V_\mathrm{td}$ (Table~\ref{tab:params}). The FCNC top-quark decays to a gauge boson plus a light up-type quark (q = u or c;  for charge conservation), $\rm t \to q\,Z$, $\rm t \to \gamma\,c$, and $\rm t \to g\,c$, occur only at the level of quantum loop corrections (Fig.~\ref{fig:top_2body_diags}), and are extremely suppressed in the SM. More precisely, they are 1-loop-, CKM-, and GIM-suppressed (with a GIM suppression factor\footnote{\label{note1}Theoretically, the b-quark running mass must be evaluated at the top pole mass scale (at which the t quark decays) in the $\rm \overline{MS}$ scheme, otherwise rates are artificially enhanced.} given by $f(m_\mathrm{b}^2/m_\mathrm{W}^2)\approx 10^{-9}$), leading to tiny amplitudes. 
In many BSM models, however, the GIM suppression can be relaxed, and one-loop diagrams mediated by new bosons may also contribute, yielding effective couplings orders of magnitude larger, and correspondingly enhancing such ultrarare branching fractions~\cite{Beneke:2000hk,Aguilar-Saavedra:2004mfd}.

\begin{table}[htpb!]
\caption{Compilation of rare two-body top quark decays. For each decay, we provide the predicted branching fraction(s) and the theoretical approach used to compute it, as well as the current experimental upper limit and that estimated for HL-LHC. The last two columns indicate whether the decay can be produced at FCC-ee/FCC-hh.
\label{tab:top_2body_decays}}
\resizebox{\textwidth}{!}{%
\input{tables/exclusive_t_decays_q_boson.tex}
}
\end{table}

\begin{figure}[htpb!]
\centering
\includegraphics[width=0.6\textwidth]{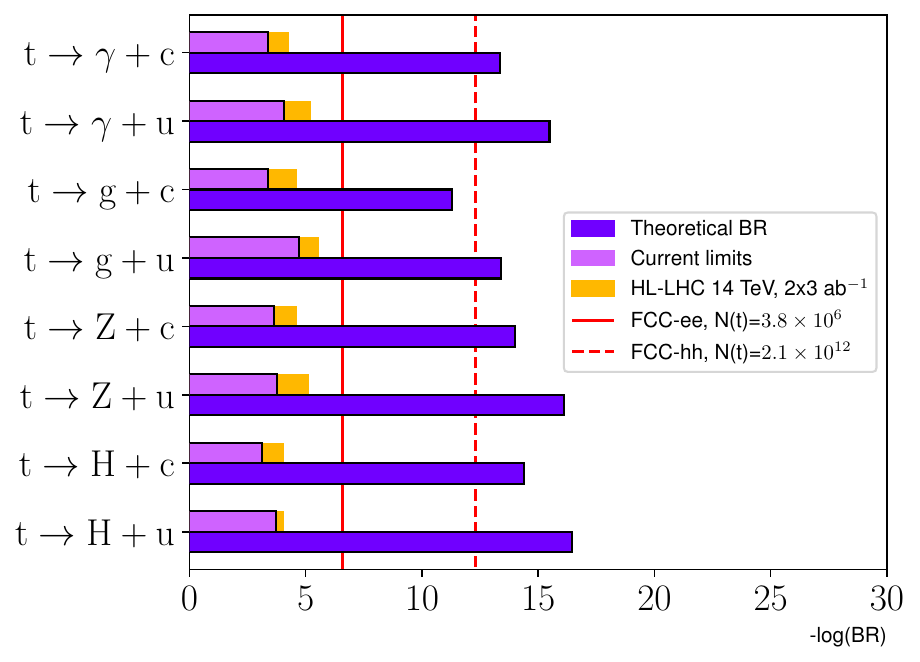}
\caption{Branching fractions (in negative log scale) of rare two-body $\rm t \to V+q,H+q$ decays (with V = gauge boson, and $\rm q = u,c$). The most recent theoretical predictions (blue bars) are compared to current experimental upper limits (violet) and expected conservative HL-LHC bounds (orange). The red vertical lines indicate the FCC-ee (solid) and FCC-hh (dashed) reach based only on the total number of top quarks expected at both facilities.
}
\label{fig:t_V_q_limits}
\end{figure}

Table~\ref{tab:top_2body_decays} lists the branching fractions of FCNC top decays, and Fig.~\ref{fig:t_V_q_limits} displays them graphically. The predicted SM rates are tiny, $\mathcal{O}(10^{-12}$--$10^{-17})$,
and very sensitive to the value of the b-quark mass used to calculate them$^{\ref{note1}}$. The decay branching fractions $\rm t \to Z\,q$, $\rm t \to \gamma \,q$, $\rm t \to g\,q$, $\rm t \to H\,q$, are further suppressed for $\rm q = u$ compared to the $\rm q = c$ case, by a CKM factor $(|V_\mathrm{ub}|/|V_\mathrm{cb}|)^2 \approx 0.008$. 
Experimentally, FCNC decays of the top quark have attracted lots of interest in the quest for BSM phenomena at the LHC, and all eight channels have been searched for. The current experimental upper limits are set in the $\mathcal{O}(10^{-4}$--$10^{-5})$ range, and will be improved by about a factor of ten at the end of the HL-LHC. In the absence of BSM effects enhancing such FCNC top decays, only the FCC-hh will reach top production rates capable of probing the $\rm t \to g\,c$ decay.

\subsection{Three-body top quark decays}

\begin{figure}[htpb!]
\centering
\includegraphics[width=0.9\textwidth]{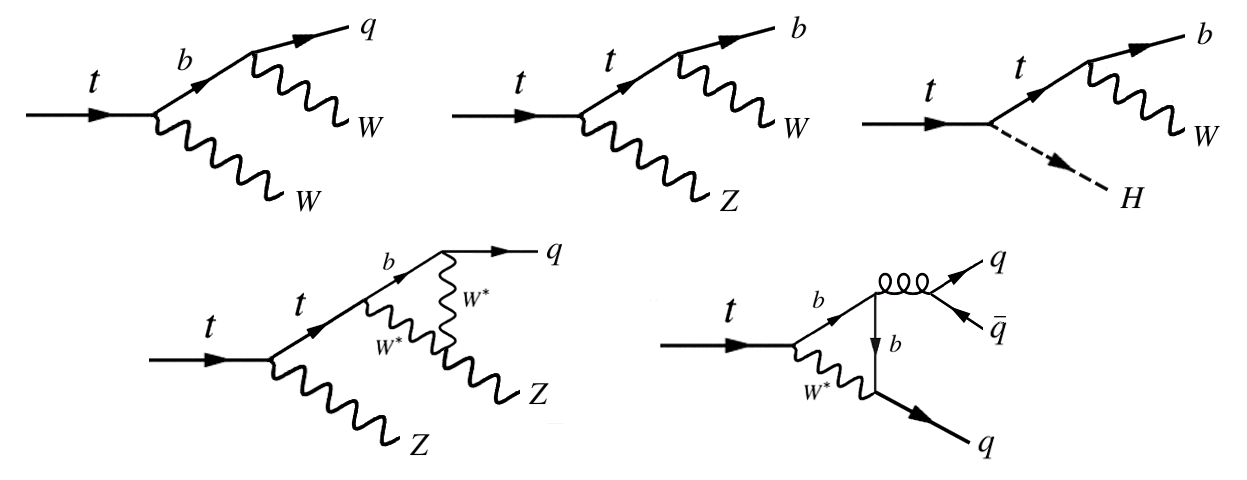}
\caption{Representative diagrams of rare three-body decays of the top quark into a pair of bosons plus a quark at tree-level (upper), and via loops into ZZq (lower left) and into three quarks (lower right). The outgoing quarks are either b, or up-type quarks ($\rm q = c,u$), by charge conservation.
\label{fig:top_3body_diags}}
\end{figure}

Thanks to its large mass, the top quark has different rare decays kinematically accessible involving the presence of multiple heavy bosons in the final state~\cite{Jenkins:1996zd,Beneke:2000hk}. The possibility of 3-body radiative decays of the top quark $\rm t \to W\,b\, X$, where X can be a Z or a Higgs boson has been considered, \eg\ in Refs.~\cite{Decker:1992wz,Mahlon:1994us,Altarelli:2000nt} (the cases $\rm X = \gamma, g$ are simply NLO real QED or QCD corrections to the dominant $\rm t W\,b$ decay, for which one needs also an energy threshold to avoid infrared/collinear divergences in the decay rates, and not considered here as they are not rare). The corresponding diagrams are shown in Fig.~\ref{fig:top_3body_diags}. Properly taking into account the finite width effects is key to compute any close-to-threshold 3-body decays with (partially) offshell particles~\cite{Altarelli:2000nt,Eilam:2006uh}. 

\begin{figure}[htpb!]
\centering
\includegraphics[width=0.6\textwidth]{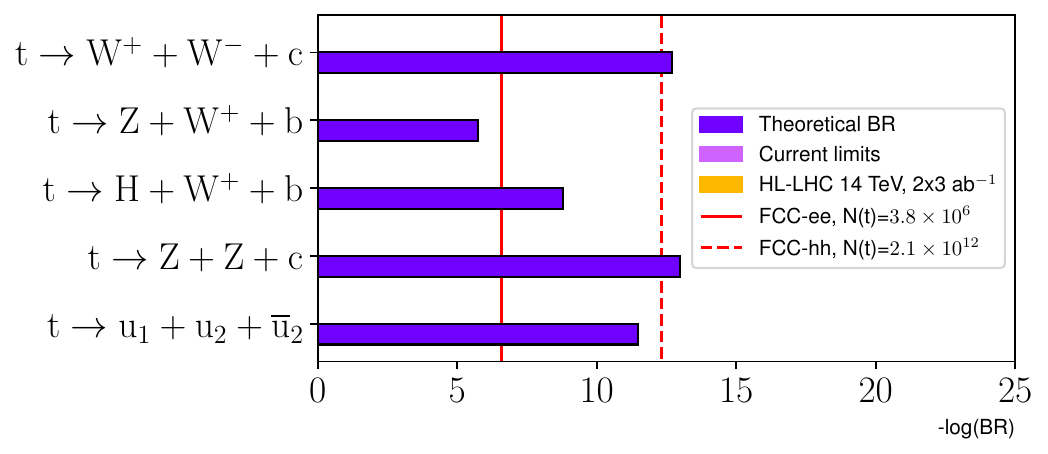}
\caption{Branching fractions (in negative log scale) of rare three-body top quark decays: Most recent theoretical predictions are shown as blue bars. 
The red vertical lines indicate the FCC-ee (solid) and FCC-hh (dashed) reach based only on the total number of top quarks expected at both facilities. }
\label{fig:top_3body_limits}
\end{figure}

Table~\ref{tab:top_3body_decays} collects the branching fractions for rare three-body decays of the top quark, and Fig.~\ref{fig:top_3body_limits} presents them in graphical form. No experimental search has been performed to date. By an amusing coincidence, the $\rm t \to Z\,W\,b$ decay is kinematically allowed because $m_\mathrm{t}\approx (m_\mathrm{W}+m_\mathrm{b})+m_\mathrm{Z}$ to within a few GeV, which can be satisfied within the $\mathcal{O}(2$~GeV) widths of the onshell electroweak bosons (Table~\ref{tab:params}). The corresponding branching fraction is $\BR(\rm t \to Z\,W\,b)\approx 2\cdot10^{-6}$ and, 
although no experimental limits exist yet from the LHC data, it could be potentially discovered at the HL-LHC, and for sure observed at FCC-ee and FCC-hh. 
The branching fraction of $\rm t \to H\, W \, b$, first evaluated in~\cite{Han:2013sea}, has been recomputed here at NLO using \mgshort\ (via $\rm t \to H\, W^*(\mu\nu)b$, with $m_\mathrm{b}(m_\mathrm{t}) = 2.6$~GeV) with the parameters listed in Table~\ref{tab:params}, and found to be $\BR = 1.6 \cdot10^{-9}$. Other decays are highly GIM-suppressed (Fig.~\ref{fig:top_3body_diags}, bottom) and have much smaller rates. The $\rm t \to ZZq$ is below $\mathcal{O}(10^{-13})$, and is likely impossible to observe anywhere unless some BSM mechanism enhances it. 
The top quark can also decay into three up-type quarks, either through the same diagrams shown in Fig.~\ref{fig:top_2body_diags} where the emitted boson further decays/splits into $\uubar$ or $\ccbar$, or through a virtual W boson exchange. This process is dominated by the gluon splitting into two up-type quarks shown in the Fig.~\ref{fig:top_3body_diags} (lower right panel), and has a decay rate of $\mathcal{O}(10^{-12})$~\cite{Cordero-Cid:2004pco} commensurate with that of the ``parent'' $\rm t \to u\,g$ two-body decay (Table~\ref{tab:top_2body_decays}).

\begin{table}[htpb!]
\caption{Compilation of rare three-body top quark decays ($\rm u_{1,2}= u,c$ quarks). For each decay, we provide the predicted branching fraction(s) and the theoretical approach used to compute it. The last two columns indicate whether the decay can be produced at FCC-ee/FCC-hh.
\label{tab:top_3body_decays}}
\resizebox{\textwidth}{!}{%
\input{tables/rate_t_decays_3_body.tex}
}
\end{table}

\subsection{Semiexclusive top quark decays into a quark plus a meson}
\label{sec:top_semiexcl}

\begin{figure}[htpb!]
\centering
\includegraphics[width=0.65\textwidth]{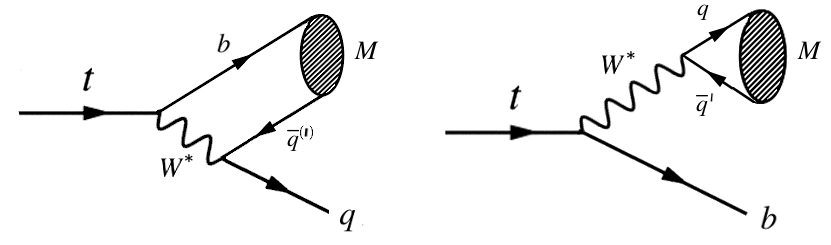}
\caption{Schematic diagrams of semiexclusive two-body decays of the top quark into a meson plus a $\rm q = c,u$ quark (left) or a b quark (right).\label{fig:top_meson_quark_diags}}
\end{figure}

There are also theoretical studies of semiexclusive top quark decays in which interactions among the decay quarks (either from the W decay, or combining the primary bottom quark with the W-decay quarks) lead to the formation of final states with one meson recoiling against a jet~\cite{dEnterria:2020ygk,Handoko:1998tz,Handoko:1999iu,Slabospitskii:2021rns}. Such decays can provide a new method to measure the top-quark mass via a two-body (jet-meson) invariant mass analysis, with different systematic uncertainties of those of the currently existing approaches~\cite{dEnterria:2020ygk}. Typical diagrams for the process are shown in Fig.~\ref{fig:top_meson_quark_diags} with the single meson recoiling against a $\rm q=u,c$ quark (right), or against a b quark (left).

\begin{figure}[htpb!]
\centering
\includegraphics[width=0.6\textwidth]{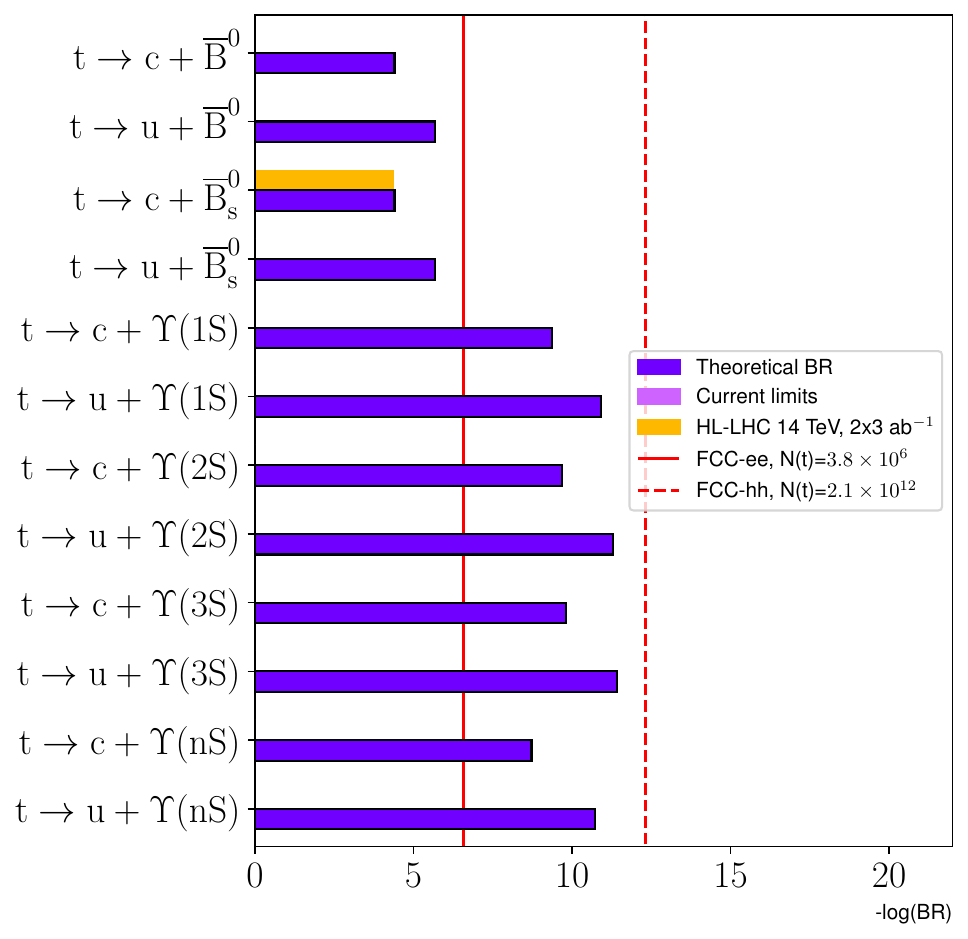}
\caption{Branching fractions (in negative log scale) of semiexclusive $\rm t \to c/u\ quark + meson$ decays. Most recent theoretical predictions are shown as blue bars. The red vertical lines indicate the FCC-ee (solid) and FCC-hh (dashed) reach based only on the total number of top quarks expected at both facilities.}
\label{fig:t_meson_cu_quark_limits}
\end{figure}

Branching fractions are listed in Table~\ref{tab:t_uc_meson_decays} for a bottom meson plus a c or u quark. The results of this table are shown in graphical form in Fig.~\ref{fig:t_meson_cu_quark_limits}. Theoretical branching fractions are in the $\mathcal{O}(10^{-5}$--$10^{-12})$ range. No experimental study has been performed to date. The combined $\rm t\to B_\mathrm{(s)}+u/c$ decays have a relatively large rate of $\BR\approx 10^{-4}$, that makes them worth an experimental search at the LHC. Dedicated studies exist that prove the feasibility to observe them via $\rm t \to b$-jet$\,+\,$c-jet at the HL-LHC~\cite{dEnterria:2020ygk}. Four channels are producible at FCC-ee, and all of them can be produced at FCC-hh. 

\begin{table}[htpb!]
\caption{
Compilation of semiexclusive decays of the top quark into a c or u quark plus a meson. For each decay, we provide the predicted branching fraction(s) and the theoretical approach used to compute it, as well as the current experimental upper limit and that estimated for HL-LHC. The last two columns indicate whether the decay can be produced at FCC-ee/FCC-hh.
\label{tab:t_uc_meson_decays}}
\resizebox{\textwidth}{!}{%
\input{tables/exclusive_t_decays_u_meson.tex}
}
\end{table}

\begin{figure}[htpb!]
\centering
\includegraphics[width=0.6\textwidth]{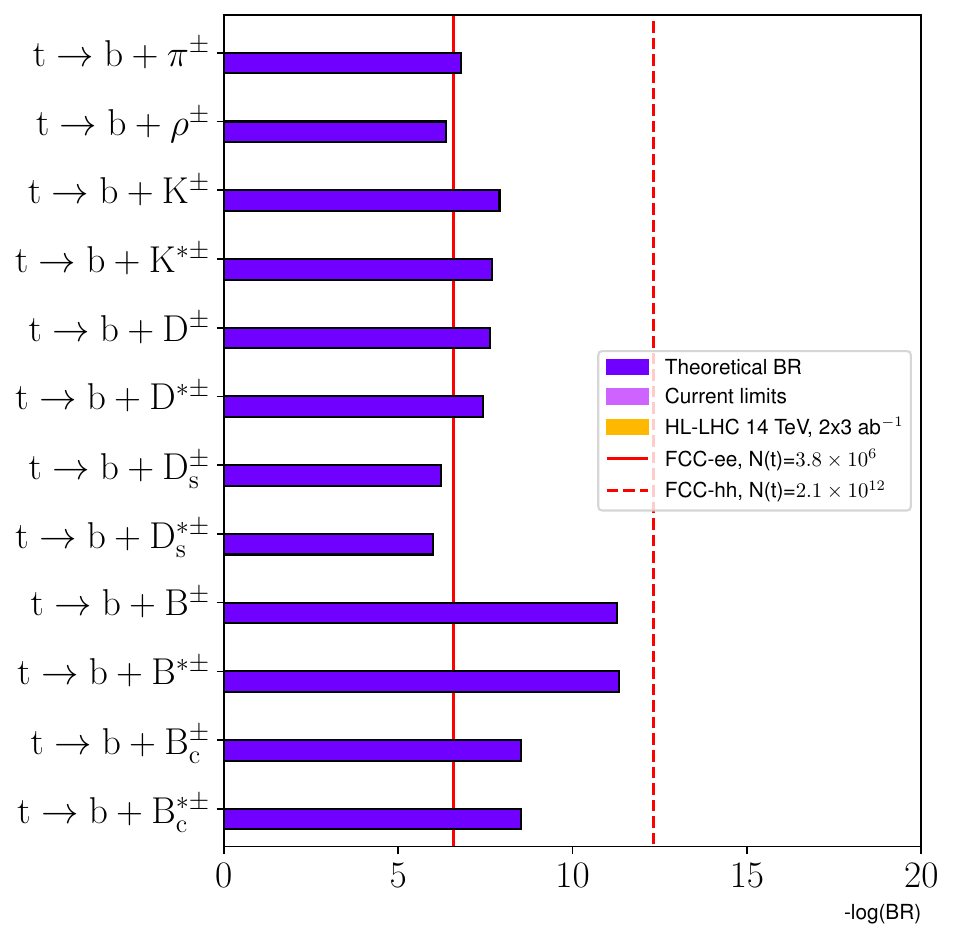}
\caption{
Branching fractions (in negative log scale) of semiexclusive $\rm t \to b\ quark + meson$ decays. The theoretical rates obtained with Eq.~(\ref{eq:t_M_b_decay}) are shown as blue bars. The red vertical lines indicate the FCC-ee (solid) and FCC-hh (dashed) reach based only on the total number of top quarks expected at both facilities.}
\label{fig:t_meson_b_quark_limits}
\end{figure}

The right diagram of Fig.~\ref{fig:top_meson_quark_diags} corresponds to the process of a top quark decaying through an offshell W boson with mass $m_\mathrm{W^*}\approx m_\mathrm{M}$ close to a mesonic resonance M that recoils against a b-quark jet.
The amplitude for the process $\rm t\to b+M^+$ is similar to that for the well-known $\rm \tau^\pm\to \pi^\pm+\nu_\tau$ and $\rm \tau^\pm\to \rho^\pm+\nu_\tau$ decays~\cite{Grossman:2015cak}
\begin{equation}
   \Gamma(\tau^\pm\to {\rm M}^\pm\nu_\tau) = S_{\rm EW}\,\frac{G_F^2 m_\tau^3}{16\pi}\,|V_{ij}|^2 f_{\rm M}^2
    \left( 1 - \frac{m_{\rm M}^2}{m_\tau^2} \right)^2 \!
    \left( 1 + b_{\rm M}\,\frac{m_{\rm M}^2}{m_\tau^2} \right) ,
\end{equation}
where $b_\mathrm{M} = 0$ for pseudoscalar mesons ($\rm M^\pm = \pi^\pm$) and $b_\mathrm{M} = 2$ for vector mesons  ($\rm M^\pm = \rho^\pm$), and  the factor $S_\text{EW}=1.0154$ includes the leading-logarithmic~\cite{Sirlin:1977sv,Marciano:1988vm} and non-logarithmic electroweak corrections~\cite{Braaten:1990ef}. For the particular $\rm t\to b+\pi^+/\rho^+$ decays, the relevant amplitudes read
\begin{eqnarray}
    i M_{\rm P}(\rm t\to b+\pi^+)&=& \frac{\pi\ \alpha(m_{\rm t})}{2 \sin^2\theta_\mathrm{W}}\frac{V_{\rm u d}f_{\rm M}}{m_{\rm W}^2}\left[\bar{u}(p_2,r)\gamma_\mu(1-\gamma^5)u(p_1,s)\right] p_3^\mu,\\
    i M_{\rm VM}(\rm t\to b+\rho^+)&=& \frac{\pi\ \alpha(m_{\rm t})}{2 \sin^2\theta_\mathrm{W}}\frac{V_{\rm u d}f_{\rm M}m_\rho}{m_{\rm W}^2}\left[\bar{u}(p_2,r)\gamma_\mu(1-\gamma^5)u(p_1,s)\right] \epsilon^{*\mu}(p_3,\lambda).
\end{eqnarray}
Here, $p_1, p_2, p_3$ are the momentum of t quark, b quark and meson, respectively. The indices $r,s$ denote the spin states of the corresponding particles, the index $\lambda$ denotes polarization state of the vector meson, $u(\overline{u})(p,r)$ is the Dirac spinor associated with the creation (annihilation) operator of a fermion with momentum $p$ and spin $r$. For the generic case of a semiexclusive decay into a meson $\rm M^\pm$ plus a b (or $\overline{\rm b}$ quark), taking the meson and b quark as massless compared to the top quark, the decay rate of both pseudoscalar and vector mesons are the same, and read
\begin{eqnarray}
\Gamma(\rm t\to b+M^\pm) 
&=&\frac{1}{8\pi m_{\rm t} } \sum_{\lambda}\frac{1}{2}\sum_{r,s}|M|^2 
=\frac {\pi\ \alpha(m_t)^2 m_{\rm t}^3 |V_{\rm q q'}|^2f_{\rm M}^2} {16 m_{\rm W}^4 \sin \theta_w^4} 
= 1.34 \cdot 10^{-5} \,\mbox{[GeV$^{-1}]$}\ |V_{\rm q q'}|^2f_{\rm M}^2,
\label{eq:t_M_b_decay}
\end{eqnarray}
where $f_\mathrm{M}$ is the decay constant of the meson, and $|V_\mathrm{qq'}|^2$ the relevant CKM matrix element, and where the last numerical equality is obtained with the parameters of Table~\ref{tab:params}. The decay $\BR(\rm t \to \pi^+ + b)$ was first estimated in Ref.~\cite{Beneke:2000hk}, but their expression differs from Eq.~(\ref{eq:t_M_b_decay}) here by a factor of 1/9~\cite{Long-ShunLu2024}. We extend here the study of semiexclusive $\rm t\to meson + b$-jet decays to include the $\rm M^\pm = \rho^\pm,\, K^\pm,\, D^\pm,\, D^\pm_{s},\, B^\pm,\, B_\mathrm{c}^\pm$ channels (and their corresponding excited states) by employing the decay constants from the recent compilation for the light mesons~\cite{Grossman:2015cak}, and from lattice calculations~\cite{Becirevic:2012ti,Colquhoun:2015oha} for the heavy ones (Table~\ref{tab:params}). The resulting $\rm t \to b+M$ branching fractions are tabulated in Table~\ref{tab:t_b_meson_decays}, and also shown in graphical form in Fig.~\ref{fig:t_meson_b_quark_limits}. Estimated theoretical branching fractions are in the $\mathcal{O}(10^{-7}$--$10^{-12})$ range. No experimental search has been performed to date at the LHC. 
About four channels are producible in the clean FCC-ee environment, and all of them will be accessible at FCC-hh.

\begin{table}[htpb!]
\caption{
Compilation of semiexclusive decays of the top quark into a b quark plus a meson. For each decay, we provide the predicted branching fraction(s) and the theoretical approach used to compute it, or derived here using Eq.~(\ref{eq:t_M_b_decay}). The last two columns indicate whether the decay can be produced at FCC-ee/FCC-hh.
\label{tab:t_b_meson_decays}}
\resizebox{\textwidth}{!}{%
\input{tables/exclusive_t_decays_b_meson.tex}
}
\end{table}

\clearpage
\section{Summary}

We have presented a comprehensive survey of the theoretical and experimental status of about 200 rare and exclusive few-body decays of the four heaviest Standard Model (SM) particles: the Higgs, the electroweak (W, Z) bosons, and the top quark. Rare decays are defined here as those having branching fractions below $\BR \approx 10^{-5}$, and we focus on those with two- to four-particles in the final state. Such decay processes remain experimentally unobserved and only upper limits have been set for about 50 of them. The study of these decay processes provides a powerful window into physics beyond the Standard Model (BSM) either directly, by probing SM suppressed or forbidden processes (such as flavour-changing neutral currents FCNC, lepton-flavour or lepton-flavour-universality violating processes, or spin-selection-rules violating decays), or indirectly as SM backgrounds to multiple exotic BSM decays (\eg\ into axion-like particles ALPs, massive gravitons, or dark photons). Additionally, such decays offer a unique opportunity to constrain the light-fermion Yukawa couplings, can provide alternative means to measure the W and/or top-quark masses, and help improve our understanding of quantum chromodynamics (QCD) factorization with small nonperturbative corrections.

First, we have systematically collected and organized in tabular form the theoretical branching fractions of abbout 200 rare decay channels, indicating the model(s) used for the calculations, while providing any existing experimental upper limits on them. Among those, we have estimated for the first time the rates of about 60 new processes including ultrarare Higgs boson decays into photons and/or neutrinos (with rates $10^{-12}$--$10^{-40}$) and into Z bosons plus gluons or photons (with rates $10^{-6}$, $10^{-9}$), radiative H and Z boson decays into leptonium states (with rates $10^{-10}$--$10^{-23}$), exclusive radiative H and Z boson quark-flavour-changing decays (with rates $10^{-14}$--$10^{-27}$), exclusive Z decays into a pair of $\phi$ mesons (with an $\mathcal{O}(10^{-12})$ rate), three-body H\,W\,b top-quark decay (with an $\mathcal{O}(10^{-10})$ rate), and semiexclusive top-quark decays into a quark plus a meson (with rates $10^{-7}$--$10^{-12}$). In addition, we have revised and updated predictions for a few other rare Z-boson and top-quark partial widths. We have also studied in detail all Z decay channels leading to a triphoton final state, and found that the sum of all relevant exclusive photon-plus-meson decays amounts to $\BR(Z\to \gamma+\mathrm{M}\to 3\gamma) = 1.8\cdot 10^{-10}$, which is about one-third of the direct rate estimated here to be $\BR(Z\to 3\gamma) = 6.4\cdot 10^{-10}$. All such decays will need to be taken into account as backgrounds for future searches of $\mathrm{Z}\to\gamma a(\gaga)$ processes, where $a$ can be a BSM particle such as spin-0 (axion-like) or spin-2 (graviton-like) object.

Secondly, the feasibility of measuring each of these unobserved decays has been estimated for p-p collisions at the High-Luminosity Large Hadron Collider (HL-LHC), as well as for $\epem$ and p-p collisions at the 
\begin{table}[htpb!]
\caption{Selection of rare and exclusive decays of the H, Z, W bosons and top quark potentially observable in pp(14~TeV) collisions at the HL-LHC. The last column indicates the approximate ratio of theoretical to our expected projected rates, $\BR(\rm th)/\BR(\rm exp)$.
\label{tab:hl_lhc_summary}}
\resizebox{\textwidth}{!}{%
\input{tables/important_channels.tex}}
\end{table}
Future Circular Collider (FCC-ee and FCC-hh).
From the number of H, Z, W bosons, and top quarks expected to be produced at the HL-LHC, FCC-ee, and FCC-hh colliders, and by statistically extrapolating the current 95\% confidence-level limits set, we provide estimates of the conservatively achievable experimental upper bounds (or observations) in those machines. Among those, in Table~\ref{tab:hl_lhc_summary} we have selected a few interesting channels that can be observed and/or deserve further study in p-p collisions at the HL-LHC.
The last column indicates the approximate ratio of theoretical to expected experimental rates, $\BR(\rm th)/\BR(\rm exp)$. Two reasons motivate this selection of channels: either (i) they have relatively large rates, $\mathcal{O}(10^{-5})$ (with a measurement having been attempted or not, to date), or (ii) they have lower rates, $\BR\lesssim 10^{-8}$, but measurements of upper bounds have been performed, and our conservative projections indicate $\BR(\rm th)/\BR(\rm exp)$ not far from unity. For the Higgs boson, we list first the rare $\rm H \to 4\gamma$ decay that has not been directly searched for at the LHC to date, but limits exist on the process $\rm H \to a(\gaga)a(\gaga)$ with two intermediate ALPs decaying into photons
~\cite{ATLAS:2015rsn,CMS:2022fyt,CMS:2022xxa,ATLAS:2023ian}. Although its rate makes its observation impossible at the LHC in the absence of enhancing BSM effects, it would not be too difficult for ATLAS/CMS to recast any current and future similar ALP searches into upper bounds on the $\rm H\to 4\gamma$ ``continuum'' decay. Next, we find that the HL-LHC can conservatively set upper bounds on the exclusive $\rm H\to \gamma\,\rho$ and $\rm H\to \gamma\,\jpsi$ decays at about four and ten times their expected SM values, respectively. Similarly, we find that HL-LHC can potentially observe $\rm Z\to \gamma\,\phi$, $\rm Z\to \gamma\,\jpsi$, and $\rm Z\to \gamma\,\Upsilon(1S)$ with ratios between the theoretical rates and projected experimental limits as large as 1/5, 1/3, and 1/4, respectively. In the W boson case, the least suppressed rare decays are $\rm W \to \gamma\,\pi,\,\gamma\,D_\mathrm{s}$, but with experimental bounds expected at about two orders-of-magnitude their SM values. Finally, the semiexclusive decay of the top quark into a charm jet and B meson (identified experimentally as a b-jet) has $\mathcal{O}(10^{-4})$ rates that make it observable at the HL-LHC~\cite{dEnterria:2020ygk}, and the three-body $\rm t \to W\,Z\,b$ decay can occur in about one out of two million top-quark decays, and a search should also be attempted. Of course, the HL-LHC channel selection of Table~\ref{tab:hl_lhc_summary} is driven by the SM rates and their potential visibility, but as explained in this work there are many other suppressed decays of the four heaviest particles that can be enhanced in many BSM scenarios, and should be an active part of target searches in the next years.\\

Finally, we have shown that a future high-luminosity $\epem$ electroweak factory, such as the FCC-ee with $6\cdot10^{12}$ Z bosons, $5\cdot10^{8}$ W bosons, $1.9\cdot10^{6}$ Higgs bosons, and $3.8\cdot10^{6}$ top quarks produced in very clean experimental conditions, can discover about half of the 200 rare decays discussed here. Eventually, the FCC-hh with huge data samples expected, can produce most of the SM decay channels listed in this work, although their experimental observation will be difficult in a challenging hadronic collider environment. We hope that this document can help guide and prioritize upcoming experimental and theoretical studies of rare and exclusive few-body decays of the most massive SM particles, as well as further motivate BSM searches, at the LHC and future colliders.\\


\paragraph*{\hspace{-0.35cm}Acknowledgments.---}Informative discussions and valuable feedback from Hua-Sheng~Shao and J.-Ph.~Lansberg are gratefully acknowledged. We thank Dao-Neng~Gao, Long-Shun Lu, Richard~Ruiz, and Bin~Yan for feedback on a previous version of the paper. Support from the EU STRONG-2020 project under the program H2020-INFRAIA-2018-1, grant agreement No.\ 824093 is acknowledged.\\

\paragraph*{\hspace{-0.35cm}Data availability statement.---} All data that support the findings of this study are included within the article.

\bibliographystyle{myutphys}
\bibliography{references}
\end{document}

%% file: tables/exclusive_H_decays_ultrarare.tex
\begin{tabular}{ll<{\hspace{-2mm}}>{\hspace{-1mm}}r<{\hspace{-6mm}}lllllcc}
\toprule
 &  &  & \multicolumn{2}{c}{} & \multicolumn{2}{c}{Exp. limits} & \multicolumn{2}{c}{Producible at} \\
 &   & \multicolumn{2}{c}{Branching fraction} & Framework & 2024 & HL-LHC & FCC-ee & FCC-hh \\
\midrule
\multirow[c]{10}{*}{$\rm H \, \to$} & $\rm \nu+\overline{\nu}$ & \ $2.0$ & $\times~ 10^{-26}$ & loop SM, \mgshort~(this work) & -- & -- & \textcolor{red}{\ding{56}} & \textcolor{red}{\ding{56}} \\
\cline{2-9}
 & \multirow[c]{2}{*}{$\rm \gamma+\nu+\overline{\nu}$} & \ $3.4$ & $\times~ 10^{-4}$ & loop SM~\cite{Sun:2013cba} & \multirow[c]{2}{*}{--} & \multirow[c]{2}{*}{--} & \multirow[c]{2}{*}{\textcolor{green}{\Checkmark}} & \multirow[c]{2}{*}{\textcolor{green}{\Checkmark}} \\
 &  & \ $3.6$ & $\times~ 10^{-4}$ & loop SM, \mgshort~(this work) &  &  &  &  \\
\cline{2-9}
 & $\rm \gamma+\gamma+\gamma~$ & \ $1.0$ & $\times~ 10^{-40}$ & loop SM, \mgshort~(this work) & -- & -- & \textcolor{red}{\ding{56}} & \textcolor{red}{\ding{56}} \\
\cline{2-9}
 & $\rm \gamma+\gamma+\gamma+\gamma$ & \ $5.4$ & $\times~ 10^{-12}$ & loop SM, \mgshort~(this work) & -- & -- & \textcolor{red}{\ding{56}} & \textcolor{red}{\ding{56}} \\
\cline{2-9}
 & \multirow[c]{2}{*}{$\rm Z+\gamma+\gamma$} & \ $2.0$ & $\times~ 10^{-9}$ & loop SM(with cuts)~\cite{Abbasabadi:2004wq} & \multirow[c]{2}{*}{--} & \multirow[c]{2}{*}{--} & \multirow[c]{2}{*}{\textcolor{red}{\ding{56}}} & \multirow[c]{2}{*}{\textcolor{green}{\Checkmark}} \\
 &  & \ $2.4$ & $\times~ 10^{-9}$ & loop SM, \mgshort~(this work) &  &  &  &  \\
\cline{2-9}
 & \multirow[c]{3}{*}{$\rm Z+g+g$} & \ $3.6$ & $\times~ 10^{-7}$ & loop SM~\cite{Abbasabadi:2008zz} & \multirow[c]{3}{*}{--} & \multirow[c]{3}{*}{--} & \multirow[c]{3}{*}{\textcolor{red}{\ding{56}}} & \multirow[c]{3}{*}{\textcolor{green}{\Checkmark}} \\
 &  & \ $5.5$ & $\times~ 10^{-7}$ & loop SM\footnote{With updated SM parameters}~\cite{Kniehl:1990yb} &  &  &  &  \\
 &  & \ $6.3$ & $\times~ 10^{-7}$ & loop SM, \mgshort~(this work) &  &  &  &  \\

\bottomrule
\end{tabular}

%% file: tables/exclusive_H_decays_gamma_meson.tex
\begin{tabular}{l<{\hspace{-4mm}}l<{\hspace{-2mm}}>{\hspace{-1mm}}lr<{\hspace{-6mm}}llccccc}
\toprule
 &  &  &  & \multicolumn{2}{c}{} & \multicolumn{2}{c}{Exp. limits} & \multicolumn{2}{c}{Producible at} \\
$\rm H\to \, \gamma$ & $+$ & $\rm M$ & \multicolumn{2}{c}{Branching fraction} & Framework & 2024 & HL-LHC & FCC-ee & FCC-hh \\
\midrule
\multirow[c]{19}{*}{$\rm H\, \to \,\gamma$} & \multirow[c]{19}{*}{+} & $\rm \rho^0$ & \ $(1.68\pm 0.08)$ & $\times~ 10^{-5}$ & SCET+LCDA~\cite{Konig:2015qat} & $<$\ $3.7$$\times~ 10^{-4}$~\cite{CMS:2024tgj} & $\lesssim5.7\times~ 10^{-5}$ & \textcolor{green}{\Checkmark} & \textcolor{green}{\Checkmark} \\

 &  & $\rm \omega$ & \ $(1.48\pm 0.08)$ & $\times~ 10^{-6}$ & SCET+LCDA~\cite{Konig:2015qat} & $<$\ $5.5$$\times~ 10^{-4}$~\cite{ATLAS:2023alf} & $\lesssim8.2\times~ 10^{-5}$ & \textcolor{green}{\Checkmark} & \textcolor{green}{\Checkmark} \\

 &  & $\rm \phi$ & \ $(2.31\pm 0.11)$ & $\times~ 10^{-6}$ & SCET+LCDA~\cite{Konig:2015qat} & $<$\ $3.0$$\times~ 10^{-4}$~\cite{CMS:2024tgj} & $\lesssim4.5\times~ 10^{-5}$ & \textcolor{green}{\Checkmark} & \textcolor{green}{\Checkmark} \\
\cline{3-10}
 &  & \multirow[c]{3}{*}{$\rm J/\psi$} & \ $(3.01\pm 0.15)$ & $\times~ 10^{-6}$ & NRQCD~(NLL)+LDME~\cite{Brambilla:2019fmu} & \multirow[c]{3}{*}{$<$\ $2.0$$\times~ 10^{-4}$~\cite{ATLAS:2022rej}} & \multirow[c]{3}{*}{$\lesssim 3.9\times~ 10^{-5}$~\cite{ATLAS:2015xkp}} & \multirow[c]{3}{*}{\textcolor{green}{\Checkmark}} & \multirow[c]{3}{*}{\textcolor{green}{\Checkmark}} \\
 &  &  & \ $\left(2.99^{+0.16}_{-0.15}\right)$ & $\times~ 10^{-6}$ & NRQCD+LCDA~\cite{Bodwin:2017wdu} &  &  &  &  \\
 &  &  & \ $(2.95\pm 0.17)$ & $\times~ 10^{-6}$ & SCET+LCDA~\cite{Konig:2015qat} &  &  &  &  \\
\cline{3-10}
 &  & $\rm \psi(2S)$ & \ $(1.03\pm 0.06)$ & $\times~ 10^{-6}$ & SCET+LCDA\footnote{private communication with author of Ref. \cite{Bodwin:2017wdu}}~\cite{CMS:2024hhg} & $<$\ $9.9$$\times~ 10^{-4}$~\cite{CMS:2024hhg} & $\lesssim1.4\times~ 10^{-4}$ & \textcolor{green}{\Checkmark} & \textcolor{green}{\Checkmark} \\
\cline{3-10}
 &  & \multirow[c]{4}{*}{$\rm \Upsilon(1S)$} & \ $3.0$ & $\times~ 10^{-8}$ & NRQCD (NLO)+LDME~\cite{Dong:2022bkd} & \multirow[c]{4}{*}{$<$\ $2.5$$\times~ 10^{-4}$~\cite{ATLAS:2022rej}} & \multirow[c]{4}{*}{$\lesssim3.8\times~ 10^{-5}$} & \multirow[c]{4}{*}{\textcolor{red}{\ding{56}}} & \multirow[c]{4}{*}{\textcolor{green}{\Checkmark}} \\
 &  &  & \ $\left(9.97^{+4.04}_{-3.03}\right)$ & $\times~ 10^{-9}$ & NRQCD~(NLL)+LDME~\cite{Brambilla:2019fmu} &  &  &  &  \\
 &  &  & \ $\left(5.22^{+2.02}_{-1.70}\right)$ & $\times~ 10^{-9}$ & NRQCD+LCDA~\cite{Bodwin:2017wdu} &  &  &  &  \\
 &  &  & \ $\left(4.61^{+1.76}_{-1.23}\right)$ & $\times~ 10^{-9}$ & SCET+LCDA~\cite{Konig:2015qat} &  &  &  &  \\
\cline{3-10}
 &  & \multirow[c]{4}{*}{$\rm \Upsilon(2S)$} & \ $1.4$ & $\times~ 10^{-8}$ & NRQCD (NLO)+LDME~\cite{Dong:2022bkd} & \multirow[c]{4}{*}{$<$\ $4.2$$\times~ 10^{-4}$~\cite{ATLAS:2022rej}} & \multirow[c]{4}{*}{$\lesssim6.4\times~ 10^{-5}$} & \multirow[c]{4}{*}{\textcolor{red}{\ding{56}}} & \multirow[c]{4}{*}{\textcolor{green}{\Checkmark}} \\
 &  &  & \ $\left(2.62^{+1.39}_{-0.91}\right)$ & $\times~ 10^{-9}$ & NRQCD~(NLL)+LDME~\cite{Brambilla:2019fmu} &  &  &  &  \\
 &  &  & \ $\left(1.42^{+0.72}_{-0.57}\right)$ & $\times~ 10^{-9}$ & NRQCD+LCDA~\cite{Bodwin:2017wdu} &  &  &  &  \\
 &  &  & \ $\left(2.34^{+0.76}_{-1.00}\right)$ & $\times~ 10^{-9}$ & SCET+LCDA~\cite{Konig:2015qat} &  &  &  &  \\
\cline{3-10}
 &  & \multirow[c]{4}{*}{$\rm \Upsilon(3S)$} & \ $1.1$ & $\times~ 10^{-8}$ & NRQCD (NLO)+LDME~\cite{Dong:2022bkd} & \multirow[c]{4}{*}{$<$\ $3.4$$\times~ 10^{-4}$~\cite{ATLAS:2022rej}} & \multirow[c]{4}{*}{$\lesssim5.2\times~ 10^{-5}$} & \multirow[c]{4}{*}{\textcolor{red}{\ding{56}}} & \multirow[c]{4}{*}{\textcolor{green}{\Checkmark}} \\
 &  &  & \ $\left(1.87^{+1.05}_{-0.69}\right)$ & $\times~ 10^{-9}$ & NRQCD~(NLL)+LDME~\cite{Brambilla:2019fmu} &  &  &  &  \\
 &  &  & \ $\left(9.1^{+4.8}_{-3.8}\right)$ & $\times~ 10^{-10}$ & NRQCD+LCDA~\cite{Bodwin:2017wdu} &  &  &  &  \\
 &  &  & \ $\left(2.13^{+0.76}_{-1.12}\right)$ & $\times~ 10^{-9}$ & SCET+LCDA~\cite{Konig:2015qat} &  &  &  &  \\

\bottomrule
\end{tabular}

%% file: tables/exclusive_H_decays_Z_meson.tex
\begin{tabular}{l<{\hspace{-4mm}}l<{\hspace{-2mm}}>{\hspace{-1mm}}lr<{\hspace{-6mm}}llccccc}
\toprule
 &  &  &  & \multicolumn{2}{c}{} & \multicolumn{2}{c}{Exp. limits} & \multicolumn{2}{c}{Producible at} \\
$\rm H\to \, Z$ & $+$ & $\rm M$ & \multicolumn{2}{c}{Branching fraction} & Framework & 2024 & HL-LHC & FCC-ee & FCC-hh \\
\midrule
\multirow[c]{25}{*}{$\rm H\, \to \,Z$} & \multirow[c]{25}{*}{+} & \multirow[c]{2}{*}{$\rm \pi^0$} & \ $(2.3\pm 0.1)$ & $\times~ 10^{-6}$ & EFT+NRQM~\cite{LHCHiggsCrossSectionWorkingGroup:2016ypw} & \multirow[c]{2}{*}{--} & \multirow[c]{2}{*}{--} & \multirow[c]{2}{*}{\textcolor{green}{\Checkmark}} & \multirow[c]{2}{*}{\textcolor{green}{\Checkmark}} \\
 &  &  & \ $(2.3\pm 0.1)$ & $\times~ 10^{-6}$ & EFT+LCDA~\cite{Alte:2016yuw} &  &  &  &  \\
\cline{3-10}
 &  & $\rm \eta$ & \ $(8.3\pm 0.9)$ & $\times~ 10^{-7}$ & EFT+LCDA~\cite{Alte:2016yuw} & -- & -- & \textcolor{green}{\Checkmark} & \textcolor{green}{\Checkmark} \\
\cline{3-10}
 &  & \multirow[c]{2}{*}{$\rm \rho^0$} & \ $(1.4\pm 0.1)$ & $\times~ 10^{-5}$ & EFT+NRQM~\cite{LHCHiggsCrossSectionWorkingGroup:2016ypw} & \multirow[c]{2}{*}{$<$\ $1.2$$\times~ 10^{-2}$~\cite{CMS:2020ggo}} & \multirow[c]{2}{*}{$\lesssim1.8\times~ 10^{-3}$} & \multirow[c]{2}{*}{\textcolor{green}{\Checkmark}} & \multirow[c]{2}{*}{\textcolor{green}{\Checkmark}} \\
 &  &  & \ $(7.19\pm 0.29)$ & $\times~ 10^{-6}$ & EFT+LCDA~\cite{Alte:2016yuw} &  &  &  &  \\
\cline{3-10}
 &  & \multirow[c]{2}{*}{$\rm \omega$} & \ $(1.6\pm 0.1)$ & $\times~ 10^{-6}$ & EFT+NRQM~\cite{LHCHiggsCrossSectionWorkingGroup:2016ypw} & \multirow[c]{2}{*}{--} & \multirow[c]{2}{*}{--} & \multirow[c]{2}{*}{\textcolor{green}{\Checkmark}} & \multirow[c]{2}{*}{\textcolor{green}{\Checkmark}} \\
 &  &  & \ $(5.6\pm 0.2)$ & $\times~ 10^{-7}$ & EFT+LCDA~\cite{Alte:2016yuw} &  &  &  &  \\
\cline{3-10}
 &  & $\rm \eta'$ & \ $(1.24\pm 0.13)$ & $\times~ 10^{-6}$ & EFT+LCDA~\cite{Alte:2016yuw} & -- & -- & \textcolor{green}{\Checkmark} & \textcolor{green}{\Checkmark} \\
\cline{3-10}
 &  & \multirow[c]{2}{*}{$\rm \phi$} & \ $(4.2\pm 0.3)$ & $\times~ 10^{-6}$ & EFT+NRQM~\cite{LHCHiggsCrossSectionWorkingGroup:2016ypw} & \multirow[c]{2}{*}{$<$\ $3.6$$\times~ 10^{-3}$~\cite{CMS:2020ggo}} & \multirow[c]{2}{*}{$\lesssim5.4\times~ 10^{-4}$} & \multirow[c]{2}{*}{\textcolor{green}{\Checkmark}} & \multirow[c]{2}{*}{\textcolor{green}{\Checkmark}} \\
 &  &  & \ $(2.42\pm 0.10)$ & $\times~ 10^{-6}$ & EFT+LCDA~\cite{Alte:2016yuw} &  &  &  &  \\
\cline{3-10}
 &  & \multirow[c]{2}{*}{$\rm \eta_c$} & \ $(1.00\pm 0.01)$ & $\times~ 10^{-5}$ & EFT+LCDA~\cite{Becirevic:2017chd} & \multirow[c]{2}{*}{--} & \multirow[c]{2}{*}{--} & \multirow[c]{2}{*}{\textcolor{green}{\Checkmark}} & \multirow[c]{2}{*}{\textcolor{green}{\Checkmark}} \\
 &  &  & \ $(1.0\pm 0.1)$ & $\times~ 10^{-5}$ & EFT+NRQM~\cite{LHCHiggsCrossSectionWorkingGroup:2016ypw} &  &  &  &  \\
\cline{3-10}
 &  & \multirow[c]{3}{*}{$\rm J/\psi$} & \ $3.4$ & $\times~ 10^{-6}$ & NRQCD~(NLO)+LMDE~\cite{Sun:2018xft} & \multirow[c]{3}{*}{$<$\ $1.9$$\times~ 10^{-3}$~\cite{CMS:2022fsq}} & \multirow[c]{3}{*}{$\lesssim 2.1\times~ 10^{-4}$~\cite{CMS:2022kdd}} & \multirow[c]{3}{*}{\textcolor{green}{\Checkmark}} & \multirow[c]{3}{*}{\textcolor{green}{\Checkmark}} \\
 &  &  & \ $(2.3\pm 0.1)$ & $\times~ 10^{-6}$ & EFT+LCDA~\cite{Alte:2016yuw} &  &  &  &  \\
 &  &  & \ $3.2$ & $\times~ 10^{-6}$ & EFT+NRQM~\cite{Gao:2014xlv} &  &  &  &  \\
\cline{3-10}
 &  & $\rm \psi(2S)$ & \ $1.5$ & $\times~ 10^{-6}$ & EFT+NRQM~\cite{Gao:2014xlv} & $<$\ $6.6$$\times~ 10^{-3}$~\cite{CMS:2022fsq} & $\lesssim1.0\times~ 10^{-3}$ & \textcolor{green}{\Checkmark} & \textcolor{green}{\Checkmark} \\
\cline{3-10}
 &  & \multirow[c]{2}{*}{$\rm \eta_b$} & \ $(2.69\pm 0.05)$ & $\times~ 10^{-5}$ & EFT+LCDA~\cite{Becirevic:2017chd} & \multirow[c]{2}{*}{--} & \multirow[c]{2}{*}{--} & \multirow[c]{2}{*}{\textcolor{green}{\Checkmark}} & \multirow[c]{2}{*}{\textcolor{green}{\Checkmark}} \\
 &  &  & \ $\left(4.739^{+0.276}_{-0.244}\right)$ & $\times~ 10^{-5}$ & EFT~(NLO)+LCDA~\cite{Zhu:2018xxs} &  &  &  &  \\
\cline{3-10}
 &  & \multirow[c]{3}{*}{$\rm \Upsilon(1S)$} & \ $1.7$ & $\times~ 10^{-5}$ & NRQCD~(NLO)+LMDE~\cite{Sun:2018xft} & \multirow[c]{3}{*}{--} & \multirow[c]{3}{*}{--} & \multirow[c]{3}{*}{\textcolor{green}{\Checkmark}} & \multirow[c]{3}{*}{\textcolor{green}{\Checkmark}} \\
 &  &  & \ $(1.54\pm 0.06)$ & $\times~ 10^{-5}$ & EFT+LCDA~\cite{Alte:2016yuw} &  &  &  &  \\
 &  &  & \ $1.7$ & $\times~ 10^{-5}$ & EFT+NRQM~\cite{Gao:2014xlv} &  &  &  &  \\
\cline{3-10}
 &  & \multirow[c]{2}{*}{$\rm \Upsilon(2S)$} & \ $(7.5\pm 0.3)$ & $\times~ 10^{-6}$ & EFT+LCDA~\cite{Alte:2016yuw} & \multirow[c]{2}{*}{--} & \multirow[c]{2}{*}{--} & \multirow[c]{2}{*}{\textcolor{green}{\Checkmark}} & \multirow[c]{2}{*}{\textcolor{green}{\Checkmark}} \\
 &  &  & \ $8.9$ & $\times~ 10^{-6}$ & EFT+NRQM~\cite{Gao:2014xlv} &  &  &  &  \\
\cline{3-10}
 &  & \multirow[c]{2}{*}{$\rm \Upsilon(3S)$} & \ $(5.63\pm 0.24)$ & $\times~ 10^{-6}$ & EFT+LCDA~\cite{Alte:2016yuw} & \multirow[c]{2}{*}{--} & \multirow[c]{2}{*}{--} & \multirow[c]{2}{*}{\textcolor{green}{\Checkmark}} & \multirow[c]{2}{*}{\textcolor{green}{\Checkmark}} \\
 &  &  & \ $6.7$ & $\times~ 10^{-6}$ & EFT+NRQM~\cite{Gao:2014xlv} &  &  &  &  \\

\bottomrule
\end{tabular}

%% file: tables/exclusive_H_decays_boson_flavouredmeson.tex
\begin{tabular}{l<{\hspace{-4mm}}l<{\hspace{-2mm}}>{\hspace{-1mm}}lr<{\hspace{-6mm}}llccccc}
\toprule
 &  &  &  & \multicolumn{2}{c}{} & \multicolumn{2}{c}{Exp. limits} & \multicolumn{2}{c}{Producible at} \\
$\rm H\to \, X$ & $+$ & $\rm M$ & \multicolumn{2}{c}{Branching fraction} & Framework & 2024 & HL-LHC & FCC-ee & FCC-hh \\
\midrule
\multirow[c]{4}{*}{$\rm H\, \to \,\gamma$} & \multirow[c]{4}{*}{+} & $\rm K^{*0}$ & \ $2.6$ & $\times~ 10^{-23}$ & EFT+LCDA~(this work) & $<$\ $2.2$$\times~ 10^{-4}$~\cite{ATLAS:2023alf} & $\lesssim3.3\times~ 10^{-5}$ & \textcolor{red}{\ding{56}} & \textcolor{red}{\ding{56}} \\

 &  & $\rm D^{*0}$ & \ $6.7$ & $\times~ 10^{-27}$ & EFT+LCDA~(this work) & $<$\ $1.0$$\times~ 10^{-3}$~\cite{ATLAS:2024dpw} & $\lesssim1.5\times~ 10^{-4}$ & \textcolor{red}{\ding{56}} & \textcolor{red}{\ding{56}} \\

 &  & $\rm B^{*0}$ & \ $8.2$ & $\times~ 10^{-16}$ & EFT+LCDA~(this work) & -- & -- & \textcolor{red}{\ding{56}} & \textcolor{red}{\ding{56}} \\

 &  & $\rm B_s^{*0}$ & \ $1.8$ & $\times~ 10^{-14}$ & EFT+LCDA~(this work) & -- & -- & \textcolor{red}{\ding{56}} & \textcolor{red}{\ding{56}} \\
\cline{1-10}
\multirow[c]{4}{*}{$\rm H\, \to \,Z$} & \multirow[c]{4}{*}{+} & $\rm K^{*0}$ & \ $2.2$ & $\times~ 10^{-25}$ & EFT+LCDA~(this work) & -- & -- & \textcolor{red}{\ding{56}} & \textcolor{red}{\ding{56}} \\

 &  & $\rm D^{*0}$ & \ $1.8$ & $\times~ 10^{-30}$ & EFT+LCDA~(this work) & -- & -- & \textcolor{red}{\ding{56}} & \textcolor{red}{\ding{56}} \\

 &  & $\rm B^{*0}$ & \ $2.4$ & $\times~ 10^{-19}$ & EFT+LCDA~(this work) & -- & -- & \textcolor{red}{\ding{56}} & \textcolor{red}{\ding{56}} \\

 &  & $\rm B_s^{*0}$ & \ $2.9$ & $\times~ 10^{-17}$ & EFT+LCDA~(this work) & -- & -- & \textcolor{red}{\ding{56}} & \textcolor{red}{\ding{56}} \\

\bottomrule
\end{tabular}

%% file: tables/exclusive_H_decays_W_meson.tex
\begin{tabular}{l<{\hspace{-4mm}}l<{\hspace{-2mm}}>{\hspace{-1mm}}lr<{\hspace{-6mm}}llccccc}
\toprule
 &  &  &  & \multicolumn{2}{c}{} & \multicolumn{2}{c}{Exp. limits} & \multicolumn{2}{c}{Producible at} \\
$\rm H\to \, W$ & $+$ & $\rm M$ & \multicolumn{2}{c}{Branching fraction} & Framework & 2024 & HL-LHC & FCC-ee & FCC-hh \\
\midrule
\multirow[c]{22}{*}{$\rm H\, \to \,W^\mp$} & \multirow[c]{22}{*}{+} & \multirow[c]{2}{*}{$\rm \pi^\pm$} & \ $(4.2\pm 0.2)$ & $\times~ 10^{-6}$ & EFT+NRQM~\cite{LHCHiggsCrossSectionWorkingGroup:2016ypw} & \multirow[c]{2}{*}{--} & \multirow[c]{2}{*}{--} & \multirow[c]{2}{*}{\textcolor{green}{\Checkmark}} & \multirow[c]{2}{*}{\textcolor{green}{\Checkmark}} \\
 &  &  & \ $(4.3\pm 0.2)$ & $\times~ 10^{-6}$ & EFT+LCDA~\cite{Alte:2016yuw} &  &  &  &  \\
\cline{3-10}
 &  & \multirow[c]{2}{*}{$\rm \rho^\pm$} & \ $(1.5\pm 0.1)$ & $\times~ 10^{-5}$ & EFT+NRQM~\cite{LHCHiggsCrossSectionWorkingGroup:2016ypw} & \multirow[c]{2}{*}{--} & \multirow[c]{2}{*}{--} & \multirow[c]{2}{*}{\textcolor{green}{\Checkmark}} & \multirow[c]{2}{*}{\textcolor{green}{\Checkmark}} \\
 &  &  & \ $(1.09\pm 0.05)$ & $\times~ 10^{-5}$ & EFT+LCDA~\cite{Alte:2016yuw} &  &  &  &  \\
\cline{3-10}
 &  & \multirow[c]{2}{*}{$\rm K^\pm$} & \ $(3.3\pm 0.1)$ & $\times~ 10^{-7}$ & EFT+NRQM~\cite{LHCHiggsCrossSectionWorkingGroup:2016ypw} & \multirow[c]{2}{*}{--} & \multirow[c]{2}{*}{--} & \multirow[c]{2}{*}{\textcolor{red}{\ding{56}}} & \multirow[c]{2}{*}{\textcolor{green}{\Checkmark}} \\
 &  &  & \ $(3.3\pm 0.1)$ & $\times~ 10^{-7}$ & EFT+LCDA~\cite{Alte:2016yuw} &  &  &  &  \\
\cline{3-10}
 &  & \multirow[c]{2}{*}{$\rm K^{*\pm}$} & \ $(4.3\pm 0.2)$ & $\times~ 10^{-7}$ & EFT+NRQM~\cite{LHCHiggsCrossSectionWorkingGroup:2016ypw} & \multirow[c]{2}{*}{--} & \multirow[c]{2}{*}{--} & \multirow[c]{2}{*}{\textcolor{red}{\ding{56}}} & \multirow[c]{2}{*}{\textcolor{green}{\Checkmark}} \\
 &  &  & \ $(5.6\pm 0.4)$ & $\times~ 10^{-7}$ & EFT+LCDA~\cite{Alte:2016yuw} &  &  &  &  \\
\cline{3-10}
 &  & \multirow[c]{2}{*}{$\rm D^\pm$} & \ $(5.8\pm 0.6)$ & $\times~ 10^{-7}$ & EFT+NRQM~\cite{LHCHiggsCrossSectionWorkingGroup:2016ypw} & \multirow[c]{2}{*}{--} & \multirow[c]{2}{*}{--} & \multirow[c]{2}{*}{\textcolor{green}{\Checkmark}} & \multirow[c]{2}{*}{\textcolor{green}{\Checkmark}} \\
 &  &  & \ $(5.6\pm 0.5)$ & $\times~ 10^{-7}$ & EFT+LCDA~\cite{Alte:2016yuw} &  &  &  &  \\
\cline{3-10}
 &  & \multirow[c]{2}{*}{$\rm D^{*\pm}$} & \ $(1.3\pm 0.1)$ & $\times~ 10^{-6}$ & EFT+NRQM~\cite{LHCHiggsCrossSectionWorkingGroup:2016ypw} & \multirow[c]{2}{*}{--} & \multirow[c]{2}{*}{--} & \multirow[c]{2}{*}{\textcolor{green}{\Checkmark}} & \multirow[c]{2}{*}{\textcolor{green}{\Checkmark}} \\
 &  &  & \ $(1.04\pm 0.14)$ & $\times~ 10^{-6}$ & EFT+LCDA~\cite{Alte:2016yuw} &  &  &  &  \\
\cline{3-10}
 &  & \multirow[c]{2}{*}{$\rm D_s^\pm$} & \ $(1.6\pm 0.1)$ & $\times~ 10^{-5}$ & EFT+NRQM~\cite{LHCHiggsCrossSectionWorkingGroup:2016ypw} & \multirow[c]{2}{*}{--} & \multirow[c]{2}{*}{--} & \multirow[c]{2}{*}{\textcolor{green}{\Checkmark}} & \multirow[c]{2}{*}{\textcolor{green}{\Checkmark}} \\
 &  &  & \ $(1.71\pm 0.11)$ & $\times~ 10^{-5}$ & EFT+LCDA~\cite{Alte:2016yuw} &  &  &  &  \\
\cline{3-10}
 &  & \multirow[c]{2}{*}{$\rm D_s^{*\pm}$} & \ $(3.5\pm 0.2)$ & $\times~ 10^{-5}$ & EFT+NRQM~\cite{LHCHiggsCrossSectionWorkingGroup:2016ypw} & \multirow[c]{2}{*}{--} & \multirow[c]{2}{*}{--} & \multirow[c]{2}{*}{\textcolor{green}{\Checkmark}} & \multirow[c]{2}{*}{\textcolor{green}{\Checkmark}} \\
 &  &  & \ $(2.51\pm 0.19)$ & $\times~ 10^{-5}$ & EFT+LCDA~\cite{Alte:2016yuw} &  &  &  &  \\
\cline{3-10}
 &  & \multirow[c]{2}{*}{$\rm B^\pm$} & \ $(1.6\pm 0.4)$ & $\times~ 10^{-10}$ & EFT+NRQM~\cite{LHCHiggsCrossSectionWorkingGroup:2016ypw} & \multirow[c]{2}{*}{--} & \multirow[c]{2}{*}{--} & \multirow[c]{2}{*}{\textcolor{red}{\ding{56}}} & \multirow[c]{2}{*}{\textcolor{green}{\Checkmark}} \\
 &  &  & \ $(1.54\pm 0.40)$ & $\times~ 10^{-10}$ & EFT+LCDA~\cite{Alte:2016yuw} &  &  &  &  \\
\cline{3-10}
 &  & \multirow[c]{2}{*}{$\rm B^{*\pm}$} & \ $(1.3\pm 0.2)$ & $\times~ 10^{-5}$ & EFT+NRQM~\cite{LHCHiggsCrossSectionWorkingGroup:2016ypw} & \multirow[c]{2}{*}{--} & \multirow[c]{2}{*}{--} & \multirow[c]{2}{*}{\textcolor{green}{\Checkmark}} & \multirow[c]{2}{*}{\textcolor{green}{\Checkmark}} \\
 &  &  & \ $(1.41\pm 0.36)$ & $\times~ 10^{-10}$ & EFT+LCDA~\cite{Alte:2016yuw} &  &  &  &  \\
\cline{3-10}
 &  & \multirow[c]{2}{*}{$\rm B_c^\pm$} & \ $(1.6\pm 0.2)$ & $\times~ 10^{-8}$ & EFT+NRQM~\cite{LHCHiggsCrossSectionWorkingGroup:2016ypw} & \multirow[c]{2}{*}{--} & \multirow[c]{2}{*}{--} & \multirow[c]{2}{*}{\textcolor{red}{\ding{56}}} & \multirow[c]{2}{*}{\textcolor{green}{\Checkmark}} \\
 &  &  & \ $(8.21\pm 0.83)$ & $\times~ 10^{-8}$ & EFT+LCDA~\cite{Alte:2016yuw} &  &  &  &  \\

\bottomrule
\end{tabular}

%% file: tables/exclusive_H_decays_boson_leptonium.tex
\begin{tabular}{l<{\hspace{-4mm}}l<{\hspace{-2mm}}>{\hspace{-1mm}}lr<{\hspace{-6mm}}llccccc}
\toprule
 &  &  &  & \multicolumn{2}{c}{} & \multicolumn{2}{c}{Exp. limits} & \multicolumn{2}{c}{Producible at} \\
$\rm H\to \, V$ & $+$ & $\rm (\ell\ell)$ & \multicolumn{2}{c}{Branching fraction} & Framework & 2024 & HL-LHC & FCC-ee & FCC-hh \\
\midrule
\multirow[c]{6}{*}{$\rm H\, \to \,\gamma$} & \multirow[c]{6}{*}{+} & \multirow[c]{2}{*}{$\rm (ee)_1$} & \ $3.5$ & $\times~ 10^{-12}$ & Eq. \eqref{eq:Gamma_H_ll_gamma}~(this work) & \multirow[c]{2}{*}{--} & \multirow[c]{2}{*}{--} & \multirow[c]{2}{*}{\textcolor{red}{\ding{56}}} & \multirow[c]{2}{*}{\textcolor{red}{\ding{56}}} \\
 &  &  & \ $1.1$ & $\times~ 10^{-11}$ & RQM~\cite{Martynenko:2024rfj} &  &  &  &  \\
\cline{3-10}
 &  & \multirow[c]{2}{*}{$\rm (\mu\mu)_1$} & \ $3.5$ & $\times~ 10^{-12}$ & Eq. \eqref{eq:Gamma_H_ll_gamma}~(this work) & \multirow[c]{2}{*}{--} & \multirow[c]{2}{*}{--} & \multirow[c]{2}{*}{\textcolor{red}{\ding{56}}} & \multirow[c]{2}{*}{\textcolor{red}{\ding{56}}} \\
 &  &  & \ $1.1$ & $\times~ 10^{-11}$ & RQM~\cite{Martynenko:2024rfj} &  &  &  &  \\
\cline{3-10}
 &  & \multirow[c]{2}{*}{$\rm (\tau\tau)_1$} & \ $2.2$ & $\times~ 10^{-12}$ & Eq. \eqref{eq:Gamma_H_ll_gamma}~(this work) & \multirow[c]{2}{*}{--} & \multirow[c]{2}{*}{--} & \multirow[c]{2}{*}{\textcolor{red}{\ding{56}}} & \multirow[c]{2}{*}{\textcolor{red}{\ding{56}}} \\
 &  &  & \ $3.5$ & $\times~ 10^{-12}$ & RQM~\cite{Martynenko:2024rfj} &  &  &  &  \\
\cline{1-10}
\multirow[c]{9}{*}{$\rm H\, \to \,Z$} & \multirow[c]{9}{*}{+} & \multirow[c]{2}{*}{$\rm (ee)_1$} & \ $5.2$ & $\times~ 10^{-13}$ & Eq. \eqref{eq:Gamma_H_ll_Z}~(this work) & \multirow[c]{2}{*}{--} & \multirow[c]{2}{*}{--} & \multirow[c]{2}{*}{\textcolor{red}{\ding{56}}} & \multirow[c]{2}{*}{\textcolor{red}{\ding{56}}} \\
 &  &  & \ $7.9$ & $\times~ 10^{-13}$ & RQM~\cite{Martynenko:2024rfj} &  &  &  &  \\
\cline{3-10}
 &  & \multirow[c]{2}{*}{$\rm (\mu\mu)_1$} & \ $5.7$ & $\times~ 10^{-13}$ & Eq. \eqref{eq:Gamma_H_ll_Z}~(this work) & \multirow[c]{2}{*}{--} & \multirow[c]{2}{*}{--} & \multirow[c]{2}{*}{\textcolor{red}{\ding{56}}} & \multirow[c]{2}{*}{\textcolor{red}{\ding{56}}} \\
 &  &  & \ $9.8$ & $\times~ 10^{-13}$ & RQM~\cite{Martynenko:2024rfj} &  &  &  &  \\
\cline{3-10}
 &  & \multirow[c]{2}{*}{$\rm (\tau\tau)_1$} & \ $1.4$ & $\times~ 10^{-11}$ & Eq. \eqref{eq:Gamma_H_ll_Z}~(this work) & \multirow[c]{2}{*}{--} & \multirow[c]{2}{*}{--} & \multirow[c]{2}{*}{\textcolor{red}{\ding{56}}} & \multirow[c]{2}{*}{\textcolor{red}{\ding{56}}} \\
 &  &  & \ $5.7$ & $\times~ 10^{-11}$ & RQM~\cite{Martynenko:2024rfj} &  &  &  &  \\
\cline{3-10}
 &  & $\rm (ee)_0$ & \ $2.7$ & $\times~ 10^{-16}$ & Eq. \eqref{eq:Gamma_H_ll_Z}~(this work) & -- & -- & \textcolor{red}{\ding{56}} & \textcolor{red}{\ding{56}} \\

 &  & $\rm (\mu\mu)_0$ & \ $1.1$ & $\times~ 10^{-14}$ & Eq. \eqref{eq:Gamma_H_ll_Z}~(this work) & -- & -- & \textcolor{red}{\ding{56}} & \textcolor{red}{\ding{56}} \\

 &  & $\rm (\tau\tau)_0$ & \ $3.2$ & $\times~ 10^{-12}$ & Eq. \eqref{eq:Gamma_H_ll_Z}~(this work) & -- & -- & \textcolor{red}{\ding{56}} & \textcolor{red}{\ding{56}} \\

\bottomrule
\end{tabular}

%% file: tables/exclusive_H_decays_meson_meson.tex
\begin{tabular}{ll<{\hspace{-4mm}}>{\hspace{-2mm}}l<{\hspace{-2mm}}>{\hspace{-1mm}}lr<{\hspace{-6mm}}llccccc}
\toprule
 &  &  &  &  & \multicolumn{2}{c}{} & \multicolumn{2}{c}{Exp. limits} & \multicolumn{2}{c}{Producible at} \\
$\rm H\, \to $ & M  & + & M & \multicolumn{2}{c}{Branching fraction} & Framework & 2024 & HL-LHC & FCC-ee & FCC-hh \\
\midrule
\multirow[c]{23}{*}{$\rm H\,\to$} & $\rm \phi$ & + & $\rm J/\psi$ & \ $1.0$ & $\times~ 10^{-9}$ & LC+LCDA~\cite{Kartvelishvili:2008tz} & -- & -- & \textcolor{red}{\ding{56}} & \textcolor{green}{\Checkmark} \\

 & \multirow[c]{6}{*}{$\rm J/\psi$} & \multirow[c]{6}{*}{+} & $\rm \eta_c$ & \ $1.9$ & $\times~ 10^{-15}$ & RQM~\cite{Martynenko:2025nsg} & -- & -- & \textcolor{red}{\ding{56}} & \textcolor{red}{\ding{56}} \\
\cline{4-11}
 &  &  & \multirow[c]{5}{*}{$\rm J/\psi$} & \ $1.7$ & $\times~ 10^{-10}$ & RQM~\cite{Faustov:2022jfk} & \multirow[c]{5}{*}{$<$\ $3.8$$\times~ 10^{-4}$~\cite{CMS:2022fsq}} & \multirow[c]{5}{*}{$\lesssim5.8\times~ 10^{-5}$} & \multirow[c]{5}{*}{\textcolor{red}{\ding{56}}} & \multirow[c]{5}{*}{\textcolor{green}{\Checkmark}} \\
 &  &  &  & \ $(5.8\text{ -- }6.0)$ & $\times~ 10^{-9}$ & NRQCD+LDME~\cite{Belov:2023iop} &  &  &  &  \\
 &  &  &  & \ $2.1$ & $\times~ 10^{-10}$ & RQM~\cite{Faustov:2023phi} &  &  &  &  \\
 &  &  &  & \ $(5.9\pm 2.3)$ & $\times~ 10^{-10}$ & NRQCD/NRCSM~\cite{Gao:2022iam} &  &  &  &  \\
 &  &  &  & \ $1.5$ & $\times~ 10^{-10}$ & LC+LCDA~\cite{Kartvelishvili:2008tz} &  &  &  &  \\
\cline{2-11}
 & \multirow[c]{2}{*}{$\rm \psi(2S)$} & \multirow[c]{2}{*}{+} & $\rm J/\psi$ & \ $\mathcal{O}(5)$ & $\times~ 10^{-11}$ & -- & $<$\ $2.1$$\times~ 10^{-3}$~\cite{CMS:2022fsq} & $\lesssim3.2\times~ 10^{-4}$ & \textcolor{red}{\ding{56}} & \textcolor{green}{\Checkmark} \\

 &  &  & $\rm \psi(2S)$ & \ $(5.1\pm 2.0)$ & $\times~ 10^{-11}$ & NRQCD/NRCSM~\cite{Gao:2022iam} & $<$\ $3.0$$\times~ 10^{-3}$~\cite{CMS:2022fsq} & $\lesssim4.5\times~ 10^{-4}$ & \textcolor{red}{\ding{56}} & \textcolor{green}{\Checkmark} \\
\cline{2-11}
 & $\rm \chi_{c1}$ & + & $\rm \eta_c$ & \ $1.9$ & $\times~ 10^{-15}$ & RQM~\cite{Martynenko:2025nsg} & -- & -- & \textcolor{red}{\ding{56}} & \textcolor{red}{\ding{56}} \\

 & $\rm h_c$ & + & $\rm J/\psi$ & \ $1.9$ & $\times~ 10^{-15}$ & RQM~\cite{Martynenko:2025nsg} & -- & -- & \textcolor{red}{\ding{56}} & \textcolor{red}{\ding{56}} \\

 & $\rm B_c^\mp$ & + & $\rm B_c^\pm$ & \ $(2.0\text{ -- }3.0)$ & $\times~ 10^{-10}$ & RQM~\cite{Belov:2021toy} & -- & -- & \textcolor{red}{\ding{56}} & \textcolor{green}{\Checkmark} \\

 & $\rm B_c^{*\mp}$ & + & $\rm B_c^{*\pm}$ & \ $(1.4\text{ -- }1.7)$ & $\times~ 10^{-10}$ & RQM~\cite{Belov:2021toy} & -- & -- & \textcolor{red}{\ding{56}} & \textcolor{green}{\Checkmark} \\
\cline{2-11}
 & \multirow[c]{7}{*}{$\rm \Upsilon(1S)$} & \multirow[c]{7}{*}{+} & \multirow[c]{2}{*}{$\rm J/\psi$} & \ $(2.7\text{ -- }3.6)$ & $\times~ 10^{-10}$ & NRQCD+LDME~\cite{Belov:2023iop} & \multirow[c]{2}{*}{--} & \multirow[c]{2}{*}{--} & \multirow[c]{2}{*}{\textcolor{red}{\ding{56}}} & \multirow[c]{2}{*}{\textcolor{green}{\Checkmark}} \\
 &  &  &  & \ $1.6$ & $\times~ 10^{-11}$ & LC+LCDA~\cite{Kartvelishvili:2008tz} &  &  &  &  \\
\cline{4-11}
 &  &  & \multirow[c]{5}{*}{$\rm \Upsilon(1S)$} & \ $1.8$ & $\times~ 10^{-10}$ & RQM~\cite{Faustov:2022jfk} & \multirow[c]{5}{*}{$<$\ $1.7$$\times~ 10^{-3}$~\cite{CMS:2022fsq}} & \multirow[c]{5}{*}{$\lesssim2.6\times~ 10^{-4}$} & \multirow[c]{5}{*}{\textcolor{red}{\ding{56}}} & \multirow[c]{5}{*}{\textcolor{green}{\Checkmark}} \\
 &  &  &  & \ $(8.5\text{ -- }9.2)$ & $\times~ 10^{-10}$ & NRQCD+LDME~\cite{Belov:2023iop} &  &  &  &  \\
 &  &  &  & \ $2.3$ & $\times~ 10^{-9}$ & RQM~\cite{Faustov:2023phi} &  &  &  &  \\
 &  &  &  & \ $(4.3\pm 0.9)$ & $\times~ 10^{-10}$ & NRQCD/NRCSM~\cite{Gao:2022iam} &  &  &  &  \\
 &  &  &  & \ $2.3$ & $\times~ 10^{-9}$ & LC+LCDA~\cite{Kartvelishvili:2008tz} &  &  &  &  \\
\cline{2-11}
 & $\rm \Upsilon(2S)$ & + & $\rm \Upsilon(2S)$ & \ $(1.0\pm 0.2)$ & $\times~ 10^{-10}$ & NRQCD/NRCSM~\cite{Gao:2022iam} & -- & -- & \textcolor{red}{\ding{56}} & \textcolor{green}{\Checkmark} \\

 & $\rm \Upsilon(3S)$ & + & $\rm \Upsilon(3S)$ & \ $(5.7\pm 1.2)$ & $\times~ 10^{-11}$ & NRQCD/NRCSM~\cite{Gao:2022iam} & -- & -- & \textcolor{red}{\ding{56}} & \textcolor{green}{\Checkmark} \\

 & $\rm \Upsilon(mS)$ & + & $\rm \Upsilon(nS)$ & \  &  & -- & $<$\ $3.5$$\times~ 10^{-4}$~\cite{CMS:2022fsq} & $\lesssim 9.2\times~ 10^{-6}$~\cite{CMS:2022kdd} & \textcolor{red}{\ding{56}} & \textcolor{red}{\ding{56}} \\

\bottomrule
\end{tabular}

%% file: tables/exclusive_Z_decays_not_gamma_meson.tex
\begin{tabular}{lr@{\hskip 2mm}llccc}
\toprule
 &  & \multicolumn{2}{c}{} & \multicolumn{2}{c}{Exp. limits} & Producible at \\
  & \multicolumn{2}{c}{Branching fraction} & Framework & 2024 & HL-LHC & FCC-ee \\
\midrule
\multirow[c]{2}{*}{$\rm Z\; \to \;g+g+g$} & \ $1.9$ & $\times~ 10^{-6}$ & NLO SM~\cite{Hopker:1993pb} & \multirow[c]{2}{*}{$<$\ $1.1$$\times~ 10^{-2}$~\cite{DELPHI:1996bav}} & \multirow[c]{2}{*}{--} & \multirow[c]{2}{*}{\textcolor{green}{\Checkmark}} \\
 & \ $1.8$ & $\times~ 10^{-6}$ & loop SM, \mgshort~(this work) &  &  &  \\
\cline{1-7}
\multirow[c]{2}{*}{$\rm Z\; \to \;g+g+\gamma$} & \ $7.0$ & $\times~ 10^{-7}$ & NLO SM\footnote{With updated SM parameters. The original result was $\BR = 1.8 \times 10^{-6}$ \cite{Laursen:1982rg}}~\cite{Laursen:1982rg} & \multirow[c]{2}{*}{--} & \multirow[c]{2}{*}{--} & \multirow[c]{2}{*}{\textcolor{green}{\Checkmark}} \\
 & \ $6.9$ & $\times~ 10^{-7}$ & loop SM, \mgshort~(this work) &  &  &  \\
\cline{1-7}
$\rm Z\; \to \;g+g+\gamma_\mathrm{soft}$ & \ $1.2$ & $\times~ 10^{-9}$ & loop SM, \mgshort~(this work) & -- & -- & \textcolor{green}{\Checkmark} \\
\cline{1-7}
\multirow[c]{2}{*}{$\rm Z\; \to \;\gamma+\gamma+\gamma$} & \ $5.4$ & $\times~ 10^{-10}$ & NLO SM~\cite{Glover:1993nv} & \multirow[c]{2}{*}{$<$\ $2.2$$\times~ 10^{-6}$~\cite{ATLAS:2015rsn}} & \multirow[c]{2}{*}{$\lesssim1.3\times~ 10^{-7}$} & \multirow[c]{2}{*}{\textcolor{green}{\Checkmark}} \\
 & \ $6.6$ & $\times~ 10^{-10}$ & loop SM, \mgshort~(this work) &  &  &  \\
\cline{1-7}
$\rm Z\; \to \;\gamma+\gamma+\gamma_\mathrm{soft}$ & \ $2.2$ & $\times~ 10^{-12}$ & loop SM, \mgshort~(this work) & $<$\ $1.5$$\times~ 10^{-5}$~\cite{CDF:2013lma} & $\lesssim2.1\times~ 10^{-7}$ & \textcolor{green}{\Checkmark} \\
\cline{1-7}
\multirow[c]{2}{*}{$\rm Z\; \to \;\gamma+\nu+\hat{\nu}$} & \ $7.2$ & $\times~ 10^{-10}$ & loop SM~\cite{Hernandez:1999xn} & \multirow[c]{2}{*}{--} & \multirow[c]{2}{*}{--} & \multirow[c]{2}{*}{\textcolor{green}{\Checkmark}} \\
 & \ $1.4$ & $\times~ 10^{-10}$ & loop SM, \mgshort~(this work) &  &  &  \\

\bottomrule
\end{tabular}

%% file: tables/exclusive_Z_decays_gamma_meson_light_quarks.tex
\begin{tabular}{l<{\hspace{-4mm}}l<{\hspace{-2mm}}>{\hspace{-1mm}}lr<{\hspace{-6mm}}llccc}
\toprule
 &  &  &  & \multicolumn{2}{c}{} & \multicolumn{2}{c}{Exp. limits} & Producible at \\
$\rm Z\,\to\, \gamma$ & $+$ & $\rm M$ & \multicolumn{2}{c}{Branching fraction} & Framework & 2024 & HL-LHC & FCC-ee \\
\midrule
\multirow[c]{7}{*}{$\rm Z\,\to \,\gamma$} & \multirow[c]{7}{*}{+} & $\rm \pi^0$ & \ $(9.8\pm 1.0)$ & $\times~ 10^{-12}$ & SCET+LCDA~\cite{Grossman:2015cak} & $<$\ $2.0$$\times~ 10^{-5}$~\cite{CDF:2013lma} & $\lesssim2.8\times~ 10^{-7}$ & \textcolor{green}{\Checkmark} \\

 &  & $\rm \eta$ & \ $(1.0\text{ -- }17.0)$ & $\times~ 10^{-10}$ & SCET+LCDA~\cite{Alte:2015dpo} & $<$\ $5.1$$\times~ 10^{-5}$~\cite{ALEPH:1991qhf} & -- & \textcolor{green}{\Checkmark} \\

 &  & $\rm \rho^0$ & \ $(4.19\pm 0.47)$ & $\times~ 10^{-9}$ & SCET+LCDA~\cite{Grossman:2015cak} & $<$\ $4.0$$\times~ 10^{-6}$~\cite{ATLAS:2017gko} & $\lesssim2.9\times~ 10^{-7}$ & \textcolor{green}{\Checkmark} \\

 &  & $\rm \omega$ & \ $(2.82\pm 0.41)$ & $\times~ 10^{-8}$ & SCET+LCDA~\cite{Grossman:2015cak} & $<$\ $3.9$$\times~ 10^{-6}$~\cite{ATLAS:2023alf} & $\lesssim5.8\times~ 10^{-7}$ & \textcolor{green}{\Checkmark} \\

 &  & $\rm \eta'$ & \ $(3.1\text{ -- }4.8)$ & $\times~ 10^{-9}$ & SCET+LCDA~\cite{Alte:2015dpo} & $<$\ $4.2$$\times~ 10^{-5}$~\cite{ALEPH:1991qhf} & -- & \textcolor{green}{\Checkmark} \\
\cline{3-9}
 &  & \multirow[c]{2}{*}{$\rm \phi$} & \ $(1.04\pm 0.12)$ & $\times~ 10^{-8}$ & SCET+LCDA~\cite{Grossman:2015cak} & \multirow[c]{2}{*}{$<$\ $7.0$$\times~ 10^{-7}$~\cite{ATLAS:2017gko}} & \multirow[c]{2}{*}{$\lesssim5.1\times~ 10^{-8}$} & \multirow[c]{2}{*}{\textcolor{green}{\Checkmark}} \\
 &  &  & \ $(1.17\pm 0.08)$ & $\times~ 10^{-8}$ & LC+LCDA~\cite{Huang:2014cxa} &  &  &  \\

\bottomrule
\end{tabular}

%% file: tables/exclusive_Z_decays_gamma_meson_charmonium.tex
\begin{tabular}{l<{\hspace{-4mm}}l<{\hspace{-2mm}}>{\hspace{-1mm}}lr<{\hspace{-6mm}}llccc}
\toprule
 &  &  &  & \multicolumn{2}{c}{} & \multicolumn{2}{c}{Exp. limits} & Producible at \\
$\rm Z\,\to\, \gamma$ & $+$ & $\rm M$ & \multicolumn{2}{c}{Branching fraction} & Framework & 2024 & HL-LHC & FCC-ee \\
\midrule
\multirow[c]{25}{*}{$\rm Z\,\to \,\gamma$} & \multirow[c]{25}{*}{+} & \multirow[c]{5}{*}{$\rm \eta_c$} & \ $\left(1.32^{+0.56}_{-0.54}\right)$ & $\times~ 10^{-8}$ & NRQCD+LDME~\cite{Wang:2023ssg} & \multirow[c]{5}{*}{--} & \multirow[c]{5}{*}{--} & \multirow[c]{5}{*}{\textcolor{green}{\Checkmark}} \\
 &  &  & \ $(9.5\pm 0.2)$ & $\times~ 10^{-9}$ & NRQCD (NNLO+NLL)~\cite{Sang:2023hjl} &  &  &  \\
 &  &  & \ $(7.42\pm 0.61)$ & $\times~ 10^{-9}$ & NRQCD~(NLO+NLL)~\cite{Chung:2019ota} &  &  &  \\
 &  &  & \ $6.6$ & $\times~ 10^{-9}$ & NRQCD+LDME~\cite{Luchinsky:2017jab} &  &  &  \\
 &  &  & \ $(9.4\pm 1.0)$ & $\times~ 10^{-9}$ & LC+LCDA~\cite{Luchinsky:2017jab} &  &  &  \\
\cline{3-9}
 &  & \multirow[c]{7}{*}{$\rm J/\psi$} & \ $\left(1.04^{+0.18}_{-0.16}\right)$ & $\times~ 10^{-7}$ & NRQCD+LDME~\cite{Wang:2023ssg} & \multirow[c]{7}{*}{$<$\ $6.0$$\times~ 10^{-7}$~\cite{CMS:2024hhg,CMS:2018gcm}} & \multirow[c]{7}{*}{$\lesssim 3.1\times~ 10^{-7}$~\cite{ATLAS:2015xkp}} & \multirow[c]{7}{*}{\textcolor{green}{\Checkmark}} \\
 &  &  & \ $\left(5.75^{+0.08}_{-0.09}\right)$ & $\times~ 10^{-8}$ & NRQCD (NNLO+NLL)~\cite{Sang:2023hjl} &  &  &  \\
 &  &  & \ $\left(8.96^{+1.51}_{-1.38}\right)$ & $\times~ 10^{-8}$ & LC+LCDA~\cite{Bodwin:2017pzj} &  &  &  \\
 &  &  & \ $4.5$ & $\times~ 10^{-8}$ & NRQCD+LDME~\cite{Luchinsky:2017jab} &  &  &  \\
 &  &  & \ $(8.8\pm 0.9)$ & $\times~ 10^{-8}$ & LC+LCDA~\cite{Luchinsky:2017jab} &  &  &  \\
 &  &  & \ $(8.02\pm 0.45)$ & $\times~ 10^{-8}$ & SCET+LCDA~\cite{Grossman:2015cak} &  &  &  \\
 &  &  & \ $(9.96\pm 1.86)$ & $\times~ 10^{-8}$ & NRQCD+LDME~\cite{Huang:2014cxa} &  &  &  \\
\cline{3-9}
 &  & $\rm \psi(2S)$ & \ $\left(4.83^{+1.02}_{-0.91}\right)$ & $\times~ 10^{-8}$ & SCET+LCDA\footnote{private communication with author of Ref. \cite{Bodwin:2017wdu}}~\cite{CMS:2024hhg} & $<$\ $1.3$$\times~ 10^{-6}$~\cite{CMS:2024hhg,CMS:2018gcm} & $\lesssim1.0\times~ 10^{-7}$ & \textcolor{green}{\Checkmark} \\
\cline{3-9}
 &  & \multirow[c]{3}{*}{$\rm \chi_{c0}$} & \ $(3.74\pm 0.05)$ & $\times~ 10^{-10}$ & NRQCD+LDME~\cite{Sang:2022erv} & \multirow[c]{3}{*}{--} & \multirow[c]{3}{*}{--} & \multirow[c]{3}{*}{\textcolor{green}{\Checkmark}} \\
 &  &  & \ $1.4$ & $\times~ 10^{-10}$ & NRQCD+LDME~\cite{Luchinsky:2017jab} &  &  &  \\
 &  &  & \ $(5.0\pm 2.0)$ & $\times~ 10^{-10}$ & LC+LCDA~\cite{Luchinsky:2017jab} &  &  &  \\
\cline{3-9}
 &  & \multirow[c]{3}{*}{$\rm \chi_{c1}$} & \ $\left(2.383^{+0.014}_{-0.017}\right)$ & $\times~ 10^{-9}$ & NRQCD+LDME~\cite{Sang:2022erv} & \multirow[c]{3}{*}{--} & \multirow[c]{3}{*}{--} & \multirow[c]{3}{*}{\textcolor{green}{\Checkmark}} \\
 &  &  & \ $8.7$ & $\times~ 10^{-10}$ & NRQCD+LDME~\cite{Luchinsky:2017jab} &  &  &  \\
 &  &  & \ $(5.6\pm 2.0)$ & $\times~ 10^{-9}$ & LC+LCDA~\cite{Luchinsky:2017jab} &  &  &  \\
\cline{3-9}
 &  & \multirow[c]{3}{*}{$\rm h_c$} & \ $\left(3.487^{+0.206}_{-0.230}\right)$ & $\times~ 10^{-9}$ & NRQCD+LDME~\cite{Sang:2022erv} & \multirow[c]{3}{*}{--} & \multirow[c]{3}{*}{--} & \multirow[c]{3}{*}{\textcolor{green}{\Checkmark}} \\
 &  &  & \ $3.0$ & $\times~ 10^{-9}$ & NRQCD+LDME~\cite{Luchinsky:2017jab} &  &  &  \\
 &  &  & \ $(1.0\pm 0.4)$ & $\times~ 10^{-8}$ & LC+LCDA~\cite{Luchinsky:2017jab} &  &  &  \\
\cline{3-9}
 &  & \multirow[c]{3}{*}{$\rm \chi_{c2}$} & \ $\left(3.38^{+0.19}_{-0.22}\right)$ & $\times~ 10^{-10}$ & NRQCD+LDME~\cite{Sang:2022erv} & \multirow[c]{3}{*}{--} & \multirow[c]{3}{*}{--} & \multirow[c]{3}{*}{\textcolor{green}{\Checkmark}} \\
 &  &  & \ $2.9$ & $\times~ 10^{-10}$ & NRQCD+LDME~\cite{Luchinsky:2017jab} &  &  &  \\
 &  &  & \ $(1.0\pm 0.4)$ & $\times~ 10^{-9}$ & LC+LCDA~\cite{Luchinsky:2017jab} &  &  &  \\

\bottomrule
\end{tabular}

%% file: tables/exclusive_Z_decays_gamma_meson_bottomonium.tex
\begin{tabular}{l<{\hspace{-4mm}}l<{\hspace{-2mm}}>{\hspace{-1mm}}lr<{\hspace{-6mm}}llccc}
\toprule
 &  &  &  & \multicolumn{2}{c}{} & \multicolumn{2}{c}{Exp. limits} & Producible at \\
$\rm Z\,\to\, \gamma$ & $+$ & $\rm M$ & \multicolumn{2}{c}{Branching fraction} & Framework & 2024 & HL-LHC & FCC-ee \\
\midrule
\multirow[c]{19}{*}{$\rm Z\,\to \,\gamma$} & \multirow[c]{19}{*}{+} & \multirow[c]{3}{*}{$\rm \eta_b$} & \ $\left(2.79^{+0.20}_{-0.19}\right)$ & $\times~ 10^{-8}$ & NRQCD+LDME~\cite{Wang:2023ssg} & \multirow[c]{3}{*}{--} & \multirow[c]{3}{*}{--} & \multirow[c]{3}{*}{\textcolor{green}{\Checkmark}} \\
 &  &  & \ $(2.43\pm 0.01)$ & $\times~ 10^{-8}$ & NRQCD (NNLO+NLL)~\cite{Sang:2023hjl} &  &  &  \\
 &  &  & \ $(2.8\pm 0.5)$ & $\times~ 10^{-8}$ & NRQCD~(NLO+NLL)~\cite{Chung:2019ota} &  &  &  \\
\cline{3-9}
 &  & \multirow[c]{6}{*}{$\rm \Upsilon(1S)$} & \ $(5.41\pm 0.39)$ & $\times~ 10^{-8}$ & NRQCD+LDME~\cite{Wang:2023ssg} & \multirow[c]{6}{*}{$<$\ $1.1$$\times~ 10^{-6}$~\cite{ATLAS:2022rej}} & \multirow[c]{6}{*}{$\lesssim1.7\times~ 10^{-7}$} & \multirow[c]{6}{*}{\textcolor{green}{\Checkmark}} \\
 &  &  & \ $(4.63\pm 0.02)$ & $\times~ 10^{-8}$ & NRQCD (NNLO+NLL)~\cite{Sang:2023hjl} &  &  &  \\
 &  &  & \ $(5.61\pm 0.29)$ & $\times~ 10^{-8}$ & NRQCD+LDME~\cite{Dong:2022ayy} &  &  &  \\
 &  &  & \ $\left(4.8^{+0.3}_{-0.2}\right)$ & $\times~ 10^{-8}$ & LC+LCDA~\cite{Bodwin:2017pzj} &  &  &  \\
 &  &  & \ $(5.39\pm 0.16)$ & $\times~ 10^{-8}$ & SCET+LCDA~\cite{Grossman:2015cak} &  &  &  \\
 &  &  & \ $(4.93\pm 0.51)$ & $\times~ 10^{-8}$ & NRQCD+LDME~\cite{Huang:2014cxa} &  &  &  \\
\cline{3-9}
 &  & \multirow[c]{2}{*}{$\rm \Upsilon(2S)$} & \ $(2.66\pm 0.31)$ & $\times~ 10^{-8}$ & NRQCD+LDME~\cite{Dong:2022ayy} & \multirow[c]{2}{*}{$<$\ $1.3$$\times~ 10^{-6}$~\cite{ATLAS:2022rej}} & \multirow[c]{2}{*}{$\lesssim2.0\times~ 10^{-7}$} & \multirow[c]{2}{*}{\textcolor{green}{\Checkmark}} \\
 &  &  & \ $\left(2.44^{+0.14}_{-0.13}\right)$ & $\times~ 10^{-8}$ & LC+LCDA~\cite{Bodwin:2017pzj} &  &  &  \\
\cline{3-9}
 &  & \multirow[c]{2}{*}{$\rm \Upsilon(3S)$} & \ $(1.93\pm 0.25)$ & $\times~ 10^{-8}$ & NRQCD+LDME~\cite{Dong:2022ayy} & \multirow[c]{2}{*}{$<$\ $2.4$$\times~ 10^{-6}$~\cite{ATLAS:2022rej}} & \multirow[c]{2}{*}{$\lesssim3.7\times~ 10^{-7}$} & \multirow[c]{2}{*}{\textcolor{green}{\Checkmark}} \\
 &  &  & \ $\left(1.88^{+0.11}_{-0.10}\right)$ & $\times~ 10^{-8}$ & LC+LCDA~\cite{Bodwin:2017pzj} &  &  &  \\
\cline{3-9}
 &  & $\rm \Upsilon(4S)$ & \ $(1.22\pm 0.13)$ & $\times~ 10^{-8}$ & SCET+LCDA~\cite{Grossman:2015cak} & -- & -- & \textcolor{green}{\Checkmark} \\

 &  & $\rm \Upsilon(nS)$ & \ $\left(9.96^{+0.28}_{-0.26}\right)$ & $\times~ 10^{-8}$ & SCET+LCDA~\cite{Grossman:2015cak} & -- & -- & \textcolor{green}{\Checkmark} \\

 &  & $\rm \chi_{b0}$ & \ $\left(2.73^{+0.05}_{-0.04}\right)$ & $\times~ 10^{-10}$ & NRQCD+LDME~\cite{Sang:2022erv} & -- & -- & \textcolor{green}{\Checkmark} \\

 &  & $\rm \chi_{b1}$ & \ $\left(1.473^{+0.010}_{-0.011}\right)$ & $\times~ 10^{-9}$ & NRQCD+LDME~\cite{Sang:2022erv} & -- & -- & \textcolor{green}{\Checkmark} \\

 &  & $\rm h_b$ & \ $\left(9.27^{+0.36}_{-0.41}\right)$ & $\times~ 10^{-10}$ & NRQCD+LDME~\cite{Sang:2022erv} & -- & -- & \textcolor{green}{\Checkmark} \\

 &  & $\rm \chi_{b2}$ & \ $\left(2.92^{+0.12}_{-0.14}\right)$ & $\times~ 10^{-10}$ & NRQCD+LDME~\cite{Sang:2022erv} & -- & -- & \textcolor{green}{\Checkmark} \\

\bottomrule
\end{tabular}

%% file: tables/exclusive_Z_decays_gamma_meson_flavoured.tex
\begin{tabular}{l<{\hspace{-4mm}}l<{\hspace{-2mm}}>{\hspace{-1mm}}lr<{\hspace{-6mm}}llccc}
\toprule
 &  &  &  & \multicolumn{2}{c}{} & \multicolumn{2}{c}{Exp. limits} & Producible at \\
$\rm Z\,\to\, \gamma$ & $+$ & $\rm M$ & \multicolumn{2}{c}{Branching fraction} & Framework & 2024 & HL-LHC & FCC-ee \\
\midrule
\multirow[c]{5}{*}{$\rm Z\,\to \,\gamma$} & \multirow[c]{5}{*}{+} & $\rm K^0$ & \ $3.3$ & $\times~ 10^{-20}$ & SCET+LCDA~(this work) & $<$\ $3.0$$\times~ 10^{-6}$~\cite{ATLAS:2024dpw} & $\lesssim4.5\times~ 10^{-7}$ & \textcolor{red}{\ding{56}} \\
\cline{3-9}
 &  & \multirow[c]{2}{*}{$\rm D^0$} & <\ $1.0$ & $\times~ 10^{-15}$ & SCET+LCDA~\cite{Grossman:2015cak,LHCb:2022kta} & \multirow[c]{2}{*}{$<$\ $4.0$$\times~ 10^{-6}$~\cite{ATLAS:2024dpw}} & \multirow[c]{2}{*}{$\lesssim6.0\times~ 10^{-7}$} & \multirow[c]{2}{*}{\textcolor{red}{\ding{56}}} \\
 &  &  & \ $1.4$ & $\times~ 10^{-25}$ & SCET+LCDA~(this work) &  &  &  \\
\cline{3-9}
 &  & $\rm B^0$ & \ $8.3$ & $\times~ 10^{-17}$ & SCET+LCDA~(this work) & -- & -- & \textcolor{red}{\ding{56}} \\

 &  & $\rm B_s^0$ & \ $2.3$ & $\times~ 10^{-15}$ & SCET+LCDA~(this work) & -- & -- & \textcolor{red}{\ding{56}} \\

\bottomrule
\end{tabular}

%% file: tables/exclusive_Z_decays_3_gammas.tex
\begin{tabular}{lllr@{\hskip 2mm}lcccc}
\toprule
$\rm Z\,\to \,\gamma$ & + & $\rm M(\gamma\gamma)$ & \multicolumn{2}{c}{Branching fraction} \\
\midrule
\multirow[c]{9}{*}{$\rm Z\,\to \,\gamma$} & \multirow[c]{9}{*}{+} & $\rm \pi^0$$(\gamma\gamma)$ & $9.7$ & $\times~ 10^{-12}$ \\
 &  & $\rm \eta$$(\gamma\gamma)$ & $6.3$ & $\times~ 10^{-11}$ \\
 &  & $\rm \eta'$$(\gamma\gamma)$ & $1.1$ & $\times~ 10^{-10}$ \\
 &  & $\rm \eta_c$$(\gamma\gamma)$ & $2.1$ & $\times~ 10^{-12}$ \\
 &  & $\rm \chi_{c0}$$(\gamma\gamma)$ & $7.6$ & $\times~ 10^{-14}$ \\
 &  & $\rm \chi_{c1}$$(\gamma\gamma)$ & $1.5$ & $\times~ 10^{-14}$ \\
 &  & $\rm \chi_{c2}$$(\gamma\gamma)$ & $9.6$ & $\times~ 10^{-14}$ \\
 &  & $\rm \chi_{b0}$$(\gamma\gamma)$ & $1.6$ & $\times~ 10^{-14}$ \\
 &  & $\rm \chi_{b2}$$(\gamma\gamma)$ & $1.6$ & $\times~ 10^{-14}$ \\
\cline{1-5}
Sum &  &  & $1.8$ & $\times~ 10^{-10}$ \\
\cline{1-5}
$Z$ & $\longrightarrow$ & $\gamma\gamma\gamma$ & $6.6$ & $\times~ 10^{-10}$ \\
\cline{1-5}
Total &  &  & $8.4$ & $\times~ 10^{-10}$ \\
\bottomrule
\end{tabular}

%% file: tables/exclusive_Z_decays_W_meson.tex
\begin{tabular}{l<{\hspace{-4mm}}l<{\hspace{-2mm}}>{\hspace{-1mm}}lr<{\hspace{-6mm}}llcccc}
\toprule
 &  &  &  & \multicolumn{2}{c}{} & \multicolumn{2}{c}{Exp. limits} & Producible at \\
$\rm Z \,\to\,W$ & $+$ & $\rm M$ & \multicolumn{2}{c}{Branching fraction} & Framework & 2024 & HL-LHC & FCC-ee \\
\midrule
\multirow[c]{12}{*}{$\rm Z\,\to \,W^\mp$} & \multirow[c]{12}{*}{+} & $\rm \pi^\pm$ & \ $(1.51\pm 0.01)$ & $\times~ 10^{-10}$ & SCET+LCDA~\cite{Grossman:2015cak} & $<$\ $7.0$$\times~ 10^{-5}$~\cite{ALEPH:1991qhf} & -- & \textcolor{green}{\Checkmark} \\

 &  & $\rm \rho^\pm$ & \ $(4.0\pm 0.1)$ & $\times~ 10^{-10}$ & SCET+LCDA~\cite{Grossman:2015cak} & $<$\ $8.3$$\times~ 10^{-5}$~\cite{ALEPH:1991qhf} & -- & \textcolor{green}{\Checkmark} \\

 &  & $\rm K^\pm$ & \ $(1.16\pm 0.01)$ & $\times~ 10^{-11}$ & SCET+LCDA~\cite{Grossman:2015cak} & -- & -- & \textcolor{green}{\Checkmark} \\

 &  & $\rm K^{*\pm}$ & \ $(1.96\pm 0.12)$ & $\times~ 10^{-11}$ & SCET+LCDA~\cite{Grossman:2015cak} & -- & -- & \textcolor{green}{\Checkmark} \\

 &  & $\rm D^\pm$ & \ $(1.99\pm 0.17)$ & $\times~ 10^{-11}$ & SCET+LCDA~\cite{Grossman:2015cak} & -- & -- & \textcolor{green}{\Checkmark} \\

 &  & $\rm D_s^\pm$ & \ $(6.04\pm 0.30)$ & $\times~ 10^{-10}$ & SCET+LCDA~\cite{Grossman:2015cak} & -- & -- & \textcolor{green}{\Checkmark} \\

 &  & $\rm D^{*\pm}$ & \ $3.8$ & $\times~ 10^{-11}$ & SCET+LCDA~(this work) & -- & -- & \textcolor{green}{\Checkmark} \\

 &  & $\rm D_s^{*\pm}$ & \ $9.9$ & $\times~ 10^{-10}$ & SCET+LCDA~(this work) & -- & -- & \textcolor{green}{\Checkmark} \\

 &  & $\rm B^\pm$ & \ $5.4$ & $\times~ 10^{-15}$ & SCET+LCDA~(this work) & -- & -- & \textcolor{red}{\ding{56}} \\

 &  & $\rm B^{*\pm}$ & \ $4.8$ & $\times~ 10^{-15}$ & SCET+LCDA~(this work) & -- & -- & \textcolor{red}{\ding{56}} \\

 &  & $\rm B_c^\pm$ & \ $3.1$ & $\times~ 10^{-12}$ & SCET+LCDA~(this work) & -- & -- & \textcolor{green}{\Checkmark} \\

 &  & $\rm B_c^{*\pm}$ & \ $3.1$ & $\times~ 10^{-12}$ & SCET+LCDA~(this work) & -- & -- & \textcolor{green}{\Checkmark} \\

\bottomrule
\end{tabular}

%% file: tables/exclusive_Z_decays_gamma_meson_leptonium.tex
\begin{tabular}{l<{\hspace{-4mm}}l<{\hspace{-2mm}}>{\hspace{-1mm}}lr<{\hspace{-6mm}}llccc}
\toprule
 &  &  &  & \multicolumn{2}{c}{} & \multicolumn{2}{c}{Exp. limits} & Producible at \\
$\rm Z\,\to\, \gamma$ & $+$ & $\rm (\ell\ell)$ & \multicolumn{2}{c}{Branching fraction} & Framework & 2024 & HL-LHC & FCC-ee \\
\midrule
\multirow[c]{6}{*}{$\rm Z\,\to \,\gamma$} & \multirow[c]{6}{*}{+} & $\rm (ee)_1$ & \ $7.3$ & $\times~ 10^{-21}$ & Eq. \eqref{eq:Zgammall1}~(this work) & -- & -- & \textcolor{red}{\ding{56}} \\

 &  & $\rm (\mu\mu)_1$ & \ $3.1$ & $\times~ 10^{-16}$ & Eq. \eqref{eq:Zgammall1}~(this work) & -- & -- & \textcolor{red}{\ding{56}} \\

 &  & $\rm (\tau\tau)_1$ & \ $8.9$ & $\times~ 10^{-14}$ & Eq. \eqref{eq:Zgammall1}~(this work) & -- & -- & \textcolor{red}{\ding{56}} \\

 &  & $\rm (ee)_0$ & \ $4.7$ & $\times~ 10^{-23}$ & Eq. \eqref{eq:Zgammall0}~(this work) & -- & -- & \textcolor{red}{\ding{56}} \\

 &  & $\rm (\mu\mu)_0$ & \ $2.0$ & $\times~ 10^{-18}$ & Eq. \eqref{eq:Zgammall0}~(this work) & -- & -- & \textcolor{red}{\ding{56}} \\

 &  & $\rm (\tau\tau)_0$ & \ $5.7$ & $\times~ 10^{-16}$ & Eq. \eqref{eq:Zgammall0}~(this work) & -- & -- & \textcolor{red}{\ding{56}} \\

\bottomrule
\end{tabular}

%% file: tables/exclusive_Z_decays_light_meson_meson.tex
\begin{tabular}{l>{\hspace{-1mm}}l<{\hspace{-3mm}}>{\hspace{-2mm}}l<{\hspace{-2mm}}>{\hspace{-1mm}}lr<{\hspace{-6mm}}llccccc}
\toprule
 &  &  &  &  & \multicolumn{2}{c}{} & \multicolumn{2}{c}{Exp. limits} & Producible at \\
$Z\,\to$ & M  & $+$ & M & \multicolumn{2}{c}{Branching fraction} & Framework & 2024 & HL-LHC & at FCC-ee \\
\midrule
\multirow[c]{4}{*}{$\rm Z\,\to$} & $\rm \pi^0$ & + & $\rm \pi^0$ & \multicolumn{2}{l}{(forbidden)} & -- & $<$\ $1.5$$\times~ 10^{-5}$~\cite{CDF:2013lma} & $\lesssim2.1\times~ 10^{-7}$ & \textcolor{red}{\ding{56}} \\

 & $\rm \pi^+$ & + & $\rm \pi^-$ & \ $1.7$ & $\times~ 10^{-12}$ & SCET+LCDA~\cite{Cheng:2018khi} & -- & -- & \textcolor{green}{\Checkmark} \\

 & $\rm K^+$ & + & $\rm K^-$ & \ $8.3$ & $\times~ 10^{-13}$ & SCET+LCDA~\cite{Cheng:2018khi} & -- & -- & \textcolor{green}{\Checkmark} \\

 & $\rm \phi$ & + & $\rm \phi$ & \ $2.1$ & $\times~ 10^{-12}$ & EFT+NRQM~(this work) & -- & -- & \textcolor{green}{\Checkmark} \\

\bottomrule
\end{tabular}

%% file: tables/exclusive_Z_decays_charmonium_meson.tex
\begin{tabular}{l>{\hspace{-1mm}}l<{\hspace{-3mm}}>{\hspace{-2mm}}l<{\hspace{-2mm}}>{\hspace{-1mm}}lr<{\hspace{-6mm}}llccccc}
\toprule
 &  &  &  &  & \multicolumn{2}{c}{} & \multicolumn{2}{c}{Exp. limits} & Producible at \\
$Z\,\to$ & M  & $+$ & M & \multicolumn{2}{c}{Branching fraction} & Framework & 2024 & HL-LHC & at FCC-ee \\
\midrule
\multirow[c]{37}{*}{$\rm Z\,\to$} & \multirow[c]{10}{*}{$\rm \eta_c$} & \multirow[c]{10}{*}{+} & \multirow[c]{3}{*}{$\rm J/\psi$} & \ $(1.5\pm 0.4)$ & $\times~ 10^{-11}$ & NRQCD/NRCSM~\cite{Gao:2022mwa} & \multirow[c]{3}{*}{--} & \multirow[c]{3}{*}{--} & \multirow[c]{3}{*}{\textcolor{green}{\Checkmark}} \\
 &  &  &  & \ $(1.8\text{ -- }2.7)$ & $\times~ 10^{-11}$ & NRQCD+LDME (NLO)~\cite{Luo:2022ugd} &  &  &  \\
 &  &  &  & \ $2.7$ & $\times~ 10^{-14}$ & NRQCD+LDME~\cite{Likhoded:2017jmx} &  &  &  \\
\cline{4-10}
 &  &  & \multirow[c]{2}{*}{$\rm \chi_{c0}$} & \ $2.3$ & $\times~ 10^{-12}$ & NRQCD+LDME~\cite{Likhoded:2017jmx} & \multirow[c]{2}{*}{--} & \multirow[c]{2}{*}{--} & \multirow[c]{2}{*}{\textcolor{green}{\Checkmark}} \\
 &  &  &  & \ $(2.3\pm 1.0)$ & $\times~ 10^{-12}$ & LC+LCDA~\cite{Likhoded:2017jmx} &  &  &  \\
\cline{4-10}
 &  &  & $\rm \chi_{c1}$ & \ $5.4$ & $\times~ 10^{-14}$ & NRQCD+LDME~\cite{Likhoded:2017jmx} & -- & -- & \textcolor{red}{\ding{56}} \\
\cline{4-10}
 &  &  & \multirow[c]{2}{*}{$\rm h_c$} & \ $2.1$ & $\times~ 10^{-13}$ & NRQCD+LDME~\cite{Likhoded:2017jmx} & \multirow[c]{2}{*}{--} & \multirow[c]{2}{*}{--} & \multirow[c]{2}{*}{\textcolor{green}{\Checkmark}} \\
 &  &  &  & \ $(1.0\pm 0.5)$ & $\times~ 10^{-12}$ & LC+LCDA~\cite{Likhoded:2017jmx} &  &  &  \\
\cline{4-10}
 &  &  & \multirow[c]{2}{*}{$\rm \chi_{c2}$} & \ $9.7$ & $\times~ 10^{-13}$ & NRQCD+LDME~\cite{Likhoded:2017jmx} & \multirow[c]{2}{*}{--} & \multirow[c]{2}{*}{--} & \multirow[c]{2}{*}{\textcolor{green}{\Checkmark}} \\
 &  &  &  & \ $(4.6\pm 2.0)$ & $\times~ 10^{-12}$ & LC+LCDA~\cite{Likhoded:2017jmx} &  &  &  \\
\cline{2-10}
 & \multirow[c]{15}{*}{$\rm J/\psi$} & \multirow[c]{15}{*}{+} & \multirow[c]{5}{*}{$\rm J/\psi$} & \ $(1.1\pm 0.3)$ & $\times~ 10^{-10}$ & NRQCD/NRCSM~\cite{Gao:2022mwa} & \multirow[c]{5}{*}{$<$\ $1.1$$\times~ 10^{-6}$~\cite{CMS:2022fsq}} & \multirow[c]{5}{*}{$\lesssim1.7\times~ 10^{-7}$} & \multirow[c]{5}{*}{\textcolor{green}{\Checkmark}} \\
 &  &  &  & \ $\left(1.11^{+0.34}_{-0.24}\right)$ & $\times~ 10^{-10}$ & NRQCD+LDME~\cite{Li:2023tzx} &  &  &  \\
 &  &  &  & \ $(1.1\text{ -- }1.3)$ & $\times~ 10^{-10}$ & NRQCD+LDME (NLO)~\cite{Luo:2022ugd} &  &  &  \\
 &  &  &  & \ $2.3$ & $\times~ 10^{-14}$ & NRQCD+LDME~\cite{Likhoded:2017jmx} &  &  &  \\
 &  &  &  & \ $2.7$ & $\times~ 10^{-11}$ & NRQCD~\cite{Bergstrom:1990bu} &  &  &  \\
\cline{4-10}
 &  &  & \multirow[c]{3}{*}{$\rm \chi_{c0}$} & \ $(1.1\text{ -- }4.1)$ & $\times~ 10^{-12}$ & NRQCD+LDME (NLO)~\cite{Luo:2022ugd} & \multirow[c]{3}{*}{--} & \multirow[c]{3}{*}{--} & \multirow[c]{3}{*}{\textcolor{green}{\Checkmark}} \\
 &  &  &  & \ $8.3$ & $\times~ 10^{-14}$ & NRQCD+LDME~\cite{Likhoded:2017jmx} &  &  &  \\
 &  &  &  & \ $(4.7\pm 2.0)$ & $\times~ 10^{-13}$ & LC+LCDA~\cite{Likhoded:2017jmx} &  &  &  \\
\cline{4-10}
 &  &  & \multirow[c]{2}{*}{$\rm \chi_{c1}$} & \ $(3.5\text{ -- }4.4)$ & $\times~ 10^{-12}$ & NRQCD+LDME (NLO)~\cite{Luo:2022ugd} & \multirow[c]{2}{*}{--} & \multirow[c]{2}{*}{--} & \multirow[c]{2}{*}{\textcolor{green}{\Checkmark}} \\
 &  &  &  & \ $3.5$ & $\times~ 10^{-15}$ & NRQCD+LDME~\cite{Likhoded:2017jmx} &  &  &  \\
\cline{4-10}
 &  &  & \multirow[c]{2}{*}{$\rm h_c$} & \ $1.5$ & $\times~ 10^{-12}$ & NRQCD+LDME~\cite{Likhoded:2017jmx} & \multirow[c]{2}{*}{--} & \multirow[c]{2}{*}{--} & \multirow[c]{2}{*}{\textcolor{green}{\Checkmark}} \\
 &  &  &  & \ $(9.5\pm 5.0)$ & $\times~ 10^{-12}$ & LC+LCDA~\cite{Likhoded:2017jmx} &  &  &  \\
\cline{4-10}
 &  &  & \multirow[c]{3}{*}{$\rm \chi_{c2}$} & \ $(9.6\text{ -- }24.8)$ & $\times~ 10^{-13}$ & NRQCD+LDME (NLO)~\cite{Luo:2022ugd} & \multirow[c]{3}{*}{--} & \multirow[c]{3}{*}{--} & \multirow[c]{3}{*}{\textcolor{green}{\Checkmark}} \\
 &  &  &  & \ $1.4$ & $\times~ 10^{-13}$ & NRQCD+LDME~\cite{Likhoded:2017jmx} &  &  &  \\
 &  &  &  & \ $(9.3\pm 4.0)$ & $\times~ 10^{-13}$ & LC+LCDA~\cite{Likhoded:2017jmx} &  &  &  \\
\cline{2-10}
 & \multirow[c]{4}{*}{$\rm \chi_{c0}$} & \multirow[c]{4}{*}{+} & \multirow[c]{2}{*}{$\rm \chi_{c1}$} & \ $7.6$ & $\times~ 10^{-14}$ & NRQCD+LDME~\cite{Likhoded:2017jmx} & \multirow[c]{2}{*}{--} & \multirow[c]{2}{*}{--} & \multirow[c]{2}{*}{\textcolor{red}{\ding{56}}} \\
 &  &  &  & \ $(1.4\pm 1.0)$ & $\times~ 10^{-12}$ & LC+LCDA~\cite{Likhoded:2017jmx} &  &  &  \\
\cline{4-10}
 &  &  & $\rm h_c$ & \ $3.5$ & $\times~ 10^{-16}$ & NRQCD+LDME~\cite{Likhoded:2017jmx} & -- & -- & \textcolor{red}{\ding{56}} \\

 &  &  & $\rm \chi_{c2}$ & \ $6.4$ & $\times~ 10^{-15}$ & NRQCD+LDME~\cite{Likhoded:2017jmx} & -- & -- & \textcolor{red}{\ding{56}} \\
\cline{2-10}
 & \multirow[c]{5}{*}{$\rm \chi_{c1}$} & \multirow[c]{5}{*}{+} & $\rm \chi_{c1}$ & \ $3.9$ & $\times~ 10^{-16}$ & NRQCD+LDME~\cite{Likhoded:2017jmx} & -- & -- & \textcolor{red}{\ding{56}} \\
\cline{4-10}
 &  &  & \multirow[c]{2}{*}{$\rm h_c$} & \ $2.9$ & $\times~ 10^{-14}$ & NRQCD+LDME~\cite{Likhoded:2017jmx} & \multirow[c]{2}{*}{--} & \multirow[c]{2}{*}{--} & \multirow[c]{2}{*}{\textcolor{red}{\ding{56}}} \\
 &  &  &  & \ $(6.1\pm 5.0)$ & $\times~ 10^{-13}$ & LC+LCDA~\cite{Likhoded:2017jmx} &  &  &  \\
\cline{4-10}
 &  &  & \multirow[c]{2}{*}{$\rm \chi_{c2}$} & \ $1.3$ & $\times~ 10^{-13}$ & NRQCD+LDME~\cite{Likhoded:2017jmx} & \multirow[c]{2}{*}{--} & \multirow[c]{2}{*}{--} & \multirow[c]{2}{*}{\textcolor{red}{\ding{56}}} \\
 &  &  &  & \ $(2.8\pm 2.0)$ & $\times~ 10^{-12}$ & LC+LCDA~\cite{Likhoded:2017jmx} &  &  &  \\
\cline{2-10}
 & \multirow[c]{2}{*}{$\rm h_c$} & \multirow[c]{2}{*}{+} & $\rm h_c$ & \ $9.9$ & $\times~ 10^{-17}$ & NRQCD+LDME~\cite{Likhoded:2017jmx} & -- & -- & \textcolor{red}{\ding{56}} \\

 &  &  & $\rm \chi_{c2}$ & \ $2.3$ & $\times~ 10^{-16}$ & NRQCD+LDME~\cite{Likhoded:2017jmx} & -- & -- & \textcolor{red}{\ding{56}} \\
\cline{2-10}
 & $\rm \chi_{c2}$ & + & $\rm \chi_{c2}$ & \ $1.3$ & $\times~ 10^{-16}$ & NRQCD+LDME~\cite{Likhoded:2017jmx} & -- & -- & \textcolor{red}{\ding{56}} \\

\bottomrule
\end{tabular}

%% file: tables/exclusive_Z_decays_bottomonium_meson.tex
\begin{tabular}{l>{\hspace{-1mm}}l<{\hspace{-3mm}}>{\hspace{-2mm}}l<{\hspace{-2mm}}>{\hspace{-1mm}}lr<{\hspace{-6mm}}llccccc}
\toprule
 &  &  &  &  & \multicolumn{2}{c}{} & \multicolumn{2}{c}{Exp. limits} & Producible at \\
$Z\,\to$ & M  & $+$ & M & \multicolumn{2}{c}{Branching fraction} & Framework & 2024 & HL-LHC & at FCC-ee \\
\midrule
\multirow[c]{10}{*}{$\rm Z\,\to$} & \multirow[c]{6}{*}{$\rm \Upsilon(1S)$} & \multirow[c]{6}{*}{+} & $\rm J/\psi$ & \ $4.6$ & $\times~ 10^{-11}$ & NRQCD~\cite{Bergstrom:1990bu} & -- & -- & \textcolor{green}{\Checkmark} \\

 &  &  & $\rm \eta_b(1S)$ & \ $(1.9\pm 0.2)$ & $\times~ 10^{-11}$ & NRQCD/NRCSM~\cite{Gao:2022mwa} & -- & -- & \textcolor{green}{\Checkmark} \\
\cline{4-10}
 &  &  & \multirow[c]{2}{*}{$\rm \Upsilon(1S)$} & \ $(6.2\pm 0.3)$ & $\times~ 10^{-11}$ & NRQCD(NLO)+LDME~\cite{Li:2024zun} & \multirow[c]{2}{*}{$<$\ $1.8$$\times~ 10^{-6}$~\cite{CMS:2022fsq}} & \multirow[c]{2}{*}{$\lesssim2.7\times~ 10^{-7}$} & \multirow[c]{2}{*}{\textcolor{green}{\Checkmark}} \\
 &  &  &  & \ $\left(4.4^{+0.6}_{-0.3}\right)$ & $\times~ 10^{-13}$ & NRQCD/NRCSM~\cite{Gao:2022mwa} &  &  &  \\
\cline{4-10}
 &  &  & $\rm \Upsilon(2S)$ & \ $(5.6\pm 0.2)$ & $\times~ 10^{-11}$ & NRQCD(NLO)+LDME~\cite{Li:2024zun} & -- & -- & \textcolor{green}{\Checkmark} \\

 &  &  & $\rm \Upsilon(3S)$ & \ $(4.1\pm 0.2)$ & $\times~ 10^{-11}$ & NRQCD(NLO)+LDME~\cite{Li:2024zun} & -- & -- & \textcolor{green}{\Checkmark} \\
\cline{2-10}
 & \multirow[c]{2}{*}{$\rm \Upsilon(2S)$} & \multirow[c]{2}{*}{+} & $\rm \Upsilon(2S)$ & \ $(1.4\pm 0.1)$ & $\times~ 10^{-11}$ & NRQCD(NLO)+LDME~\cite{Li:2024zun} & -- & -- & \textcolor{green}{\Checkmark} \\

 &  &  & $\rm \Upsilon(3S)$ & \ $(1.9\pm 0.1)$ & $\times~ 10^{-11}$ & NRQCD(NLO)+LDME~\cite{Li:2024zun} & -- & -- & \textcolor{green}{\Checkmark} \\
\cline{2-10}
 & $\rm \Upsilon(3S)$ & + & $\rm \Upsilon(3S)$ & \ $(6.8\pm 0.3)$ & $\times~ 10^{-12}$ & NRQCD(NLO)+LDME~\cite{Li:2024zun} & -- & -- & \textcolor{green}{\Checkmark} \\

 & $\rm \Upsilon(mS)$ & + & $\rm \Upsilon(nS)$ & \ $2.1$ & $\times~ 10^{-12}$ & NRQCD~\cite{Bergstrom:1990bu} & $<$\ $3.9$$\times~ 10^{-7}$~\cite{CMS:2022fsq} & $\lesssim5.9\times~ 10^{-8}$ & \textcolor{green}{\Checkmark} \\

\bottomrule
\end{tabular}

%% file: tables/exclusive_W_decays_gamma_meson.tex
\begin{tabular}{l<{\hspace{-3mm}}>{\hspace{-2mm}}l<{\hspace{-2mm}}>{\hspace{-1mm}}lr<{\hspace{-6mm}}llccccc}
\toprule
 &  &  &  & \multicolumn{2}{c}{} & \multicolumn{2}{c}{Exp. limits} & \multicolumn{2}{c}{Producible at} \\
$\rm W^\pm \, \to \, \gamma$ & + & $\rm M$ & \multicolumn{2}{c}{Branching fraction} & Framework & 2024 & HL-LHC & FCC-ee & FCC-hh \\
\midrule
\multirow[c]{14}{*}{$\rm W^\pm \, \to \,\gamma$} & \multirow[c]{14}{*}{+} & $\rm \pi^\pm$ & \ $(4.0\pm 0.8)$ & $\times~ 10^{-9}$ & SCET+LCDA~\cite{Grossman:2015cak} & $<$\ $1.9$$\times~ 10^{-6}$~\cite{ATLAS:2023jfq,CMS:2020oqe} & $\lesssim2.9\times~ 10^{-7}$ & \textcolor{green}{\Checkmark} & \textcolor{green}{\Checkmark} \\

 &  & $\rm \rho^\pm$ & \ $(8.74\pm 1.91)$ & $\times~ 10^{-9}$ & SCET+LCDA~\cite{Grossman:2015cak} & $<$\ $5.2$$\times~ 10^{-6}$~\cite{ATLAS:2023jfq} & $\lesssim7.9\times~ 10^{-7}$ & \textcolor{green}{\Checkmark} & \textcolor{green}{\Checkmark} \\

 &  & $\rm K^\pm$ & \ $(3.25\pm 0.69)$ & $\times~ 10^{-10}$ & SCET+LCDA~\cite{Grossman:2015cak} & $<$\ $1.7$$\times~ 10^{-6}$~\cite{ATLAS:2023jfq} & $\lesssim2.6\times~ 10^{-7}$ & \textcolor{red}{\ding{56}} & \textcolor{green}{\Checkmark} \\

 &  & $\rm K^{*\pm}$ & \ $(4.78\pm 1.15)$ & $\times~ 10^{-10}$ & SCET+LCDA~\cite{Grossman:2015cak} & -- & -- & \textcolor{red}{\ding{56}} & \textcolor{green}{\Checkmark} \\

 &  & $\rm D^\pm$ & \ $\left(1.38^{+0.51}_{-0.33}\right)$ & $\times~ 10^{-9}$ & SCET+LCDA~\cite{Grossman:2015cak} & -- & -- & \textcolor{red}{\ding{56}} & \textcolor{green}{\Checkmark} \\
\cline{3-10}
 &  & \multirow[c]{3}{*}{$\rm D_s^\pm$} & \ $4.7$ & $\times~ 10^{-9}$ & NRQCD+LDME~\cite{Bagdatova:2023etj} & \multirow[c]{3}{*}{$<$\ $6.5$$\times~ 10^{-4}$~\cite{LHCb:2022kta}} & \multirow[c]{3}{*}{$\lesssim1.2\times~ 10^{-5}$} & \multirow[c]{3}{*}{\textcolor{green}{\Checkmark}} & \multirow[c]{3}{*}{\textcolor{green}{\Checkmark}} \\
 &  &  & \ $3.4$ & $\times~ 10^{-9}$ & LC+LCDA~\cite{Bagdatova:2023etj} &  &  &  &  \\
 &  &  & \ $\left(3.66^{+1.49}_{-0.85}\right)$ & $\times~ 10^{-8}$ & SCET+LCDA~\cite{Grossman:2015cak} &  &  &  &  \\
\cline{3-10}
 &  & \multirow[c]{2}{*}{$\rm D_s^{*\pm}$} & \ $8.9$ & $\times~ 10^{-10}$ & NRQCD+LDME~\cite{Bagdatova:2023etj} & \multirow[c]{2}{*}{--} & \multirow[c]{2}{*}{--} & \multirow[c]{2}{*}{\textcolor{red}{\ding{56}}} & \multirow[c]{2}{*}{\textcolor{green}{\Checkmark}} \\
 &  &  & \ $3.4$ & $\times~ 10^{-9}$ & LC+LCDA~\cite{Bagdatova:2023etj} &  &  &  &  \\
\cline{3-10}
 &  & \multirow[c]{3}{*}{$\rm B^\pm$} & \ $\left(2.616^{+3.146}_{-1.330}\right)$ & $\times~ 10^{-12}$ & HQET+LCDA~\cite{Beneke:2023nmj} & \multirow[c]{3}{*}{--} & \multirow[c]{3}{*}{--} & \multirow[c]{3}{*}{\textcolor{red}{\ding{56}}} & \multirow[c]{3}{*}{\textcolor{green}{\Checkmark}} \\
 &  &  & \ $\left(1.99^{+2.49}_{-0.82}\right)$ & $\times~ 10^{-12}$ & SCET+LCDA\footnote{updated inputs from \cite{Beneke:2023nmj} for the result from \cite{Grossman:2015cak}}~\cite{Beneke:2023nmj} &  &  &  &  \\
 &  &  & \ $\left(1.55^{+0.79}_{-0.60}\right)$ & $\times~ 10^{-12}$ & SCET+LCDA~\cite{Grossman:2015cak} &  &  &  &  \\
\cline{3-10}
 &  & $\rm B_c^\pm$ & \ $1.3$ & $\times~ 10^{-13}$ & SCET+LCDA~(this work) & -- & -- & \textcolor{red}{\ding{56}} & \textcolor{green}{\Checkmark} \\

\bottomrule
\end{tabular}

%% file: tables/exclusive_W_decays_meson_meson.tex
\begin{tabular}{ll<{\hspace{-3mm}}>{\hspace{-2mm}}l<{\hspace{-2mm}}>{\hspace{-1mm}}lr<{\hspace{-6mm}}llccccc}
\toprule
 &  &  &  &  & \multicolumn{2}{c}{} & \multicolumn{2}{c}{Exp. limits} & \multicolumn{2}{c}{Producible at} \\
$\rm W^\pm\, \to $ & M  & + & M & \multicolumn{2}{c}{Branching fraction} & Framework & 2024 & HL-LHC & FCC-ee & FCC-hh \\
\midrule
\multirow[c]{41}{*}{$\rm W^\pm\, \to $} & \multirow[c]{4}{*}{$\rm \eta_c$} & \multirow[c]{4}{*}{+} & \multirow[c]{2}{*}{$\rm D_s^\pm$} & \ $2.1$ & $\times~ 10^{-12}$ & NRCSM+LCDA~\cite{Luchinsky:2021amp} & \multirow[c]{2}{*}{--} & \multirow[c]{2}{*}{--} & \multirow[c]{2}{*}{\textcolor{red}{\ding{56}}} & \multirow[c]{2}{*}{\textcolor{green}{\Checkmark}} \\
 &  &  &  & \ $\left(1.31^{+0.34}_{-0.22}\right)$ & $\times~ 10^{-11}$ & LC+LCDA~\cite{Luchinsky:2021amp} &  &  &  &  \\
\cline{4-11}
 &  &  & \multirow[c]{2}{*}{$\rm D_s^{*\pm}$} & \ $3.0$ & $\times~ 10^{-12}$ & NRCSM+LCDA~\cite{Luchinsky:2021amp} & \multirow[c]{2}{*}{--} & \multirow[c]{2}{*}{--} & \multirow[c]{2}{*}{\textcolor{red}{\ding{56}}} & \multirow[c]{2}{*}{\textcolor{green}{\Checkmark}} \\
 &  &  &  & \ $\left(1.48^{+0.36}_{-0.22}\right)$ & $\times~ 10^{-11}$ & LC+LCDA~\cite{Luchinsky:2021amp} &  &  &  &  \\
\cline{2-11}
 & \multirow[c]{7}{*}{$\rm J/\psi$} & \multirow[c]{7}{*}{+} & $\rm D^\pm$ & \ $\mathcal{O}($ & $ 10^{-13}$) & LC+$\delta$ appr.~(this work) & -- & -- & \textcolor{red}{\ding{56}} & \textcolor{green}{\Checkmark} \\
\cline{4-11}
 &  &  & \multirow[c]{3}{*}{$\rm D_s^\pm$} & \ $2.6$ & $\times~ 10^{-12}$ & NRQCD+LDME~\cite{Bagdatova:2023etj} & \multirow[c]{3}{*}{--} & \multirow[c]{3}{*}{--} & \multirow[c]{3}{*}{\textcolor{red}{\ding{56}}} & \multirow[c]{3}{*}{\textcolor{green}{\Checkmark}} \\
 &  &  &  & \ $\left(1.8^{+0.4}_{-0.2}\right)$ & $\times~ 10^{-11}$ & LC+LCDA~\cite{Luchinsky:2021amp} &  &  &  &  \\
 &  &  &  & \ $2.1$ & $\times~ 10^{-12}$ & NRCSM+LCDA~\cite{Luchinsky:2021amp} &  &  &  &  \\
\cline{4-11}
 &  &  & \multirow[c]{3}{*}{$\rm D_s^{*\pm}$} & \ $3.0$ & $\times~ 10^{-12}$ & NRCSM+LCDA~\cite{Luchinsky:2021amp} & \multirow[c]{3}{*}{--} & \multirow[c]{3}{*}{--} & \multirow[c]{3}{*}{\textcolor{red}{\ding{56}}} & \multirow[c]{3}{*}{\textcolor{green}{\Checkmark}} \\
 &  &  &  & \ $1.7$ & $\times~ 10^{-12}$ & NRQCD+LDME~\cite{Bagdatova:2023etj} &  &  &  &  \\
 &  &  &  & \ $\left(2.03^{+0.46}_{-0.24}\right)$ & $\times~ 10^{-11}$ & LC+LCDA~\cite{Luchinsky:2021amp} &  &  &  &  \\
\cline{2-11}
 & \multirow[c]{2}{*}{$\rm \psi(2S)$} & \multirow[c]{2}{*}{+} & $\rm D_s^\pm$ & \ $5.1$ & $\times~ 10^{-11}$ & NRQCD+LDME~\cite{Bagdatova:2023etj} & -- & -- & \textcolor{red}{\ding{56}} & \textcolor{green}{\Checkmark} \\

 &  &  & $\rm D_s^{*\pm}$ & \ $7.4$ & $\times~ 10^{-12}$ & NRQCD+LDME~\cite{Bagdatova:2023etj} & -- & -- & \textcolor{red}{\ding{56}} & \textcolor{green}{\Checkmark} \\
\cline{2-11}
 & \multirow[c]{6}{*}{$\rm \chi_{c0}$} & \multirow[c]{6}{*}{+} & \multirow[c]{3}{*}{$\rm D_s^\pm$} & \ $9.4$ & $\times~ 10^{-14}$ & NRQCD+LDME~\cite{Bagdatova:2023etj} & \multirow[c]{3}{*}{--} & \multirow[c]{3}{*}{--} & \multirow[c]{3}{*}{\textcolor{red}{\ding{56}}} & \multirow[c]{3}{*}{\textcolor{green}{\Checkmark}} \\
 &  &  &  & \ $4.7$ & $\times~ 10^{-14}$ & NRCSM+LCDA~\cite{Luchinsky:2021amp} &  &  &  &  \\
 &  &  &  & \ $\left(7.1^{+3.5}_{-3.1}\right)$ & $\times~ 10^{-13}$ & LC+LCDA~\cite{Luchinsky:2021amp} &  &  &  &  \\
\cline{4-11}
 &  &  & \multirow[c]{3}{*}{$\rm D_s^{*\pm}$} & \ $1.2$ & $\times~ 10^{-13}$ & NRQCD+LDME~\cite{Bagdatova:2023etj} & \multirow[c]{3}{*}{--} & \multirow[c]{3}{*}{--} & \multirow[c]{3}{*}{\textcolor{red}{\ding{56}}} & \multirow[c]{3}{*}{\textcolor{green}{\Checkmark}} \\
 &  &  &  & \ $8.1$ & $\times~ 10^{-14}$ & NRCSM+LCDA~\cite{Luchinsky:2021amp} &  &  &  &  \\
 &  &  &  & \ $\left(8.0^{+3.7}_{-3.1}\right)$ & $\times~ 10^{-13}$ & LC+LCDA~\cite{Luchinsky:2021amp} &  &  &  &  \\
\cline{2-11}
 & \multirow[c]{6}{*}{$\rm \chi_{c1}$} & \multirow[c]{6}{*}{+} & \multirow[c]{3}{*}{$\rm D_s^\pm$} & \ $2.0$ & $\times~ 10^{-13}$ & NRQCD+LDME~\cite{Bagdatova:2023etj} & \multirow[c]{3}{*}{--} & \multirow[c]{3}{*}{--} & \multirow[c]{3}{*}{\textcolor{red}{\ding{56}}} & \multirow[c]{3}{*}{\textcolor{green}{\Checkmark}} \\
 &  &  &  & \ $2.9$ & $\times~ 10^{-13}$ & NRCSM+LCDA~\cite{Luchinsky:2021amp} &  &  &  &  \\
 &  &  &  & \ $\left(7.83^{+3.40}_{-3.05}\right)$ & $\times~ 10^{-12}$ & LC+LCDA~\cite{Luchinsky:2021amp} &  &  &  &  \\
\cline{4-11}
 &  &  & \multirow[c]{3}{*}{$\rm D_s^{*\pm}$} & \ $2.0$ & $\times~ 10^{-13}$ & NRQCD+LDME~\cite{Bagdatova:2023etj} & \multirow[c]{3}{*}{--} & \multirow[c]{3}{*}{--} & \multirow[c]{3}{*}{\textcolor{red}{\ding{56}}} & \multirow[c]{3}{*}{\textcolor{green}{\Checkmark}} \\
 &  &  &  & \ $4.0$ & $\times~ 10^{-13}$ & NRCSM+LCDA~\cite{Luchinsky:2021amp} &  &  &  &  \\
 &  &  &  & \ $\left(8.83^{+3.50}_{-3.06}\right)$ & $\times~ 10^{-12}$ & LC+LCDA~\cite{Luchinsky:2021amp} &  &  &  &  \\
\cline{2-11}
 & \multirow[c]{4}{*}{$\rm h_c$} & \multirow[c]{4}{*}{+} & \multirow[c]{2}{*}{$\rm D_s^\pm$} & \ $1.4$ & $\times~ 10^{-13}$ & NRCSM+LCDA~\cite{Luchinsky:2021amp} & \multirow[c]{2}{*}{--} & \multirow[c]{2}{*}{--} & \multirow[c]{2}{*}{\textcolor{red}{\ding{56}}} & \multirow[c]{2}{*}{\textcolor{green}{\Checkmark}} \\
 &  &  &  & \ $\left(2.13^{+0.97}_{-0.83}\right)$ & $\times~ 10^{-12}$ & LC+LCDA~\cite{Luchinsky:2021amp} &  &  &  &  \\
\cline{4-11}
 &  &  & \multirow[c]{2}{*}{$\rm D_s^{*\pm}$} & \ $2.0$ & $\times~ 10^{-13}$ & NRCSM+LCDA~\cite{Luchinsky:2021amp} & \multirow[c]{2}{*}{--} & \multirow[c]{2}{*}{--} & \multirow[c]{2}{*}{\textcolor{red}{\ding{56}}} & \multirow[c]{2}{*}{\textcolor{green}{\Checkmark}} \\
 &  &  &  & \ $\left(2.4^{+1.1}_{-0.9}\right)$ & $\times~ 10^{-12}$ & LC+LCDA~\cite{Luchinsky:2021amp} &  &  &  &  \\
\cline{2-11}
 & \multirow[c]{6}{*}{$\rm \chi_{c2}$} & \multirow[c]{6}{*}{+} & \multirow[c]{3}{*}{$\rm D_s^\pm$} & \ $3.9$ & $\times~ 10^{-14}$ & NRQCD+LDME~\cite{Bagdatova:2023etj} & \multirow[c]{3}{*}{--} & \multirow[c]{3}{*}{--} & \multirow[c]{3}{*}{\textcolor{red}{\ding{56}}} & \multirow[c]{3}{*}{\textcolor{green}{\Checkmark}} \\
 &  &  &  & \ $9.6$ & $\times~ 10^{-14}$ & NRCSM+LCDA~\cite{Luchinsky:2021amp} &  &  &  &  \\
 &  &  &  & \ $\left(1.42^{+0.62}_{-0.53}\right)$ & $\times~ 10^{-12}$ & LC+LCDA~\cite{Luchinsky:2021amp} &  &  &  &  \\
\cline{4-11}
 &  &  & \multirow[c]{3}{*}{$\rm D_s^{*\pm}$} & \ $3.9$ & $\times~ 10^{-14}$ & NRQCD+LDME~\cite{Bagdatova:2023etj} & \multirow[c]{3}{*}{--} & \multirow[c]{3}{*}{--} & \multirow[c]{3}{*}{\textcolor{red}{\ding{56}}} & \multirow[c]{3}{*}{\textcolor{green}{\Checkmark}} \\
 &  &  &  & \ $1.4$ & $\times~ 10^{-13}$ & NRCSM+LCDA~\cite{Luchinsky:2021amp} &  &  &  &  \\
 &  &  &  & \ $\left(1.6^{+0.7}_{-0.6}\right)$ & $\times~ 10^{-12}$ & LC+LCDA~\cite{Luchinsky:2021amp} &  &  &  &  \\
\cline{2-11}
 & \multirow[c]{2}{*}{$\rm B_s^0$} & \multirow[c]{2}{*}{+} & $\rm B_c^\pm$ & \ $2.5$ & $\times~ 10^{-12}$ & NRQCD+LDME~\cite{Bagdatova:2023etj} & -- & -- & \textcolor{red}{\ding{56}} & \textcolor{green}{\Checkmark} \\

 &  &  & $\rm B_c^{*\pm}$ & \ $2.0$ & $\times~ 10^{-12}$ & NRQCD+LDME~\cite{Bagdatova:2023etj} & -- & -- & \textcolor{red}{\ding{56}} & \textcolor{green}{\Checkmark} \\
\cline{2-11}
 & \multirow[c]{2}{*}{$\rm B_s^*$} & \multirow[c]{2}{*}{+} & $\rm B_c^\pm$ & \ $2.7$ & $\times~ 10^{-12}$ & NRQCD+LDME~\cite{Bagdatova:2023etj} & -- & -- & \textcolor{red}{\ding{56}} & \textcolor{green}{\Checkmark} \\

 &  &  & $\rm B_c^{*\pm}$ & \ $2.7$ & $\times~ 10^{-12}$ & NRQCD+LDME~\cite{Bagdatova:2023etj} & -- & -- & \textcolor{red}{\ding{56}} & \textcolor{green}{\Checkmark} \\
\cline{2-11}
 & \multirow[c]{2}{*}{$\rm \Upsilon(1S)$} & \multirow[c]{2}{*}{+} & $\rm B^\pm$ & \ $\mathcal{O}$ & $\ (10^{-16})$ & LC+$\delta$ appr.~(this work) & -- & -- & \textcolor{red}{\ding{56}} & \textcolor{red}{\ding{56}} \\

 &  &  & $\rm B_c^\pm$ & \ $\mathcal{O}$ & $\ (10^{-14})$ & LC+$\delta$ appr.~(this work) & -- & -- & \textcolor{red}{\ding{56}} & \textcolor{red}{\ding{56}} \\

\bottomrule
\end{tabular}

%% file: tables/exclusive_t_decays_q_boson.tex
\begin{tabular}{ll<{\hspace{-3mm}}>{\hspace{-2mm}}l<{\hspace{-2mm}}>{\hspace{-1mm}}lr<{\hspace{-6mm}}llccccc}
\toprule
 &  &  &  &  & \multicolumn{2}{c}{} & \multicolumn{2}{c}{Exp. limits} & \multicolumn{2}{c}{Producible at} \\
$\rm t\;\to $ & V & $+$ & q & \multicolumn{2}{c}{Branching fraction} & Framework & 2024 & HL-LHC & FCC-ee & FCC-hh \\
\midrule
\multirow[c]{15}{*}{t$\;\to$} & \multirow[c]{4}{*}{$\rm \gamma$} & \multirow[c]{4}{*}{+} & \multirow[c]{2}{*}{$\rm c$} & \ $(4.55\pm 0.23)$ & $\times~ 10^{-14}$ & loop SM~\cite{Balaji:2020qjg} & \multirow[c]{2}{*}{$<$\ $4.0$$\times~ 10^{-4}$~\cite{ATLAS:2019mke}} & \multirow[c]{2}{*}{$\lesssim 5.2\times~ 10^{-5}$~\cite{Cerri:2018ypt}} & \multirow[c]{2}{*}{\textcolor{red}{\ding{56}}} & \multirow[c]{2}{*}{\textcolor{red}{\ding{56}}} \\
 &  &  &  & \ $\left(4.6^{+2.0}_{-1.0}\right)$ & $\times~ 10^{-14}$ & loop SM~\cite{Aguilar-Saavedra:2004mfd} &  &  &  &  \\
\cline{4-11}
 &  &  & \multirow[c]{2}{*}{$\rm u$} & \ $(3.26\pm 0.34)$ & $\times~ 10^{-16}$ & loop SM~\cite{Balaji:2020qjg} & \multirow[c]{2}{*}{$<$\ $8.9$$\times~ 10^{-5}$~\cite{ATLAS:2019mke}} & \multirow[c]{2}{*}{$\lesssim 6.1\times~ 10^{-6}$~\cite{Cerri:2018ypt}} & \multirow[c]{2}{*}{\textcolor{red}{\ding{56}}} & \multirow[c]{2}{*}{\textcolor{red}{\ding{56}}} \\
 &  &  &  & \ $3.7$ & $\times~ 10^{-16}$ & loop SM~\cite{Aguilar-Saavedra:2004mfd} &  &  &  &  \\
\cline{2-11}
 & \multirow[c]{5}{*}{$\rm g$} & \multirow[c]{5}{*}{+} & \multirow[c]{3}{*}{$\rm c$} & \ $(5.31\pm 0.27)$ & $\times~ 10^{-12}$ & loop SM~\cite{Balaji:2020qjg} & \multirow[c]{3}{*}{$<$\ $4.1$$\times~ 10^{-4}$~\cite{CMS:2016uzc}} & \multirow[c]{3}{*}{$\lesssim 2.3\times~ 10^{-5}$~\cite{Cerri:2018ypt}} & \multirow[c]{3}{*}{\textcolor{red}{\ding{56}}} & \multirow[c]{3}{*}{\textcolor{green}{\Checkmark}} \\
 &  &  &  & \ $5.7$ & $\times~ 10^{-12}$ & loop SM~\cite{Eilam:2006uh} &  &  &  &  \\
 &  &  &  & \ $\left(4.6^{+3.0}_{-1.0}\right)$ & $\times~ 10^{-12}$ & loop SM~\cite{Aguilar-Saavedra:2004mfd} &  &  &  &  \\
\cline{4-11}
 &  &  & \multirow[c]{2}{*}{$\rm u$} & \ $(3.81\pm 0.34)$ & $\times~ 10^{-14}$ & loop SM~\cite{Balaji:2020qjg} & \multirow[c]{2}{*}{$<$\ $2.0$$\times~ 10^{-5}$~\cite{CMS:2016uzc}} & \multirow[c]{2}{*}{$\lesssim 2.7\times~ 10^{-6}$~\cite{Cerri:2018ypt}} & \multirow[c]{2}{*}{\textcolor{red}{\ding{56}}} & \multirow[c]{2}{*}{\textcolor{red}{\ding{56}}} \\
 &  &  &  & \ $3.7$ & $\times~ 10^{-14}$ & loop SM~\cite{Aguilar-Saavedra:2004mfd} &  &  &  &  \\
\cline{2-11}
 & \multirow[c]{2}{*}{$\rm Z$} & \multirow[c]{2}{*}{+} & $\rm c$ & \ $1.0$ & $\times~ 10^{-14}$ & loop SM~\cite{Aguilar-Saavedra:2004mfd} & $<$\ $2.4$$\times~ 10^{-4}$~\cite{ATLAS:2018zsq} & $\lesssim 2.3\times~ 10^{-5}$~\cite{Liu:2020bem} & \textcolor{red}{\ding{56}} & \textcolor{red}{\ding{56}} \\

 &  &  & $\rm u$ & \ $8.0$ & $\times~ 10^{-17}$ & loop SM~\cite{Aguilar-Saavedra:2004mfd} & $<$\ $1.7$$\times~ 10^{-4}$~\cite{ATLAS:2018zsq} & $\lesssim 7.3\times~ 10^{-6}$~\cite{Liu:2020bem} & \textcolor{red}{\ding{56}} & \textcolor{red}{\ding{56}} \\
\cline{2-11}
 & \multirow[c]{4}{*}{$\rm H$} & \multirow[c]{4}{*}{+} & \multirow[c]{2}{*}{$\rm c$} & \ $\left(4.19^{+1.09}_{-0.86}\right)$ & $\times~ 10^{-15}$ & loop SM~\cite{Altmannshofer:2019ogm} & \multirow[c]{2}{*}{$<$\ $7.3$$\times~ 10^{-4}$~\cite{CMS:2021hug}} & \multirow[c]{2}{*}{$\lesssim 8.5\times~ 10^{-5}$~\cite{ATLAS:2016qxw}} & \multirow[c]{2}{*}{\textcolor{red}{\ding{56}}} & \multirow[c]{2}{*}{\textcolor{red}{\ding{56}}} \\
 &  &  &  & \ $3.0$ & $\times~ 10^{-15}$ & loop SM~\cite{Aguilar-Saavedra:2004mfd} &  &  &  &  \\
\cline{4-11}
 &  &  & \multirow[c]{2}{*}{$\rm u$} & \ $\left(3.66^{+1.15}_{-0.97}\right)$ & $\times~ 10^{-17}$ & loop SM~\cite{Altmannshofer:2019ogm} & \multirow[c]{2}{*}{$<$\ $1.9$$\times~ 10^{-4}$~\cite{CMS:2021hug}} & \multirow[c]{2}{*}{$\lesssim 8.5\times~ 10^{-5}$~\cite{ATLAS:2016qxw}} & \multirow[c]{2}{*}{\textcolor{red}{\ding{56}}} & \multirow[c]{2}{*}{\textcolor{red}{\ding{56}}} \\
 &  &  &  & \ $2.0$ & $\times~ 10^{-17}$ & loop SM~\cite{Aguilar-Saavedra:2004mfd} &  &  &  &  \\

\bottomrule
\end{tabular}

%% file: tables/rate_t_decays_3_body.tex
\begin{tabular}{ll<{\hspace{-2mm}}>{\hspace{-1mm}}l<{\hspace{-2mm}}>{\hspace{-1mm}}l<{\hspace{-2mm}}>{\hspace{-1mm}}l<{\hspace{-2mm}}>{\hspace{-1mm}}lr<{\hspace{-6mm}}llccccc}
\toprule
 &  &  &  &  &  &  & \multicolumn{2}{c}{} & \multicolumn{2}{c}{Exp. limits} & \multicolumn{2}{c}{Producible at} \\
$\rm t\,\to$ & X & + & Y & +  & Z & \multicolumn{2}{c}{Branching fraction} & Framework & 2024 & HL-LHC & FCC-ee & FCC-hh \\
\midrule
\multirow[c]{9}{*}{t$\, \to$} & \multirow[c]{2}{*}{$\rm W^+$} & \multirow[c]{2}{*}{+} & \multirow[c]{2}{*}{$\rm W^-$} & \multirow[c]{2}{*}{+} & \multirow[c]{2}{*}{$\rm c$} & \ $2.0$ & $\times~ 10^{-13}$ & LO SM~\cite{Bar-Shalom:2005ldb} & \multirow[c]{2}{*}{--} & \multirow[c]{2}{*}{--} & \multirow[c]{2}{*}{\textcolor{red}{\ding{56}}} & \multirow[c]{2}{*}{\textcolor{red}{\ding{56}}} \\
 &  &  &  &  &  & \ $1.0$ & $\times~ 10^{-13}$ & LO SM~\cite{Beneke:2000hk} &  &  &  &  \\
\cline{2-13}
 & \multirow[c]{3}{*}{$\rm Z$} & \multirow[c]{3}{*}{+} & \multirow[c]{3}{*}{$\rm W^+$} & \multirow[c]{3}{*}{+} & \multirow[c]{3}{*}{$\rm b$} & \ $1.8$ & $\times~ 10^{-6}$ & LO SM~\cite{Papaefstathiou:2017xuv} & \multirow[c]{3}{*}{--} & \multirow[c]{3}{*}{--} & \multirow[c]{3}{*}{\textcolor{green}{\Checkmark}} & \multirow[c]{3}{*}{\textcolor{green}{\Checkmark}} \\
 &  &  &  &  &  & \ $2.0$ & $\times~ 10^{-6}$ & LO SM~\cite{Altarelli:2000nt} &  &  &  &  \\
 &  &  &  &  &  & \ $\left(5.4^{+4.7}_{-2.0}\right)$ & $\times~ 10^{-7}$ & LO SM~\cite{Mahlon:1998fr} &  &  &  &  \\
\cline{2-13}
 & \multirow[c]{2}{*}{$\rm H$} & \multirow[c]{2}{*}{+} & \multirow[c]{2}{*}{$\rm W^+$} & \multirow[c]{2}{*}{+} & \multirow[c]{2}{*}{$\rm b$} & \ $1.8$ & $\times~ 10^{-9}$ & LO SM~\cite{Han:2013sea} & \multirow[c]{2}{*}{--} & \multirow[c]{2}{*}{--} & \multirow[c]{2}{*}{\textcolor{red}{\ding{56}}} & \multirow[c]{2}{*}{\textcolor{green}{\Checkmark}} \\
 &  &  &  &  &  & \ $1.6$ & $\times~ 10^{-9}$ & loop SM, \mgshort~(this work) &  &  &  &  \\
\cline{2-13}
 & $\rm Z $ & + & $\rm Z$ & + & $\rm c$ & $<$\ $1.0$ & $\times~ 10^{-13}$ & LO SM~\cite{Bar-Shalom:2005ldb} & -- & -- & \textcolor{red}{\ding{56}} & \textcolor{red}{\ding{56}} \\

 & $\rm u_1$ & + & $\rm u_2$ & + & $\rm \overline{u}_2$ & \ $3.4$ & $\times~ 10^{-12}$ & loop SM~\cite{Cordero-Cid:2004pco} & -- & -- & \textcolor{red}{\ding{56}} & \textcolor{green}{\Checkmark} \\

\bottomrule
\end{tabular}

%% file: tables/exclusive_t_decays_u_meson.tex
\begin{tabular}{ll<{\hspace{-3mm}}>{\hspace{-2mm}}l<{\hspace{-2mm}}>{\hspace{-1mm}}lr<{\hspace{-6mm}}llccccc}
\toprule
 &  &  &  &  & \multicolumn{2}{c}{} & \multicolumn{2}{c}{Exp. limits} & \multicolumn{2}{c}{Producible at} \\
$\rm t\;\to $ & M & $+$ & q & \multicolumn{2}{c}{Branching fraction} & Framework & 2024 & HL-LHC & FCC-ee & FCC-hh \\
\midrule
\multirow[c]{16}{*}{t$\;\to$} & \multirow[c]{2}{*}{$\rm \overline{B}^0$} & \multirow[c]{2}{*}{+} & $\rm c$ & \ $\left(2.1^{+2.1}_{-1.1}\right)$ & $\times~ 10^{-6}$ & NRQCD+LDME~\cite{dEnterria:2020ygk} & -- & -- & \textcolor{green}{\Checkmark} & \textcolor{green}{\Checkmark} \\

 &  &  & $\rm u$ & \ $\left(4.0^{+4.0}_{-2.0}\right)$ & $\times~ 10^{-5}$ & NRQCD+LDME~\cite{dEnterria:2020ygk} & -- & -- & \textcolor{green}{\Checkmark} & \textcolor{green}{\Checkmark} \\
\cline{2-11}
 & \multirow[c]{2}{*}{$\rm \overline{B}^0_s$} & \multirow[c]{2}{*}{+} & $\rm c$ & \ $\left(4.0^{+4.0}_{-2.0}\right)$ & $\times~ 10^{-5}$ & NRQCD+LDME~\cite{dEnterria:2020ygk} & \  & $\lesssim 4.0\times~ 10^{-5}$~\cite{dEnterria:2020ygk} & \textcolor{green}{\Checkmark} & \textcolor{green}{\Checkmark} \\

 &  &  & $\rm u$ & \ $\left(2.1^{+2.1}_{-1.1}\right)$ & $\times~ 10^{-6}$ & NRQCD+LDME~\cite{dEnterria:2020ygk} & -- & -- & \textcolor{green}{\Checkmark} & \textcolor{green}{\Checkmark} \\
\cline{2-11}
 & \multirow[c]{4}{*}{$\rm \Upsilon(1S)$} & \multirow[c]{4}{*}{+} & \multirow[c]{3}{*}{$\rm c$} & \ $4.3$ & $\times~ 10^{-10}$ & NRQCD+CSM~\cite{Slabospitskii:2021rns} & \multirow[c]{3}{*}{--} & \multirow[c]{3}{*}{--} & \multirow[c]{3}{*}{\textcolor{red}{\ding{56}}} & \multirow[c]{3}{*}{\textcolor{green}{\Checkmark}} \\
 &  &  &  & \ $(1.0\text{ -- }1.5)$ & $\times~ 10^{-9}$ & NRQCD+LDME~\cite{dEnterria:2020ygk} &  &  &  &  \\
 &  &  &  & \ $(6.4\pm 1.3)$ & $\times~ 10^{-10}$ & NRQCD+COM~\cite{Handoko:1999iu} &  &  &  &  \\
\cline{4-11}
 &  &  & $\rm u$ & \ $(1.0\text{ -- }1.5)$ & $\times~ 10^{-11}$ & NRQCD+LDME~\cite{dEnterria:2020ygk} & -- & -- & \textcolor{red}{\ding{56}} & \textcolor{green}{\Checkmark} \\
\cline{2-11}
 & \multirow[c]{3}{*}{$\rm \Upsilon(2S)$} & \multirow[c]{3}{*}{+} & \multirow[c]{2}{*}{$\rm c$} & \ $2.1$ & $\times~ 10^{-10}$ & NRQCD+CSM~\cite{Slabospitskii:2021rns} & \multirow[c]{2}{*}{--} & \multirow[c]{2}{*}{--} & \multirow[c]{2}{*}{\textcolor{red}{\ding{56}}} & \multirow[c]{2}{*}{\textcolor{green}{\Checkmark}} \\
 &  &  &  & \ $(1.7\text{ -- }5.3)$ & $\times~ 10^{-10}$ & NRQCD+LDME~\cite{dEnterria:2020ygk} &  &  &  &  \\
\cline{4-11}
 &  &  & $\rm u$ & \ $(1.7\text{ -- }5.3)$ & $\times~ 10^{-12}$ & NRQCD+LDME~\cite{dEnterria:2020ygk} & -- & -- & \textcolor{red}{\ding{56}} & \textcolor{green}{\Checkmark} \\
\cline{2-11}
 & \multirow[c]{3}{*}{$\rm \Upsilon(3S)$} & \multirow[c]{3}{*}{+} & \multirow[c]{2}{*}{$\rm c$} & \ $1.6$ & $\times~ 10^{-10}$ & NRQCD+CSM~\cite{Slabospitskii:2021rns} & \multirow[c]{2}{*}{--} & \multirow[c]{2}{*}{--} & \multirow[c]{2}{*}{\textcolor{red}{\ding{56}}} & \multirow[c]{2}{*}{\textcolor{green}{\Checkmark}} \\
 &  &  &  & \ $(2.7\text{ -- }3.8)$ & $\times~ 10^{-10}$ & NRQCD+LDME~\cite{dEnterria:2020ygk} &  &  &  &  \\
\cline{4-11}
 &  &  & $\rm u$ & \ $(2.7\text{ -- }3.8)$ & $\times~ 10^{-12}$ & NRQCD+LDME~\cite{dEnterria:2020ygk} & -- & -- & \textcolor{red}{\ding{56}} & \textcolor{green}{\Checkmark} \\
\cline{2-11}
 & \multirow[c]{2}{*}{$\rm \Upsilon(nS)$} & \multirow[c]{2}{*}{+} & $\rm c$ & \ $\left(1.9^{+0.2}_{-0.1}\right)$ & $\times~ 10^{-9}$ & NRQCD+LDME~\cite{dEnterria:2020ygk} & -- & -- & \textcolor{red}{\ding{56}} & \textcolor{green}{\Checkmark} \\

 &  &  & $\rm u$ & \ $\left(1.9^{+0.2}_{-0.1}\right)$ & $\times~ 10^{-11}$ & NRQCD+LDME~\cite{dEnterria:2020ygk} & -- & -- & \textcolor{red}{\ding{56}} & \textcolor{green}{\Checkmark} \\

\bottomrule
\end{tabular}

%% file: tables/exclusive_t_decays_b_meson.tex
\begin{tabular}{ll<{\hspace{-3mm}}>{\hspace{-2mm}}l<{\hspace{-2mm}}>{\hspace{-1mm}}lr<{\hspace{-6mm}}llccccc}
\toprule
 &  &  &  &  & \multicolumn{2}{c}{} & \multicolumn{2}{c}{Exp. limits} & \multicolumn{2}{c}{Producible at} \\
$\rm t\;\to $ & M & $+$ & b & \multicolumn{2}{c}{Branching fraction} & Framework & 2024 & HL-LHC & FCC-ee & FCC-hh \\
\midrule
\multirow[c]{14}{*}{t$\;\to$} & \multirow[c]{2}{*}{$\rm \pi^\pm$} & \multirow[c]{2}{*}{+} & \multirow[c]{2}{*}{$\rm b$} & \ $1.6$ & $\times~ 10^{-7}$ & Eq. \eqref{eq:t_M_b_decay}~(this work) & \multirow[c]{2}{*}{--} & \multirow[c]{2}{*}{--} & \multirow[c]{2}{*}{\textcolor{red}{\ding{56}}} & \multirow[c]{2}{*}{\textcolor{green}{\Checkmark}} \\
 &  &  &  & \ $2.0$ & $\times~ 10^{-8}$ & EFT+LCDA~\cite{Beneke:2000hk} &  &  &  &  \\
\cline{2-11}
 & $\rm \rho^\pm$ & + & $\rm b$ & \ $4.3$ & $\times~ 10^{-7}$ & Eq. \eqref{eq:t_M_b_decay}~(this work) & -- & -- & \textcolor{green}{\Checkmark} & \textcolor{green}{\Checkmark} \\

 & $\rm K^\pm$ & + & $\rm b$ & \ $1.2$ & $\times~ 10^{-8}$ & Eq. \eqref{eq:t_M_b_decay}~(this work) & -- & -- & \textcolor{red}{\ding{56}} & \textcolor{green}{\Checkmark} \\

 & $\rm K^{*\pm}$ & + & $\rm b$ & \ $2.1$ & $\times~ 10^{-8}$ & Eq. \eqref{eq:t_M_b_decay}~(this work) & -- & -- & \textcolor{red}{\ding{56}} & \textcolor{green}{\Checkmark} \\

 & $\rm D^\pm$ & + & $\rm b$ & \ $2.3$ & $\times~ 10^{-8}$ & Eq. \eqref{eq:t_M_b_decay}~(this work) & -- & -- & \textcolor{red}{\ding{56}} & \textcolor{green}{\Checkmark} \\

 & $\rm D^{*\pm}$ & + & $\rm b$ & \ $3.7$ & $\times~ 10^{-8}$ & Eq. \eqref{eq:t_M_b_decay}~(this work) & -- & -- & \textcolor{red}{\ding{56}} & \textcolor{green}{\Checkmark} \\
\cline{2-11}
 & \multirow[c]{2}{*}{$\rm D_s^\pm$} & \multirow[c]{2}{*}{+} & \multirow[c]{2}{*}{$\rm b$} & \ $6.0$ & $\times~ 10^{-7}$ & Eq. \eqref{eq:t_M_b_decay}~(this work) & \multirow[c]{2}{*}{--} & \multirow[c]{2}{*}{--} & \multirow[c]{2}{*}{\textcolor{green}{\Checkmark}} & \multirow[c]{2}{*}{\textcolor{green}{\Checkmark}} \\
 &  &  &  & \ $1.0$ & $\times~ 10^{-7}$ & EFT+LCDA~\cite{Beneke:2000hk} &  &  &  &  \\
\cline{2-11}
 & $\rm D_s^{*\pm}$ & + & $\rm b$ & \ $9.8$ & $\times~ 10^{-7}$ & Eq. \eqref{eq:t_M_b_decay}~(this work) & -- & -- & \textcolor{green}{\Checkmark} & \textcolor{green}{\Checkmark} \\

 & $\rm B^\pm$ & + & $\rm b$ & \ $5.3$ & $\times~ 10^{-12}$ & Eq. \eqref{eq:t_M_b_decay}~(this work) & -- & -- & \textcolor{red}{\ding{56}} & \textcolor{green}{\Checkmark} \\

 & $\rm B^{*\pm}$ & + & $\rm b$ & \ $4.7$ & $\times~ 10^{-12}$ & Eq. \eqref{eq:t_M_b_decay}~(this work) & -- & -- & \textcolor{red}{\ding{56}} & \textcolor{green}{\Checkmark} \\

 & $\rm B_c^\pm$ & + & $\rm b$ & \ $3.1$ & $\times~ 10^{-9}$ & Eq. \eqref{eq:t_M_b_decay}~(this work) & -- & -- & \textcolor{red}{\ding{56}} & \textcolor{green}{\Checkmark} \\

 & $\rm B_c^{*\pm}$ & + & $\rm b$ & \ $3.1$ & $\times~ 10^{-9}$ & Eq. \eqref{eq:t_M_b_decay}~(this work) & -- & -- & \textcolor{red}{\ding{56}} & \textcolor{green}{\Checkmark} \\

\bottomrule
\end{tabular}

%% file: tables/important_channels.tex
\begin{tabular}{ll<{\hspace{-3mm}}>{\hspace{-2mm}}l<{\hspace{-2mm}}>{\hspace{-2mm}}l<{\hspace{-2mm}}>{\hspace{-2mm}}l<{\hspace{-2mm}}>{\hspace{-1mm}}lr<{\hspace{-6mm}}ccccclccc}
\toprule
 &  &  &  &  &  &  &  & \multicolumn{2}{c}{Exp. limits} &  \\
 &   &    &      &       &        & \multicolumn{2}{c}{Branching fraction} & 2023 & HL-LHC & $\BR(\rm th)/\BR(\rm exp)$ \\
\midrule
\multirow[c]{7}{*}{H$~\to$} & \!\!\!\!$\gaga\gaga$ & & & & & \ $5.4$ & $\times~ 10^{-12}$ & -- & -- & --\\
\cline{2-11}
 & \multirow[c]{2}{*}{$\rm \gamma$} & \multirow[c]{2}{*}{$+$} & $\rm \rho^0$ &  &  & \ $(1.68\pm 0.08)$ & $\times~ 10^{-5}$ & $<$\ $3.7$$\times~ 10^{-4}$~\cite{CMS:2024tgj} & $\lesssim5.7\times~ 10^{-5}$ & $\sim$ 1/4 \\

 &  &  & $\rm J/\psi$ &  &  & \ $(2.95\pm 0.17)$ & $\times~ 10^{-6}$ & $<$\ $2.0$$\times~ 10^{-4}$~\cite{ATLAS:2022rej} & $\lesssim 3.9\times~ 10^{-5}$~\cite{ATLAS:2015xkp} & $\sim$ 1/10 \\
\cline{2-11}
 & \multirow[c]{2}{*}{$\rm W^\mp$} & \multirow[c]{2}{*}{$+$} & $\rm \rho^\pm$ &  &  & \ $(1.5\pm 0.1)$ & $\times~ 10^{-5}$ & -- & -- & -- \\

 &  &  & $\rm D_s^{*\pm}$ &  &  & \ $(3.5\pm 0.2)$ & $\times~ 10^{-5}$ & -- & -- & -- \\
\cline{2-11}
 & \multirow[c]{2}{*}{$\rm Z$} & \multirow[c]{2}{*}{$+$} & $\rm \rho^0$ &  &  & \ $(1.4\pm 0.1)$ & $\times~ 10^{-5}$ & $<$\ $1.2$$\times~ 10^{-2}$~\cite{CMS:2020ggo} & $\lesssim1.8\times~ 10^{-3}$ & $\sim$ 1/100 \\

 &  &  & $\rm \Upsilon(1S)$ &  &  & \ $1.7$ & $\times~ 10^{-5}$ & -- & -- & -- \\
\cline{1-11}
\multirow[c]{5}{*}{Z$~\to$} & \multirow[c]{5}{*}{$\rm \gamma$} & \multirow[c]{5}{*}{$+$} & $\rm \rho^0$ &  &  & \ $(4.19\pm 0.47)$ & $\times~ 10^{-9}$ & $<$\ $4.0$$\times~ 10^{-6}$~\cite{ATLAS:2017gko} & $\lesssim2.9\times~ 10^{-7}$ & $\sim$ 1/70 \\

 &  &  & $\rm \omega$ &  &  & \ $(2.82\pm 0.41)$ & $\times~ 10^{-8}$ & $<$\ $3.9$$\times~ 10^{-6}$~\cite{ATLAS:2023alf} & $\lesssim5.8\times~ 10^{-7}$ & $\sim$ 1/20 \\

 &  &  & $\rm \phi$ &  &  & \ $(1.17\pm 0.08)$ & $\times~ 10^{-8}$ & $<$\ $7.0$$\times~ 10^{-7}$~\cite{ATLAS:2017gko} & $\lesssim5.0\times~ 10^{-8}$ & $\sim$ 1/5 \\

 &  &  & $\rm J/\psi$ &  &  & \ $(9.96\pm 1.86)$ & $\times~ 10^{-8}$ & $<$\ $6.0$$\times~ 10^{-7}$~\cite{CMS:2024hhg} & $\lesssim 3.1\times~ 10^{-7}$~\cite{ATLAS:2015xkp} & $\sim$ 1/3 \\

 &  &  & $\rm \Upsilon(1S)$ &  &  & \ $(4.93\pm 0.51)$ & $\times~ 10^{-8}$ & $<$\ $1.1$$\times~ 10^{-6}$~\cite{ATLAS:2018xfc} & $\lesssim1.7\times~ 10^{-7}$ & $\sim$ 1/4 \\
\cline{1-11}
\multirow[c]{2}{*}{W$~\to$} & \multirow[c]{2}{*}{$\rm \gamma$} & \multirow[c]{2}{*}{$+$} & $\rm \pi^\pm$ &  &  & \ $(4.0\pm 0.8)$ & $\times~ 10^{-9}$ & $<$\ $1.9$$\times~ 10^{-6}$~\cite{ATLAS:2023jfq,CMS:2020oqe} & $\lesssim2.9\times~ 10^{-7}$ & $\sim$ 1/70 \\

 &  &  & $\rm D_s^\pm$ &  &  & \ $\left(3.66^{+1.49}_{-0.85}\right)$ & $\times~ 10^{-8}$ & $<$\ $6.5$$\times~ 10^{-4}$~\cite{LHCb:2022kta} & $\lesssim1.2\times~ 10^{-5}$ & $\sim$ 1/300 \\
\cline{1-11}
\multirow[c]{2}{*}{~~t$~~\to$} & $\rm c$ & $+$ & $\rm \overline{B}^0_s$ &  &  & \ $\left(4.0^{+4.0}_{-2.0}\right)$ & $\times~ 10^{-5}$ & --  & $\lesssim 4.0\times~ 10^{-5}$~\cite{dEnterria:2020ygk} & $\sim$ 1/1 \\

 & W & $+$ & Z & \!\!\!\! \!\!\!\! $+$~~~b &  &  $1.8$ & $\times~ 10^{-6}$ & -- & -- & --\\

\bottomrule
\end{tabular}

%% file: references.bib
@article{Bauer:2001yt,
    author = "Bauer, Christian W. and Pirjol, Dan and Stewart, Iain W.",
    title = "{Soft collinear factorization in effective field theory}",
    eprint = "hep-ph/0109045",
    archivePrefix = "arXiv",
    reportNumber = "UCSD-PTH-01-15",
    doi = "10.1103/PhysRevD.65.054022",
    journal = "Phys. Rev. D",
    volume = "65",
    pages = "054022",
    year = "2002"
}

@article{Georgi:1993mps,
    author = "Georgi, H.",
    title = "{Effective field theory}",
    doi = "10.1146/annurev.ns.43.120193.001233",
    journal = "Ann. Rev. Nucl. Part. Sci.",
    volume = "43",
    pages = "209--252",
    year = "1993"
}

@article{Neubert:1993mb,
    author = "Neubert, Matthias",
    title = "{Heavy quark symmetry}",
    eprint = "hep-ph/9306320",
    archivePrefix = "arXiv",
    reportNumber = "SLAC-PUB-6263",
    doi = "10.1016/0370-1573(94)90091-4",
    journal = "Phys. Rept.",
    volume = "245",
    pages = "259--396",
    year = "1994"
}

@article{Glashow:1970gm,
    author = "Glashow, S. L. and Iliopoulos, J. and Maiani, L.",
    title = "{Weak Interactions with Lepton-Hadron Symmetry}",
    doi = "10.1103/PhysRevD.2.1285",
    journal = "Phys. Rev. D",
    volume = "2",
    pages = "1285--1292",
    year = "1970"
}

@article{Perez:2003ad,
    author = "Perez, M. A. and Tavares-Velasco, G. and Toscano, J. J.",
    title = "{New physics effects in rare Z decays}",
    eprint = "hep-ph/0305227",
    archivePrefix = "arXiv",
    doi = "10.1142/S0217751X04017100",
    journal = "Int. J. Mod. Phys. A",
    volume = "19",
    pages = "159--178",
    year = "2004"
}

@article{Huang:2014cxa,
    author = "Huang, Ting-Chung and Petriello, Frank",
    title = "{Rare exclusive decays of the Z-boson revisited}",
    eprint = "1411.5924",
    archivePrefix = "arXiv",
    primaryClass = "hep-ph",
    doi = "10.1103/PhysRevD.92.014007",
    journal = "Phys. Rev. D",
    volume = "92",
    pages = "014007",
    year = "2015"
}

@article{Sang:2023hjl,
    author = "Sang, Wen-Long and Yang, De-Shan and Zhang, Yu-Dong",
    title = "{Z-boson radiative decays to an S-wave quarkonium at NNLO and NLL accuracy}",
    eprint = "2302.06439",
    archivePrefix = "arXiv",
    primaryClass = "hep-ph",
    doi = "10.1103/PhysRevD.108.014021",
    journal = "Phys. Rev. D",
    volume = "108",
    pages = "014021",
    year = "2023"
}

@article{Sang:2022erv,
    author = "Sang, Wen-Long and Yang, De-Shan and Zhang, Yu-Dong",
    title = "{Z boson radiative decays to a P-wave quarkonium at NNLO and LL accuracy}",
    eprint = "2208.10118",
    archivePrefix = "arXiv",
    primaryClass = "hep-ph",
    doi = "10.1103/PhysRevD.106.094023",
    journal = "Phys. Rev. D",
    volume = "106",
    pages = "094023",
    year = "2022"
}

@article{Gao:2022mwa,
    author = "Gao, Dao-Neng and Gong, Xi",
    title = "{Note on rare Z-boson decays to double heavy quarkonia}",
    eprint = "2208.12652",
    archivePrefix = "arXiv",
    primaryClass = "hep-ph",
    reportNumber = "USTC-ICTS/PCFT-22-25",
    doi = "10.1088/1674-1137/acb7d1",
    journal = "Chin. Phys. C",
    volume = "47",
    pages = "043106",
    year = "2023"
}

@article{Grossman:2015cak,
    author = {Grossman, Yuval and K\"onig, Matthias and Neubert, Matthias},
    title = "{Exclusive radiative decays of W and Z bosons in QCD factorization}",
    eprint = "1501.06569",
    archivePrefix = "arXiv",
    primaryClass = "hep-ph",
    reportNumber = "MITP-15-002",
    doi = "10.1007/JHEP04(2015)101",
    journal = "JHEP",
    volume = "04",
    pages = "101",
    year = "2015"
}

@article{Alte:2015dpo,
    author = {Alte, Stefan and K\"onig, Matthias and Neubert, Matthias},
    title = "{Exclusive radiative Z-boson decays to mesons with flavor-singlet components}",
    eprint = "1512.09135",
    archivePrefix = "arXiv",
    primaryClass = "hep-ph",
    reportNumber = "MITP-15-117",
    doi = "10.1007/JHEP02(2016)162",
    journal = "JHEP",
    volume = "02",
    pages = "162",
    year = "2016"
}

@article{dEnterria:2022alo,
    author = "d'Enterria, David and Perez-Ramos, Redamy and Shao, Hua-Sheng",
    title = "{Ditauonium spectroscopy}",
    eprint = "2204.07269",
    archivePrefix = "arXiv",
    primaryClass = "hep-ph",
    doi = "10.1140/epjc/s10052-022-10831-x",
    journal = "Eur. Phys. J. C",
    volume = "82",
    pages = "923",
    year = "2022"
}

@article{Bodwin:2017pzj,
    author = "Bodwin, Geoffrey T. and Chung, Hee Sok and Ee, June-Haak and Lee, Jungil",
    title = "{Z-boson decays to a vector quarkonium plus a photon}",
    eprint = "1709.09320",
    archivePrefix = "arXiv",
    primaryClass = "hep-ph",
    doi = "10.1103/PhysRevD.97.016009",
    journal = "Phys. Rev. D",
    volume = "97",
    pages = "016009",
    year = "2018"
}

@article{Bodwin:2017wdu,
    author = "Bodwin, Geoffrey T. and Chung, Hee Sok and Ee, June-Haak and Lee, Jungil",
    title = "{Addendum: New approach to the resummation of logarithms in Higgs-boson decays to a vector quarkonium plus a photon [Phys. Rev. D 95, 054018 (2017)]}",
    eprint = "1710.09872",
    archivePrefix = "arXiv",
    primaryClass = "hep-ph",
    doi = "10.1103/PhysRevD.96.116014",
    journal = "Phys. Rev. D",
    volume = "96",
    pages = "116014",
    year = "2017"
}

@article{Wang:2018rjg,
    author = "Wang, Jun-Zhang and Sun, Zhi-Feng and Liu, Xiang and Matsuki, Takayuki",
    title = "{Higher bottomonium zoo}",
    eprint = "1802.04938",
    archivePrefix = "arXiv",
    primaryClass = "hep-ph",
    doi = "10.1140/epjc/s10052-018-6372-1",
    journal = "Eur. Phys. J. C",
    volume = "78",
    pages = "915",
    year = "2018"
}

@article{Alte:2016yuw,
    author = {Alte, Stefan and K\"onig, Matthias and Neubert, Matthias},
    title = "Exclusive Weak Radiative {Higgs} Decays in the Standard Model and Beyond",
    eprint = "1609.06310",
    archivePrefix = "arXiv",
    primaryClass = "hep-ph",
    reportNumber = "MITP-16-070",
    doi = "10.1007/JHEP12(2016)037",
    journal = "JHEP",
    volume = "12",
    pages = "037",
    year = "2016"
}

@article{Isidori:2013cla,
    author = "Isidori, Gino and Manohar, Aneesh V. and Trott, Michael",
    title = "Probing the nature of the {Higgs}-like boson via {$\rm H \to V \mathcal{F}$} decays",
    eprint = "1305.0663",
    archivePrefix = "arXiv",
    primaryClass = "hep-ph",
    reportNumber = "CERN-PH-TH-2013-159, CERN-PH-TH-159",
    doi = "10.1016/j.physletb.2013.11.054",
    journal = "Phys. Lett. B",
    volume = "728",
    pages = "131--135",
    year = "2014"
}

@article{Faustov:2022jfk,
author = "Faustov, R. N. and Martynenko, A. P. and Martynenko, F. A.",
title = "{Relativistic corrections to paired production of charmonium and bottomonium in decays of the Higgs boson}",
eprint = "2209.12321",
archivePrefix = "arXiv",
primaryClass = "hep-ph",
doi = "10.1103/PhysRevD.107.056002",
journal = "Phys. Rev. D",
volume = "107",
pages = "056002",
year = "2023"
}

@article{Gao:2022iam,
    author = "Gao, Dao-Neng and Gong, Xi",
    title = "{Higgs boson decays into a pair of heavy vector quarkonia}",
    eprint = "2203.00514",
    archivePrefix = "arXiv",
    primaryClass = "hep-ph",
    reportNumber = "USTC-ICTS/PCFT-22-05",
    doi = "10.1016/j.physletb.2022.137243",
    journal = "Phys. Lett. B",
    volume = "832",
    pages = "137243",
    year = "2022"
}

@article{Kartvelishvili:2008tz,
    author = "Kartvelishvili, V. and Luchinsky, A. V. and Novoselov, A. A.",
    title = "{Double vector quarkonia production in exclusive Higgs boson decays}",
    eprint = "0810.0953",
    archivePrefix = "arXiv",
    primaryClass = "hep-ph",
    doi = "10.1103/PhysRevD.79.114015",
    journal = "Phys. Rev. D",
    volume = "79",
    pages = "114015",
    year = "2009"
}

@article{Belov:2021toy,
    author = "Belov, I. N. and Berezhnoy, A. V. and Dorokhov, A. E. and Likhoded, A. K. and Martynenko, A. P. and Martynenko, F. A.",
    title = "{Higgs boson decay to paired $\rm B_c$: Relativistic and one-loop corrections}",
    eprint = "2105.02207",
    archivePrefix = "arXiv",
    primaryClass = "hep-ph",
    doi = "10.1016/j.nuclphysa.2021.122285",
    journal = "Nucl. Phys. A",
    volume = "1015",
    pages = "122285",
    year = "2021"
}

@article{Jiang:2015pah,
    author = "Jiang, Jun and Qiao, Cong-Feng",
    title = "{$\rm B_c$ Production in Higgs boson decays}",
    eprint = "1512.01327",
    archivePrefix = "arXiv",
    primaryClass = "hep-ph",
    doi = "10.1103/PhysRevD.93.054031",
    journal = "Phys. Rev. D",
    volume = "93",
    pages = "054031",
    year = "2016"
}

@article{Luo:2022ugd,
    author = "Luo, Xuan and Fu, Hai-Bing and Tian, Hai-Jiang and Li, Cong",
    title = "{Next-to-leading-order QCD correction to the exclusive double charmonium production via Z decays}",
    eprint = "2209.08802",
    archivePrefix = "arXiv",
    primaryClass = "hep-ph",
    month = "9",
    year = "2022"
}

@article{Likhoded:2017jmx,
    author = "Likhoded, A. K. and Luchinsky, A. V.",
    title = "{Double charmonia production in exclusive Z boson decays}",
    eprint = "1712.03108",
    archivePrefix = "arXiv",
    primaryClass = "hep-ph",
    doi = "10.1142/S0217732318500785",
    journal = "Mod. Phys. Lett. A",
    volume = "33",
    pages = "1850078",
    year = "2018"
}

@article{Belov:2023iop,
    author = "Belov, I. N. and Berezhnoy, A. V. and Leshchenko, E. A. and Likhoded, A. K.",
    title = "{QCD one-loop correction to Higgs boson decay into quarkonium pairs}",
    eprint = "2304.11620",
    archivePrefix = "arXiv",
    primaryClass = "hep-ph",
    doi = "10.1103/PhysRevD.108.036013",
    journal = "Phys. Rev. D",
    volume = "108",
    pages = "036013",
    year = "2023"
}

@article{Beneke:2023nmj,
    author = "Beneke, Martin and Finauri, Gael and Vos, K. Keri and Wei, Yanbing",
    title = "{QCD light-cone distribution amplitudes of heavy mesons from boosted HQET}",
    eprint = "2305.06401",
    archivePrefix = "arXiv",
    primaryClass = "hep-ph",
    reportNumber = "TUM-HEP-1455/23, Nikhef-2023-003",
    doi = "10.1007/JHEP09(2023)066",
    journal = "JHEP",
    volume = "09",
    pages = "066",
    year = "2023"
}

@article{Konig:2015qat,
    author = {K\"onig, Matthias and Neubert, Matthias},
    title = "Exclusive Radiative {Higgs} Decays as Probes of Light-Quark {Yukawa} Couplings",
    eprint = "1505.03870",
    archivePrefix = "arXiv",
    primaryClass = "hep-ph",
    reportNumber = "MITP-15-031",
    doi = "10.1007/JHEP08(2015)012",
    journal = "JHEP",
    volume = "08",
    pages = "012",
    year = "2015"
}

@article{Luchinsky:2021amp,
    author = "Luchinsky, A. V. and Likhoded, A. K.",
    title = "{Charmonia production in $\rm W \to (c\bar c) D_s^{(*)}$ decays}",
    doi = "10.1142/S0217732321500589",
    eprint = "1801.08998",
    journal = "Mod. Phys. Lett. A",
    volume = "36",
    pages = "2150058",
    year = "2021"
}

@article{Eilam:2006uh,
    author = "Eilam, Gad and Frank, Mariana and Turan, Ismail",
    title = "{Rare decay of the top $\rm t \to cgg$ in the standard model}",
    eprint = "hep-ph/0601151",
    archivePrefix = "arXiv",
    reportNumber = "CUMQ-HEP-138",
    doi = "10.1103/PhysRevD.73.053011",
    journal = "Phys. Rev. D",
    volume = "73",
    pages = "053011",
    year = "2006"
}

@article{Balaji:2020qjg,
    author = "Balaji, Shyam",
    title = "{$CP$ asymmetries in the rare top decays $\rm t\to c\gamma$ and $\rm t\to c g$}",
    eprint = "2009.03315",
    archivePrefix = "arXiv",
    primaryClass = "hep-ph",
    doi = "10.1103/PhysRevD.102.113010",
    journal = "Phys. Rev. D",
    volume = "102",
    pages = "113010",
    year = "2020"
}

@article{dEnterria:2020ygk,
    author = "d'Enterria, David and Shao, Hua-Sheng",
    title = "{Rare two-body decays of the top quark into a bottom meson plus an up or charm quark}",
    eprint = "2005.08102",
    archivePrefix = "arXiv",
    primaryClass = "hep-ph",
    doi = "10.1007/JHEP07(2020)127",
    journal = "JHEP",
    volume = "07",
    pages = "127",
    year = "2020"
}

@article{Papaefstathiou:2017xuv,
    author = "Papaefstathiou, Andreas and Tetlalmatzi-Xolocotzi, Gilberto",
    title = "{Rare top quark decays at a 100 TeV proton-proton collider: $\rm t \to bWZ$ and $\rm t\to hc$}",
    eprint = "1712.06332",
    archivePrefix = "arXiv",
    primaryClass = "hep-ph",
    reportNumber = "NIKHEF-2017-069, MCNET-17-24",
    doi = "10.1140/epjc/s10052-018-5701-8",
    journal = "Eur. Phys. J. C",
    volume = "78",
    pages = "214",
    year = "2018"
}

@article{Cordero-Cid:2004pco,
    author = "Cordero-Cid, Adriana and Hernandez, J. M. and Tavares-Velasco, G. and Toscano, J. J.",
    title = "{Rare top quark decay $\rm t \to u_{1} \bar{u}_{2} u_{2}$ in the standard model}",
    eprint = "hep-ph/0411188",
    archivePrefix = "arXiv",
    doi = "10.1103/PhysRevD.73.094005",
    journal = "Phys. Rev. D",
    volume = "73",
    pages = "094005",
    year = "2006"
}

@inproceedings{Mahlon:1998fr,
    author = "Mahlon, Gregory",
    title = "{Theoretical expectations in radiative top decays}",
    booktitle = "{Thinkshop on Top Quark Physics for Run II}",
    eprint = "hep-ph/9810485",
    archivePrefix = "arXiv",
    reportNumber = "MCGILL-98-31",
    month = "10",
    year = "1998"
}

@article{Bar-Shalom:2005ldb,
    author = "Bar-Shalom, Shaouly and Eilam, Gad and Frank, Mariana and Turan, Ismail",
    title = "{Width effects on near threshold decays of the top quark $\rm t \to cWW, cZZ$ and of neutral Higgs bosons}",
    eprint = "hep-ph/0506167",
    archivePrefix = "arXiv",
    reportNumber = "MADPH-05-1425, CUMQ-HEP-134, WM-05-108",
    doi = "10.1103/PhysRevD.72.055018",
    journal = "Phys. Rev. D",
    volume = "72",
    pages = "055018",
    year = "2005"
}

@inproceedings{Beneke:2000hk,
    author = "Beneke, M. and others",
    title = "{Top quark physics}",
    booktitle = "{Workshop on Standard Model Physics (and more) at the LHC (First Plenary Meeting)}",
    eprint = "hep-ph/0003033",
    archivePrefix = "arXiv",
    reportNumber = "CERN-TH-2000-100, FERMILAB-CONF-00-398-T",
    doi = "10.5170/CERN-2000-004.419",
    pages = "419--529",
    month = "3",
    year = "2000"
}

@article{Aguilar-Saavedra:2004mfd,
    author = "Aguilar-Saavedra, J. A.",
    editor = "del Aguila, F. and Pittau, R. and Djouadi, A. and Papadopoulos, C. G.",
    title = "{Top flavor-changing neutral interactions: Theoretical expectations and experimental detection}",
    eprint = "hep-ph/0409342",
    archivePrefix = "arXiv",
    journal = "Acta Phys. Polon. B",
    volume = "35",
    pages = "2695--2710",
    year = "2004"
}

@article{Bagdatova:2023etj,
    author = "Bagdatova, Alsu G. and Baranov, Sergey P. and Sakharov, Alexander S.",
    title = "{Study of exclusive two-body W decays with fully reconstructible kinematics}",
    eprint = "2303.02136",
    archivePrefix = "arXiv",
    primaryClass = "hep-ph",
    doi = "10.1142/S0217732323500736",
    journal = "Mod. Phys. Lett. A",
    volume = "38",
    pages = "2350073",
    year = "2023"
}

@article{Bodwin:2013gca,
    author = "Bodwin, Geoffrey T. and Petriello, Frank and Stoynev, Stoyan and Velasco, Mayda",
    title = "{Higgs boson decays to quarkonia and the $\rm H\bar{c}c$  coupling}",
    eprint = "1306.5770",
    archivePrefix = "arXiv",
    primaryClass = "hep-ph",
    doi = "10.1103/PhysRevD.88.053003",
    journal = "Phys. Rev. D",
    volume = "88",
    pages = "053003",
    year = "2013"
}

@article{Li:2023tzx,
author = "Li, Cong and Sun, Zhan and Zhang, Gui-Yuan",
title = "{Analysis of double-$\jpsi$ production in Z decay at next-to-leading-order QCD accuracy}",
eprint = "2307.08376",
archivePrefix = "arXiv",
primaryClass = "hep-ph",
doi = "10.1007/JHEP10(2023)120",
journal = "JHEP",
volume = "10",
pages = "120",
year = "2023"
}

@article{Dong:2022ayy,
    author = "Dong, Hongxin and Sun, Peng and Yan, Bin and Yuan, C. -P.",
    title = "{Probing the Zbb anomalous couplings via exclusive Z boson decay}",
    eprint = "2201.11635",
    archivePrefix = "arXiv",
    primaryClass = "hep-ph",
    reportNumber = "LA-UR-22-20027, MSUHEP-22-001",
    doi = "10.1016/j.physletb.2022.137076",
    journal = "Phys. Lett. B",
    volume = "829",
    pages = "137076",
    year = "2022"
}

@article{Faustov:2023phi,
    author = "Faustov, R. N. and Martynenko, A. P. and Martynenko, F. A.",
    title = "Pair Quarkonium Production in {Higgs} Boson Decay",
    doi = "10.1134/S1547477123030287",
    journal = "Phys. Part. Nucl. Lett.",
    volume = "20",
    pages = "368--371",
    year = "2023"
}

@article{Handoko:1999iu,
    author = "Handoko, L. T. and Qiao, Cong-Feng",
    title = "{$\rm t \to c \Upsilon$ within and beyond the Standard Model}",
    eprint = "hep-ph/9907375",
    archivePrefix = "arXiv",
    reportNumber = "DESY-99-096, LFTMLIPI-139901",
    doi = "10.1088/0954-3899/27/7/302",
    journal = "J. Phys. G",
    volume = "27",
    pages = "1391--1404",
    year = "2001"
}

@article{Slabospitskii:2021rns,
    author = "Slabospitskii, Sergei",
    title = "{Top-quark decay into $\Upsilon$-meson}",
    eprint = "2107.04398",
    archivePrefix = "arXiv",
    primaryClass = "hep-ph",
    doi = "10.1016/j.physletb.2021.136704",
    journal = "Phys. Lett. B",
    volume = "822",
    pages = "136704",
    year = "2021"
}

@article{Altarelli:2000nt,
    author = "Altarelli, Guido and Conti, L. and Lubicz, V.",
    title = "{The $\rm t \to WZ b$ decay in the standard model: A Critical reanalysis}",
    eprint = "hep-ph/0010090",
    archivePrefix = "arXiv",
    reportNumber = "CERN-TH-2000-170, RM3-TH-00-12",
    doi = "10.1016/S0370-2693(00)01333-2",
    journal = "Phys. Lett. B",
    volume = "502",
    pages = "125--132",
    year = "2001"
}

@article{ATLAS:2018zsq,
    author = "Aaboud, M. and others",
    collaboration = "ATLAS",
    title = "{Search for flavour-changing neutral current top-quark decays $\rm t\to qZ$ in proton-proton collisions at $\sqrt{s}=13$ TeV with the ATLAS detector}",
    eprint = "1803.09923",
    archivePrefix = "arXiv",
    primaryClass = "hep-ex",
    reportNumber = "CERN-EP-2018-018",
    doi = "10.1007/JHEP07(2018)176",
    journal = "JHEP",
    volume = "07",
    pages = "176",
    year = "2018"
}

@article{Luchinsky:2017jab,
    author = "Luchinsky, A. V.",
    title = "Leading order {NRQCD} and Light-Cone Analysis of Exclusive Charmonia Production in Radiative {Z}-boson Decays",
    eprint = "1706.04091",
    archivePrefix = "arXiv",
    primaryClass = "hep-ph",
    month = "6",
    year = "2017"
}

@article{CMS:2018gcm,
    author = "Sirunyan, Albert M and others",
    collaboration = "CMS",
    title = "{Search for rare decays of Z and Higgs bosons to $\jpsi$ and a photon in proton-proton collisions at $\sqrt{s} =$ 13 TeV}",
    eprint = "1810.10056",
    archivePrefix = "arXiv",
    primaryClass = "hep-ex",
    reportNumber = "CMS-SMP-17-012, CERN-EP-2018-250",
    doi = "10.1140/epjc/s10052-019-6562-5",
    journal = "Eur. Phys. J. C",
    volume = "79",
    pages = "94",
    year = "2019"
}

@article{CMS:2022fsq,
    author = "Tumasyan, Armen and others",
    collaboration = "CMS",
    title = "{Search for Higgs boson decays into Z and $\jpsi$ and for Higgs and Z boson decays into $\jpsi$ or $\Upsilon$ pairs in pp collisions at $\sqrt{s} =13$~TeV}",
    eprint = "2206.03525",
    archivePrefix = "arXiv",
    primaryClass = "hep-ex",
    reportNumber = "CMS-HIG-20-008, CERN-EP-2022-074",
    doi = "10.1016/j.physletb.2022.137534",
    journal = "Phys. Lett. B",
    volume = "842",
    pages = "137534",
    year = "2023"
}

@article{Gao:2014xlv,
    author = "Gao, Dao-Neng",
    title = "{A note on Higgs decays into Z boson and $\jpsi(\Upsilon)$}",
    eprint = "1406.7102",
    archivePrefix = "arXiv",
    primaryClass = "hep-ph",
    reportNumber = "USTC-ICTS-14-10",
    doi = "10.1016/j.physletb.2014.09.019",
    journal = "Phys. Lett. B",
    volume = "737",
    pages = "366--368",
    year = "2014"
}

@article{ATLAS:2018xfc,
    author = "Aaboud, Morad and others",
    collaboration = "ATLAS",
    title = "{Searches for exclusive Higgs and Z boson decays into $\jpsi\gamma$, $\psi(2S)\gamma$, and $\Upsilon(nS)\gamma$ at $\sqrt{s}=13$ TeV with the ATLAS detector}",
    eprint = "1807.00802",
    archivePrefix = "arXiv",
    primaryClass = "hep-ex",
    reportNumber = "CERN-EP-2018-154",
    doi = "10.1016/j.physletb.2018.09.024",
    journal = "Phys. Lett. B",
    volume = "786",
    pages = "134--155",
    year = "2018"
}

@article{LHCb:2022kta,
    author = "Aaij, R. and others",
    collaboration = "LHCb,",
    title = "{Search for the rare decays $\rm W^+ \to D^+_s\gamma$ and $\rm Z \to D^0\gamma$ at LHCb}",
    eprint = "2212.07120",
    archivePrefix = "arXiv",
    primaryClass = "hep-ex",
    reportNumber = "CERN-EP-2022-252, LHCb-PAPER-2022-033",
    doi = "10.1088/1674-1137/aceae9",
    journal = "Chin. Phys. C",
    volume = "47",
    pages = "093002",
    year = "2023"
}

@article{Guberina:1980dc,
    author = "Guberina, B. and Kuhn, Johann H. and Peccei, R. D. and Ruckl, R.",
    title = "Rare Decays of the {Z$^0$}",
    reportNumber = "MPI-PAE/PTh 10/80",
    doi = "10.1016/0550-3213(80)90287-4",
    journal = "Nucl. Phys. B",
    volume = "174",
    pages = "317--334",
    year = "1980"
}

@article{Arnellos:1981gy,
    author = "Arnellos, Lampros and Marciano, William J. and Parsa, Zohreh",
    title = "Radiative Decays {$\rm W^\pm \to \rho^\pm \gamma$ and $\rm Z^0 \to \rho^0 \gamma$}",
    reportNumber = "PRINT-81-0619 (NORTHWESTERN)",
    doi = "10.1016/0550-3213(82)90496-5",
    journal = "Nucl. Phys. B",
    volume = "196",
    pages = "378--393",
    year = "1982"
}

@article{CDF:1998kzn,
    author = "Abe, F. and others",
    collaboration = "CDF",
    title = "{Search for the rare decay $\rm W^\pm \to D_s^{\pm} \gamma$ in $\rm p\bar{p}$ collisions at $\sqrt{s} = 1.8$ TeV}",
    reportNumber = "FERMILAB-PUB-98-110-E",
    doi = "10.1103/PhysRevD.58.091101",
    journal = "Phys. Rev. D",
    volume = "58",
    pages = "091101",
    year = "1998"
}

@article{ATLAS:2023alf,
    author = "Aad, Georges and others",
    collaboration = "ATLAS",
    title = "{Search for exclusive Higgs and Z boson decays to $\omega\gamma$ and Higgs boson decays to $K^{*}\gamma$ with the ATLAS detector}",
    eprint = "2301.09938",
    archivePrefix = "arXiv",
    primaryClass = "hep-ex",
    reportNumber = "CERN-EP-2022-288",
    doi = "10.1016/j.physletb.2023.138292",
    journal = "Phys. Lett. B",
    volume = "847",
    pages = "138292",
    year = "2023"
}

@article{ATLAS:2022rej,
    author = "Aad, Georges and others",
    collaboration = "ATLAS",
    title = "{Searches for exclusive Higgs and Z boson decays into a vector quarkonium state and a photon using 139~fb$^{-1}$ of ATLAS $\sqrt{s}=13$~TeV proton\textendash{}proton collision data}",
    eprint = "2208.03122",
    archivePrefix = "arXiv",
    primaryClass = "hep-ex",
    reportNumber = "CERN-EP-2022-128",
    doi = "10.1140/epjc/s10052-023-11869-1",
    journal = "Eur. Phys. J. C",
    volume = "83",
    pages = "781",
    year = "2023"
}

@unpublished{CMS:2022kdd,
    author = "Tumasyan, Armen and others",
    collaboration = "CMS",
    title = "{Search for rare Higgs boson decays with mesons at the HL-LHC}",
    note = "CMS-PAS-FTR-21-009",
    year = "2022"
}

@article{CMS:2020ggo,
    author = "Sirunyan, Albert M and others",
    collaboration = "CMS",
    title = "{Search for decays of the 125 GeV Higgs boson into a Z boson and a $\rho$ or $\phi$ meson}",
    eprint = "2007.05122",
    archivePrefix = "arXiv",
    primaryClass = "hep-ex",
    reportNumber = "CMS-HIG-19-012, CERN-EP-2020-120, FERMILAB-PUB-20-340-CMS",
    doi = "10.1007/JHEP11(2020)039",
    journal = "JHEP",
    volume = "11",
    pages = "039",
    year = "2020"
}

@article{CMS:2019wch,
    author = "Sirunyan, Albert M and others",
    collaboration = "CMS",
    title = "{Search for Higgs and Z boson decays to $\jpsi$ or $\Upsilon$ pairs in the four-muon final state in proton-proton collisions at $\sqrts = 13$~TeV}",
    eprint = "1905.10408",
    archivePrefix = "arXiv",
    primaryClass = "hep-ex",
    reportNumber = "CMS-HIG-18-025, CERN-EP-2019-082",
    doi = "10.1016/j.physletb.2019.134811",
    journal = "Phys. Lett. B",
    volume = "797",
    pages = "134811",
    year = "2019"
}

@article{ATLAS:2017gko,
    author = "Aaboud, M. and others",
    collaboration = "ATLAS",
    title = "{Search for exclusive Higgs and Z boson decays to $\phi\gamma$ and $\rho\gamma$ with the ATLAS detector}",
    eprint = "1712.02758",
    archivePrefix = "arXiv",
    primaryClass = "hep-ex",
    reportNumber = "CERN-EP-2017-273",
    doi = "10.1007/JHEP07(2018)127",
    journal = "JHEP",
    volume = "07",
    pages = "127",
    year = "2018"
}

@article{Brambilla:2019fmu,
    author = "Brambilla, Nora and Chung, Hee Sok and Lai, Wai Kin and Shtabovenko, Vladyslav and Vairo, Antonio",
    title = "{Order $v^4$ corrections to Higgs boson decay into $\jpsi + \gamma$}",
    eprint = "1907.06473",
    archivePrefix = "arXiv",
    primaryClass = "hep-ph",
    reportNumber = "TUM-EFT 123/19, P3H-19-016, TTP19-019",
    doi = "10.1103/PhysRevD.100.054038",
    journal = "Phys. Rev. D",
    volume = "100",
    pages = "054038",
    year = "2019"
}

@article{Chung:2019ota,
    author = "Chung, Hee Sok and Ee, June-Haak and Kang, Daekyoung and Kim, U-Rae and Lee, Jungil and Wang, Xiang-Peng",
    title = "Pseudoscalar Quarkonium$+\gamma$ Production at {NLL+NLO} accuracy",
    eprint = "1906.03275",
    archivePrefix = "arXiv",
    primaryClass = "hep-ph",
    doi = "10.1007/JHEP10(2019)162",
    journal = "JHEP",
    volume = "10",
    pages = "162",
    year = "2019"
}

@unpublished{TheATLAScollaboration:2013nbo,
    author = "Aad, Georges and others",
    collaboration = "ATLAS",
    title = "{Sensitivity of ATLAS at HL-LHC to flavour changing neutral currents in top quark decays $\rm t \to cH$, with $\rm H \to \gamma\gamma$}",
    note = "ATL-PHYS-PUB-2013-012",
    year = "2013"
}

@unpublished{ATLAS:2015xkp,
    author = "Aad, Georges and others",
    collaboration = "ATLAS",
    title = "{Search for the Standard Model Higgs and Z boson decays to $\jpsi\,\gamma$: HL-LHC projections}",
    note = "ATL-PHYS-PUB-2015-043",
    year = "2015"
}

@article{CDF:2013lma,
    author = "Aaltonen, Timo Antero and others",
    collaboration = "CDF",
    title = "First Search for Exotic {Z} boson Decays into Photons and Neutral Pions in Hadron Collisions",
    eprint = "1311.3282",
    archivePrefix = "arXiv",
    primaryClass = "hep-ex",
    reportNumber = "FERMILAB-PUB-13-509-E",
    doi = "10.1103/PhysRevLett.112.111803",
    journal = "Phys. Rev. Lett.",
    volume = "112",
    pages = "111803",
    year = "2014"
}

@article{ALEPH:1991qhf,
    author = "Decamp, D. and others",
    collaboration = "ALEPH",
    title = "{Searches for new particles in Z decays using the ALEPH detector}",
    reportNumber = "CERN-PPE-91-149",
    doi = "10.1016/0370-1573(92)90177-2",
    journal = "Phys. Rept.",
    volume = "216",
    pages = "253--340",
    year = "1992"
}

@article{Fael:2018ktm,
    author = "Fael, Matteo and Mannel, Thomas",
    title = "{On the decays $\rm B \to K^{(*)} + $ leptonium}",
    eprint = "1803.08880",
    archivePrefix = "arXiv",
    primaryClass = "hep-ph",
    reportNumber = "SI-HEP-2018-12, QFET-2018-06",
    doi = "10.1016/j.nuclphysb.2018.05.015",
    journal = "Nucl. Phys. B",
    volume = "932",
    pages = "370--384",
    year = "2018"
}

@article{Korchin:2013ifa,
    author = "Korchin, Alexander Yu. and Kovalchuk, Vladimir A.",
    title = "{Polarization effects in the Higgs boson decay to $\rm \gamma Z$ and test of $CP$ and $CPT$ symmetries}",
    eprint = "1303.0365",
    archivePrefix = "arXiv",
    primaryClass = "hep-ph",
    doi = "10.1103/PhysRevD.88.036009",
    journal = "Phys. Rev. D",
    volume = "88",
    pages = "036009",
    year = "2013"
}

@article{DELPHI:1996bav,
    author = "Abreu, P. and others",
    collaboration = "DELPHI",
    title = "{An Upper limit for Br(Z$^0$ $\to$ g g g) from symmetric three jet Z$^0$ hadronic decays}",
    reportNumber = "CERN-PPE-96-131",
    doi = "10.1016/S0370-2693(96)01450-5",
    journal = "Phys. Lett. B",
    volume = "389",
    pages = "405--415",
    year = "1996"
}

@article{ATLAS:2019mke,
    author = "Aad, Georges and others",
    collaboration = "ATLAS",
    title = "{Search for flavour-changing neutral currents in processes with one top quark and a photon using 81 fb$^{-1}$ of pp collisions at $\sqrt{s} = 13$ TeV with the ATLAS experiment}",
    eprint = "1908.08461",
    archivePrefix = "arXiv",
    primaryClass = "hep-ex",
    reportNumber = "CERN-EP-2019-155",
    doi = "10.1016/j.physletb.2019.135082",
    journal = "Phys. Lett. B",
    volume = "800",
    pages = "135082",
    year = "2020"
}

@article{CMS:2021hug,
    author = "Tumasyan, A. and others",
    collaboration = "CMS",
    title = "Search for Flavor-Changing Neutral Current Interactions of the Top Quark and {Higgs} boson in Final States with Two Photons in Proton-Proton Collisions at $\sqrt{s}=13$~{TeV}",
    eprint = "2111.02219",
    archivePrefix = "arXiv",
    primaryClass = "hep-ex",
    reportNumber = "CMS-TOP-20-007, CERN-EP-2021-201",
    doi = "10.1103/PhysRevLett.129.032001",
    journal = "Phys. Rev. Lett.",
    volume = "129",
    pages = "032001",
    year = "2022"
}

@article{CMS:2020oqe,
    author = "Sirunyan, Albert M and others",
    collaboration = "CMS",
    title = "{Search for the rare decay of the W boson into a pion and a photon in proton-proton collisions at $\sqrts=13$~TeV}",
    eprint = "2011.06028",
    archivePrefix = "arXiv",
    primaryClass = "hep-ex",
    reportNumber = "CMS-SMP-20-008, CERN-EP-2020-197",
    doi = "10.1016/j.physletb.2021.136409",
    journal = "Phys. Lett. B",
    volume = "819",
    pages = "136409",
    year = "2021"
}

@article{ATLAS:2015rsn,
    author = "Aad, Georges and others",
    collaboration = "ATLAS",
    title = "{Search for new phenomena in events with at least three photons collected in pp collisions at $\sqrt{s}$ = 8 TeV with the ATLAS detector}",
    eprint = "1509.05051",
    archivePrefix = "arXiv",
    primaryClass = "hep-ex",
    reportNumber = "CERN-PH-EP-2015-187",
    doi = "10.1140/epjc/s10052-016-4034-8",
    journal = "Eur. Phys. J. C",
    volume = "76",
    pages = "210",
    year = "2016"
}

@article{Mangano:2014xta,
    author = "Mangano, Michelangelo and Melia, Tom",
    title = "{Rare exclusive hadronic W decays in a $\rm t\bar{t}$ environment}",
    eprint = "1410.7475",
    archivePrefix = "arXiv",
    primaryClass = "hep-ph",
    doi = "10.1140/epjc/s10052-015-3482-x",
    journal = "Eur. Phys. J. C",
    volume = "75",
    pages = "258",
    year = "2015"
}

@article{Melia:2016knk,
    author = "Melia, Tom",
    editor = "Aguilar-Ben\'\i{}tez, M and Fuster, J and Mart\'\i{}-Garc\'\i{}a, S and Santamar\'\i{}a, A",
    title = "{Exclusive hadronic W decay: $\rm W \to \pi \gamma$ and $\rm W \to \pi^+ \pi^+ \pi^-$}",
    doi = "10.1016/j.nuclphysbps.2015.09.341",
    journal = "Nucl. Part. Phys. Proc.",
    volume = "273-275",
    pages = "2102--2106",
    year = "2016"
}

@article{Laursen:1982rg,
    author = "Laursen, Morten L. and Samuel, Mark A. and Tupper, Gary B. and Sen, Achin",
    title = "{$\rm Z^0$} Decay Into Two Gluons and a Photon for Massive Quarks",
    reportNumber = "PRINT-82-0618 (OKLAHOMA-STATE)",
    doi = "10.1103/PhysRevD.27.196",
    journal = "Phys. Rev. D",
    volume = "27",
    pages = "196",
    year = "1983"
}

@article{Hopker:1993pb,
    author = "Hopker, Roland and van der Bij, J. J.",
    title = "{Z$^0$ decay into three gluons}",
    reportNumber = "FREIBURG-THEP-93-3",
    doi = "10.1103/PhysRevD.49.3779",
    journal = "Phys. Rev. D",
    volume = "49",
    pages = "3779--3782",
    year = "1994"
}

@article{Glover:1993nv,
    author = "Glover, E. W. Nigel and Morgan, A. G.",
    title = "{Z boson decay into photons}",
    reportNumber = "DTP-93-4, DTP-93-04",
    doi = "10.1007/BF01650444",
    journal = "Z. Phys. C",
    volume = "60",
    pages = "175--180",
    year = "1993"
}

@article{ATLAS:2023jfq,
    author = "Aad, Georges and others",
    collaboration = "ATLAS",
    title = "Search for the Exclusive {W} boson Hadronic Decays {$\rm W^{\pm}\to \pi^{\pm}\gamma$, $\rm W^{\pm}\to K^{\pm}\gamma$ and $\rm W^{\pm}\to\rho^{\pm}\gamma$ with the ATLAS} Detector",
    eprint = "2309.15887",
    archivePrefix = "arXiv",
    primaryClass = "hep-ex",
    reportNumber = "CERN-EP-2023-211",
    doi = "10.1103/PhysRevLett.133.161804",
    journal = "Phys. Rev. Lett.",
    volume = "133",
    pages = "161804",
    year = "2024"
}

@article{Hernandez:1999xn,
    author = "Hernandez, J. M. and Perez, M. A. and Tavares-Velasco, G. and Toscano, J. J.",
    title = "{Decay $\rm Z \to \nu\bar{\nu}\gamma$ in the standard model}",
    eprint = "hep-ph/9903391",
    archivePrefix = "arXiv",
    doi = "10.1103/PhysRevD.60.013004",
    journal = "Phys. Rev. D",
    volume = "60",
    pages = "013004",
    year = "1999"
}

@article{dEnterria:2023yao,
    author = "d'Enterria, David and Shao, Hua-Sheng",
    title = "{Prospects for ditauonium discovery at colliders}",
    eprint = "2302.07365",
    archivePrefix = "arXiv",
    primaryClass = "hep-ph",
    doi = "10.1016/j.physletb.2023.137960",
    journal = "Phys. Lett. B",
    volume = "842",
    pages = "137960",
    year = "2023"
}

@article{Lepage:1979zb,
    author = "Lepage, G. Peter and Brodsky, Stanley J.",
    title = "{Exclusive Processes in Quantum Chromodynamics: Evolution Equations for Hadronic Wave Functions and the Form-Factors of Mesons}",
    reportNumber = "SLAC-PUB-2343",
    doi = "10.1016/0370-2693(79)90554-9",
    journal = "Phys. Lett. B",
    volume = "87",
    pages = "359--365",
    year = "1979"
}

@article{Lepage:1980fj,
    author = "Lepage, G. Peter and Brodsky, Stanley J.",
    title = "{Exclusive Processes in Perturbative Quantum Chromodynamics}",
    reportNumber = "SLAC-PUB-2478",
    doi = "10.1103/PhysRevD.22.2157",
    journal = "Phys. Rev. D",
    volume = "22",
    pages = "2157",
    year = "1980"
}

@article{Efremov:1978rn,
    author = "Efremov, A. V. and Radyushkin, A. V.",
    title = "Asymptotical Behavior of Pion Electromagnetic Form-Factor in {QCD}",
    reportNumber = "JINR-E2-11983",
    doi = "10.1007/BF01032111",
    journal = "Theor. Math. Phys.",
    volume = "42",
    pages = "97--110",
    year = "1980"
}

@article{Efremov:1979qk,
    author = "Efremov, A. V. and Radyushkin, A. V.",
    title = "Factorization and Asymptotical Behavior of Pion Form-Factor in {QCD}",
    reportNumber = "JINR-P2-12900",
    doi = "10.1016/0370-2693(80)90869-2",
    journal = "Phys. Lett. B",
    volume = "94",
    pages = "245--250",
    year = "1980"
}

@article{Chernyak:1983ej,
    author = "Chernyak, V. L. and Zhitnitsky, A. R.",
    title = "Asymptotic Behavior of Exclusive Processes in {QCD}",
    reportNumber = "IYF-83-103, IYF-83-104, IYF-83-105, IYF-83-106, IYF-83-107, IYF-83-108",
    doi = "10.1016/0370-1573(84)90126-1",
    journal = "Phys. Rept.",
    volume = "112",
    pages = "173",
    year = "1984"
}

@article{Azzi:2019yne,
    author = "Azzi, P. and others",
    editor = "Dainese, Andrea and Mangano, Michelangelo and Meyer, Andreas B. and Nisati, Aleandro and Salam, Gavin and Vesterinen, Mika Anton",
    title = "{Report from Working Group 1: Standard Model Physics at the HL-LHC and HE-LHC}",
    eprint = "1902.04070",
    archivePrefix = "arXiv",
    primaryClass = "hep-ph",
    reportNumber = "CERN-LPCC-2018-03",
    doi = "10.23731/CYRM-2019-007.1",
    journal = "CERN Yellow Rep. Monogr.",
    volume = "7",
    pages = "1--220",
    year = "2019"
}

@article{CMS:2012qbp,
    author = "Chatrchyan, Serguei and others",
    collaboration = "CMS",
    title = "Observation of a New Boson at a Mass of 125 {GeV with the CMS Experiment at the LHC}",
    eprint = "1207.7235",
    archivePrefix = "arXiv",
    primaryClass = "hep-ex",
    reportNumber = "CMS-HIG-12-028, CERN-PH-EP-2012-220",
    doi = "10.1016/j.physletb.2012.08.021",
    journal = "Phys. Lett. B",
    volume = "716",
    pages = "30--61",
    year = "2012"
}

@article{ATLAS:2012yve,
    author = "Aad, Georges and others",
    collaboration = "ATLAS",
    title = "{Observation of a new particle in the search for the Standard Model Higgs boson with the ATLAS detector at the LHC}",
    eprint = "1207.7214",
    archivePrefix = "arXiv",
    primaryClass = "hep-ex",
    reportNumber = "CERN-PH-EP-2012-218",
    doi = "10.1016/j.physletb.2012.08.020",
    journal = "Phys. Lett. B",
    volume = "716",
    pages = "1--29",
    year = "2012"
}

@article{FCC:2018evy,
    author = "Abada, A. and others",
    collaboration = "FCC",
    title = "{FCC-ee: The Lepton Collider}: {Future Circular Collider Conceptual Design Report Volume 2}",
    reportNumber = "CERN-ACC-2018-0057",
    doi = "10.1140/epjst/e2019-900045-4",
    journal = "Eur. Phys. J. ST",
    volume = "228",
    pages = "261--623",
    year = "2019"
}

@article{FCC:2018vvp,
    author = "Abada, A. and others",
    collaboration = "FCC",
    title = "{FCC-hh: The Hadron Collider}: {Future Circular Collider Conceptual Design Report Volume 3}",
    reportNumber = "CERN-ACC-2018-0058",
    doi = "10.1140/epjst/e2019-900087-0",
    journal = "Eur. Phys. J. ST",
    volume = "228",
    pages = "755--1107",
    year = "2019"
}

@unpublished{FCC:2023,
    author = "Grojean, C. and Janot, P. and Mangano, M. and others",
    collaboration = "FCC-ee",
    title = "{FCC-ee Midterm report: Physics and Experiments}",
    note = "CERN-FCC-2023-000i",
    year = "2023"
}

@article{Alwall:2014hca,
    author = "Alwall, J. and Frederix, R. and Frixione, S. and Hirschi, V. and Maltoni, F. and Mattelaer, O. and Shao, H. -S. and Stelzer, T. and Torrielli, P. and Zaro, M.",
    title = "{The automated computation of tree-level and next-to-leading order differential cross sections, and their matching to parton shower simulations}",
    eprint = "1405.0301",
    archivePrefix = "arXiv",
    primaryClass = "hep-ph",
    reportNumber = "CERN-PH-TH-2014-064, CP3-14-18, LPN14-066, MCNET-14-09, ZU-TH-14-14",
    doi = "10.1007/JHEP07(2014)079",
    journal = "JHEP",
    volume = "07",
    pages = "079",
    year = "2014"
}

@article{Frederix:2018nkq,
    author = "Frederix, R. and Frixione, S. and Hirschi, V. and Pagani, D. and Shao, H. -S. and Zaro, M.",
    title = "{The automation of next-to-leading order electroweak calculations}",
    eprint = "1804.10017",
    archivePrefix = "arXiv",
    primaryClass = "hep-ph",
    reportNumber = "Nikhef/2018-015, TUM-HEP-1138/18, NIKHEF-2018-015, TUM-HEP-1138-18",
    doi = "10.1007/JHEP11(2021)085",
    journal = "JHEP",
    volume = "07",
    pages = "185",
    year = "2018",
    note = "[Erratum: JHEP 11 (2021) 085]"
}

@article{CEPCStudyGroup:2018ghi,
    author = "Dong, Mingyi and others",
    editor = "Guimar\~aes da Costa, Jo\~ao Barreiro and others",
    collaboration = "CEPC Study Group",
    title = "{CEPC Conceptual Design Report: Volume 2 - Physics \& Detector}",
    eprint = "1811.10545",
    archivePrefix = "arXiv",
    primaryClass = "hep-ex",
    reportNumber = "IHEP-CEPC-DR-2018-02, IHEP-EP-2018-01, IHEP-TH-2018-01",
    month = "11",
    year = "2018"
}

@article{dEnterria:2021ljz,
    author = "d'Enterria, David",
    title = "{Collider constraints on axion-like particles}",
    booktitle = "{Workshop on Feebly Interacting Particles}",
    eprint = "2102.08971",
    archivePrefix = "arXiv",
    primaryClass = "hep-ex",
    year = "2021"
}

@article{Bauer:2018uxu,
    author = "Bauer, Martin and Heiles, Mathias and Neubert, Matthias and Thamm, Andrea",
    title = "{Axion-Like Particles at Future Colliders}",
    eprint = "1808.10323",
    archivePrefix = "arXiv",
    primaryClass = "hep-ph",
    reportNumber = "CERN-TH-2018-199, MITP/18-075",
    doi = "10.1140/epjc/s10052-019-6587-9",
    journal = "Eur. Phys. J. C",
    volume = "79",
    pages = "74",
    year = "2019"
}

@article{dEnterria:2023npy,
    author = "d'Enterria, David and Tamlihat, Malak Ait and Schoeffel, Laurent and Shao, Hua-Sheng and Tayalati, Yahya",
    title = "{Collider constraints on massive gravitons coupling to photons}",
    eprint = "2306.15558",
    archivePrefix = "arXiv",
    primaryClass = "hep-ph",
    doi = "10.1016/j.physletb.2023.138237",
    journal = "Phys. Lett. B",
    volume = "846",
    pages = "138237",
    year = "2023"
}

@article{Manohar:1983md,
    author = "Manohar, Aneesh and Georgi, Howard",
    title = "{Chiral quarks and the Nonrelativistic Quark Model}",
    reportNumber = "HUTP-83/A042a",
    doi = "10.1016/0550-3213(84)90231-1",
    journal = "Nucl. Phys. B",
    volume = "234",
    pages = "189--212",
    year = "1984"
}

@article{CMS:2019vaj,
    author = "Sirunyan, Albert M and others",
    collaboration = "CMS",
    title = "{Search for W boson decays to three charged pions}",
    eprint = "1901.11201",
    archivePrefix = "arXiv",
    primaryClass = "hep-ex",
    reportNumber = "CMS-SMP-18-009, CERN-EP-2019-001",
    doi = "10.1103/PhysRevLett.122.151802",
    journal = "Phys. Rev. Lett.",
    volume = "122",
    pages = "151802",
    year = "2019"
}

@article{ATLAS:2018jqi,
    author = "Aaboud, Morad and others",
    collaboration = "ATLAS",
    title = "{Search for top-quark decays $\rm t \to Hq$ with 36 fb$^{-1}$ of pp collision data at $\sqrt{s}=13$ TeV with the ATLAS detector}",
    eprint = "1812.11568",
    archivePrefix = "arXiv",
    primaryClass = "hep-ex",
    reportNumber = "CERN-EP-2018-295",
    doi = "10.1007/JHEP05(2019)123",
    journal = "JHEP",
    volume = "05",
    pages = "123",
    year = "2019"
}

@article{Curtin:2014cca,
    author = "Curtin, David and Essig, Rouven and Gori, Stefania and Shelton, Jessie",
    title = "Illuminating Dark Photons with High-Energy Colliders",
    eprint = "1412.0018",
    archivePrefix = "arXiv",
    primaryClass = "hep-ph",
    reportNumber = "YITP-SB-14-49",
    doi = "10.1007/JHEP02(2015)157",
    journal = "JHEP",
    volume = "02",
    pages = "157",
    year = "2015"
}

@article{Agrawal:2021dbo,
    author = "Agrawal, Prateek and others",
    title = "{Feebly-interacting particles: FIPs 2020 workshop report}",
    eprint = "2102.12143",
    archivePrefix = "arXiv",
    primaryClass = "hep-ph",
    doi = "10.1140/epjc/s10052-021-09703-7",
    journal = "Eur. Phys. J. C",
    volume = "81",
    pages = "1015",
    year = "2021"
}

@article{LHCHiggsCrossSectionWorkingGroup:2016ypw,
    author = "de Florian, D. and others",
    collaboration = "LHC Higgs Cross Section Working Group",
    title = "{Handbook of LHC Higgs Cross Sections: 4. Deciphering the Nature of the Higgs Sector}",
    eprint = "1610.07922",
    archivePrefix = "arXiv",
    primaryClass = "hep-ph",
    reportNumber = "CERN-2017-002-M, CERN-2017-002",
    doi = "10.23731/CYRM-2017-002",
    volume = "2/2017",
    month = "10",
    year = "2016"
}

@article{Djouadi:2018xqq,
    author = "Djouadi, Abdelhak and Kalinowski, Jan and Muehlleitner, Margarete and Spira, Michael",
    collaboration = "HDECAY",
    title = "{HDECAY: Twenty${++}$ years after}",
    eprint = "1801.09506",
    archivePrefix = "arXiv",
    primaryClass = "hep-ph",
    reportNumber = "LPT-ORSAY-18-04, CERN-TH-2017-262, LPT-Orsay-18-04, KA-TP-03-2018, PSI-PR-18-02",
    doi = "10.1016/j.cpc.2018.12.010",
    journal = "Comput. Phys. Commun.",
    volume = "238",
    pages = "214--231",
    year = "2019"
}

@article{Mangano:2016jyj,
    author = "Mangano, M. L. and others",
    title = "{Physics at a 100 TeV pp Collider: Standard Model Processes}",
    eprint = "1607.01831",
    archivePrefix = "arXiv",
    primaryClass = "hep-ph",
    reportNumber = "CERN-TH-2016-112",
    doi = "10.23731/CYRM-2017-003.1",
    month = "7",
    year = "2016"
}

@article{ALEPH:2013dgf,
    author = "Schael, S. and others",
    collaboration = "LEP Electroweak Working Group",
    title = "{Electroweak measurements in electron-positron collisions at W-boson-pair energies at LEP}",
    eprint = "1302.3415",
    archivePrefix = "arXiv",
    primaryClass = "hep-ex",
    reportNumber = "CERN-PH-EP-2013-022",
    doi = "10.1016/j.physrep.2013.07.004",
    journal = "Phys. Rept.",
    volume = "532",
    pages = "119--244",
    year = "2013"
}

@article{ALEPH:2005ab,
    author = "Schael, S. and others",
    collaboration = "LEP/SLD Electroweak Working Groups",
    title = "{Precision electroweak measurements on the Z resonance}",
    eprint = "hep-ex/0509008",
    archivePrefix = "arXiv",
    reportNumber = "SLAC-R-774",
    doi = "10.1016/j.physrep.2005.12.006",
    journal = "Phys. Rept.",
    volume = "427",
    pages = "257--454",
    year = "2006"
}

@article{Kagan:2014ila,
    author = "Kagan, Alexander L. and Perez, Gilad and Petriello, Frank and Soreq, Yotam and Stoynev, Stoyan and Zupan, Jure",
    title = "{Exclusive window onto Higgs Yukawa couplings}",
    eprint = "1406.1722",
    archivePrefix = "arXiv",
    primaryClass = "hep-ph",
    doi = "10.1103/PhysRevLett.114.101802",
    journal = "Phys. Rev. Lett.",
    volume = "114",
    pages = "101802",
    year = "2015"
}

@article{Beneke:2000ry,
    author = "Beneke, M. and Buchalla, G. and Neubert, M. and Sachrajda, Christopher T.",
    title = "{QCD factorization for exclusive, nonleptonic B meson decays: General arguments and the case of heavy light final states}",
    eprint = "hep-ph/0006124",
    archivePrefix = "arXiv",
    reportNumber = "CERN-TH-2000-159, CLNS-00-1675, PITHA-00-06, SHEP-00-06",
    doi = "10.1016/S0550-3213(00)00559-9",
    journal = "Nucl. Phys. B",
    volume = "591",
    pages = "313--418",
    year = "2000"
}

@article{Perez:2015lra,
    author = "Perez, Gilad and Soreq, Yotam and Stamou, Emmanuel and Tobioka, Kohsaku",
    title = "{Prospects for measuring the Higgs boson coupling to light quarks}",
    eprint = "1505.06689",
    archivePrefix = "arXiv",
    primaryClass = "hep-ph",
    doi = "10.1103/PhysRevD.93.013001",
    journal = "Phys. Rev. D",
    volume = "93",
    pages = "013001",
    year = "2016"
}

@article{CMS:2016uzc,
    author = "Khachatryan, Vardan and others",
    collaboration = "CMS",
    title = "{Search for anomalous Wtb couplings and flavour-changing neutral currents in t-channel single top quark production in pp collisions at $\sqrt{s} =$ 7 and 8 TeV}",
    eprint = "1610.03545",
    archivePrefix = "arXiv",
    primaryClass = "hep-ex",
    reportNumber = "CMS-TOP-14-007, CERN-EP-2016-207",
    doi = "10.1007/JHEP02(2017)028",
    journal = "JHEP",
    volume = "02",
    pages = "028",
    year = "2017"
}

@article{Liu:2020bem,
    author = "Liu, Yao-Bei and Moretti, Stefano",
    title = "{Probing tqZ anomalous couplings in the trilepton signal at the HL-LHC, HE-LHC and FCC-hh}",
    eprint = "2010.05148",
    archivePrefix = "arXiv",
    primaryClass = "hep-ph",
    doi = "10.1088/1674-1137/abe0c0",
    journal = "Chin. Phys. C",
    volume = "45",
    pages = "043110",
    year = "2021"
}

@article{Cerri:2018ypt,
    author = "Cerri, A. and others",
    editor = "Dainese, Andrea and Mangano, Michelangelo and Meyer, Andreas B. and Nisati, Aleandro and Salam, Gavin and Vesterinen, Mika Anton",
    title = "{Report from Working Group 4}: {Opportunities in Flavour Physics at the HL-LHC and HE-LHC}",
    eprint = "1812.07638",
    archivePrefix = "arXiv",
    primaryClass = "hep-ph",
    reportNumber = "CERN-LPCC-2018-06",
    doi = "10.23731/CYRM-2019-007.867",
    journal = "CERN Yellow Rep. Monogr.",
    volume = "7",
    pages = "867--1158",
    year = "2019"
}

@article{Belle-II:2018jsg,
    author = "Altmannshofer, W. and others",
    editor = "Kou, E. and Urquijo, P.",
    collaboration = "Belle-II",
    title = "{The Belle II Physics Book}",
    eprint = "1808.10567",
    archivePrefix = "arXiv",
    primaryClass = "hep-ex",
    reportNumber = "KEK Preprint 2018-27, BELLE2-PUB-PH-2018-001, FERMILAB-PUB-18-398-T, JLAB-THY-18-2780, INT-PUB-18-047, UWThPh 2018-26",
    doi = "10.1093/ptep/ptz106",
    journal = "PTEP",
    volume = "2019",
    pages = "123C01",
    year = "2019",
    note = "[Erratum: PTEP 2020 (2020) 029201]"
}

@article{LHCb:2018roe,
    author = "Aaij, Roel and others",
    collaboration = "LHCb",
    title = "{Physics case for an LHCb Upgrade II - Opportunities in flavour physics, and beyond, in the HL-LHC era}",
    eprint = "1808.08865",
    archivePrefix = "arXiv",
    primaryClass = "hep-ex",
    reportNumber = "LHCB-PUB-2018-009, CERN-LHCC-2018-027",
    month = "8",
    year = "2018"
}

@article{Bodwin:1994jh,
    author = "Bodwin, Geoffrey T. and Braaten, Eric and Lepage, G. Peter",
    title = "{Rigorous QCD analysis of inclusive annihilation and production of heavy quarkonium}",
    eprint = "hep-ph/9407339",
    archivePrefix = "arXiv",
    reportNumber = "ANL-HEP-PR-94-24, FERMILAB-PUB-94-073-T, NUHEP-TH-94-5",
    doi = "10.1103/PhysRevD.55.5853",
    journal = "Phys. Rev. D",
    volume = "51",
    pages = "1125--1171",
    year = "1995",
    note = "[Erratum: Phys. Rev. D 55 (1997) 5853]"
}

@article{Jones:2020bvu,
    author = "Jones, E. and Murray, W. J.",
    title = "{Mass biases in exclusive radiative hadronic decays of W bosons at the LHC}",
    eprint = "2009.01073",
    archivePrefix = "arXiv",
    primaryClass = "hep-ex",
    doi = "10.1088/1367-2630/ac3572",
    journal = "New J. Phys.",
    volume = "23",
    pages = "113035",
    year = "2021"
}

@article{Ebert:2002pp,
    author = "Ebert, D. and Faustov, R. N. and Galkin, V. O.",
    title = "{Properties of heavy quarkonia and $B_c$ mesons in the relativistic quark model}",
    eprint = "hep-ph/0210381",
    archivePrefix = "arXiv",
    reportNumber = "HU-EP-02-45",
    doi = "10.1103/PhysRevD.67.014027",
    journal = "Phys. Rev. D",
    volume = "67",
    pages = "014027",
    year = "2003"
}

@article{Bergstrom:1990bu,
    author = "Bergstrom, L. and Robinett, R. W.",
    title = "{On the rare decays $\rm Z \to VV$ and $\rm Z \to VP$}",
    reportNumber = "PSU/TH/61, USITP-90-02",
    doi = "10.1103/PhysRevD.41.3513",
    journal = "Phys. Rev. D",
    volume = "41",
    pages = "3513",
    year = "1990"
}

@article{Doroshenko:1987nj,
    author = "Doroshenko, M. N. and Kartvelishvili, V. G. and Chikovani, E. G. and Esakiya, Sh. M.",
    title = "{Vector quarkonium in decays of heavy Higgs particles}",
    journal = "Yad. Fiz.",
    volume = "46",
    pages = "864--868",
    year = "1987"
}

@article{Chen:2022wit,
    author = "Chen, Long-Bin and Li, Hai Tao and Wang, Jian and Wang, Yefan",
    title = "{Analytic result for the top-quark width at next-to-next-to-leading order in QCD}",
    eprint = "2212.06341",
    archivePrefix = "arXiv",
    primaryClass = "hep-ph",
    doi = "10.1103/PhysRevD.108.054003",
    journal = "Phys. Rev. D",
    volume = "108",
    pages = "054003",
    year = "2023"
}

@article{Handoko:1998tz,
    author = "Handoko, L. T.",
    title = "{Branching ratio and CP violation in semiinclusive flavor changing top decays}",
    eprint = "hep-ph/9803262",
    archivePrefix = "arXiv",
    reportNumber = "LFTMLIPI-139801, HUPD-9801, LFTMLIPI-139801",
    doi = "10.1007/BF03545794",
    journal = "Nuovo Cim. A",
    volume = "111",
    pages = "1275",
    year = "1998"
}

@article{Becirevic:2012ti,
    author = "Becirevic, Damir and Lubicz, Vittorio and Sanfilippo, Francesco and Simula, Silvano and Tarantino, Cecilia",
    title = "{D-meson decay constants and a check of factorization in non-leptonic B-decays}",
    eprint = "1201.4039",
    archivePrefix = "arXiv",
    primaryClass = "hep-lat",
    reportNumber = "LPT-11-119, RM3-TH-21-1, RM3-TH-12-1",
    doi = "10.1007/JHEP02(2012)042",
    journal = "JHEP",
    volume = "02",
    pages = "042",
    year = "2012"
}

@article{Colquhoun:2015oha,
    author = "Colquhoun, B. and Davies, C. T. H. and Dowdall, R. J. and Kettle, J. and Koponen, J. and Lepage, G. P. and Lytle, A. T.",
    collaboration = "HPQCD",
    title = "{B-meson decay constants: a more complete picture from full lattice QCD}",
    eprint = "1503.05762",
    archivePrefix = "arXiv",
    primaryClass = "hep-lat",
    doi = "10.1103/PhysRevD.91.114509",
    journal = "Phys. Rev. D",
    volume = "91",
    pages = "114509",
    year = "2015"
}

@article{Chapon:2020heu,
    author = "Chapon, Emilien and others",
    title = "{Prospects for quarkonium studies at the high-luminosity LHC}",
    eprint = "2012.14161",
    archivePrefix = "arXiv",
    primaryClass = "hep-ph",
    reportNumber = "MIT-CTP/5231, JLAB-THY-20-3240",
    doi = "10.1016/j.ppnp.2021.103906",
    journal = "Prog. Part. Nucl. Phys.",
    volume = "122",
    pages = "103906",
    year = "2022"
}

@article{Lansberg:2019adr,
    author = "Lansberg, Jean-Philippe",
    title = "{New Observables in Inclusive Production of Quarkonia}",
    eprint = "1903.09185",
    archivePrefix = "arXiv",
    primaryClass = "hep-ph",
    doi = "10.1016/j.physrep.2020.08.007",
    journal = "Phys. Rept.",
    volume = "889",
    pages = "1--106",
    year = "2020"
}

@article{Shifman:1978bx,
    author = "Shifman, Mikhail A. and Vainshtein, A. I. and Zakharov, Valentin I.",
    title = "{QCD and Resonance Physics. Theoretical Foundations}",
    reportNumber = "ITEP-73-1978, ITEP-80-1978",
    doi = "10.1016/0550-3213(79)90022-1",
    journal = "Nucl. Phys. B",
    volume = "147",
    pages = "385--447",
    year = "1979"
}

@article{Colangelo:2000dp,
    author = "Colangelo, Pietro and Khodjamirian, Alexander",
    editor = "Shifman, M. and Ioffe, Boris",
    title = "{QCD sum rules, a modern perspective}",
    eprint = "hep-ph/0010175",
    archivePrefix = "arXiv",
    reportNumber = "CERN-TH-2000-296, BARI-TH-2000-394",
    doi = "10.1142/9789812810458_0033",
    pages = "1495--1576",
    month = "10",
    year = "2000"
}

@article{Boughezal:2016wmq,
    author = "Boughezal, Radja and Campbell, John M. and Ellis, R. Keith and Focke, Christfried and Giele, Walter and Liu, Xiaohui and Petriello, Frank and Williams, Ciaran",
    title = "{Color singlet production at NNLO in MCFM}",
    eprint = "1605.08011",
    archivePrefix = "arXiv",
    primaryClass = "hep-ph",
    reportNumber = "FERMILAB-PUB-16-120-T, IPPP-16-32",
    doi = "10.1140/epjc/s10052-016-4558-y",
    journal = "Eur. Phys. J. C",
    volume = "77",
    pages = "7",
    year = "2017"
}

@article{NNPDF:2017mvq,
    author = "Ball, Richard D. and others",
    collaboration = "NNPDF",
    title = "{Parton distributions from high-precision collider data}",
    eprint = "1706.00428",
    archivePrefix = "arXiv",
    primaryClass = "hep-ph",
    reportNumber = "EDINBURGH-2017-08, NIKHEF-2017-006, OUTP-17-04P, TIF-UNIMI-2017-3, CAVENDISH-HEP-17-06, CERN-TH-2017-077, Edinburgh 2017/08,
  Nikhef/2017-006, OUTP-17-04P,TIF-UNIMI-2017-3",
    doi = "10.1140/epjc/s10052-017-5199-5",
    journal = "Eur. Phys. J. C",
    volume = "77",
    pages = "663",
    year = "2017"
}

@article{Aranda:2020tqw,
    author = "Aranda, J. I. and Gonz\'alez-Estrada, G. and Monta\~no, J. and Ram\'\i{}rez-Zavaleta, F. and Tututi, E. S.",
    title = "{Revisiting the rare $\rm H\to q_iq_j$ decays in the standard model}",
    eprint = "2009.07166",
    archivePrefix = "arXiv",
    primaryClass = "hep-ph",
    doi = "10.1088/1361-6471/abb44d",
    journal = "J. Phys. G",
    volume = "47",
    pages = "125001",
    year = "2020"
}

@article{Benitez-Guzman:2015ana,
    author = "Benitez-Guzm\'an, L. G. and Garc\'\i{}a-Jim\'enez, I. and L\'opez-Osorio, M. A. and Mart\'\i{}nez-Pascual, E. and Toscano, J. J.",
    title = "{Revisiting the flavor changing neutral current Higgs decays $\rm H\;\to \;{q}_{i}{q}_{j}$ in the Standard Model}",
    eprint = "1506.02718",
    archivePrefix = "arXiv",
    primaryClass = "hep-ph",
    doi = "10.1088/0954-3899/42/8/085002",
    journal = "J. Phys. G",
    volume = "42",
    pages = "085002",
    year = "2015"
}

@article{Hirschi:2015iia,
    author = "Hirschi, Valentin and Mattelaer, Olivier",
    title = "{Automated event generation for loop-induced processes}",
    eprint = "1507.00020",
    archivePrefix = "arXiv",
    primaryClass = "hep-ph",
    reportNumber = "IPPP-15-35, DCPT-15-70, MCNET-15-14",
    doi = "10.1007/JHEP10(2015)146",
    journal = "JHEP",
    volume = "10",
    pages = "146",
    year = "2015"
}

@article{Kamenik:2023hvi,
    author = "Kamenik, Jernej F. and Korajac, Arman and Szewc, Manuel and Tammaro, Michele and Zupan, Jure",
    title = "{Flavor-violating Higgs and Z boson decays at a future circular lepton collider}",
    eprint = "2306.17520",
    archivePrefix = "arXiv",
    primaryClass = "hep-ph",
    doi = "10.1103/PhysRevD.109.L011301",
    journal = "Phys. Rev. D",
    volume = "109",
    pages = "L011301",
    year = "2024"
}

@article{Pokraka:2017ore,
    author = "Pokraka, Andrzej and Czarnecki, Andrzej",
    title = "{Parapositronium can decay into three photons}",
    eprint = "1707.09466",
    archivePrefix = "arXiv",
    primaryClass = "hep-ph",
    reportNumber = "ALBERTA-THY-19-17",
    doi = "10.1103/PhysRevD.96.093002",
    journal = "Phys. Rev. D",
    volume = "96",
    pages = "093002",
    year = "2017"
}

@article{Bernreuther:1981ah,
    author = "Bernreuther, W. and Nachtmann, O.",
    title = "{Weak Interaction Effects in Positronium}",
    reportNumber = "HD-THEP-81-7",
    doi = "10.1007/BF01545680",
    journal = "Z. Phys. C",
    volume = "11",
    pages = "235",
    year = "1981"
}

@article{CMS:2022fyt,
    author = "Tumasyan, Armen and others",
    collaboration = "CMS",
    title = "{Search for exotic Higgs boson decays $\rm H\to a a\to 4\gamma$ with events containing two merged diphotons in proton-proton collisions at $\sqrt{s}$ = 13~TeV}",
    eprint = "2209.06197",
    archivePrefix = "arXiv",
    primaryClass = "hep-ex",
    reportNumber = "CMS-HIG-21-016, CERN-EP-2022-151",
    doi = "10.1103/PhysRevLett.131.101801",
    journal = "Phys. Rev. Lett.",
    volume = "131",
    pages = "101801",
    year = "2023"
}

@article{CMS:2022xxa,
    author = "Tumasyan, Armen and others",
    collaboration = "CMS",
    title = "{Search for the exotic decay of the Higgs boson into two light pseudoscalars with four photons in the final state in proton-proton collisions at $ \sqrt{s} $ = 13 TeV}",
    eprint = "2208.01469",
    archivePrefix = "arXiv",
    primaryClass = "hep-ex",
    reportNumber = "CMS-HIG-21-003, CERN-EP-2022-095",
    doi = "10.1007/JHEP07(2023)148",
    journal = "JHEP",
    volume = "07",
    pages = "148",
    year = "2023"
}

@article{Curtin:2013fra,
    author = "Curtin, David and others",
    title = "{Exotic decays of the 125 GeV Higgs boson}",
    eprint = "1312.4992",
    archivePrefix = "arXiv",
    primaryClass = "hep-ph",
    reportNumber = "YITP-13-47, PITT-PACC-1314",
    doi = "10.1103/PhysRevD.90.075004",
    journal = "Phys. Rev. D",
    volume = "90",
    pages = "075004",
    year = "2014"
}

@article{Zhu:2018xxs,
    author = "Zhu, Rong-Fei and Feng, Tai-Fu and Zhang, Hai-Bin",
    title = "{The QCD corrections of the process $\rm h \to \eta_{b}Z$}",
    doi = "10.1142/S0217732318300082",
    journal = "Mod. Phys. Lett. A",
    volume = "33",
    pages = "1830008",
    year = "2018"
}

@article{Sun:2018xft,
    author = "Sun, Qing-Feng and Wang, An-Min",
    title = "{Next-to-leading order QCD corrections to the decay of Higgs to vector meson and Z boson}",
    eprint = "1801.10528",
    archivePrefix = "arXiv",
    primaryClass = "hep-ph",
    doi = "10.1088/1674-1137/42/3/033105",
    journal = "Chin. Phys. C",
    volume = "42",
    pages = "033105",
    year = "2018"
}

@article{Becirevic:2017chd,
    author = "Be\v{c}irevi\'c, Damir and Meli\'c, Bla\v{z}enka and Patra, Monalisa and Sumensari, Olcyr",
    title = "{Seeking a CP-odd Higgs boson via $h\to \eta_{c,b}\ell^+\ell^-$}",
    eprint = "1705.01112",
    archivePrefix = "arXiv",
    primaryClass = "hep-ph",
    reportNumber = "LPT-Orsay-17-18",
    doi = "10.1103/PhysRevD.97.015008",
    journal = "Phys. Rev. D",
    volume = "97",
    pages = "015008",
    year = "2018"
}

@article{Landau:1948kw,
    author = "Landau, L. D.",
    title = "{On the angular momentum of a system of two photons}",
    doi = "10.1016/B978-0-08-010586-4.50070-5",
    journal = "Dokl. Akad. Nauk SSSR",
    volume = "60",
    pages = "207--209",
    year = "1948"
}

@article{Yang:1950rg,
    author = "Yang, Chen-Ning",
    title = "{Selection Rules for the Dematerialization of a Particle Into Two Photons}",
    doi = "10.1103/PhysRev.77.242",
    journal = "Phys. Rev.",
    volume = "77",
    pages = "242--245",
    year = "1950"
}

@article{CMS:2023eos,
author = "Hayrapetyan, Aram and others",
collaboration = "CMS",
title = "{Search for an exotic decay of the Higgs boson into a Z boson and a pseudoscalar particle in proton-proton collisions at $\sqrts = 13$~TeV}",
eprint = "2311.00130",
archivePrefix = "arXiv",
primaryClass = "hep-ex",
reportNumber = "CMS-HIG-22-003, CERN-EP-2023-223",
doi = "10.1016/j.physletb.2024.138582",
journal = "Phys. Lett. B",
volume = "852",
pages = "138582",
year = "2024"
}

@article{LEPWorkingGroupforHiggsbosonsearches:2003ing,
    author = "Barate, R. and others",
    collaboration = "LEP Working Group for Higgs boson searches, ALEPH, DELPHI, L3, OPAL",
    title = "{Search for the standard model Higgs boson at LEP}",
    eprint = "hep-ex/0306033",
    archivePrefix = "arXiv",
    reportNumber = "CERN-EP-2003-011",
    doi = "10.1016/S0370-2693(03)00614-2",
    journal = "Phys. Lett. B",
    volume = "565",
    pages = "61--75",
    year = "2003"
}

@inproceedings{dEnterria:2017jmj,
    author = "d'Enterria, David and Rebello Teles, Patricia and Martins, Daniel E.",
    title = "{Measurements of $\gamma \gamma \to \textrm{Higgs}$ and $\gamma \gamma \to W^{+}W^{-}$ in $e^{+}e^{-}$ collisions at the Future Circular Collider}",
    booktitle = "{17th conference on Elastic and Diffractive Scattering}",
    eprint = "1712.07023",
    archivePrefix = "arXiv",
    primaryClass = "hep-ph",
    month = "12",
    year = "2017"
}

@article{Mandal:2023lhp,
    author = "Mandal, Rusa and Nandi, Soumitra and Ray, Ipsita",
    title = "{Constraining inverse moment of B-meson distribution amplitude using Lattice QCD data}",
    eprint = "2308.07033",
    archivePrefix = "arXiv",
    primaryClass = "hep-ph",
    doi = "10.1016/j.physletb.2023.138345",
    journal = "Phys. Lett. B",
    volume = "848",
    pages = "138345",
    year = "2024"
}

@article{Khodjamirian:2020hob,
    author = "Khodjamirian, Alexander and Mandal, Rusa and Mannel, Thomas",
    title = "{Inverse moment of the B$_{s}$-meson distribution amplitude from QCD sum rule}",
    eprint = "2008.03935",
    archivePrefix = "arXiv",
    primaryClass = "hep-ph",
    reportNumber = "SI-HEP-2020-20, P3H-20-039",
    doi = "10.1007/JHEP10(2020)043",
    journal = "JHEP",
    volume = "10",
    pages = "043",
    year = "2020"
}

@article{Lu:2021ttf,
    author = "Lu, Long-Sheng",
    title = "{Factorization of radiative leptonic D-meson decay with sub-leading power corrections}",
    eprint = "2104.01562",
    archivePrefix = "arXiv",
    primaryClass = "hep-ph",
    doi = "10.1088/1674-1137/abf5c9",
    journal = "Chin. Phys. C",
    volume = "45",
    pages = "073101",
    year = "2021"
}

@article{FlavourLatticeAveragingGroupFLAG:2021npn,
    author = "Aoki, Y. and others",
    collaboration = "Flavour Lattice Averaging Group (FLAG)",
    title = "{FLAG Review 2021}",
    eprint = "2111.09849",
    archivePrefix = "arXiv",
    primaryClass = "hep-lat",
    reportNumber = "CERN-TH-2021-191, JLAB-THY-21-3528, FERMILAB-PUB-21-620-SCD-T",
    doi = "10.1140/epjc/s10052-022-10536-1",
    journal = "Eur. Phys. J. C",
    volume = "82",
    pages = "869",
    year = "2022"
}

@article{Lee:2005gza,
    author = "Lee, Seung J. and Neubert, Matthias",
    title = "{Model-independent properties of the B-meson distribution amplitude}",
    eprint = "hep-ph/0509350",
    archivePrefix = "arXiv",
    reportNumber = "CLNS-05-1925",
    doi = "10.1103/PhysRevD.72.094028",
    journal = "Phys. Rev. D",
    volume = "72",
    pages = "094028",
    year = "2005"
}

@article{Altmannshofer:2019ogm,
    author = "Altmannshofer, Wolfgang and Maddock, Brian and Tuckler, Douglas",
    title = "Rare Top Decays as Probes of Flavorful {Higgs} Bosons",
    eprint = "1904.10956",
    archivePrefix = "arXiv",
    primaryClass = "hep-ph",
    doi = "10.1103/PhysRevD.100.015003",
    journal = "Phys. Rev. D",
    volume = "100",
    pages = "015003",
    year = "2019"
}

@article{Azhothkaran:2020ipl,
    author = "Azhothkaran, Bhaghyesh and Nilakanthan, V. K.",
    title = "Decay Constants of {S}-Wave Heavy Quarkonia",
    doi = "10.1007/s10773-020-04474-5",
    journal = "Int. J. Theor. Phys.",
    volume = "59",
    pages = "2016--2028",
    year = "2020"
}

@article{VanRoyen:1967nq,
    author = "Van Royen, R. and Weisskopf, V. F.",
    title = "Hadron Decay Processes and the Quark Model",
    doi = "10.1007/BF02823542",
    journal = "Nuovo Cim. A",
    volume = "50",
    pages = "617--645",
    year = "1967",
    note = "[Erratum: Nuovo Cim. A 51 (1967) 583]"
}

@article{ATLAS:2023ian,
    author = "Aad, Georges and others",
    collaboration = "ATLAS",
    title = "{Search for short- and long-lived axion-like particles in $\rm H\to a a \to 4\gamma$ decays with the ATLAS experiment at the LHC}",
    eprint = "2312.03306",
    archivePrefix = "arXiv",
    primaryClass = "hep-ex",
    reportNumber = "CERN-EP-2023-202",
    doi = "10.1140/epjc/s10052-024-12979-0",
    journal = "Eur. Phys. J. C",
    volume = "84",
    pages = "742",
    year = "2024"
}

@unpublished{ATLAS:2016qxw,
    author = "Aad, Georges and others",
    collaboration = "ATLAS",
    title = "{Expected sensitivity of ATLAS to FCNC top quark decays $\rm t \to Zu$ and $\rm t \to Hq$ at the High Luminosity LHC}",
    note = "ATL-PHYS-PUB-2016-019",
    year = "2016"
}

@article{CMS:2013rmy,
    author = "Chatrchyan, Serguei and others",
    collaboration = "CMS",
    title = "{Search for a Higgs boson decaying into a Z and a photon in pp collisions at $\sqrt{s}$ = 7 and 8 TeV}",
    eprint = "1307.5515",
    archivePrefix = "arXiv",
    primaryClass = "hep-ex",
    reportNumber = "CMS-HIG-13-006, CERN-PH-EP-2013-113",
    doi = "10.1016/j.physletb.2013.09.057",
    journal = "Phys. Lett. B",
    volume = "726",
    pages = "587--609",
    year = "2013"
}

@article{Jenkins:1996zd,
    author = "Jenkins, Elizabeth Ellen",
    title = "The Rare top decays {$\rm t \to b W^{+} Z$ and $\rm t \to c W^{+} W^{-}$}",
    eprint = "hep-ph/9612211",
    archivePrefix = "arXiv",
    reportNumber = "UCSD-PTH-96-28",
    doi = "10.1103/PhysRevD.56.458",
    journal = "Phys. Rev. D",
    volume = "56",
    pages = "458--466",
    year = "1997"
}

@article{Wang:2023ssg,
author = "Wang, Guang-Yu and Zheng, Xu-Chang and Wu, Xing-Gang and Xu, Guang-Zhi",
title = "{Z-boson decays into S-wave quarkonium plus a photon up to $\mathcal{O}(\alpha_s v^2)$ corrections}",
eprint = "2312.08796",
archivePrefix = "arXiv",
primaryClass = "hep-ph",
doi = "10.1103/PhysRevD.109.074004",
journal = "Phys. Rev. D",
volume = "109",
pages = "074004",
year = "2024"
}

@article{Nemenov:2001smx,
    author = "Nemenov, Leonid",
    title = "{Elementary Relativistic Atoms}",
    doi = "10.1007/3-540-45395-4_12",
    journal = "Lect. Notes Phys.",
    volume = "570",
    pages = "223--245",
    year = "2001"
}

@article{Dong:2022bkd,
    author = "Dong, Hongxin and Sun, Peng and Yan, Bin",
    title = "{Probing the $\rm H\to\gaga$ coupling via Higgs boson exclusive decay into quarkonia plus a photon at the HL-LHC}",
    eprint = "2208.05153",
    archivePrefix = "arXiv",
    primaryClass = "hep-ph",
    doi = "10.1103/PhysRevD.106.095013",
    journal = "Phys. Rev. D",
    volume = "106",
    pages = "095013",
    year = "2022"
}

@article{Mahlon:1994us,
    author = "Mahlon, Gregory and Parke, Stephen J.",
    title = "{Finite width effects in top quark decays}",
    eprint = "hep-ph/9412250",
    archivePrefix = "arXiv",
    reportNumber = "FERMILAB-PUB-94-406-T",
    doi = "10.1016/0370-2693(95)00083-W",
    journal = "Phys. Lett. B",
    volume = "347",
    pages = "394--398",
    year = "1995"
}

@article{Decker:1992wz,
    author = "Decker, Roger and Nowakowski, Marek and Pilaftsis, Apostolos",
    title = "{Dominant three-body decays of a heavy Higgs and top quark}",
    eprint = "hep-ph/9301283",
    archivePrefix = "arXiv",
    reportNumber = "MZ-TH-92-25, TTP-92-23",
    doi = "10.1007/BF01565067",
    journal = "Z. Phys. C",
    volume = "57",
    pages = "339--348",
    year = "1993"
}

@article{Han:2013sea,
    author = "Han, Tao and Ruiz, Richard",
    title = "{Higgs} bosons from top quark decays",
    eprint = "1312.3324",
    archivePrefix = "arXiv",
    primaryClass = "hep-ph",
    reportNumber = "PITT-PACC-1313",
    doi = "10.1103/PhysRevD.89.074045",
    journal = "Phys. Rev. D",
    volume = "89",
    pages = "074045",
    year = "2014"
}

@article{Sun:2013cba,
    author = "Sun, Yi and Gao, Dao-Neng",
    title = "{Higgs decays to $\gamma$ and invisible particles in the standard model}",
    eprint = "1310.8404",
    archivePrefix = "arXiv",
    primaryClass = "hep-ph",
    reportNumber = "USTC-ICTS-13-14",
    doi = "10.1103/PhysRevD.89.017301",
    journal = "Phys. Rev. D",
    volume = "89",
    pages = "017301",
    year = "2014"
}

@article{Ma:1979px,
    author = "Ma, Ernest and Pramudita, A.",
    title = "Flavor Changing Effective Neutral Current Couplings in the {Weinberg--Salam} Model",
    reportNumber = "UH-511-372-79",
    doi = "10.1103/PhysRevD.22.214",
    journal = "Phys. Rev. D",
    volume = "22",
    pages = "214",
    year = "1980"
}

@article{ATLAS:2024dpw,
    author = "Aad, Georges and others",
    collaboration = "ATLAS",
    title = "{Searches for exclusive Higgs boson decays into D$^*\gamma$ and Z boson decays into D$^0\gamma$ and K$_s^0\gamma$ in pp collisions at $\sqrts = 13$~TeV with the ATLAS detector}",
    eprint = "2402.18731",
    archivePrefix = "arXiv",
    primaryClass = "hep-ex",
    reportNumber = "CERN-EP-2024-037",
    doi = "10.1016/j.physletb.2024.138762",
    journal = "Phys. Lett. B",
    volume = "855",
    pages = "138762",
    year = "2024"
}

@article{Braaten:2001bf,
    author = "Braaten, Eric and Jia, Yu and Mehen, Thomas",
    title = "{B production asymmetries in perturbative QCD}",
    eprint = "hep-ph/0108201",
    archivePrefix = "arXiv",
    doi = "10.1103/PhysRevD.66.034003",
    journal = "Phys. Rev. D",
    volume = "66",
    pages = "034003",
    year = "2002"
}

@phdthesis{Konig:2018wuf,
    author = {K\"onig, Matthias},
    title = "{Effective field theories in the standard model and beyond}",
    doi = "10.25358/openscience-4471",
    school = "Mainz U.",
    year = "2018"
}

@article{Bodwin:2007fz,
    author = "Bodwin, Geoffrey T. and Chung, Hee Sok and Kang, Daekyoung and Lee, Jungil and Yu, Chaehyun",
    title = "{Improved determination of color-singlet nonrelativistic QCD matrix elements for S-wave charmonium}",
    eprint = "0710.0994",
    archivePrefix = "arXiv",
    primaryClass = "hep-ph",
    reportNumber = "ANL-HEP-PR-07-48",
    doi = "10.1103/PhysRevD.77.094017",
    journal = "Phys. Rev. D",
    volume = "77",
    pages = "094017",
    year = "2008"
}

@article{Eichten:1995ch,
    author = "Eichten, Estia J. and Quigg, Chris",
    title = "{Quarkonium wave functions at the origin}",
    eprint = "hep-ph/9503356",
    archivePrefix = "arXiv",
    reportNumber = "FERMILAB-PUB-95-045-T, CLNS-95-1329",
    doi = "10.1103/PhysRevD.52.1726",
    journal = "Phys. Rev. D",
    volume = "52",
    pages = "1726--1728",
    year = "1995"
}

@article{CMS:2024tgj,
    author = "Hayrapetyan, Aram and others",
    collaboration = "CMS",
    title = "{Search for the Higgs boson decays to a $\rho^0$, $\phi$, or K$^{*0}$ meson and a photon in proton-proton collisions at $\sqrt{s}$ = 13 TeV}",
    eprint = "2410.18289",
    archivePrefix = "arXiv",
    primaryClass = "hep-ex",
    reportNumber = "CMS-HIG-23-005, CERN-EP-2024-252",
    month = "10",
    year = "2024"
}

@article{Martynenko:2024rfj,
    author = "Martynenko, F. A. and Martynenko, A. P. and Eskin, A. V.",
    title = "{Production of dileptonic bound states in the Higgs boson decay}",
    eprint = "2405.00829",
    archivePrefix = "arXiv",
    primaryClass = "hep-ph",
    doi = "10.1103/PhysRevD.110.056016",
    journal = "Phys. Rev. D",
    volume = "110",
    pages = "056016",
    year = "2024"
}

@article{RBC-UKQCD:2008mhs,
    author = "Allton, C. and others",
    collaboration = "RBC-UKQCD",
    title = "Physical Results from 2+1 Flavor Domain Wall {QCD and SU(2)} Chiral Perturbation Theory",
    eprint = "0804.0473",
    archivePrefix = "arXiv",
    primaryClass = "hep-lat",
    reportNumber = "BNL-HET-08-5, EDINBURGH-2008-06, RBRC-730, SHEP-0812, BNL-HET-08/5, CU-TP-1182, Edinburgh 2008/06, KEK-TH-1232, RBRC-730,
  SHEP-0812",
    doi = "10.1103/PhysRevD.78.114509",
    journal = "Phys. Rev. D",
    volume = "78",
    pages = "114509",
    year = "2008"
}

@article{Pullin:2021ebn,
    author = "Pullin, Ben and Zwicky, Roman",
    title = "{Radiative decays of heavy-light mesons and the $ {f}_{H,{H}^{\ast },{H}_1}^{(T)} $ decay constants}",
    eprint = "2106.13617",
    archivePrefix = "arXiv",
    primaryClass = "hep-ph",
    reportNumber = "CP3-Origins-2020-13 DNRF90",
    doi = "10.1007/JHEP09(2021)023",
    journal = "JHEP",
    volume = "09",
    pages = "023",
    year = "2021"
}

@article{Puhan:2023ekt,
    author = "Puhan, Satyajit and Sharma, Shubham and Kaur, Navpreet and Kumar, Narinder and Dahiya, Harleen",
    title = "{T-even TMDs for the spin-0 pseudo-scalar mesons upto twist-4 using light-front formalism}",
    eprint = "2310.03464",
    archivePrefix = "arXiv",
    primaryClass = "hep-ph",
    doi = "10.1007/JHEP02(2024)075",
    journal = "JHEP",
    volume = "02",
    pages = "075",
    year = "2024"
}

@article{Wang:2017bgv,
    author = "Wang, Wei and Xu, Ji and Yang, Deshan and Zhao, Shuai",
    title = "{Relativistic corrections to light-cone distribution amplitudes of S-wave B$_{c}$ mesons and heavy quarkonia}",
    eprint = "1706.06241",
    archivePrefix = "arXiv",
    primaryClass = "hep-ph",
    doi = "10.1007/JHEP12(2017)012",
    journal = "JHEP",
    volume = "12",
    pages = "012",
    year = "2017"
}

@article{Berezhnoy:2016etd,
    author = "Berezhnoy, A. V. and Likhoded, A. K. and Onishchenko, A. I. and Poslavsky, S. V.",
    title = "{Next-to-leading order QCD corrections to paired B$_c$ production in $\epem$ annihilation}",
    eprint = "1610.00354",
    archivePrefix = "arXiv",
    primaryClass = "hep-ph",
    doi = "10.1016/j.nuclphysb.2016.12.013",
    journal = "Nucl. Phys. B",
    volume = "915",
    pages = "224--242",
    year = "2017"
}

@article{Liao:2024sgv,
    author = "Liao, Qi-Li and Jiang, Jun",
    title = "{Production of higher excited quarkonium pair at the super Z factory}",
    doi = "10.1088/1674-1137/ad3c2e",
    journal = "Chin. Phys. C",
    volume = "48",
    pages = "073102",
    year = "2024"
}

@article{Liao:2022tqx,
    author = "Liao, Qi-Li and Jiang, Jun and Zhao, Yu-Han",
    title = "{Production of double heavy quarkonia at super Z factory}",
    eprint = "2206.06123",
    archivePrefix = "arXiv",
    primaryClass = "hep-ph",
    doi = "10.1140/epjc/s10052-023-11174-x",
    journal = "Eur. Phys. J. C",
    volume = "83",
    pages = "22",
    year = "2023"
}

@article{Kniehl:1990yb,
    author = "Kniehl, Bernd A.",
    title = "{The Higgs boson decay H$ \to $Z $g g$}",
    reportNumber = "MAD-PH-563",
    doi = "10.1016/0370-2693(90)90360-I",
    journal = "Phys. Lett. B",
    volume = "244",
    pages = "537--540",
    year = "1990"
}

@article{Abbasabadi:2008zz,
    author = "Abbasabadi, Ali and Repko, Wayne W.",
    title = "{Higgs boson decay into two gluons and a Z boson}",
    doi = "10.1007/s10773-007-9590-0",
    journal = "Int. J. Theor. Phys.",
    volume = "47",
    pages = "1490--1496",
    year = "2008"
}

@article{Nishijima:1951,
    author = {Nishijima, K.},
    title = "{Generalized Furry's Theorem for Closed Loops}",
    journal = {Prog. Theor. Phys.},
    volume = {6},
    pages = {614-615},
    year = {1951},
    doi = {10.1143/ptp/6.4.614},
    eprint = {https://academic.oup.com/ptp/article-pdf/6/4/614/5419315/6-4-614.pdf},
}

@misc{Long-ShunLu2024,
  author = "Lu, Long-Shun",
  date = "2024",
  howpublished = "private communication, and \href{https://inspirehep.net/literature/2863306}{arXiv:2412.19173 [hep-ph]}"
}

@article{CMS:2024hhg,
    author = "Hayrapetyan, Aram and others",
    collaboration = "CMS",
    title = "{Search for rare decays of the Z and Higgs bosons to a J/$\psi$ or $\psi$(2S) meson and a photon in proton-proton collisions at $\sqrt{s}$ = 13 TeV}",
    eprint = "2411.15000",
    archivePrefix = "arXiv",
    primaryClass = "hep-ex",
    reportNumber = "CMS-SMP-22-012, CERN-EP-2024-253",
    month = "11",
    year = "2024"
}

@article{Abbasabadi:2004wq,
    author = "Abbasabadi, Ali and Repko, Wayne W.",
    title = "{A Note on the rare decay of a Higgs boson into photons and a Z boson}",
    eprint = "hep-ph/0411152",
    archivePrefix = "arXiv",
    doi = "10.1103/PhysRevD.71.017304",
    journal = "Phys. Rev. D",
    volume = "71",
    pages = "017304",
    year = "2005"
}

@article{Sirlin:1977sv,
    author = "Sirlin, A.",
    title = "Current Algebra Formulation of Radiative Corrections in Gauge Theories and the Universality of the Weak Interactions",
    reportNumber = "NYU-TR12-77",
    doi = "10.1103/RevModPhys.50.573",
    journal = "Rev. Mod. Phys.",
    volume = "50",
    pages = "573",
    year = "1978",
    note = "[Erratum: Rev.Mod.Phys. 50, 905 (1978)]"
}

@article{Marciano:1988vm,
    author = "Marciano, W. J. and Sirlin, A.",
    title = "Electroweak Radiative Corrections to tau Decay",
    doi = "10.1103/PhysRevLett.61.1815",
    journal = "Phys. Rev. Lett.",
    volume = "61",
    pages = "1815--1818",
    year = "1988"
}

@article{Braaten:1990ef,
    author = "Braaten, Eric and Li, Chong-Sheng",
    title = "{Electroweak radiative corrections to the semihadronic decay rate of the tau lepton}",
    reportNumber = "NUHEP-TH-90-30",
    doi = "10.1103/PhysRevD.42.3888",
    journal = "Phys. Rev. D",
    volume = "42",
    pages = "3888--3891",
    year = "1990"
}

@article{dEnterria:2025rjj,
    author = "d'Enterria, David and Le, Van Dung",
    title = "{Rare few-body decays of the Standard Model Higgs boson}",
    journal = "Sci. Post. Comm. Reps.",
    eprint = "2508.00466",
    archivePrefix = "arXiv",
    primaryClass = "hep-ph",
    year = "2026"
}

@article{Dicus:1975cz,
    author = "Dicus, Duane A.",
    title = "An Estimate of the Rate of the Rare Decay $\pi^0 \to 3\gamma$",
    reportNumber = "CPT-258-TEXAS, ORO-3992-214",
    doi = "10.1103/PhysRevD.12.2133",
    journal = "Phys. Rev. D",
    volume = "12",
    pages = "2133",
    year = "1975"
}

@article{ParticleDataGroup:2024cfk,
    author = "Navas, S. and others",
    collaboration = "Particle Data Group",
    title = "{Review of particle physics}",
    doi = "10.1103/PhysRevD.110.030001",
    journal = "Phys. Rev. D",
    volume = "110",
    pages = "030001",
    year = "2024"
}

@article{Li:2024zun,
    author = "Li, Cong and Jiang, Ying-Zhao and Sun, Zhan",
    title = "{Studies of Z boson decay into double $\Upsilon$ mesons at the NLO QCD accuracy}",
    eprint = "2407.19418",
    archivePrefix = "arXiv",
    primaryClass = "hep-ph",
    doi = "10.1103/PhysRevD.110.054018",
    journal = "Phys. Rev. D",
    volume = "110",
    pages = "054018",
    year = "2024"
}

@article{Cheng:2018khi,
    author = "Cheng, Shan and Qin, Qin",
    title = "{$Z \to \pi^+\pi^-, K^+K^-$: A touchstone of the perturbative QCD approach}",
    eprint = "1810.10524",
    archivePrefix = "arXiv",
    primaryClass = "hep-ph",
    reportNumber = "SI-HEP-2018-30, QFET-2018-19",
    doi = "10.1103/PhysRevD.99.016019",
    journal = "Phys. Rev. D",
    volume = "99",
    pages = "016019",
    year = "2019"
}

@article{Martynenko:2025nsg,
    author = "Martynenko, F. A. and Martynenko, A. P. and Eskin, A. V.",
    title = "{Production of heavy quark bound states in rare exclusive decays of Higgs boson}",
    eprint = "2505.04530",
    archivePrefix = "arXiv",
    primaryClass = "hep-ph",
    month = "5",
    year = "2025"
}

@misc{Tammaro2025,
  author = "Tammaro, Michele",
  date = "2025",
  howpublished = "private communication"
}

@misc{MartynenkoPrivateComm,
  author = "Martynenko, A. P.",
  date = "2025",
  howpublished = "private communication"
}
